\documentclass[12pt,a4paper,openright,oneside,titlepage]{book}

\pdfoutput=1

\usepackage[utf8]{inputenc}
\usepackage[english]{babel}

\usepackage[T1]{fontenc}
\usepackage{tgpagella}    				
\usepackage{mathrsfs}
\usepackage{bm} 							
\usepackage{lmodern}      				

\usepackage{revsymb}
\usepackage{textcomp}

\usepackage[titles]{tocloft}      		
\usepackage[sf,medium]{titlesec}  		
\usepackage[Sonny]{fncychap}
\usepackage{fancyhdr}
\usepackage[hang,flushmargin]{footmisc} 	

\let\origdoublepage\cleardoublepage
\newcommand{\clearemptydoublepage}{%
  \clearpage
  {\pagestyle{empty}\origdoublepage}%
}
\let\cleardoublepage\clearemptydoublepage

\renewcommand{\chaptermark}[1]{\markboth{\textbf{#1}}{}} 
\fancypagestyle{plain}
{
\fancyhead{}
\fancyfoot{}

}
\usepackage[svgnames,dvipsnames]{xcolor}
\usepackage{pgf}
\usepackage{tikz}
\usepackage{graphicx}
\usepackage[hang,font=footnotesize,labelsep=period,labelfont=bf]{caption}
\usepackage[font=footnotesize,subrefformat=parens]{subcaption}
\usepackage[crop=pdfcrop]{pstool}
\usepackage{ytableau}

\usetikzlibrary{calc,arrows,trees}
\tikzstyle{Dot}=[circle,draw=black,fill=black!50,minimum size=4pt,inner sep=0pt]
\tikzstyle{Node}=[circle,draw=black!50,fill=YellowGreen!30,inner sep=0pt,minimum size=6mm]

\usepackage{amsmath,amsfonts,amssymb}
\usepackage{cases} 						
\usepackage{wasysym} 					
\usepackage{mathtools}

\usepackage[square,authoryear]{natbib}
\usepackage[colorlinks=true, urlcolor=BrickRed, citecolor=BrickRed, linkcolor=BrickRed]{hyperref} 								
\usepackage{makeidx}

\let\vect\vec
\renewcommand{\vec}[1]{\mbox{\boldmath$#1$}}
\newcommand{\scrvec}[1]{\mbox{\boldmath\scriptsize$#1$}}
\newcommand{\al}{\alpha}

\newcommand{\tr}{{\rm tr}\,}
\newcommand{\trnorm}[1]{\left|\!\left| #1 \right|\!\right|_1}
\newcommand{\norm}[1]{\parallel\! #1 \!\parallel}
\newcommand{\id}{\openone}
\newcommand{\idd}{{\openone}}
\newcommand{\aver}[1]{\langle #1 \rangle}
\newcommand{\ket}[1]{\left|{#1}\right\rangle}
\newcommand{\bra}[1]{\left\langle{#1}\right|}
\newcommand{\braket}[2]{\langle{#1}|{#2}\rangle}
\newcommand{\ketbrad}[1]{\left|{#1}\rangle\!\langle{#1}\right|}
\newcommand{\ketbra}[2]{\left|{#1}\rangle\!\langle{#2}\right|}
\newcommand{\mean}[1]{\langle{#1}\rangle}
\newcommand{\expect}[3]{\left\langle{#1}\right|\!{#2}\!\left|{#3}\right\rangle}
\newcommand{\PUA}{Q}
\newcommand{\PME}{P_{\rm e}}
\newcommand{\na}{n_A}
\newcommand{\nb}{n_B}
\newcommand{\nc}{n_C}
\newcommand{\rank}[1]{{\rm rank}\left(#1\right)}
\newcommand{\half}{\mbox{\footnotesize${1\over2}$}}
\newcommand\xyZ[3]{\mbox{\tiny $\kern-.3em(\!#1\kern-.2em #2\!)\kern-.14em #3\!$}}
\newcommand\Xyz[3]{\mbox{\tiny $\kern-.3em #1\kern-.14em(\! #2\kern-.14em #3\kern-.14em)\!$}}

\newcommand{\Quote}[3]{
\begin{tabular}[t]{p{.5\textwidth} p{.45\textwidth}}
& {\footnotesize #1} \\
& {\footnotesize \hfill ---\emph{#2}} \\
& {\footnotesize \hfill #3} \\
& 
\end{tabular}
}

\renewcommand{\title}{\bf Dealing with ignorance: \\universal discrimination, learning and quantum correlations}
\renewcommand{\author}{Gael Sent\'{i}s Herrera}
\renewcommand{\date}{\today}

\newcommand{\ackname}{Agradecimientos}

\newcommand{\introname}{Prolegomenon}
\newcommand{\chnametwo}{Fundamentals}
\newcommand{\chnamethree}{Discrimination of quantum states}
\newcommand{\chnamefour}{Programmable quantum state discrimination}
\newcommand{\chnamefive}{Quantum learning of qubit states}
\newcommand{\chnamesix}{Quantum learning of coherent states}
\newcommand{\chnameseven}{Decomposition of quantum measurements}
\newcommand{\conclusions}{Outlook}

\begin{document}

\frontmatter

\pagestyle{empty}

\begin{center}
{\Huge \title}\\

\vspace{2cm}

Tesis\\
del programa de Doctorado en F\'{i}sica de la\\
Universitat Aut\`{o}noma de Barcelona

\vspace{1cm}

{\bf \Large \author}\\

\vspace{0.5cm}

\textit{
Departament de F\'{i}sica Te\`{o}rica: Informaci\'{o} i Fen\`{o}mens Qu\`{a}ntics\\
Universitat Autònoma de Barcelona, 08193 Bellaterra (Barcelona)
}\\

\vspace{2cm}

escrita bajo la direcci\'{o}n del\\

\vspace{0.5cm}

{Dr. Ramon Mu\~{n}oz Tapia}\\

\vspace{2cm}

\includegraphics[scale=.4]{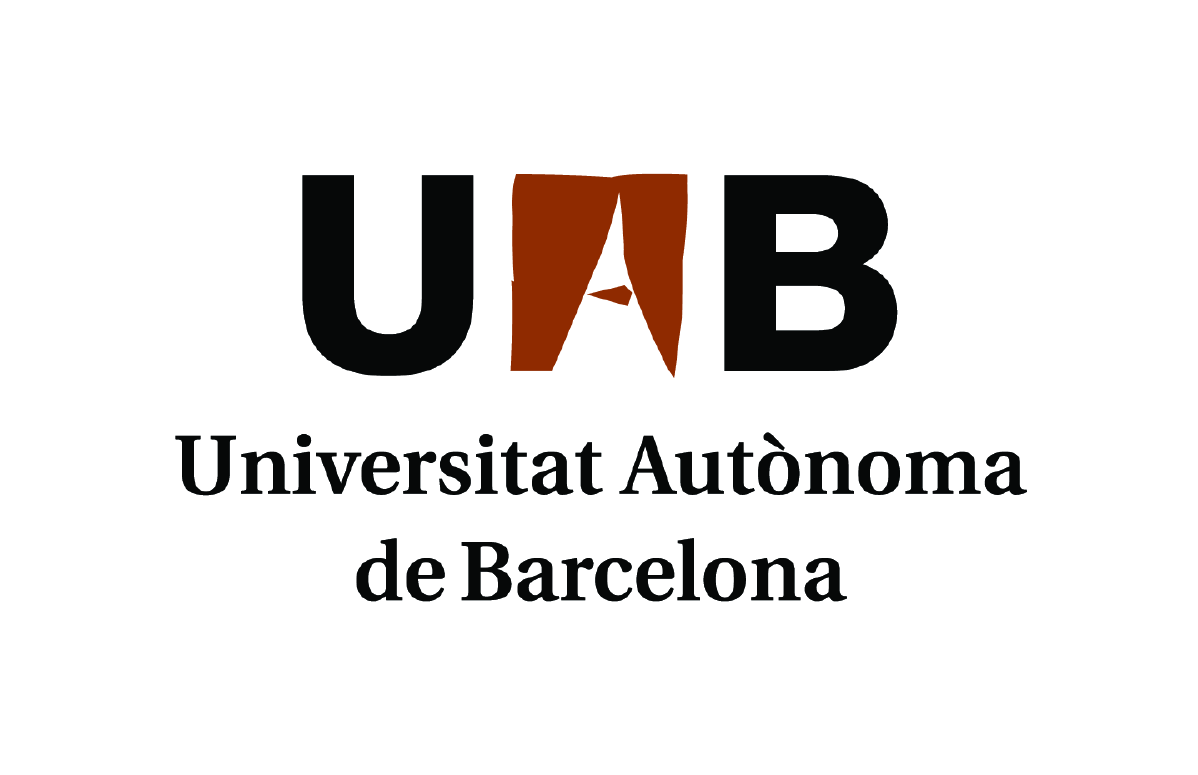}

\vspace{2cm}

Bellaterra, febrero de 2014

\end{center}

\phantom{X}
\vfill


\noindent Copyleft \textcopyleft$\;$2014 Gael Sent\'{i}s Herrera \href{gsentis@ifae.es}{\tt <gsentis@ifae.es>}\\

\noindent This work is licensed under a Creative Commons Attribution-NonCommercial-ShareAlike 4.0 International License. You are free to copy, communicate and adapt this work, as long as your use is not for commercial purposes, any derivative works are licensed under this license (or similar license to this one) and you attribute Gael Sent\'{i}s. The full license can be found at \href{http://creativecommons.org/licenses/by-nc-sa/4.0/}{\texttt{http://creativecommons.org/licenses/by-nc-sa/4.0/}}\\

\noindent Cover design: Adri\'{a}n Miguel Delgado \href{sesimple@gmail.com}{\tt <sesimple@gmail.com>}
\cleardoublepage

\phantom{X}
\vspace{3cm}
\begin{flushright}
\emph{A mis padres.}
\end{flushright}

\pagestyle{fancy}

\chapter*{\ackname}
\addcontentsline{toc}{chapter}{\ackname}
\chaptermark{\ackname}
\markboth{\ackname}{\ackname}

Este documento representa la culminaci\'{o}n de cuatro a\~{n}os de trabajo. Mi m\'{a}s profundo agradecimiento se lo debo a mi supervisor, Ramon Mu\~{n}oz Tapia, quien me ha guiado durante todos y cada uno de estos a\~{n}os.
Su entusiasmo y su intuici\'{o}n cient\'{i}fica me han inspirado desde el primer d\'{i}a a crecer como investigador. Junto a Ramon, agradezco especialmente a John Calsamiglia Costa y a Emili Bagan Capella, mis mentores intelectuales. Entre los tres conforman un equipo de investigaci\'{o}n \'{u}nico, potente y muy c\'{a}lido, del que muy orgullosamente me he sentido parte en este tiempo.

Agradezco tambi\'{e}n a todos los actuales y anteriores miembros del Grupo d'Informaci\'{o} Qu\`{a}ntica que han hecho posible el desarrollo de mi doctorado en el mejor ambiente imaginable: en lista no exhaustiva, los de los primeros tiempos, Anna Sanpera, Marià Baig, Julia Stasińska, Simone Paganelli, Gabriele de Chiara, Julio de Vicente, Bernat Gendra, Elio Ronco, Mart\'{i} Cuquet; y los reci\'{e}n llegados, Mariona Moreno, Rub\'{e}n Quesada, Andreas Winter, Marcus Huber, Alex Monràs, Claude Klöckl, Milan Mosonyi, Stefan Baeumi. De una forma u otra, todos han dejado su huella en este texto. En particular agradezco a Mart\'{i} las pol\'{e}micas de mediod\'{i}a y nuestra aventura en pol\'{i}tica universitaria; a Elio, compa\~{n}ero de despacho desde buen principio, con quien fue un placer probar la docencia universitaria; y a Bernat, colaborador, sparring intelectual y gran amigo, que me acompa\~{n}a desde hace a\'{u}n m\'{a}s tiempo.

I'm also in high debt with Stephen Bartlett and Andrew Doherty, from The University of Sydney, and with Gerardo Adesso and Madalin Gu\c{t}\u{a}, from The University of Nottingham, for their hospitality during my research stays in Australia and UK. They have gifted me with an invaluable scientific experience.

Al otro lado de la valla universitaria han estado los grandes amigos que, a la vez que yo, se aventuraban en sus respectivos doctorados. Marien, Arnim, Marta, Pere, Mar\'{i}a, Sara, Elena, con quienes debat\'{i} hasta la saciedad la intensidad, las angustias y las alegr\'{i}as del mundo de la investigaci\'{o}n. Los que se aventuraban en sus otros proyectos pero compart\'{i}an mesa de bar, imaginario colectivo y reposo, Marina, Jan, Guillem, H\'{e}ctor, Jose. Los de siempre, que me recordaban que nuestro hogar siempre estar\'{a} tambi\'{e}n en Tenerife, Fran, Christian y Adri\'{a}n (al que, adem\'{a}s, agradezco enormemente haber dise\~{n}ado la portada). A todos ellos les doy mi profunda gratitud por todos los momentos compartidos. Y, en especial, a Elena, que me ha comprendido y acompa\~{n}ado m\'{a}s profundamente que nadie.



\tableofcontents

\listoffigures


\mainmatter

\chapter{\introname}



\Quote
{
%
``We balance probabilities and choose the most likely. It is the scientific use of the imagination.''
}
{Sir Arthur Conan Doyle}
{The Hound of the Baskervilles}



\section{Introduction}

During World War II, allied forces devoted much effort to determine the extent of German military production, specially of the brand-new Panzer IV and V tanks in the times preceding D-Day. They really wanted to have an idea of how many tanks they would encounter in battlefield, for the success of an invasion crucially depended on it. The intelligence services had gathered some information, namely espionage data 
of German factories' output, aerial photographies and tank counts at previous contests. Reports indicated contradictory and huge production capabilities, between 1000 and 1500 tanks \emph{per month}. Not happy with these estimates, the allies asked statisticians to see whether their accuracy could be improved.

Only two sources of information were enough to produce incredibly accurate estimates: 
the number of tanks captured in battlefield, and their serial numbers.
With these, statisticians estimated that an average of 246 tanks were being produced per month between 1940 and 1942, while intelligence services reported a number of 1400. When, after the war, the actual German records were captured, they showed a production number of 245 tanks per month for those three years\footnote{These numbers were obtained from \citep{Brodie1947}.}. How could the statisticians be so close?

Say the total number of tanks produced in a particular month is $N$. Among $k$ captured tanks, the highest serial number turns out to be $m$. The statisticians assumed that the Germans
had numbered their tanks sequentially (and they did), hence they applied the following reasoning.
The first consequence of this assumption is that, at least, $m$ tanks were produced.
If only one tank is observed, a fairly reasonable guess of $N$ would be to double its serial number, as it is more likely that $m$ falls in the middle of the sequence of the $N$ tanks rather than in the extremes. But this is a long shot, and more precision comes with more serial numbers.
The probability that the highest serial number is $m$ in a series of $k$ out of $N$ tanks is given by the number of ways that $k-1$ tanks could have all serial numbers up to $m-1$, divided by all the possible series of $k$ tanks. Mathematically, this is expressed as
\begin{equation}\label{ch1/tanks pm}
p(m|N,k) = \frac{\binom{m-1}{k-1}}{\binom{N}{k}} \,.
\end{equation}
According to this probability, the mean value of $m$ is $\bar{m}=(N+1)k/(k+1)$. Then, \emph{assuming} that the observed $m$ coincides with $\bar{m}$, one can propose the estimator $\hat{N}=m+m/k-1$. Intuitively, this is just the highest serial number plus the average gap between serial numbers. Without going any further, this is the technique that the statisticians used to come up with the number $246$.
It is, though, a particular way of handling available information and uncertainty,
and certainly not the only possible approach.

There is an alternative solution to this problem that, involving different assumptions, 
accounts for how \emph{our} knowledge is modified when more data becomes available. 
This solution aims at obtaining the whole probability distribution of the number of tanks $p(N|m,k)$ [that is the inverse of Eq.~\eqref{ch1/tanks pm}], thus it goes beyond just giving an estimate.

\begin{figure}[t]
\begin{center}
\includegraphics[scale=0.9]{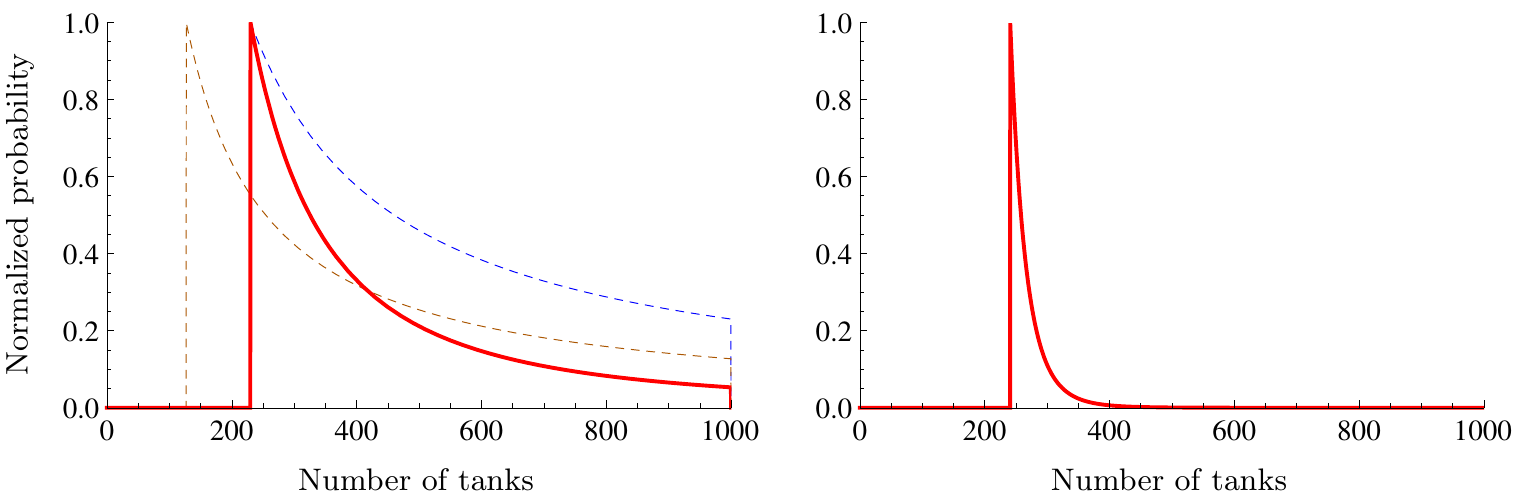}
\end{center}
\caption[Probabilities for the number of German tanks in WWII]{(left) Normalized probability of the total number $N$ of tanks when the first tank found is numbered 230 (blue dashed curve) or 127 (brown dashed curve), and when the two tanks are taken into account (red solid curve). To ease presentation, a maximum of 1000 tanks is assumed.\\
(right) Normalized probability when a series of 10 tanks is observed, with a highest serial number of 241.}\label{ch1/fig:fig1}
\end{figure}

Before any tank is found, we know nothing about $N$. We can represent this complete ignorance as a uniform probability for any value of $N$ (maybe up to a reasonable maximum, but this is not important).
Now, say one tank is found with the serial number 230. Then, two facts and one assumption comprise \emph{our} state of knowledge: 
$N$ is at least 230 (fact), the \emph{a priori} probability of that number appearing was $1/N$ (fact\footnote{As long as we keep the problem in its simplest form, e.g., not taking into account that older tanks have a greater probability to be found.}), and, as said before, any number $N$ of tanks was equally probable (assumption).
The composition of these three pieces of information yields a probability distribution for $N$, represented by the blue dashed curve in Fig.~\ref{ch1/fig:fig1}. The most likely number of tanks is $N=230$, 
but numbers around 900 still have a lot of probability,
so we better wait for more data. Say another tank is found, this time with serial number 127. 
A similar probability distribution represents this new information (brown dashed curve).
It could seem that this does not tells us anything new, since we already know that there are at least 230 tanks, but the combination of the old and the new evidence, which, roughly speaking, amounts to multiply the two distributions, 
is much more eloquent.
The red solid curve on the left side of Fig.~\ref{ch1/fig:fig1} represents our updated state of knowledge after taking into account the second tank.
It is still peaked at 230 tanks, but now the greater numbers are significantly suppressed.

Observing more tanks means a greater concentration of the probability near the peak value: the right side of Fig.~\ref{ch1/fig:fig1} shows the probability distribution for $N$ given a series of 10 tanks, where the highest serial number is 241; from this relatively small amount of data we have been able to localise $N$ around a mean value of 270, with a standard deviation of $\simeq$ 30 tanks.
For arbitrary $k$ and $m$ (given $k>2$), the probability distribution is
\begin{equation}
p(N|m,k) = \frac{k-1}{k} \frac{\binom{m-1}{k-1}}{\binom{N}{k}} \,,
\end{equation}
peaked at $N=m$ and with a mean value \mbox{$\bar{N}=\frac{(m-1)(k-1)}{k-2}$}. Although this reasoning follows a fundamentally different route than the first above, when $k$ is large enough, both $\bar{N}$ and the estimator $\hat{N}$ computed before converge. What this means is that, despite we started from strong---and different!---assumptions in both approaches ($m=\bar{m}$ in the first, and an equal chance of any total number of tanks in the second), their effect in the final result fades away as more data 
arrives.

The two methods used to solve the ``German tank problem'', paradigms of statistics, attempt to provide useful answers in uncertain scenarios. 
Their fundamentals are rooted in different interpretations of information, but they share a common feature: in front of uncertainty, they build on assumptions. Both use, in a way or another---but, maybe, the second method is more explicit---, what we \emph{think} is reasonable, what we \emph{know} beforehand, what we \emph{expect} to observe. 
Statistics gives us a lesson: 
any prediction we may make necessarily passes first through us, subjective observers of an uncertain world, and, ``despite'' that, we are able to predict with relative success. 
Well enough said by Pierre-Simon Laplace, ``probability theory is nothing but common sense reduced to calculation.''

The theory that, perhaps, best advocates the importance of the observer as an active agent in the generation of knowledge is quantum mechanics. The building block of the theory is the quantum state, a mathematical entity that does not differ too much from any of the curves in Fig.~\ref{ch1/fig:fig1}, that is, a representation of what one knows and does not know about a particular quantum system. Quantum mechanics, in contrast to its classical counterpart, is thus an intrinsically probabilistic theory, where uncertainty is considered to be a fundamental property of nature, and, moreover, where the act of observation is an intrusive process that necessarily disturbs what is being observed. In a quantum context, the concepts ``information'' and ``uncertainty'' adopt new meanings, and the role of the observer is inseparable from any experiment. 
Statistics arises as the main tool we have to make predictions about the---quantum---world. The example of the German tanks showed the importance of considering our state of knowledge in an uncertain situation---our certainties and our ignorance---as a crucial part of statistical analysis. In a nutshell, this thesis takes the lesson into the analysis of quantum information processes.

%
%
%

In the remainder of this Chapter, I summarize the main results of my research. Chapter \ref{ch2_essentials} starts by giving the reader a philosophical hint on the jumble of interpretations of probability to choose thereafter one of them, a Bayesian view. 
Then, I introduce some fundamental concepts in quantum theory widely used throughout the whole document, such as quantum states and quantum measurements. 
In Chapter~\ref{ch3_whatisknown}, I describe the main framework in which my research is situated, that is the problem of discriminating between quantum states. I focus on binary discrimination problems. I give an overview of the basics of the topic, starting from its classical analogue: 
the problem of distinguishing two probability distributions.
Although the Chapter reviews known results, 
in Section~\ref{ch3/sec:errormargin} I present an alternative derivation of the discrimination with an error margin that can be more directly generalized to encompass the setting discussed in Section~\ref{ch4/sec:margins}.
Chapters from \ref{ch4_pqsd} to \ref{ch7_povms} comprise the body of results that I have obtained during my PhD. 
The dissertation finalizes with an outlook on future work, followed by the bibliography.
\\

\section{Summary of results}

\subsection*{Programmable quantum state discrimination}

The central topic of this thesis is quantum state discrimination, a fundamental primitive in quantum statistics where one has to correctly identify the state of a system that is in one of two possible states.
The usual approach to the problem considers that the possible states are \emph{known}.
By contrast, a programmable discrimination machine performs this task when the pair of possible states is completely unknown.
The machine is visualized as a device with one data and two program ports, each fed with a number of identically prepared qubits---the data and the programs---, and it aims at correctly identifying the data state with one of the two program states. The machine is thus designed to work for \emph{every} possible pair of states.
In the first part of Chapter~\ref{ch4_pqsd}, I derive the optimal performance of programmable discrimination machines for general qubit states when an arbitrary number of copies of program and data states are available.
Two scenarios are considered: one in which the purity of the possible states is \emph{a priori} known, and the fully universal one where the machine operates over generic mixed states of unknown purity. 
Analytical results are found for both the unambiguous and minimum-error discrimination strategies. This allows to calculate the asymptotic performance of programmable discrimination machines when a large number of copies are provided and to recover the standard state discrimination and state comparison values as different limiting cases.
These results are reported in

\begin{quote}
{\sc G. Sent\'{i}s, E. Bagan, J. Calsamiglia, and R. Mu\~{n}oz Tapia}, ``Multicopy programmable discrimination of general qubit states'', \emph{Physical Review A} {\bf 82}, 042312 (2010); {\bf 83}, 039909(E) (2011).
\end{quote}


In the second part of the Chapter, 
I generalize the problem 
by allowing an error margin.
This generalized scheme has the unambiguous and the minimum-error schemes as extremal cases, when the error margin is set to zero or it is sufficiently large, respectively. Analytical results are given in the two situations where the margin is imposed on the average error probability---weak condition---or it is imposed separately on the two probabilities of assigning the state of the data to the wrong program---strong condition. It is a general feature of the proposed scheme that the success probability rises sharply as soon as a small error margin is allowed, thus providing a significant gain over the unambiguous scheme while still having high confidence results.
The contents of this second part are published in
\\

\begin{quote}
{\sc G. Sent\'{i}s, E. Bagan, J. Calsamiglia, and R. Mu\~{n}oz Tapia}, ``Programmable discrimination with an error margin'', \emph{Physical Review A} {\bf 88}, 052304 (2013).\\
\end{quote}


\subsection*{Quantum learning of qubit states}

In Chapter~\ref{ch5_learning}, by taking a closer look to the structure of the optimal measurement in programmable discrimination, I introduce a quantum learning machine for binary classification of qubit states that does not require a quantum memory. I show that this machine performs with the minimum-error rate allowed by quantum mechanics, that is, the one provided by a programmable machine, for \emph{any} size of the training set. 
This result is 
robust under (an arbitrary amount of) noise and under (statistical) variations in the composition of the training set, provided it is large enough. 
Such learning machine can be used an arbitrary number of times without retraining. 
Its required classical memory grows only logarithmically with the number of training qubits, while its excess risk decreases as the inverse of this number, and twice as fast as the excess risk of an ``estimate-and-discriminate'' machine, which estimates the (unknown) states of the training qubits and classifies the data qubit with a discrimination protocol tailored to the obtained estimates.
These results are reported in
\\

\begin{quote}
{\sc G. Sent\'{i}s, J. Calsamiglia, R. Mu\~{n}oz Tapia, and E. Bagan}, ``Quantum learning without quantum memory'', \emph{Scientific Reports} {\bf 2}, 708 (2012).\\
\end{quote}


\subsection*{Quantum learning of coherent states}

Chapter~\ref{ch6_learningcv} extends the learning concepts presented in Chapter~\ref{ch5_learning} to the domain of continuous-variables systems in a particular setting.
Using a simple model of a classical memory, consisting in an array of cells with two possible reflectivities, I propose a readout scheme that uses an imperfect coherent light source to illuminate each cell and retrieves the stored binary information by determining the state of the reflected signal. Assuming that a number of extra modes coming from the same source are at one's disposal, I show that a fully quantum processing of the signal together with the extra modes provides better results than any strategy that first tries to diminish the incomplete knowledge of the source specifications by estimating the amplitude of the extra modes, and then determines the state of the signal based on the obtained estimate. In particular, I prove this for any Gaussian estimation measurement, and I conjecture that this is the case for any local strategy based on a simple example.
A quantum-enhanced readout of a classical memory is thus observed when using classically correlated coherent signals and the value of their amplitude is not completely determined.
The results of this Chapter will be reported in 
\\

\begin{quote}
{\sc G. Sent\'{i}s, G. Adesso, and M. Gu\c{t}\u{a}}, ``Quantum reading with coherent light'', in preparation.\\
\end{quote}



\subsection*{Decomposition of quantum measurements}

The thesis closes with a study of a transversal character: the convex structure of quantum measurements. Present in all previous chapters as solutions of particular optimization problems, generalized quantum measurements, or, more accurately, their mathematical representations, form a convex set. This means that, if a certain measurement belongs to the inner region of the convex set, it is actually implementable as a convex combination of other measurements. The statistics reproduced by the original measurement is identical to the one reproduced by any of its decompositions.
In Chapter~\ref{ch7_povms}, I design an efficient and constructive algorithm to decompose any generalized quantum measurement into a convex combination of extremal measurements (i.e., measurements that cannot be decomposed as combinations of other measurements). 
I show that, if one allows for a classical post-processing step, only extremal rank-1 positive operator-valued measures are needed. For a measurement with $N$ elements on a $d$-dimensional space, the algorithm will decompose it into at most $(N-1)d+1$ extremals, whereas the best previously known upper bound scaled as $d^2$. 
Since the decomposition is not unique, I show how to tailor the algorithm to provide particular types of decompositions that exhibit some desired property.
%
%
This work is published in
\\

\begin{quote}
{\sc G. Sent\'{i}s, B. Gendra, S. D. Bartlett, and A. C. Doherty}, ``Decomposition of any quantum measurement into extremals'', \emph{Journal of Physics A: Mathematical and Theoretical} {\bf 46}, 375302 (2013).\\
\end{quote}


\chapter{\chnametwo}

\label{ch2_essentials}

\Quote
{``What exactly qualifies some physical systems to play the role of `measurer'? Was the wavefunction of the world waiting to jump for thousands of millions of years until a single-celled living creature appeared? Or did it have to wait a little longer, for some better qualified system \ldots with a PhD?''}
{John Stewart Bell}
{Against `Measurement'}

This Chapter primarily aims to provide
working definitions of key concepts in quantum mechanics that will be used extensively throughout this dissertation, such as probability distributions, quantum states and quantum measurements. A deep understanding of such concepts is an arduous quest with a variety of ends, for it belongs ultimately to the realms of interpretation and philosophy, and it is certainly not the purpose of this introduction to cover these matters in full. 
However, it is both fascinating and beneficial to 
examine the conceptual background where the statistical problems posed in the following chapters lie. This Chapter starts sketching the viewpoint considered here, that is the Bayesian interpretation of probability, what comes with it, and which are its alternatives, to detail thereafter the mathematical definitions and formalism later used.


\section{Epistemology of probability}\label{ch2/sec:epistemology}

We constantly handle probabilities in our everyday lives. We make estimations when we lack certainty, we make decisions based on statements that include expressions like ``better odds'', ``more probable'', or ``less likely''. 
We invoke common sense and probability to give a rational justification to our actions, 
yet the definition of \emph{probability}, or, more accurately, its interpretation\footnote{For an account of the mainstream interpretations of probability, see \citep{Gillies2000}.}, is far from consensus. 
A probability theory aspires to provide the procedure one should follow in facing any nondeterministic problem if one wants to be \emph{rational}, but that rationality comes in accordance with the interpretation of probability that the theory assumes. 
Choosing one particular theory carries unavoidably an epistemological compromise, namely a specific answer to the question: what \emph{is} a probability, and what does it tell us about reality? 

In modern statistics we can distinguish two major schools of thought that address such a question: frequentism and Bayesianism. However, the first attempt of a formal answer dates from 1812 and is attributed to Laplace's principle of indifference\footnote{This denomination was actually coined much later by John M. Keynes \citep{Keynes1921}.}. In his \emph{Th\'{e}orie analytique des probabilit\'{e}s}, Laplace wrote
\begin{quote}
{\footnotesize 
The theory of chance consists in reducing all the events of the same kind to a certain number of cases equally possible, that is to say, to such as we may be equally undecided about in regard to their existence, and in determining the number of cases favorable to the event whose probability is sought. The ratio of this number to that of all the cases possible is the measure of this probability, which is thus simply a fraction whose numerator is the number of favorable cases and whose denominator is the number of all the cases possible.
}
\end{quote}
The principle simply prescribes the use of the ``uniform prior probability distribution'' of all possible cases when no evidence indicates otherwise. That is to say, if I roll a die that I'm convinced is unbiased, I should assign a probability $1/6$ to each face appearing (needless to say, the principle fails at assessing any problem with no natural symmetry). This is recognized nowadays as Bayesian thinking.
In Laplace's treatise one finds no justification, for him was just common sense, but it actually implies a definite interpretative viewpoint: it locates the essence of probability in the perception of the observer, linking it with a personal belief. 
In a more recent language, the principle of indifference corresponds to the simplest noninformative prior, that is the---in principle---least compromising assumption one can make over uncertain future phenomena. But an assumption nonetheless.

Frequentism appeared in the scene as a strong critique to intuitive arguments of this sort. The felt necessity to deprive probability of any trace of subjectivism rendered what William Feller calls ``the statistical, or empirical, attitude towards probability'', initiated mainly by the contributions of Ronald A. Fisher and Richard E. von Mises \citep{Feller1950}. The frequentist standpoint conceives the probability of an event as the \emph{relative frequency} of this event happening in an infinite number of trials. Aseptic and strictly empirical. 
The frequentist methods present certain 
difficulties\footnote{See e.g. \citep{Howson2006} for a critique of frequentism in statistics.} that need not be reviewed here, but one main shortage worth remarking arises from the very definition of probability just exposed: probabilities are discussed only in relation to well-defined repeatable random experiments, hence situations that are nonrepeatable are out of the question. A typical example used to highlight this fact is the impossibility for a frequentist statistician to say anything about the probability of the Sun exploding tomorrow. One might argue that statistical inference over an ``imaginary'' ensemble of realizations of such an experiment would still be possible, but then isn't that quite the same as a subjective opinion, a human choice?

The other major approach to probability theory is Bayesianism \citep{Bernardo1994}, and it is the point of view taken in 
this thesis.
The idea, roughly speaking, is that probabilities represent \emph{degrees of belief}, and thus are intrinsically connected to an \emph{agent}, that is the individual who makes probability assignments to events. A probability is, then, a state of knowledge: it summarizes what the agent does and does not know about a given situation, i.e., it is an evaluation of his uncertainty. 
Its numerical value represents a measure of the willingness of the agent to make a bet in favor of the event in question.
In a more formal fashion, the probability of a certain hypothesis $H$, given some background information $S$, is defined as the plausibility $P(H|S)$ that the agent gives to $H$. It verifies the properties
\begin{eqnarray}
0 \leqslant P(H|S) \leqslant 1 \,, \\
P(H|S) + P(\neg H|S) = 1 \,,
\end{eqnarray}
where $\neg H$ means the negation of $H$. The plausibility $P(H|S)$ receives the more common name of \emph{prior}. In the acquisition of new evidence $E$, the prior is updated according to Bayes' rule
\begin{equation}
P(H|E,S) = \frac{P(H|S) P(E|H,S)}{P(E|S)} \,.
\end{equation}
%

Now, on a more ontologic note, there are also theories that confer \mbox{\emph{being}}---additionally to \emph{meaning}---to these notions of probability, both in the frequentist and the Bayesian perspectives. 
The common goal is to answer the second part of the question posed at the beginning of this Section: what does a probability tell us about reality? 
From the frequentist side, an attempt to explain the emergence of stable relative frequencies in nature can be found, for instance, in Karl Popper's \emph{propensity theory}\footnote{See \citep{Popper1982} or, for a more recent version of the theory, \citep{Gillies2000}.}. This theory establishes that probabilities (frequencies) are to be understood as objective tendencies of experimental situations to produce some outcomes over others. Knowledge of such ``physical properties''\footnote{Be an example of to which extent probability was regarded as a physical feature in \mbox{pre-Bayesian} theories the case of Richard E. von Mises, who even refers to probability theory as a field of theoretical physics, much as like classical mechanics or optics.} of systems is then accessible only through multiple repetitions of the experiment. This way of thinking would make sense of single-case probability attributions, which can be very appealing for solving the pressing need of an objectivistic approach to statistics---specially in intrinsically indeterministic theories like quantum mechanics---, 
but it is a somewhat \mbox{ad hoc} way of giving frequencies a scent of physical reality that is not even falsifiable, 
to put it in Popper's own terms, not to mention it carries the difficulties and critiques of the frequentist approach.

The Bayesian approach, as presented before, is strongly grounded in subjectivism. It is an exclusively epistemological approach, with no ontological endeavors. To consider probabilities plainly as degrees of belief of a decision making agent, and operating from this starting point on a logical base, together with Bayes' rule, receives the name of subjective (or personalist) Bayesianism\footnote{Subjective Bayesianism was born with the works of philosophers \citep{DeFinetti1931} and \citep{Ramsey1931}. For an accessible introduction, see \citep{Jeffrey2004}.}. This posture situates probabilities 
in the agent's mind, while leaving not a tiny bit of separated, objective essence in whatever the probabilities refer to. As a consequence, assuming one or another prior probability distribution is up to the agent's taste and consideration, in the sense that there is no ``right'' choice (of course, there may still be ``unreasonable'' choices. But, again, according to other's judgement. Not all that objective). Opposing this view there is objective Bayesianism \citep{Jaynes2003}, which supports that there is a unique rational probability that one ought to assign for any uncertain event\footnote{The discussion about true or right values for Bayesian probabilities originates with David Lewis' principal principle, and his notion of \emph{objective chance} \citep{Lewis1980}.}. It is the hope of this standpoint that a way could be found to elucidate these ``right'' probabilities, sustained by logical analysis alone. But, as for now, it is generally acknowledged that no one has succeeded in such enterprise.

From these lines onwards I will assume the subjective Bayesian viewpoint on probabilities. Therefore, no ontologic forethought will be made but, instead, a purely information-theoretic one.
This will prove to be not an inconsequential choice. Quantum mechanics, as a probabilistic theory in its essence, demands a take on the interpretation of probabilities from the very definition of its building block---the quantum state---and much further beyond, shaping accordingly the questions we ask and the way we observe.\\


\section{The quantum state}

With all this said about probabilities, I will simply identify the states of quantum systems with Bayesian probability distributions. That is to say, a quantum state is nothing more than the mathematical object we use to represent our degree of uncertainty about a particular quantum system.

To illustrate this idea, imagine we are given a quantum system prepared in a certain state. We know nothing about the preparation procedure, but we are said the state of the system is either $\ketbrad{\psi_1}$ or $\ketbrad{\psi_2}$ with probabilities $\eta_1$ and $\eta_2=1-\eta_1$, respectively. For \emph{us}, the state of the system, that is our state of knowledge, is then represented by the weighted superposition of the two possibilities $\rho = \eta_1\ketbrad{\psi_1}+\eta_2\ketbrad{\psi_2}$. In general, $\rho$ is called a \emph{density operator} and, as such, it stands for a quantum state. A density operator acts on the 
Hilbert space of the system, 
and fulfils the properties
\begin{eqnarray}
\rho &\geqslant& 0 \; \nonumber\\
\tr \rho &=& 1 \;, 
\end{eqnarray}
i.e., its matrix representation ought to have nonnegative eigenvalues (hence be Hermitian) and be normalized. If the density operator is a one-dimensional projector, i.e., it is of the form $\rho=\ketbrad{\psi}$, the state is said to be \emph{pure}. Otherwise, higher-rank density operators are said to be \emph{mixed} states. The density operator is also commonly known as \emph{density matrix}. I will use both terms interchangeably.

Pure states correspond to states of maximal knowledge, whereas mixed states correspond to less than maximal knowledge \citep{Blum1996,Fuchs1996}. This assertion is evident in the above example, in which we end up with a mixed state because the lack of knowledge about the preparation procedure forces a probabilistic description of the state of the system. This also arises when one has maximal knowledge of a bipartite system, that is when one describes its state with a pure state $\ketbrad{\psi}$ on some tensor-product Hilbert space $\mathcal{H}_1\otimes \mathcal{H}_2$. Quantum mechanics then establishes that one's knowledge of a subsystem shall be less than maximal. Indeed, the state of subsystem 1 is obtained through a partial trace operation over $\mathcal{H}_2$. Let be $\{\ket{u_i}\!\ket{v_j}\}$ a basis for $\mathcal{H}_1\otimes 	\mathcal{H}_2$; then
\begin{equation}
\ket{\psi} = \sum_{i,j} c_{ij} \ket{u_i}\!\ket{v_j} \,,
\end{equation}
and the state of subsystem 1 is
\begin{equation}
\rho = \tr\!_2 \ketbrad{\psi} = \sum_k \braket{v_k}{\psi}\!\braket{\psi}{v_k} = \sum_{i,j} c_{ij} c_{ij}^* \ketbrad{u_i} \,,
\end{equation}
i.e., a mixed state. One obtains a similar result for the state of subsystem 2.

In general, a density matrix admits infinitely many decompositions as a combination of pure states. Two ensembles of pure states $\sum_i \eta_i \ketbrad{\psi_i}$ and $\sum_i \kappa_i \ketbrad{\varphi_i}$ represent the same density matrix if its elements are connected by a unitary transformation $U$, such that
\begin{equation}
\sqrt{\eta_i} \ket{\psi_i} = \sum_j U_{ij} \sqrt{\kappa_j}\ket{\varphi_i} \,.
\end{equation}
These representations of mixed states in terms of ensembles of pure states do not immediately give an idea of how much ``less than maximal'' is the knowledge that they represent. Being able to compare mixed states in regards to their ``mixedness'' is of fundamental importance for many applications in quantum information. For two-dimensional systems there is a simple and useful way of expressing a mixed state that tells us explicitly how much mixed it is. Certainly, any mixed state $\rho$ can be expressed as
\begin{equation}\label{ch2/mixed}
\rho = r \ketbrad{\psi} + \frac{1-r}{2} \openone \,,
\end{equation}
i.e., a weighted combination of $\openone$, the identity operator on the two-dimensional Hilbert space of the system, and some pure state $\ketbrad{\psi}$. The weight $r$ is referred to as the \emph{purity} of $\rho$, in the sense that it signifies the degree of mixture between an object of maximal knowledge---the pure state---and the complete absence of it---the identity operator.\footnote{The parameter $r$ gives an idea of how close is $\rho$ to a pure state. This type decomposition exists for two-dimensional systems because there are only two possible ranks for $\rho$: it is either rank 1 (pure) or full rank (mixed), hence every mixed state can be expressed as Eq~\eqref{ch2/mixed} dictates. For $d>2$, mixed states with intermediate ranks are possible and the measure of ``mixedness'' turns subtler. In general, the answer to the question of whether a certain state $\rho_1$ is more mixed than another state $\rho_2$ is provided by the majorization relation between the eigenvalue sequences of $\rho_1$ and $\rho_2$.}


Now that pure and mixed states have been defined, a clarification is in order. Maximal knowledge shall not be misinterpreted as deterministic knowledge. The fact that I know with certainty that the state of a system is $\ketbrad{\psi}$ does not mean that I would get a deterministic result---some prefixed value---if I measure it. As it will become clear in Section \ref{ch2/sec:measurement}, the measurement outcomes would still be probabilistic. The ``maximal'' in maximal knowledge means ``to the extent that we are allowed by quantum mechanics''. And then, one can rise the following question: even though intrinsically probabilistic, if a pure state is the maximal state of knowledge of a quantum system we can aim for, should not we identify it with a \emph{property} of the system itself? Should not we attribute physical reality to the mathematical object $\ket{\psi}$? This question is as old as the quantum theory. Without entering into much detail, let me just say that, as it happens with probability theories, there is no definite answer and an alluring debate around what someone has referred to as $\psi$-ontology keeps going on. Extensions of subjective Bayesianism (see Section \ref{ch2/sec:epistemology}) into the quantum realm are, for instance, the Deutsch-Wallace variant of the many-worlds interpretation of quantum mechanics \citep{Deutsch1999,Wallace2007}, and ``Quantum Bayesianism'' \citep{Caves2002,Fuchs2010}, an interpretation of quantum theory that is cautious enough to not relate quantum states to physical properties at all. Perhaps the most extreme version of the information-theoretic approach to this matter was worded by John Wheeler in his ``it from bit'' thesis \citep{Wheeler1990}:
\begin{quote}
It from bit symbolizes the idea that every item of the physical world has at bottom—at a very deep bottom, in most instances—an immaterial source and explanation; that what we call reality arises in the last analysis from the posing of yes-no questions and the registering of equipment-evoked responses; in short, that all things physical are information-theoretic in origin and this is a participatory universe.
\end{quote}
Of course, one can also find arguments in favor of the opposed school, that is the idea of pure states being, indeed, physical properties of systems \citep{Pusey2011}. The discussion is all but settled.\\


\subsection*{The Bloch sphere}

Quantum states of two-dimensional systems, a.k.a. qubits, find a particularly useful geometrical representation in the so called \emph{Bloch sphere} picture. This representation will be used extensively in the remaining chapters of the dissertation.

\begin{figure}[t]
\begin{center}
\includegraphics[scale=1.2]{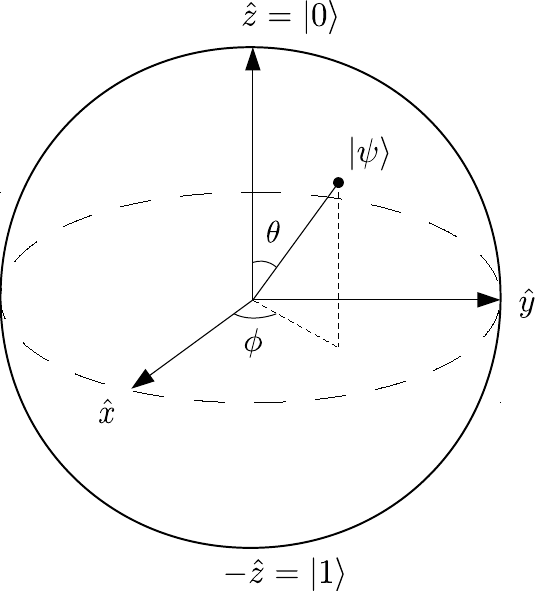}
\end{center}
\caption[The Bloch sphere]{The Bloch sphere.}\label{ch2/fig:fig1}
\end{figure}

Any pure state $\ket{\psi}$ of a two-dimensional system can be written in the computational basis, that is the basis formed by the orthogonal vectors $\ket{0}$ and $\ket{1}$, as $\ket{\psi} = \alpha \ket{0} + \beta \ket{1}$, where $\alpha$ and $\beta$ are complex numbers. Since only the relative phase between $\alpha$ and $\beta$ has any physical meaning, $\alpha$ can be taken to be real. The normalization condition $\braket{\psi}{\psi}=1$ leaves two free parameters to specify the state. In particular, one can choose the parametrization to be
\begin{equation}
\ket{\psi} = \cos \frac{\theta}{2} \ket{0} + e^{i \phi}\sin \frac{\theta}{2} \ket{1}\,,
\end{equation}
where $0\leqslant \theta < \pi$ and $0 \leqslant \phi < 2\pi$. The pair of angles $\{\theta,\phi\}$ fully determines the state $\ket{\psi}$, and, interpreted as spherical coordinates, specifies a point \mbox{$\vec{v}=(\sin\theta\cos\phi,\sin\theta\sin\phi,\cos\theta)$} in the surface of a unit 2-sphere (see Fig.~\ref{ch2/fig:fig1}). Thus, this surface represents the set of all pure states for a qubit. 

In a general way, any qubit density matrix $\rho$ can be written in the compact form
\begin{equation}\label{ch2/bloch}
\rho = \frac{\openone + r\,\vec{v}\cdot\vec{\sigma}}{2} \,,
\end{equation}
where $\vec{v}$ is the so-called \emph{Bloch vector} of the state ($|\vec{v}|=1$), $r$ is its purity, and $\vec{\sigma}=(\sigma_x,\sigma_y,\sigma_z)$ is the vector of the Hermitian, traceless Pauli matrices. As it could already be seen from Eq.~\eqref{ch2/mixed}, taking the value $r=0$ in Eq.~\eqref{ch2/bloch} yields the completely mixed state $\openone/2$, whereas $r=1$ leaves us with a rank-1 density matrix, i.e., a pure state. In the Bloch sphere picture, mixed states correspond to interior points of the sphere at a distance $r<1$ from the origin of coordinates.


\section{The quantum measurement}\label{ch2/sec:measurement}

The measurement process in quantum mechanics has been a controversial subject of study since the very origins of the theory. Two facts justify the difficulty: quantum indeterminism reveals itself upon measuring quantum systems, and, moreover, the state of the system appears to change abruptly right after the process, an experimental observation that is captured by the wave function collapse postulate of quantum mechanics. The way measurement theory is presented in standard quantum mechanics textbooks is as follows:

\begin{itemize}
\item 
Physical quantities that can be measured are formally represented by self-adjoint operators acting on the state Hilbert space called \emph{observables}. Upon measuring some observable $A$, only its eigenvalues can be observed as measurement outcomes. Say $A$ has the spectral decomposition $A=\sum_i \lambda_i \ketbrad{a_i}$. Then, the average value of $A$ when measured in some state $\rho$ is
\begin{equation}\label{ch2/meanA}
\mean{A} = \sum_i \lambda_i p(\lambda_i|\rho) = \sum_i \lambda_i \tr (\ketbrad{a_i} \rho)\,,
\end{equation}
where $p(\lambda_i|\rho)$ is the probability of obtaining the eigenvalue $\lambda_i$ as the outcome of the measurement over $\rho$. Eq.~\eqref{ch2/meanA} is just the weighted sum of the outcomes with their probabilities of occurrence, which result from projection operations on the state $\rho$. Hence the measurement of $A$ can be completely characterized as a \emph{projective measurement}, specified by a set of orthogonal projectors $\{\Pi_i \equiv \ketbrad{a_i}\}$, and its associated set of outcomes $\{\lambda_i\}$.
\item
After outcome $\lambda_i$ has been obtained, the state of the system---instantly!---becomes
\begin{equation}\label{ch2/poststate}
\rho_i = \frac{\Pi_i \rho \Pi_i}{\tr(\Pi_i \rho)} = \ketbrad{a_i} \,,
\end{equation}
where the last equality holds in this case because $\Pi_i$ is a rank-1 projector. This is the wave function collapse postulate. As it is evident from Eq.~\eqref{ch2/poststate}, if the same measurement $\{\Pi_i\}$ is applied to the posterior state $\rho_i$, the same $i$th outcome will be obtained. This repeatability is a feature of projective measurements, and its experimental verification is what caused the need to include this rather hard-to-swallow postulate in the earlier formulations of quantum theory.\footnote{No physicist is comfortable with abrupt phenomena. Some modern formulations as, for instance, Quantum Bayesianism, consider this ``spooky'' collapse simply as an update of the measurer's knowledge about the state of the system, nothing to do with a physical process.}
\end{itemize}

A projective measurement as the one described also receives the name of \emph{Projection-Valued Measure} (PVM), or \emph{von Neumann measurement}. Generically, this type of measurement includes any complete set of orthogonal, not necessarily \mbox{rank-1}, projectors $\{\Pi_i\}$ 
over the state space.
However, quantum mechanics allows for a more general measurement framework. Dropping the requirements for the elements of the measurement to be orthogonal and to be projectors, one is left with a set of \emph{positive semidefinite} operators $\{E_i\}$, i.e., self-adjoint operators with nonnegative eigenvalues, usually denoted
\begin{equation}\label{ch2/povm1}
E_i \geqslant 0 \,,
\end{equation}
that fulfil a completeness relation of the sort
\begin{equation}\label{ch2/povm2}
\sum_i E_i = \openone \,.
\end{equation}
A set of operators that verify these two conditions is called a \emph{Positive Operator-Valued Measure}, or POVM \citep{Helstrom1976}. In such a description of a measurement, the outcomes are not necessarily related to an eigenvalue of some observable but are just a label, one for each element of the set $\{E_i\}$. A picture that may resemble a POVM is that of a machine with a pilot light for each possible outcome. The machine accepts a quantum state $\rho$ as input, measures it, and blinks one of the lights. The probability of obtaining the outcome $i$, also referred to as the signalling of the element $E_i$, is given by
\begin{equation}
p(i) = \tr (E_i \rho)\,.
\end{equation}
Conditions \eqref{ch2/povm1} and \eqref{ch2/povm2} guarantee $p(i) \geqslant 0, \forall i$ and $\sum_i p(i) = 1$, respectively, rendering $p(i)$ a proper probability distribution of the outcomes. In contrast to PVMs, the POVM elements $E_i$ need not commute with each other. Also, POVMs are not repeatable.

The POVM framework is particularly useful in situations in which all that matters is the measurement device itself, i.e., both when the state of the system after the measurement is irrelevant\footnote{The post-measurement state will depend on the particular implementation of the POVM, for which there is no unique procedure.}, and---as said---when there is no interest in measuring a physical quantity but in the occurrence of certain outcomes. In other words, when the only thing one cares about is the probability distribution of the outcomes. Moreover, there are questions for which PVMs simply do not provide the best answer\footnote{A clear example will be presented in Section~\ref{ch3/sec:unambiguous}: the optimal measurement needed for unambiguous discrimination of two qubits needs three outcomes, despite the Hilbert space of the states is two-dimensional.}.

But this is a mathematical framework, and the measurements performed in a laboratory are physical after all! Some observable has to be observed, because that is the only thing we can observe. A very relevant result in the field is Neumark's dilation theorem\footnote{Alternatively spelled as Naimark's dilation theorem.} \citep{Peres1990}, which states that every POVM can be realized as a 
PVM over an auxiliary system---or \emph{ancilla} \citep{Helstrom1976}---correlated with the original system. Specifically, a $d$-dimensional system can be measured with a POVM with $n>d$ outcomes by performing a repeatable---projective---measurement over an $n$-dimensional ancilla.
%
%
This result allows us to set up the optimization problems considered here, in which we optimize some figure of merit over all possible quantum measurements, by focusing solely on sets of operators $\{E_i\}$ fulfilling the POVM conditions \eqref{ch2/povm1} and \eqref{ch2/povm2}.

\chapter{\chnamethree}

\label{ch3_whatisknown}

\Quote
{``En todas las ficciones, cada vez que un hombre se enfrenta con diversas alternativas, opta por una y elimina las otras; en la del casi inextricable Ts'ui P\^{e}n, opta---simult\'{a}neamente---por todas.''}
{Jorge Luis Borges}
{El jard\'{i}n de senderos que se bifurcan}

Quantum information is all about the processing of information that is encoded in the state of a quantum system \citep{Nielsen2000}.
But then, after the processing part has taken place, the information has to be read out, or, in other words, the state of the system has to be determined in some sense\footnote{Maybe not necessarily be \emph{completely} determined, depending on the task at hand. In any case, some attribute of it has to be extracted through a measurement.}.
There exists a variety of ways to do so, highly dependent on what type of information one is interested in and what one knows already about the state.
In particular, 
when the state is determined by selecting one among a number of possible states or hypotheses, 
one refers to the task as \emph{quantum state discrimination}. 
Orthogonal states are relatively straightforward to discriminate. If one counts with the knowledge of the various hypotheses, one can in principle discriminate perfectly among them. This is not so when the possible states are nonorthogonal. In such a case, errors will be unavoidable and the discrimination protocol shall be designed to satisfy some other optimality criteria. Designing such protocols has proven to be highly nontrivial and case-specific, the reason for which such a basic decision problem has received great attention by the quantum information community in the last 
decades\footnote{The fundamentals of quantum state discrimination were pioneered in \citep{Helstrom1976}. For a historical review on the topic, see \citep{Chefles2000}. For a more recent review, see \citep{Herzog2004a}.}.

The subsequent chapters of this dissertation (with the exception of Chapter \ref{ch7_povms}) start from quantum state discrimination problems arising in various settings, with the common denominator of the lack of classical information about the hypotheses.
It is the purpose of this Chapter to provide general background and definitions for the 
task of discriminating between \emph{known} quantum states, and set a basis upon which to build the more specific cases treated next. Also, other state determination tasks such as estimation and comparison of quantum states will be outlined.
\\



\section{The unknown quantum state}\label{ch3/sec:theunknown}

Chapter \ref{ch2_essentials} presented quantum states as probability distributions, and probability distributions as states of knowledge of an agent about some physical system. Also, it was said that measurements over the system may provide the agent with new evidence, and his state of knowledge be hence updated via Bayes' rule.
Generically, every information processing task can be depicted in an scenario involving \emph{two} agents: the first agent follows a certain processing protocol and prepares some quantum state, which is then sent to the second agent, who has to determine it through a measurement. The first agent may be referred to as \emph{sender}, \emph{preparator}, or even just \emph{source}. The second agent would be the \emph{receiver}, \emph{measurer}, or, very often, \emph{us}. In state determination problems the preparation step has been already carried out, hence the role of the second agent, that is the measurement process, is the central object of analysis.

One may think that the fact of whether there \emph{is} or there \emph{is not} an actual agent sending the state is of no importance as far as the measurer is concerned, for the only thing he should care about is the arrival of the state. However, under the Bayesian framework, the presence of a sender resolves in some way what it may look as a mere linguistic conundrum---but it is actually more than that\footnote{Besides its rightful epistemologic relevance in regards to the consistency of the Bayesian view of probabilities, the conundrum has led to mathematical theorems of paramount importance such as the quantum version of the de Finetti representation theorem.}:  
what do physicists refer to with the ubiquitous concept of an \emph{unknown quantum state} that the measurer shall unravel? Indeed, if quantum states are, in the end, states of knowledge of an agent, then how can there be an \emph{unknown} quantum state at all? Its very existence implies that it should be known, if not by the measurer, by someone else! Incorporating a sender to the scene sorts out this apparent contradiction in the sense that we, as measurers, may simply assume that he knows the preparation procedure and, therefore, the state we are commissioned to determine. In short, we are just accessing the state of knowledge of the sender through measurements on the system. This assumption may look somewhat artificial in some settings, for instance in quantum state tomography\footnote{The objective of this task is to determine an unknown state $\rho$ that some source is believed to be repeatedly preparing, which in turn characterizes it. The concept of a man-in-the-box that owns the state of knowledge $\rho$, placed inside the source, seems ridiculous. Fortunately, such an elaboration is not necessary at all. The problem and its solution are well posed in \citep{Fuchs2004}, and briefly outlined in Section~\ref{ch3/sec:othertasks}.}. 
For the time being, however, let this simple picture help to sketch the type of state determination tasks that this treatise addresses.

Let me begin with a simple binary decision problem. Imagine that the sender prepares a quantum system in some state and sends it to us, the receivers. The sender does not tell us which of two possible preparation procedures has been carried out, only that it has been selected by tossing a fair coin. Heads corresponds to the first preparation procedure, which yields the quantum state $\rho_1$, whereas tails corresponds to the second procedure, which outputs some other quantum state $\rho_2$ (the descriptions $\rho_1$ and $\rho_2$ are known). Now our task begins, that is to decide which procedure has taken place. With the piece of information that the sender has provided, our state of knowledge regarding the system has become
\begin{equation}\label{ch3/ourrho}
\rho = \frac{1}{2} \rho_1 + \frac{1}{2} \rho_2 \,.
\end{equation}
To aid in our decision we perform a measurement on the system with two outcomes, $1$ and $2$. The information gained in the measurement process is then used to make a guess: if the outcome $1$ is obtained we will say that the first procedure was selected, hence that the prepared state for the sender was $\rho_1$, and equivalently for the outcome $2$ and the state $\rho_2$. In general, there exists the possibility of making a wrong guess\footnote{I will extensively comment on this in Section \ref{ch3/sec:quantumstuff}. As for now, it is enough to consider that an erroneous guess may happen.}, and we want to engineer the measurement to minimize that chance as much as possible using the 
knowledge
we have available, that is the description of the hypothetical states $\rho_1$ and $\rho_2$ together with the fact that the coin is fair.

The described task is a particular instance of \emph{quantum state discrimination}. Now, two remarks are in order: 

The first remark is that no reference to either true or false states has been made whatsoever. Both $\rho_1$ and $\rho_2$ are states of knowledge owned by the sender, and $\rho$ is the state of knowledge owned by us before the measurement takes place (hence, at that time, two different descriptions of the same system coexist). After we measure the system we bet for one of the two preparation procedures, i.e., we bet on a past---deterministic---event: the sender's choice. It is in this sense that we may make a mistake\footnote{And if there were no sender, no one would be able to tell us that we are wrong!}. The focus here is completely upon our subjective expectation for this mistake happening.

The second remark is that the prior information we count on greatly influences the task itself. As it is obvious, the form of the state $\rho$ in Eq.~\eqref{ch3/ourrho} is a direct product of both the fairness of the coin and the two given hypotheses. If any of this information were different, $\rho$ would change and so would our measurement strategy. But there is more:

The nature of the prior information even determines the questions we might expect to answer by measuring the system. As an example, for a number of hypotheses greater than two, we may end up with a problem with no explicit optimal solution. Such settings fall under the category of \emph{\mbox{multihypothesis} quantum state discrimination} problems, in which only special cases are solvable. Taking this to the limit, in the case of complete absence of prior information we are forced to assume that the received system can be in \emph{any} state of its Hilbert space. Under these circumstances the set of possible states is infinite, and there is no realistic measurement with infinite outcomes to associate with each possibility, hence discrimination becomes nonsensical. We might then expect to answer a different question, that is which state most closely resembles the actual state. This task receives the name of \emph{quantum state estimation} and takes a 
rather
different approach. Lastly, imagine a variation of the setting in which the sender prepares two states, and tells us that they are either equal or different to each other. In such case the task is referred to as \emph{quantum state comparison}.

Starting from the scheme of two agents just exposed, I will cover in the next sections the specifics of quantum state discrimination, for which I begin with its classical analogue: discrimination of probability distributions. I will leave the discussion of estimation and comparison of quantum states to the final section of the Chapter.
\\

\section{Discrimination of probability distributions}\label{ch3/sec:classicalstuff}

One of the most fundamental problems in statistical decision theory is that of choosing between two possible explanations or models. It is called \emph{hypothesis testing}\footnote{See, e.g., \citep{Hoel1971} for an introduction on the topic.}. 
Say a medical test is designed to determine if a patient is healthy (hypothesis $H_1$\footnote{Also referred to as the ``null'' hypothesis in the topic jargon.}) or it has contracted some disease (hypothesis $H_2$).
The decision is made in view of the data obtained by the test, which produces a binary result ($i=1,2$). There are two types of errors involved: the rejection of a true $H_1$ and the acceptance of a false $H_1$, happening with probabilities $p(2|H_1)\equiv p_1(2)$ and $p(1|H_2)\equiv p_2(1)$, respectively. In general these two types of errors do not have to be treated on equal footing, since diagnosing the disease to a healthy patient may not have the same consequences as failing to detect a true disease. It would be desirable to design a test that minimizes both errors, but this is typically not possible since a reduction of one of them is tied to an increase of the other. The Bayesian-like approach to the problem consists in minimizing the average of the errors
\begin{equation}
\eta_1\,p(2|H_1) + \eta_2\,p(1|H_2) \,,
\end{equation}
with respect to some prior state of knowledge (encapsulated in the distribution $\{\eta_1,\eta_2\}$ for the \emph{a priori} probabilities of occurrence of each hypothesis). In this context, such approach is known as symmetric hypothesis testing.

Taking this medical example to more abstract grounds, the problem becomes that of discriminating two possible probability distributions $p_1(i)$ and $p_2(i)$, $i=1,\ldots,n$, by means of \emph{one} sampling. We, the discriminators, must infer the identity of the probability distribution with the smallest probability of error in average, based solely on the drawn sample and the \emph{a priori} probabilities $\eta_1$ and $\eta_2$. A reasonable candidate for the best strategy to accomplish this task is to just bet for the distribution that provides the outcome of the sampling with the largest \emph{posterior} probability, i.e., to use the Bayes decision function\footnote{It is not only reasonable but also optimal, in the sense that any other decision function provides a greater probability of error in average. A simple proof can be found, for instance, in \citep{Fuchs1996}.}. Given the outcome $i$, the posterior probability for the probability distribution $p_1(i)$ to be true is given by Bayes' rule
\begin{equation}
p(1|i) = \frac{\eta_1 p_1(i)}{p(i)} = \frac{\eta_1 p_1(i)}{\eta_1 p_1(i)+\eta_2 p_2(i)} \,,
\end{equation}
and equivalently for $p(2|i)$, where $p(i)$ is the total probability for the outcome $i$ to come up in the sampling. The Bayes decision function simply becomes
\begin{equation}
\delta(i) = 
\begin{cases}
1 & {\rm if} \quad \eta_1 p_1(i) > \eta_2 p_2(i) \\
2 & {\rm if} \quad \eta_1 p_1(i) < \eta_2 p_2(i) \,, \\
{\rm anything} & {\rm if} \quad \eta_1 p_1(i) = \eta_2 p_2(i)
\end{cases}
\end{equation}
where the value of $\delta(i)$ indicates the bet in an obvious way. With this strategy, the probability of a wrong guess $i$ is given by the minimum of the conditional probabilities, i.e. $\min \,\{p(1|i),p(2|i)\}$. This allows to concisely write the average probability of error according to Bayes decision function---hereafter simply called the minimum probability of error---as
\begin{eqnarray}
P_{\rm e} &=& \sum_{i=1}^n p(i) \min \{p(1|i),p(2|i)\} \nonumber \\
&=& \sum_{i=1}^n \min \{\eta_1 p_1(i), \eta_2 p_2(i)\} \,. \label{ch3/pe_1sample}
\end{eqnarray}

Note that $P_{\rm e}$ explicitly depends not only on the distributions to be discriminated, but also on our subjective prior state of knowledge $\{\eta_1,\eta_2\}$. As it was pointed out in Section~\ref{ch2/sec:epistemology}, prior-dependence is neither a shortage nor a strength, but a hard-coded characteristic of Bayesian statistics. 
One only needs to take this dependence into account when drawing conclusions from Bayesian analysis.

The value of $P_{\rm e}$ is intuitively related to how distinguishable $p_1(i)$ is from $p_2(i)$. Obviously, the more distinguishable, the less errors we make in identifying them. Unfortunately, although $P_{\rm e}$ has a clear operational interpretation and it is easily computable, it fails at quantifying the distinguishability of probability distributions. 
The reason for this is that it is not monotonous under the increase of the number of samplings. Indeed, Eq.~\eqref{ch3/pe_1sample} was derived for one sampling, but nothing prevented us in principle from sampling the distribution more times before making our guess. And if so, it may happen that a pair of probability distributions provides a smaller $P_{\rm e}$ than another pair when sampling once, while being the other way around if we allow the decision to be based on two samples\footnote{Examples that illustrate such situation can be found in \citep{Cover2006}.}.

It is desirable to overcome this limitation, i.e., to find a function that does not depend explicitly on the number of samplings. A reason to do so is that such function will yield a proper distinguishability measure for probability distributions in the context of decision problems. In addition, such figure will build a notion of \emph{distance} between probability distributions. 
The answer gets revealed in taking a closer look to the multiple sampling case.
\\

\subsection{The Chernoff bound}\label{ch3/sec:classicalchernoff}

Let us now sample the distribution $N$ times before making a guess. The set of possible outcomes (the sample space) is the $N$-fold Cartesian product of $\{1,2,\ldots,n\}$. Denote a particular set of $N$ outcomes as 
\begin{equation}
i^{(N)} = (i_1,i_2,\ldots,i_N) \in \{1,2,\ldots,n\}^{\times N}\,. 
\end{equation}
The two probability distributions for a given sequence $i^{(N)}$ are
\begin{equation}
p_1\left(i^{(N)}\right) = p_1(i_1) p_1(i_2) \cdots p_1(i_N) \,,
\end{equation}
and
\begin{equation}
p_2\left(i^{(N)}\right) = p_2(i_1) p_2(i_2) \cdots p_2(i_N) \,.
\end{equation}
Now, using the inequality
\begin{equation}
\min \{a,b\} \leqslant a^s b^{1-s} \,,\quad s\in [0,1] \,,
\end{equation}
that holds for any two positive numbers $a$ and $b$, the probability of error can be written as
\begin{eqnarray}
P_{\rm e}(N) &=& \sum_{i^{(N)}} \min \left\{\eta_1 p_1\left(i^{(N)}\right),\eta_2 p_2\left(i^{(N)}\right)\right\} \nonumber\\
&\leqslant & \eta_1^s \eta_2^{1-s} \sum_{i^{(N)}} \left( \prod_{k=1}^N p_1(i_k)^s p_2(i_k)^{1-s} \right) \nonumber\\
&=& \eta_1^s \eta_2^{1-s} \prod_{k=1}^N \left( \sum_{i_k=1}^n p_1(i_k)^s p_2(i_k)^{1-s} \right) \nonumber\\
&=& \eta_1^s \eta_2^{1-s} \left( \sum_{i=1}^n p_1(i)^s p_2(i)^{1-s} \right)^N \,.
\end{eqnarray}
The bound becomes even tighter when taking the minimum over $s$, that is
\begin{equation}\label{ch3/classicalchernoffbound}
P_{\rm e}(N) \leqslant \min_{s\in [0,1]} \eta_1^s \eta_2^{1-s} \left( \sum_{i=1}^n p_1(i)^s p_2(i)^{1-s} \right)^N \,.
\end{equation}
This is the \emph{Chernoff bound} \citep{Chernoff1952}. 

\begin{figure}[t]
\begin{center}
\includegraphics[scale=1.2]{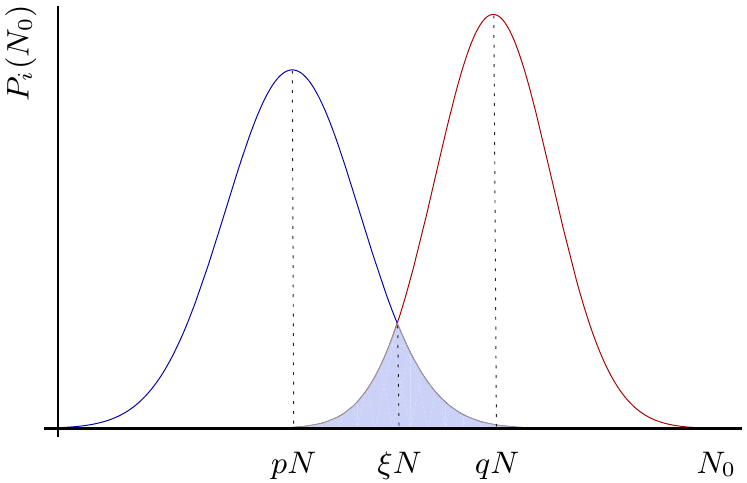}
\end{center}
\caption[Probability distributions of a biased coin and the Chernoff distance]{The probability distribution of a result of $N_0$ ``heads'' is represented for a biased coin that can be of types, 1 or 2. When $N$ is large, the curves approach Gaussians centred at $pN$ and $qN$, respectively, where $p$ ($q$) is the bias of coin 1 (2). The filled area corresponds to the error probability in distinguishing the two distributions.}\label{ch3/fig:fig1}
\end{figure}

This is a specially remarkable upper bound for the optimal $P_{\rm e}(N)$ because it is actually attained in the asymptotic limit $N\to\infty$\footnote{The proof for the attainability of the Chernoff bound is more involved and shall not be reproduced here. It can be found in \citep{Cover2006}.}. At an intuitive level, it is clear that the probability of error goes to zero as $N$ increases. It turns out that the shape of this decrease asymptotically approaches an exponential function, and, moreover, the exact rate exponent is fixed through Eq.~\eqref{ch3/classicalchernoffbound}, i.e.,
\begin{equation}
P_{\rm e}(N\to\infty) \sim e^{-N C(p_1,p_2)} \,,
\end{equation}
with
\begin{equation}\label{ch3/classicalchernoffdistance}
C(p_1,p_2) \equiv -\log \min_{s\in [0,1]} \sum_{i=1}^n p_1(i)^s p_2(i)^{1-s} \,.
\end{equation}
The exponent $C(p_1,p_2)$ is known as the \emph{Chernoff distance}. For the special case of measurements with two outcomes (i.e., $n=2$), the meaning of the Chernoff distance can be easily pinned down. This is the case of a biased coin tossed $N$ times, with two possible probability distributions for the outcomes, $p_1=\{p,1-p\}$ and $p_2=\{q,1-q\}$. A result of $N_0$ ``heads'' out of $N$ tosses, according to $p_1$, occurs with probability
\begin{equation}
P_1(N_0) = \binom{N}{N_0} p^{N_0} (1-p)^{N-N_0} \,,
\end{equation}
whereas, according to $p_2$, occurs with probability $P_2(N_0)$, defined as $P_1(N_0)$ but with $p$ replaced by $q$. In the limit of large $N$ these distributions approach Gaussians centred at $pN$ and $qN$, respectively.
Let $\xi$ be the fraction of ``heads'' above which one must decide in favor of $p_1$. That is, if $N_0 \geqslant \xi N$ one accepts the distribution $p_1$, whereas if $N_0 < \xi N$ one accepts $p_2$. The main contribution to the error probability in the asymptotic regime is due to cases in which $N_0 = \xi N$, i.e., by events that occur with the same probability for both hypotheses (see Fig.~\ref{ch3/fig:fig1}). It can be proven that 
\begin{equation}
-\lim_{N\to\infty} \frac{\log P_1(\xi N)}{N} = C(p_1,p_2) 
\end{equation}
(the same limit holds for $P_2$). This means that the Chernoff distance, defined as in Eq.~\eqref{ch3/classicalchernoffdistance} for the case of $n=2$, is exactly the exponent of the asymptotic probability of such events, and thus of the asymptotic error probability.

The Chernoff distance thus allows to properly compare pairs of probability distributions in regards to their distinguishability, in the sense of the error probability inherent to the task of discriminating among them\footnote{The error probability is just one way to define a notion of distinguishability, in this case through a decision problem. There is a variety of figures to assess how much distinguishable are two probability distributions, namely the mutual information, the statistical overlap or fidelity, or the Kullback-Leibler information, although none of them as clearly defined in an operational sense as the probability of error. For a compendium of distinguishability measures, both classical and quantum, see \citep{Fuchs1996}.}. 
Going to the asymptotic limit $N\to\infty$ is the way to get rid of $N$-dependent results, obtaining a quantity that depends solely on the pair of probability distributions, thus related to some relative property of them. Furthermore, note that even the prior dependence has disappeared in Eq.~\eqref{ch3/classicalchernoffdistance}. 
All these nice properties will hold in the quantum version of the Chernoff bound (see Section \ref{ch3/sec:quantumchernoff}), together with additional benefits of a purely quantum nature.
\\

\section{Discrimination of quantum states}\label{ch3/sec:quantumstuff}

One can think of the classical probability distributions in the previous Section as arising from some kind of fixed quantum measurement $\mathcal{E}=\{E_i\}$ performed over a quantum system which state is either $\rho_1$ or $\rho_2$, i.e.,
\begin{equation}
p_1(i) = \tr (E_i \rho_1) \,,\quad p_2(i) = \tr (E_i \rho_2) \,.
\end{equation}
The problem of discriminating quantum states is essentially different to that of discriminating probability distributions in that, in the latter case, the measurement procedure is fixed. 
The process of sampling the probability distributions 
is simply not under discussion,
since it just consists in randomly picking a value of $i$ (e.g., tossing a coin or rolling a die). Then, given the outcome, one optimizes the guessing part, i.e., one chooses optimally the Bayes decision function to indicate a guess. In the quantum analogue, outcomes are generated by applying a---not predetermined---\mbox{measurement $\mathcal{E}$}. The particular $\mathcal{E}$ used is up to the measurer's choice, and it will directly influence the probabilities of the observed outcomes. 
It is an extra freedom of the problem. Given a set of quantum states among which one has to discriminate, one then needs to optimize the two parts of the process: the measurement \emph{and} the guess. This combination is summed up neatly by the POVM formalism (recall Section \ref{ch2/sec:measurement}). Generically, a POVM, that is a set of semidefinite positive operators $\{E_i\}$ such that $\sum_i E_i = \openone$, will have as many elements as possible answers the observer may give. In other words, the occurrence of every outcome is directly associated with a particular answer (a different one, in principle) regarding the identity of the unknown state. Hence the optimization of the ``measurement and guess'' process boils down conveniently to optimize over all possible POVMs $\mathcal{E}$.

The other genuinely quantum feature that makes the task of discriminating quantum states both challenging and interesting is the fact that two nonorthogonal quantum states cannot be discriminated perfectly \citep{Nielsen2000}. The proof is very simple. Suppose that two pure states $\ket{\psi_1}$ and $\ket{\psi_2}$ are nonorthogonal, and that there is a measurement $\mathcal{E}=\{E_1,E_2\}$ that distinguishes them perfectly. This is mathematically represented by
\begin{eqnarray}
p_1(1)&=&\tr (E_1 \ketbrad{\psi_1}) = 1 \,,\label{ch3/nonorthogonal0}\\
p_2(2)&=&\tr (E_2 \ketbrad{\psi_2}) = 1 \,.\label{ch3/nonorthogonal1}
\end{eqnarray}
Because $\mathcal{E}$ is a POVM, the completeness relation $E_1 + E_2 = \openone$ holds. This guarantees that the probabilities add up to one, namely $p_1(1)+p_1(2)=1$ and $p_2(1)+p_2(2)=1$. 
%
Due to Eq.~\eqref{ch3/nonorthogonal0}, it must happen that $p_1(2)=\tr (E_2 \ketbrad{\psi_1}) = 0$. Now, suppose the decomposition
\begin{equation}
\ket{\psi_2} = \alpha \ket{\psi_1} + \beta \ket{\varphi} \,,
\end{equation} 
where w.l.o.g. $\alpha$ can be chosen such that $0\leqslant \alpha \leqslant 1$, $\braket{\psi_1}{\varphi}=0$, $|\alpha|^2+|\beta|^2=1$ by normalization, and $|\beta|<1$ since $\braket{\psi_1}{\psi_2}>0$. This means that
\begin{equation}
p_2(2)=\tr (E_2 \ketbrad{\psi_2}) = |\beta|^2 \expect{\varphi}{E_2}{\varphi} \leqslant |\beta|^2 < 1 \,,
\end{equation}
which contradicts Eq.~\eqref{ch3/nonorthogonal1}. The second last inequality follows from %
\begin{equation}
\expect{\varphi}{E_2}{\varphi} \leqslant \sum_i \expect{\varphi}{E_i}{\varphi} = \braket{\varphi}{\varphi} = 1 \,.
\end{equation}

Quantum indeterminism places in this way its footprint onto the discrimination problem. In other words, if the states to be discriminated are nonorthogonal, even if they are pure, errors will be unavoidable. Now, we may deal with these errors in different ways. The beauty of quantum discrimination resides in that we may tune the measurement $\mathcal{E}$ to meet different requisites in a certain discrimination task, existing essentially two types of approaches in what errors are concerned: \emph{minimum-error discrimination} and \emph{unambiguous discrimination}. In a nutshell, the former allows for errors while enforces a guess after each measurement, whereas the latter sets a zero-error condition in guesses by allowing for some chance of abstaining to make a guess. Additionally, there exists a third approach that interpolates between the two extremes: \emph{discrimination with error margins}. The following three sections review the basics of each approach, with the focus placed over the discrimination between two hypotheses. A brief comment on the extension to more than two hypotheses will be made in Section~\ref{ch3/sec:othertasks}.
\\




\subsection{Minimum-error discrimination}\label{ch3/sec:minimumerror}

Carl W. Helstrom pioneered the study of discrimination problems in quantum mechanics in \citep{Helstrom1976} within the context of hypothesis testing, introduced in Section~\ref{ch3/sec:classicalstuff}, but applied to quantum states.
The scenario is a particular instance of the paradigm of two agents outlined in Section~\ref{ch3/sec:theunknown}.
A sender prepares a quantum system in either the state $\rho_1$ or the state $\rho_2$ (pure or mixed), with \emph{a priori} probabilities $\eta_1$ and $\eta_2$, and sends it to us. Our task is to identify the state of the system by applying some measurement $\mathcal{E}=\{E_1,E_2\}$ and making a guess according to the obtained outcome: we shall guess that the state was $\rho_1$ if the outcome $1$ is obtained, whereas the outcome $2$ would indicate us to guess $\rho_2$. The problem consists in finding the optimal strategy, that is the optimal 
two-outcome POVM
$\mathcal{E}$, that accomplishes the task while minimizing the average probability of error. 

For an arbitrary $\mathcal{E}$, the average probability of error is 
\begin{equation}\label{ch3/pe_arbitraryE}
P_{\rm e}(\mathcal{E}) = \eta_1 \tr(E_2\rho_1) + \eta_2 \tr(E_1\rho_2) \,,
\end{equation}
that is the probability of obtaining the outcome $2$ when the state was $\rho_1$ times its \emph{a priori} probability, plus a similar term for $\rho_2$. Using the fact that $E_2=\openone-E_1$, Eq.~\eqref{ch3/pe_arbitraryE} becomes
\begin{eqnarray}
P_{\rm e}(\mathcal{E}) &=& \eta_1 \tr\left[(\openone-E_1\right)\rho_1] + \eta_2 \tr(E_1\rho_2) \nonumber\\
&=& \eta_1 + \eta_2 \tr(E_1\rho_2) - \eta_1 \tr(E_1\rho_1) \nonumber\\
&=& \eta_1 + \tr\left(E_1 \Gamma \right) \,,\label{ch3/pe_arbitraryE0}
\end{eqnarray}
where
\begin{equation}
\Gamma = \eta_2\rho_2-\eta_1\rho_1
\end{equation}
is the so-called \emph{Helstrom matrix}. Note that, if $E_1=\openone-E_2$ is used instead, one obtains the similar expression 
\begin{equation}\label{ch3/pe_arbitraryE1}
P_{\rm e}(\mathcal{E})=\eta_2-\tr(E_2\Gamma)\,.
\end{equation}

The minimum-error probability is just
\begin{equation}
P_{\rm e} \equiv P_{\rm e}(\mathcal{E}^*) = \min_\mathcal{E} P_{\rm e}(\mathcal{E})\,,
\end{equation}
and the (optimal) POVM $\mathcal{E}^*=\{E_1^*,E_2^*\}$ that accomplishes it is named \emph{Helstrom measurement}. The explicit expression for $P_{\rm e}$ was originally derived in \citep{Helstrom1976}, although a simpler and more insightful method can be found, e.g., in \citep{Herzog2004a}. It works as follows. First, note that $\Gamma$ can have, in general, positive as well as negative and zero eigenvalues. Let its spectral decomposition be
\begin{equation}\label{ch3/spectralgamma}
\Gamma = \sum_{k=1}^d \gamma_k \ketbrad{\varphi_k}\,,
\end{equation}
where $d$ is the dimension of the Hilbert space of the system. Without loss of generality one can order the eigenvalues $\gamma_k$ as
\begin{eqnarray}
\gamma_k < 0 &{\rm for}& 1 \leqslant k < k_0 \,,\nonumber\\
\gamma_k > 0 &{\rm for}& k_0 \leqslant k \leqslant D \,,\nonumber\\
\gamma_k = 0 &{\rm for}& D < k \leqslant d \,.
\end{eqnarray}
Plugging Eq.~\eqref{ch3/spectralgamma} into Eq.~\eqref{ch3/pe_arbitraryE0} one has
\begin{equation}\label{ch3/pe_provingE}
P_{\rm e}(\mathcal{E}) = \eta_1 + \sum_{k=1}^d \gamma_k \expect{\varphi_k}{E_1}{\varphi_k}\,.
\end{equation}
The constraint $0\leqslant\expect{\varphi_k}{E_1}{\varphi_k}\leqslant 1$ holds, since $\tr(E_1\rho)$ must be a probability for any $\rho$. It immediately follows that the optimal POVM element $E_1^*$, that is the one that minimizes Eq.~\eqref{ch3/pe_provingE}, must verify $\expect{\varphi_k}{E_1^*}{\varphi_k} = 1$ when $\gamma_k < 0$, and $\expect{\varphi_k}{E_1^*}{\varphi_k} = 0$ when $\gamma_k > 0$. Hence the elements of $\mathcal{E}^*$ can be written as
\begin{equation}\label{ch3/optimalE}
E_1^* = \sum_{k=1}^{k_0-1} \ketbrad{\varphi_k} \,,\quad E_2^* = \openone - E_1^* = \sum_{k=k_0}^d \ketbrad{\varphi_k} \,.
\end{equation}
The projectors onto the eigenstates of $\Gamma$ associated with the eigenvalues $\gamma_k=0$ appear in $E_2^*$ to complete the identity operator, but this is an arbitrary choice. They may be shared in any way between $E_1^*$ and $E_2^*$, for it has no effect on the value of $P_{\rm e}$.

Summing up, the optimal measurement operators $E_1^*$ and $E_2^*$ are projectors onto the orthogonal subspaces of negative and positive eigenvalues of the Helstrom matrix $\Gamma$, respectively. The projector onto the subspace of zero eigenvalues of $\Gamma$, needed to fulfil the completeness relation $E_1^*+E_2^*=\openone$, may be chosen in any way. Interestingly, if there are no negative eigenvalues, the measurement operators turn to be $E_1^*=0$ and $E_2^*=\openone$. This situation corresponds to the optimal strategy being to always guess that the state is $\rho_2$, i.e., there is no need to measure the system at all (an equivalent situation arises when there are no positive eigenvalues). 
It is worth noting that these cases may occur only for mixed states and extreme values of their priors. Indeed, a direct-guess strategy can only be optimal when a measurement is incapable of providing any extra information besides what one already knows, i.e., when the states $\rho_1$ and $\rho_2$ are so noisy that the knowledge encapsulated in the priors $\eta_1$ and $\eta_2$ is greater than what any measurement might extract.
Plugging Eq.~\eqref{ch3/optimalE} into Eqs.~\eqref{ch3/pe_arbitraryE0} and \eqref{ch3/pe_arbitraryE1}, one finds
\begin{equation}
P_{\rm e} = \eta_1 - \sum_{k=1}^{k_0-1} |\gamma_k| = \eta_2 - \sum_{k=k_0}^D |\gamma_k| \,.
\end{equation}
Taking the sum of these two alternative forms of $P_{\rm e}$ and using $\eta_1+\eta_2=1$ leads to
\begin{equation}
P_{\rm e} = \frac{1}{2} \left(1-\sum_k |\gamma_k|\right) = \frac{1}{2} \left(1-\tr|\Gamma|\right)\,.
\end{equation}
This is the well-known Helstrom formula for the minimum-error probability in discriminating $\rho_1$ and $\rho_2$, more commonly written as
\begin{equation}\label{ch3/helstrom}
P_{\rm e} = \frac{1}{2} \left( 1- \trnorm{\eta_1\rho_1-\eta_2\rho_2}\right)\,,
\end{equation}
where $\trnorm{A}=\tr|A|=\sqrt{A^\dagger A}$ is the \emph{trace norm} operation.

The form of Eq.~\eqref{ch3/helstrom} becomes much simpler in the special case of pure states, that is when $\rho_1=\ketbrad{\psi_1}$ and $\rho_2=\ketbrad{\psi_2}$:
%
\begin{equation}\label{ch3/helstrompure}
P_{\rm e} = \frac{1}{2} \left( 1 - \sqrt{1-4\eta_1\eta_2|\braket{\psi_1}{\psi_2}|^2} \right) \,.
\end{equation}
When the states are orthogonal, that is $\braket{\psi_1}{\psi_2}=0$, the discrimination can be done perfectly and $P_{\rm e}=0$. In contrast, if $\braket{\psi_1}{\psi_2}=1$, that is the case of indistinguishable states, the error probability depends only on the \emph{a priori} knowledge contained in the probabilities $\eta_1$ and $\eta_2$. When $\eta_1=\eta_2=1/2$, one has $P_{\rm e}=1/2$ since one can do no more than guessing randomly one of the states.

\begin{figure}[t]
\begin{center}
\includegraphics[scale=0.6]{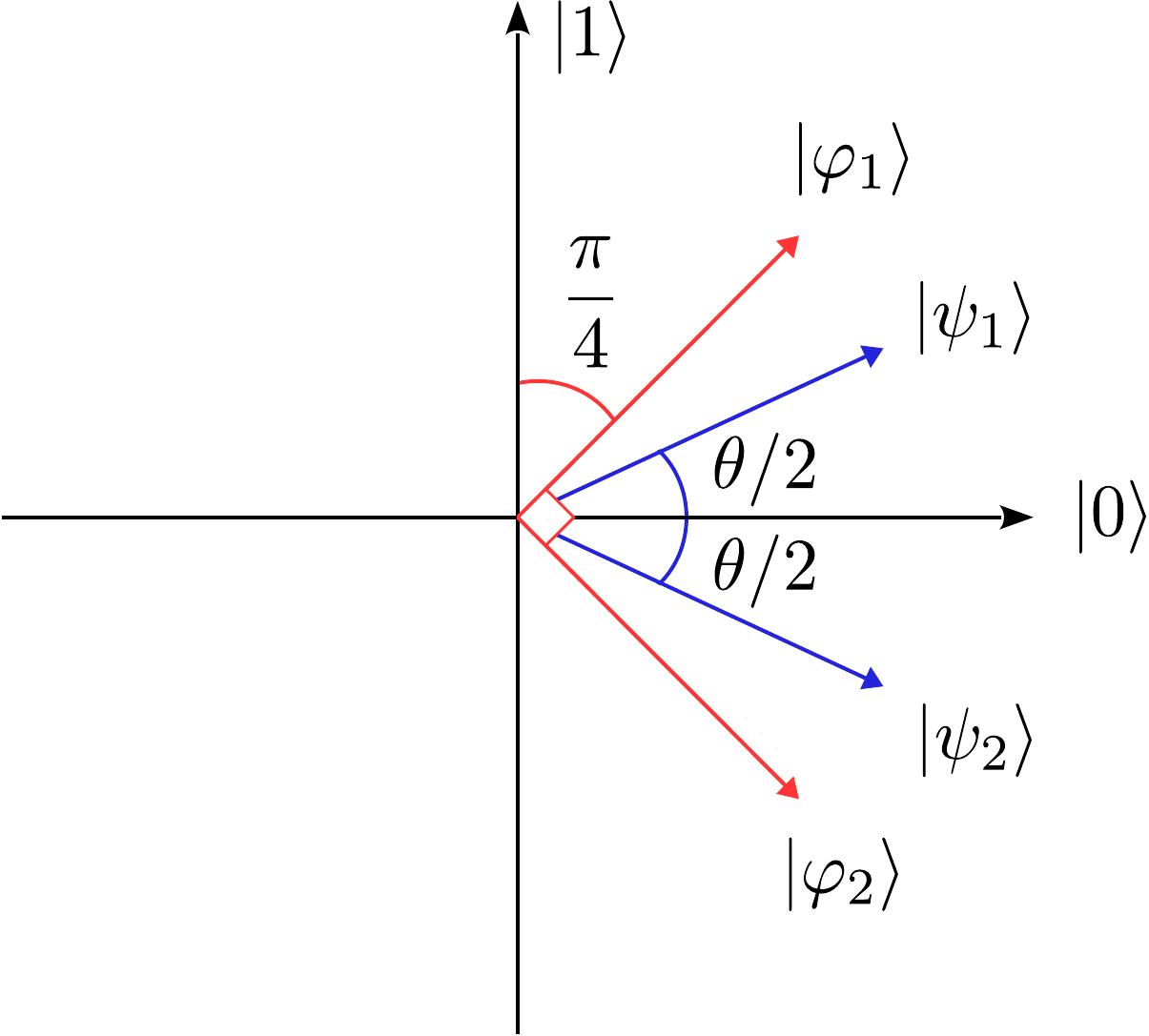}
\end{center}
\caption[Optimal POVM vectors for minimum-error discrimination]{Optimal orientation of the POVM vectors $\ket{\varphi_i}$ with respect to the states $\ket{\psi_i}$ for minimum-error discrimination.}\label{ch3/fig:fig2}
\end{figure}

It is worth mentioning that, for pure states, the matrix $\Gamma$ has rank 2 and, consequently, it has only one positive and one negative eigenvalue. Thus everything can be considered to happen in the two-dimensional subspace \mbox{$\mathcal{S}={\rm span}\{\ket{\psi_1},\ket{\psi_2}\}$}, just as if $\ket{\psi_1}$ and $\ket{\psi_2}$ were qubit states. This simplification allows for the simple and useful geometrical representation of the states and the POVM elements as vectors in a plane. Let $\{\ket{0},\ket{1}\}$ be an orthonormal basis of $\mathcal{S}$. Then, we can always write the states as
\begin{equation}\label{ch3/minerr_vecpsi}
\ket{\psi_i} = \cos\frac{\theta}{2}\ket{0} - (-1)^i \sin\frac{\theta}{2}\ket{1} \,,\quad i=1,2 \,,
\end{equation}
where $0\leqslant\theta < \pi/2$ and $|\braket{\psi_1}{\psi_2}|=\cos\theta$. Similarly, since $\mathcal{S}$ is two-dimensional, the POVM elements $E_i$ need to be one-dimensional orthogonal projectors, i.e., $E_i=\ketbrad{\varphi_i}$ for $i=1,2$, with
\begin{eqnarray}
\ket{\varphi_1} &=& \cos\frac{\phi}{2}\ket{0} + \sin\frac{\phi}{2}\ket{1} \,,\label{ch3/minerr_vecphi0}\\
\ket{\varphi_2} &=& \cos\frac{\pi-\phi}{2}\ket{0} - \sin\frac{\pi-\phi}{2}\ket{1} \,,\label{ch3/minerr_vecphi1}
\end{eqnarray}
and $\theta\leqslant\phi < \pi-\theta$.
The optimization procedure consists in finding the optimal orientation of the pair of orthogonal vectors $\ket{\varphi_i}$, i.e., the optimal angle $\phi$, such that $P_e(\mathcal{E})$, as defined in Eq.~\eqref{ch3/pe_arbitraryE}, is minimized. When the \emph{a priori} probabilities are equal, the optimal orientation is symmetric with respect to the states $\ket{\psi_i}$, that is an angle $\phi=\pi/4$ (see Fig.~\ref{ch3/fig:fig2}). When $\eta_1>\eta_2$, one just has to rotate the pair of vectors $\ket{\varphi_i}$ clockwise such that the overlap $\braket{\varphi_1}{\psi_1}$ increases (and $\braket{\varphi_2}{\psi_2}$ decreases accordingly). Such an increase translates into a greater probability of detection $\tr(E_1\rho_1)$. The reverse situation occurs when $\eta_1<\eta_2$. The optimal angles in these asymmetrical cases are trivially obtained from Eqs.~\eqref{ch3/pe_arbitraryE}, \eqref{ch3/minerr_vecpsi}, \eqref{ch3/minerr_vecphi0} and \eqref{ch3/minerr_vecphi1}, and the corresponding minimum-error probability is given by Eq.~\eqref{ch3/helstrompure}.
\\

\subsection{Unambiguous discrimination}\label{ch3/sec:unambiguous}

The minimum-error approach to the discrimination problem considered in the previous Section assumes by default a nonzero chance for erroneous guesses if the states to discriminate are nonorthogonal. There, a solution is considered optimal if this chance is minimized. However, there might be cases in which errors cannot be tolerated under any circumstances. Can one still say something about the identity of the unknown quantum state under such restriction? This question was first addressed by Ivanovic for the case of discriminating between two possible pure states $\rho_1=\ketbrad{\psi_1}$ and $\rho_2=\ketbrad{\psi_2}$ \citep{Ivanovic1987}\footnote{Historically, unambiguous discrimination was introduced first for pure states, and only recently some results for mixed states have appeared. Just the opposite as minimum-error discrimination, that started from the general case of two mixed states, and from which the pure states cases are derived.}.

The type of measurements described for minimum-error discrimination can be used to produce an outcome with no errors associated. Take a projective two-outcome measurement with elements $E_i=\ketbrad{\varphi_i}$ ($i=1,2$) defined through Eqs.~\eqref{ch3/minerr_vecphi0} and \eqref{ch3/minerr_vecphi1}, and set the extreme angle $\phi=\pi-\theta$. This angle makes the vector $\ket{\varphi_1}$ orthogonal to $\ket{\psi_2}$. The consequence is that the operator $E_1$ never ``clicks'' whenever the state is $\rho_2$, i.e., $\tr(E_1\rho_2)=0$. Thus, if the outcome $1$ is obtained, one can guess \emph{with certainty} that the state was $\rho_1$. Unfortunately, for this value of $\phi$ it also happens that $\ket{\varphi_2}$ is parallel to $\ket{\psi_2}$ and hence nonorthogonal to $\ket{\psi_1}$. This means that the outcome $2$ is not as reliable as the outcome $1$, for it will occur with some nonzero probability both if the state is $\rho_1$ and if it is $\rho_2$. The other extreme angle $\phi=\theta$ yields the reverse situation, in which outcome $2$ is error free and outcome $1$ is uncertain. Ivanovic proposed that, provided several copies of the unknown state, a series of these two measurements can be used to give conclusive guesses, at the expense of not making any guess if an unreliable outcome is obtained. A year later Dieks unified this sequence of measurements under a single POVM realization \citep{Dieks1988}, and Peres proved that such POVM is optimal in the sense that it provides a minimum probability of inconclusive outcomes \citep{Peres1988}.

\begin{figure}[t]
\begin{center}
\includegraphics[scale=0.6]{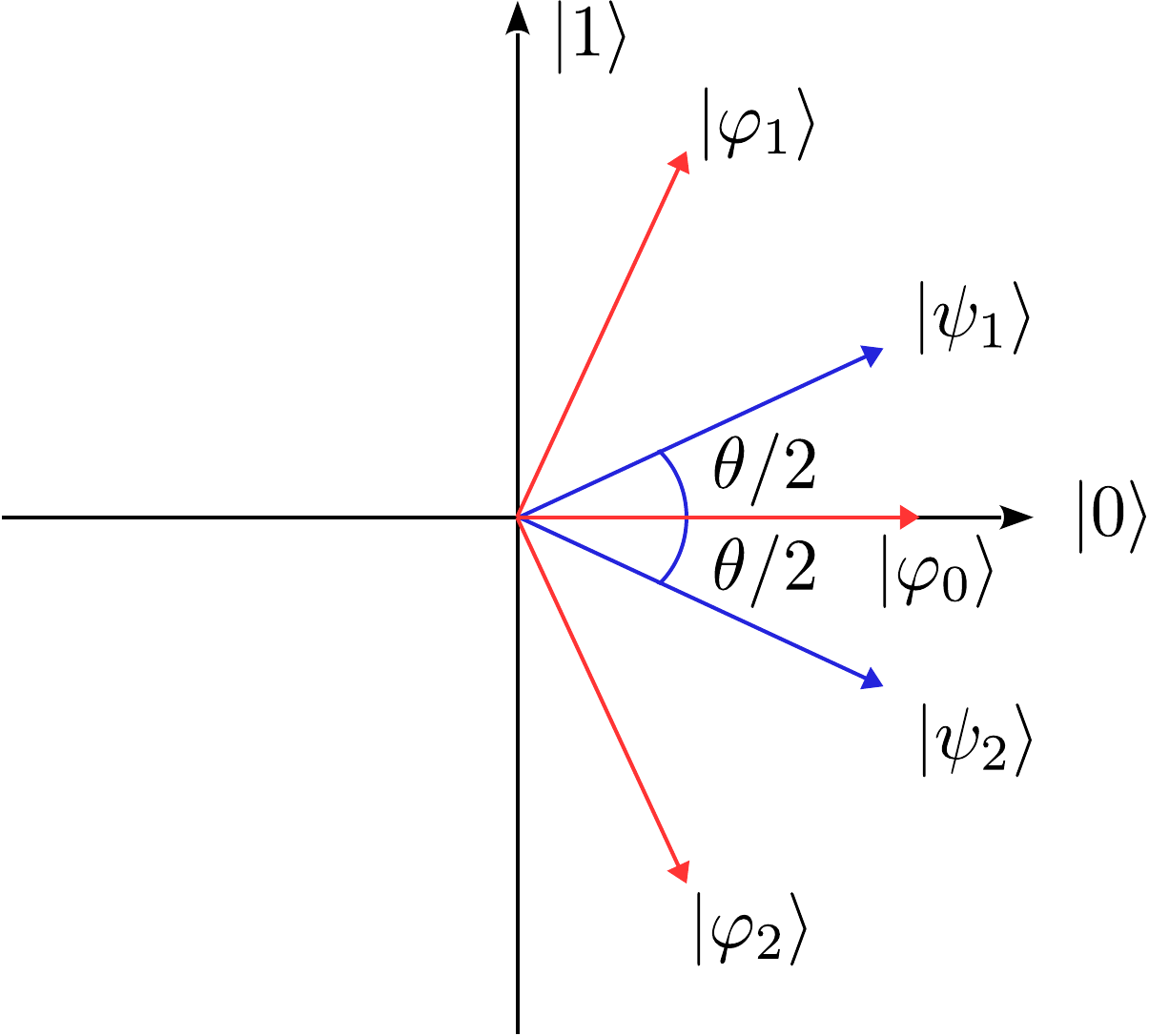}
\end{center}
\caption[Optimal POVM vectors for unambiguous discrimination]{Optimal orientation of the POVM with respect to the states $\ket{\psi_i}$ for unambiguous discrimination. The vectors associated to the POVM elements $E_1$ and $E_2$ are $\ket{\varphi_1}=|\psi_2^\perp\rangle$ and $\ket{\varphi_2}=|\psi_1^\perp\rangle$, respectively. The inconclusive element is $E_0\sim\ketbrad{\varphi_0}$.}\label{ch3/fig:fig3}
\end{figure}

The optimal solution for unambiguously discriminating two equally probable arbitrary pure states is known as the Ivanovic-Dieks-Peres (IDP) result. It invokes the use of a POVM with \emph{three}\footnote{Unambiguous discrimination is one example of a task which optimal solution requires the more general POVM formulation of quantum measurements, for it needs to overcome the limitation that von Neumann measurements impose to the number of outcomes---that of being equal to the dimension of the Hilbert space spanned by the states.} elements $\mathcal{E}=\{E_1,E_2,E_0\}$. The element $E_1$ should identify with certainty the state as $\rho_1$, the element $E_2$ should identify it as $\rho_2$ also with certainty, and the element $E_0$ completes the POVM and represents an inconclusive outcome. 
This is to say, the measurer learns nothing from such outcome about the identity of the state and thus he abstains from giving an answer. 
The unambiguous guessing requirement is mathematically represented by the condition
\begin{equation}
\tr(E_1\rho_2)=\tr(E_2\rho_1)=0 \,.
\end{equation}
This condition enforces the POVM elements to be of the form
\begin{eqnarray}
E_1 &=& \mu_1 |\psi_2^\perp\rangle\!\langle\psi_2^\perp| \,,\label{ch3/UA_E0}\\[.5em]
E_2 &=& \mu_2 |\psi_1^\perp\rangle\!\langle\psi_1^\perp| \,,\label{ch3/UA_E1}\\[.5em]
E_0 &=& \openone - E_1 - E_2 \,,\label{ch3/UA_E?}
\end{eqnarray}
where $\mu_1$ and $\mu_2$ are two coefficients yet to be determined by optimality, and $|\psi_i^\perp\rangle$ stands for a vector orthogonal to $\ket{\psi_i}$. One must now realize two facts. On the one hand, for equally probable states the probabilities of outcomes $1$ and $2$ should be equal by symmetry, hence one can safely assume that  $\mu_1=\mu_2=\mu$. On the other hand, since $\mathcal{E}$ is a POVM its elements must be semidefinite positive, i.e., the conditions $\mu \geqslant 0$ and $E_0\geqslant 0$ must hold. The latter can be assured by using the decomposition $|\psi_2^\perp\rangle = c |\psi_1^\perp\rangle + \sqrt{1-c^2}\ket{\psi_1}$, where $c=|\braket{\psi_1}{\psi_2}|$, to diagonalize $E_0$ and impose positivity for its eigenvalues. This leads to the condition
\begin{equation}
\mu \leqslant \frac{1}{1+c} \,.
\end{equation}
The probability of obtaining an inconclusive outcome is defined as
\begin{equation}
Q = \tr(E_0\rho) = \openone - \tr(E_1\rho) - \tr(E_2\rho) = 1-\mu(1-c^2) \,,
\end{equation}
where $\rho = \rho_1/2 + \rho_2/2$. Note that $\mathcal{E}$ is fully determined by the parameter $\mu$. The only thing left to do is to choose $\mu$ such that $Q$ is minimized. This happens for the maximum value $\mu = 1/(1+c)$, and yields the minimum probability of inconclusive results
\begin{equation}\label{ch3/UA_p?}
Q = c \,,
\end{equation}
and consequently the maximum probability of successful unambiguous discrimination
\begin{equation}\label{ch3/UA_ps}
P_s = \tr(E_1\rho_1)+\tr(E_2\rho_2) = 1-c \,.
\end{equation}
Eqs.~\eqref{ch3/UA_p?} and \eqref{ch3/UA_ps} comprise the IDP result. The graphical representation of the optimal $\mathcal{E}$ for unambiguous discrimination of equally-probable pure states is depicted in Fig.~\ref{ch3/fig:fig3}.

This result was generalized by Jaeger and Shimony to the case of arbitrary prior probabilities $\eta_1$ and $\eta_2$ \citep{Jaeger1995}. The bottom line of their argument is that a three-outcome POVM as described by Eqs.~(\ref{ch3/UA_E0}-\ref{ch3/UA_E?}) is optimal for unambiguous discrimination, but only when it exists, and it does not so in the whole range of values for the prior probabilities. The existence of such POVM is determined by whether the detection probabilities that it generates are valued between 0 and 1. Using $\eta_1+\eta_2=1$, it is not difficult to show that the POVM exists in the range
\begin{equation}
\frac{c^2}{1+c^2} \leqslant \eta_1 \leqslant \frac{1}{1+c^2} \,.
\end{equation}
Above this range, the optimal POVM turns out to be the first two-outcome projective measurement described at the beginning of this Section, which elements $\{E_1,E_0\}$ either identify with certainty the state $\rho_1$ or produce an inconclusive answer. Below this range, the optimal POVM is the second one described, with elements $\{E_2,E_0\}$. The general solution for arbitrary $\eta_1$ and $\eta_2$ renders the optimal inconclusive probability
\begin{equation}\label{ch3/UA_eta1eta2}
Q = 
\begin{cases}
\eta_1 + \eta_2 c^2 &{\rm if} \quad \eta_1 < \frac{c^2}{1+c^2} \\
2\sqrt{\eta_1\eta_2}c &{\rm if} \quad \frac{c^2}{1+c^2} \leqslant \eta_1 \leqslant \frac{1}{1+c^2} \\
\eta_1 c^2 + \eta_2 &{\rm if} \quad \frac{1}{1+c^2} < \eta_1
\end{cases} \,.
\end{equation}

The IDP result obtained in the late 80's, in addition to Jaeger and Shimony's generalization in 1995, completely solve the problem of unambiguous discrimination of two pure states. Results related to mixed states appeared much later. A reason for this delay may be attributed to the following common statement, phrased, e.g., by Fiur\'{a}\v{s}ek and Je\v{z}ek as: ``[\ldots] it is known that one cannot unambiguously discriminate mixed states (the reason is that the IDP scheme does not work for linearly dependent states).'' \citep{Fiurasek2003}. Indeed, the IDP method cannot be straightforwardly generalized, or, more precisely, it does not apply to general \emph{full-rank} mixed states. This is so because in such case both hypotheses have the same support\footnote{The support of a state, described by a density matrix, is defined as the subspace spanned by its eigenvectors associated to nonzero eigenvalues.}, hence a measurement operator cannot project onto a subspace that is orthogonal to the support of only one hypothesis, which is the trick that allows to conclusively say that the true hypothesis is the other one when the corresponding outcome is obtained. It is possible, however, to unambiguously discriminate mixed states which do not have the same support. Along this line are, for instance, the tasks of unambiguous discrimination between sets of states or \emph{unambiguous filtering} \citep{Sun2002}, \emph{state comparison} (see Section \ref{ch3/sec:othertasks}) and unambiguous programmable state discrimination, also known as \emph{unambiguous identification} (see Chapter \ref{ch4_pqsd}). While these tasks have case-specific solutions, results of a more general nature can be found in \citep{Rudolph2003,Herzog2005,Raynal2006}.
\\

\subsection{Discrimination with an error margin}\label{ch3/sec:errormargin}

Unambiguous and minimum-error discrimination are the two extremes of a more general scheme. Intuitively, if the unambiguous scheme is relaxed by tolerating some error rate, the success probability can be increased. Likewise, by allowing some rate of inconclusive answers in the minimum-error scheme, the reliability of the answers can also be increased. These relaxations of the zero-error condition (unambiguous scheme) and the always-guess condition (minimum-error scheme) yield two different parametrizations of the same unified approach to the problem. In the former case, the discrimination protocol is optimized for a \emph{fixed rate of inconclusive outcomes} $Q$\footnote{Analytical solutions for simple cases, numerical solutions and useful bounds were derived in \citep{Chefles1998,Zhang1999,Fiurasek2003,Eldar2003}, and a general method for converting the problem into a standard minimum-error discrimination between some stochastically transformed states was recently obtained in \citep{Bagan2012}. The techniques derived there were also successfully applied to quantum state estimation with post-processing in \citep{Gendra2012,Gendra2013}.}. In the latter, the optimal protocol is derived for a given \emph{error margin} $r$ that the probability of error must not exceed\footnote{This scheme was first considered in \citep{Touzel2007} for projective measurements. The solution for pure states allowing generalized measurements was derived in \mbox{\citep{Hayashi2008,Sugimoto2009}}.}. In both cases the optimization is carried out by maximizing the probability of success, and both are equivalent ways to connect smoothly the unambiguous and the minimum-error extremes.

These general scenarios cover many practical situations, in which only a limited rate of inconclusive answers is affordable, or a certain low error rate is tolerable. Also, cases of linearly dependent states or full-rank mixed states, where unambiguous discrimination is not possible, are in principle tractable under this general scheme, providing a way to increase the success probability over that provided by minimum-error discrimination.

In this Section, I describe the unified scheme for pure states in terms of an error margin. The results that follow were first obtained in \citep{Hayashi2008,Sugimoto2009}, but I present them here in a simpler way\footnote{The remaining of this Section follows closely the first part of \citep{Sentis2013c}.}.

Consider two pure nonorthogonal states $\rho_1=\ketbrad{\psi_1}$, $\rho_2=\ketbrad{\psi_2}$ as hypotheses of a standard two-state discrimination problem, where for simplicity we assign equal \emph{a priori} probabilities to each state. 
The discrimination with an error margin protocol can be thought of as a generalized  measurement on the system, described 
by the POVM 
$\mathcal{E}=\{E_1,E_2,E_0\}$, where, as in Section~\ref{ch3/sec:unambiguous}, the operator $E_1$ ($E_2$) is associated to the statement ``the measured state is $\rho_1$ ($\rho_2$)'', whereas  $E_0$ is associated to the inconclusive answer
 or abstention. 
The overall success, error and inconclusive probabilities are
\begin{eqnarray}
P_{\rm s} &=& \frac{1}{2} \left[ \tr (E_1 \rho_1) + \tr(E_2 \rho_2) \right]\,,\\[.5em]
P_{\rm e} &=& \frac{1}{2} \left[ \tr (E_2 \rho_1) + \tr(E_1 \rho_2) \right]\,,\\[.5em]
Q &=& \frac{1}{2} \left[ \tr (E_0 \rho_1) + \tr(E_0 \rho_2) \right]\,,
\end{eqnarray}
respectively. The relation $P_{\rm s}+P_{\rm e}+Q=1$ is guaranteed by the POVM condition~$E_0+E_1+E_2=\openone$.
The optimal discrimination  with  an error  margin  protocol is obtained by maximizing 
the success probability $P_{\rm s}$ over any possible POVM~$\mathcal{E}$ that satisfies that certain errors occur with a probability not exceeding the given margin. Generically,  these conditions imply a nonvanishing value of  the inconclusive  probability $Q$.

We consider two error margin conditions: {\em weak} and {\em strong}. The weak condition states that the \emph{average} error probability cannot exceed a margin,~i.e.,
\begin{equation}\label{ch3/weakgeneral}
P_{\rm e} = \frac{1}{2} \left[ \tr (E_2 \rho_1) + \tr(E_1 \rho_2) \right] \leqslant  r \,.
\end{equation}
The strong condition imposes a margin on the probabilities of misidentifying \emph{each} possible state, i.e.,
\begin{eqnarray}
p(\rho_2|E_1)&=&\frac{\tr (E_1\rho_2)}{\tr (E_1\rho_1)+\tr (E_1\rho_2)} \leqslant r \label{ch3/strong1}\, ,\\[.5em]
p(\rho_1|E_2)&=&\frac{\tr (E_2\rho_1)}{\tr (E_2\rho_1)+\tr (E_2\rho_2)} \leqslant r \label{ch3/strong2}\, ,
\end{eqnarray}
where $p(\rho_2|E_1)$ and $p(\rho_1|E_2)$ are the probabilities that the state identified as $\rho_1$ is actually $\rho_2$ and the other way around, respectively. 
The strong condition  is obviously 
more restrictive, 
as it sets a margin on both types of errors separately. However, as we will see, the two conditions are directly related: the strong one just corresponds to the weak one with a tighter error margin~\citep{Sugimoto2009}.
Note that both error margin schemes have the unambiguous (when~$r=0$) and the minimum-error schemes (when~$r$ is large enough) as extremal cases. 
We will denote by $r_c$ the critical margin above which the success probability does not increase and thus coincides with that of (the unrestricted) minimum-error discrimination.

\begin{figure}[t]
\begin{center}
\includegraphics[scale=.4]{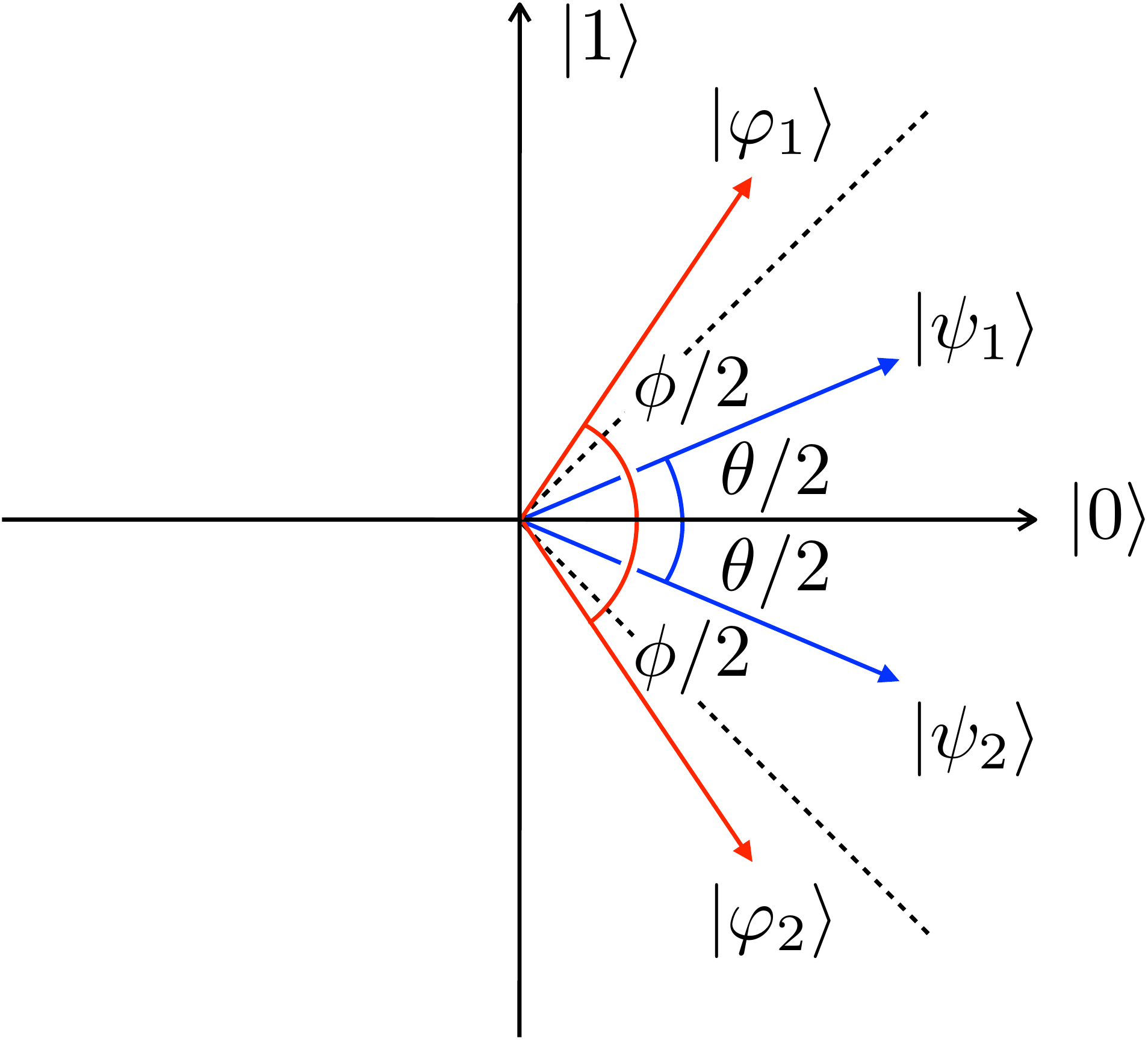}
\end{center}
\caption[Parametrization of the POVM vectors for discrimination with an error margin]{Parametrization of the states $|\psi_1\rangle$, $|\psi_2\rangle$, $|\varphi_1\rangle$ and $|\varphi_2\rangle$ as in Eqs.~(\ref{ch3/the_states}) and~(\ref{ch3/the_POVM}). The dashed lines, at an angle of $\pi/4$ with respect to the horizontal axis, represent the limit of minimum-error discrimination.}\label{ch3/fig:fig4}
\end{figure}

For the weak condition, it is straightforward to obtain the maximum success probability
by taking into account that the corresponding error probability must saturate the margin condition~\eqref{ch3/weakgeneral} for~$r \leqslant r_c$, namely $P_{\rm e}=r$. Furthermore, the symmetry of the problem dictates that $\tr (E_1 \rho_1)=\tr (E_2 \rho_2)=P_{\rm s}$ and~\mbox{$\tr (E_1 \rho_2)=\tr (E_2 \rho_1)=P_{\rm e}$}. Without loss of generality (see Fig.~\ref{ch3/fig:fig4}) and as in Section \ref{ch3/sec:minimumerror}, we can use the parametrization \eqref{ch3/minerr_vecpsi} in terms of a single angle for the input states, i.e.,
\begin{equation}
\ket{\psi_i}=\cos{\theta\over2}\;\ket{0}-(-1)^{i}\sin{\theta\over2}\;\ket{1}\,, \quad i=1,2\,,
\label{ch3/the_states}
\end{equation}
where $0\leqslant \theta < \pi/2$.
The POVM elements can be as well written as \mbox{$E_i=\mu \ketbrad{\varphi_i}$} for $i=1,2$, with
\begin{equation}
\ket{\varphi_i}=\cos{\phi\over2}\;\ket{0}-(-1)^{i}\sin{\phi\over2}\;\ket{1}\,, \quad
{\pi\over2}\leqslant\phi < \pi
\label{ch3/the_POVM}
\end{equation}
(in contrast to Eqs.~\eqref{ch3/minerr_vecphi0} and \eqref{ch3/minerr_vecphi1}, $E_1$ and $E_2$ need not be orthogonal, since in this case there is a third POVM element).
The POVM condition implies $E_0=\openone -E_1-E_2$, and the optimal value of~$\mu$ is fixed by the extremal value of the inequality $E_0\geqslant 0$. One obtains $\mu=1/(1-\cos\phi)\leqslant 1$ and finally the symmetry conditions fix~$\phi$ to be
%
\begin{equation}\label{ch3/optimal-phi}
\tan\frac{\phi}{2}=
\begin{cases}
\displaystyle\frac{\sqrt{1+c}}{\sqrt{1-c}+2\sqrt{r}} & \text{if} \quad 0\leqslant r\leqslant r_c \, , \\[1em]
1 & \text{if} \quad r_c\leqslant r\leqslant 1 \, ,
\end{cases}
\end{equation}
where  $c=|\braket{\psi_1}{\psi_2}| = \cos \theta$ is the overlap of the states~$\ket{\psi_1}$ and~$\ket{\psi_2}$.
Note that in the unambiguous limit, $r=0$,  the POVM elements~$E_1$ and~$E_2$ are orthogonal
to the states~$\ket{\psi_2}$ and~$\ket{\psi_1}$, respectively.
In the other extreme case, when the error margin coincides with, or is larger than, the minimum error,~\mbox{$r\geqslant r_c$}, one has $E_0=0$ (no abstention) and $E_1$ becomes orthogonal to $E_2$, i.e.,~$\phi=\pi/2$. In this range the measurement becomes of von Neumann type and the first case in Eq.~\eqref{ch3/optimal-phi} implies 
\begin{equation}\label{ch3/critical-r}
r_c = \frac{1}{2} \left( 1-\sqrt{1-c^{2}} \right) .
\end{equation}
Taking into account Eq.~\eqref{ch3/optimal-phi}, the optimal success probability reads 
%
\begin{equation}\label{ch3/weak}
P_{\rm s}^{W}(r) =
\begin{cases}
\left( \sqrt{r} + \sqrt{1-c}\, \right)^{2} & \text{if} \quad 0\leqslant r\leqslant r_c \, , \\[1em]
\frac{1}{2} \left( 1+\sqrt{1-c^{2}} \,\right) & \text{if} \quad r_c\leqslant r\leqslant 1 \, ,
\end{cases}
\end{equation}
%
%
%
where the superscript $W$ reminds that weak margin condition has been used.
This result was derived in~\citep{Hayashi2008} and its generalization to arbitrary prior probabilities in~\citep{Sugimoto2009} (also in~\citep{Bagan2012}, by fixing an inconclusive rate $Q$ instead of an error margin).
Note that the POVM $\mathcal{E}$ is fully determined by the angle $\phi$, which in turn is fully determined by the margin $r$ through Eq.~\eqref{ch3/optimal-phi}.

\begin{figure}[t]
\begin{center}
\includegraphics[scale=1.3]{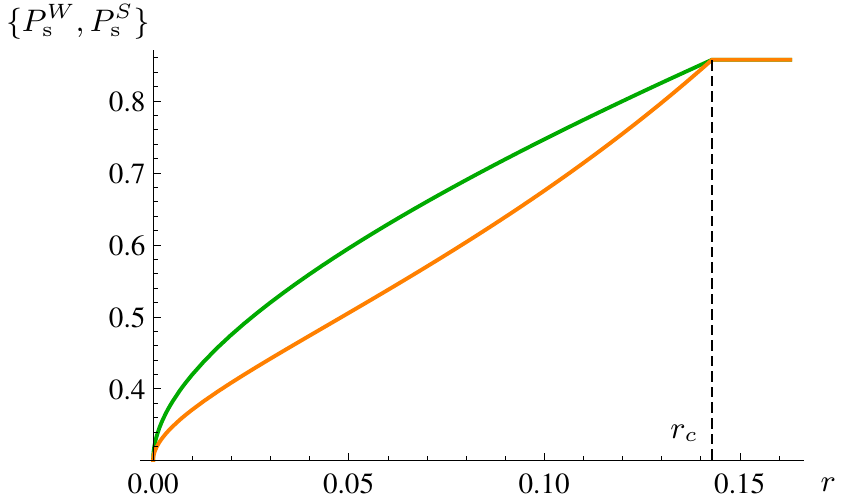}
\end{center}
\caption[Probability of success for weak and strong error margins]{The success probabilities for weak and strong error margins, $P_{\rm s}^W(r)$ (green) and $P_{\rm s}^S(r)$ (orange), for two pure states with overlap $c=0.7$. The critical margin is $r_c \simeq 0.143$. The two probabilities coincide for the extreme margins of unambiguous ($r=0$) and minimum-error discrimination ($r=r_c$).}\label{ch3/fig:fig5}
\end{figure}

The optimal success probability under the strong condition can be obtained along the same lines of the weak case, but it will prove more convenient to use the connection between both conditions to derive it directly from Eq.~\eqref{ch3/weak}.
Let us denote by  $r^S$ and $r^W$ the error margin of the strong and weak condition, respectively. 
From the symmetry of the problem, 
Eqs.~\eqref{ch3/strong1} and~\eqref{ch3/strong2} can be written in the form of a weak condition with  a margin $r^W$ as
%
%
\begin{equation}
P_{\rm e} \leqslant r^S(P_{\rm e}+P_{\rm s}) \equiv r^W .
\end{equation}
%
Hence, if $\mathcal{E}$ is the optimal POVM for a strong margin $r^S$, it is also optimal for the weak margin $r^W$, where $P_{\rm e}=r^W$ and $P_{\rm s}=P_{\rm s}^W(r^W)$ is given by Eq.~\eqref{ch3/weak}.
In terms of the success probability, the relation between $r^W$ and $r^S$ reads
\begin{equation}\label{ch3/mwms1}
 r^S=\frac{r^W}{P_{\rm s}^W(r^W)+r^W}\, .
\end{equation}
%
%
By solving for $r^W$ and substituting into Eq.~\eqref{ch3/weak} one derives the success probability for a given $r^S$, which we denote by $P_{\rm s}^S(r^S)$. For
the function $P_{\rm s}^S$ one readily obtains
\begin{equation}\label{ch3/strong}
P_{\rm s}^{S}(r) =
\begin{cases}
\displaystyle \left( \frac{\sqrt{1-r}}{\sqrt{r}-\sqrt{1-r}} \right)^{2}(1-c) & \text{if} \quad 0\leqslant r\leqslant r_c \,, \\[1em]
\frac{1}{2} \left( 1+\sqrt{1-c^{2}} \,\right) &  \text{if} \quad r_c\leqslant r\leqslant 1 \,,
\end{cases}
\end{equation}
%
in agreement with~\citep{Hayashi2008}. Note that the critical margin is the same for both the weak and the strong conditions, i.e., $r_c^W=r_c^S=r_c$. 
Indeed, beyond the critical point inconclusive results are excluded by optimality ($Q=0$ and $P_{\rm s}+P_{\rm e}=1$) and thus there is no difference between the two types of conditions. As in the weak case, there is a correspondence between the angle $\phi$ and $r^S$, thus $\mathcal{E}$ can also be parametrized in terms of the strong margin:
%
\begin{equation}\label{ch3/phi-strong}
\tan\frac{\phi}{2}=
\begin{cases}
\displaystyle
\frac{\sqrt{1-r^S}-\sqrt{r^S}}{\sqrt{1-r^S}+\sqrt{r^S}}\,{\sqrt{1+c}\over\sqrt{1-c}}
& \text{if} \quad 0\leqslant r\leqslant r_c \, , \\[1em]
1 & \text{if} \quad r_c\leqslant r\leqslant 1 \, .
\end{cases}
\end{equation}
Note that an ambiguity arises for $c=1$, as $\phi=\pi$ and then $E_1$ and $E_2$ become 
equal
to one another, independently of the value of $r^S$. Note also that for $r^S=0$ and~$r^S=r_c$ the values of $\phi$ for both, weak and strong conditions, coincide (see Fig.~\ref{ch3/fig:fig5}).
\\

\section{The many copies paradigm}

As decisions in classical hypothesis testing may be based on more than one sampling of the unknown probability distribution (see Section~\ref{ch3/sec:classicalstuff}), the discrimination of quantum states may be supported by more than one measurement of the unknown state. However, after the first measurement the state of a quantum system changes irremediably, hence a second measurement over the same system---if the first was optimal---would  
give no aid in the identification of
the original state\footnote{Although a second observer, with no knowledge about the result of the first measurement, could still ``scavenge'' information about the state that was previously measured \citep{Rapcan2011}.}. This is why a number of \emph{copies} of the system, all 
prepared
in the same unknown quantum state, is typically considered as a resource in quantum state discrimination tasks.

Formally, one considers that $N$ independent and identically-distributed (i.i.d.) states are provided. Such an ensemble of systems is described by a big $d^N$-dimensional Hilbert space $\mathcal{H}^{\otimes N}$, where $\mathcal{H}$ is the $d$-dimensional Hilbert space of each individual system.
If the state of each copy is either $\rho_1$ or $\rho_2$, then one just has to discriminate the global states $\rho_i\otimes\rho_i\otimes\ldots\otimes\rho_i\equiv\rho_i^{\otimes N}$, $i=1,2$, where $\otimes$ is the direct Kronecker product of the density matrices.

It is in the possible measurements that are at one's disposal where quantum discrimination differs the most from its classical counterpart, for quantum mechanics allows for sophisticated measurements on all $N$ systems at once.
Such \emph{collective} measurements typically outperform any strategy based on individual measurements of each copy \citep{Peres1991}, although there are cases in which they give no advantage.
The question of whether a collective measurement strategy is necessary to achieve optimal performance represents the crux of many works in quantum state discrimination.
A paradigmatic example for which this is true can be found in the context of unambiguous discrimination \citep{Chefles2001}: a set of linearly dependent states---thus not unambiguously distinguishable---can be made linearly independent if enough copies of the states are provided; one can then unambiguously determine the collective state of the set of systems through a collective measurement. On the other hand, in binary minimum-error discrimination, the optimal performance is achievable through \emph{local operations and classical communication}\footnote{This denomination stands for any strategy consisting of sequential adaptive measurements performed on each system: the result of measuring the first system determines the measurement to be used in the second, and so on.} (LOCC) if the states are pure \citep{Acin2005}, but not if they are mixed \citep{Calsamiglia2010,Higgins2011}.

The POVM formalism covers all possible measurements, thus any measurement for discriminating $\rho_1^{\otimes N}$ and $\rho_2^{\otimes N}$ can still be characterized by a two- or a three-outcome POVM just as in Sections~\ref{ch3/sec:minimumerror}, \ref{ch3/sec:unambiguous} and \ref{ch3/sec:errormargin}, but which elements $E_i$ now operate over the total Hilbert space $\mathcal{H}^{\otimes N}$. It is then straightforward to generalize the Helstrom formula for single-copy minimum-error discrimination, that is Eq.~\eqref{ch3/helstrom}, to the $N$-copy case: following identical steps, one simply obtains
%
\begin{equation}\label{ch3/helstromN}
P_{\rm e} (N) = \frac{1}{2} \left( 1-\trnorm{\eta_1 \rho_1^{\otimes N} - \eta_2 \rho_2^{\otimes N}} \right) \,.
\end{equation}
Note that the derivation of this formula imposes no additional constraints over the operators $E_i$ (apart from the POVM conditions), hence the measurement that achieves the limit~\eqref{ch3/helstromN} is, in principle, a collective one. Although the problem is formally solved, the computational cost of the trace norm grows exponentially with $N$. General analytical results for arbitrary $N$ and arbitrary states are scarce, existing only bounds for $P_{\rm e}(N)$ \citep{Audenaert2012}. 
The remaining of the Section is devoted to present two results that enable tractable analytical expressions of $P_{\rm e}(N)$ in special cases. The first is a mathematical tool that will prove useful in Chapters~\ref{ch4_pqsd} and \ref{ch5_learning} for obtaining analytical results when the number of copies is kept finite. The second concerns the asymptotic expression $P_e(N\to\infty)$.
\\

\subsection{Irreducible representations and block decomposition.}\label{ch3/sec:blockdecomposition}

The purpose of this Section is to present a particular decomposition of density operators of multicopy systems. It was introduced in \citep{Vidal1999,Cirac1999} within the context of estimation and purification of qubits, respectively, and later applied to the full estimation of qubit mixed states in \citep{Bagan2006}. Although here I will focus on qubit systems ($d=2$), it is straightforward to extend the decomposition to systems of dimension $d>2$ by including the irreducible representations of ${\rm SU}(d)$ in the formalism.

A set of $N$ qubit systems in the state $\rho$ is represented by the density operator $\rho^{\otimes N}$. This operator is invariant under the permutation of any pair of qubits, thus invariant under the action of the symmetric group $S_N$. One may use the group $S_N$ to write $\rho^{\otimes N}$ in the basis of the ${\rm SU}(2)$ invariant subspaces of $\left({\bf \frac{1}{2}}\right)^{\otimes N}$ [bold characters stand for the irreducible representations of ${\rm SU}(2)$], in a similar way as it is used to obtain the Clebsch-Gordan decomposition in ${\rm SU}(2)$. The relation between the tensor-product (decoupled) representation and that of the invariant subspaces (coupled) is
%
%
\begin{equation}
\left({\bf 1\over2}\right)^{\otimes N} = \bigoplus_{j,\alpha} {\rm \bf j^{(\alpha)}} \,,
\end{equation}
where $j=0\,(1/2), \ldots , J=N/2$ for even (odd) $N$, and $\alpha$ labels the different equivalent irreducible representations $\rm\bf j$, i.e., $\alpha=1,\ldots,\nu_j$, where $\nu_j$ is the multiplicity of $\rm\bf j$.
The density operator $\rho^{\otimes N}$, written in the invariant subspaces basis, has the block-diagonal form
\begin{equation}\label{ch3/rhoblock1}
\rho^{\otimes N} = \bigoplus_{j,\alpha} \rho_j^{(\alpha)} \,,
\end{equation}
where $\rho_j^{(\alpha)}$ represents the block associated to the subspace $\rm\bf j^{(\alpha)}$. 

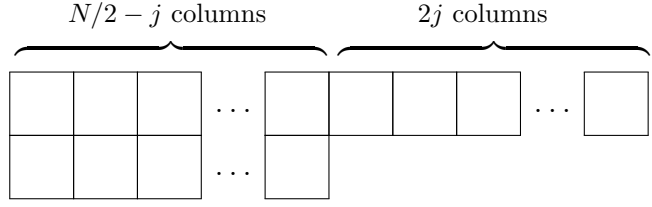
\begin{figure}[t]
\setlength{\unitlength}{1cm}
\begin{picture}(8,3)
\ytableausetup{mathmode,boxsize=2em}
\put(2.5,1){
\begin{ytableau}
\phantom. & \phantom. & \phantom. & \none[\dots] & \phantom. & \phantom. & \phantom. & \phantom. & \none[\dots] & \phantom. \\
\phantom. & \phantom. & \phantom. & \none[\dots] & \phantom.
\end{ytableau}
}
\put(2.7,2){$\overbrace{\hspace{4.1cm}}$}
\put(6.95,2){$\overbrace{\hspace{4.1cm}}$}
\put(3.4,2.5){\footnotesize $N/2-j$ columns}
\put(8,2.5){\footnotesize $2j$ columns}
\end{picture}
\caption{A generic Young diagram with $N$ boxes.}\label{ch3/fig:fig6}
\end{figure}

The explicit form of the blocks can be easily obtained by analysing the Young diagrams that can be constructed with $N$ boxes, one for each qubit. 
There will be as many different $\rm\bf j$ as Young diagrams\footnote{Given a Young diagram, the value of the associated label $\alpha$ corresponds to a specific Young tableau for that diagram (see below). As the explicit form of $\rho_j^{(\alpha)}$ does not depend on $\alpha$, one only needs to focus on Young diagrams for now.}. A particular $\rm\bf j$ corresponds to a diagram with $N/2-j$ double-box and $2j$ single-box columns (see Fig.~\ref{ch3/fig:fig6}), where each of the former is associated to a fully-antisymmetric two-qubit state or \emph{singlet}, and the remaining to a fully-symmetric state of $2j$ qubits. This means that the matrix $\rho_j^{(\alpha)}$ has dimension $2j+1$, and each singlet contributes a multiplicative factor $\det \rho$ to it.

Let $r$ be the purity of the state $\rho$ and $\vect{v}$ its Bloch vector. Let $\{\ket{j,m,\alpha}\}$ be a basis of the subspace $\rm\bf j^{(\alpha)}$ (in analogy to the angular momentum basis), constructed from the computational basis $\{\ket{0},\ket{1}\}$ of a single qubit. If $\vect{v}=\hat{z}$, the matrix $\rho_j^{(\alpha)}$ is diagonal in the basis $\{\ket{j,m,\alpha}\}$ and its expression is easily deduced. Since
\begin{equation}
\det \rho = \frac{1-r^2}{4} \,,
\end{equation}
one can write $\rho_j^{(\alpha)}$ as
\begin{align}\label{ch3/rhoblock_diagform}
\rho_j^{(\alpha)} = \left(\frac{1-r^2}{4}\right)^{N/2-j} \sum_{m=-j}^j \left(\frac{1-r}{2}\right)^{j-m} \left(\frac{1+r}{2}\right)^{j+m} \ketbrad{j,m,\alpha} \,.
\end{align}
For an arbitrary direction $\vect{v}$, it suffices to rotate the basis elements $\ket{j,m,\alpha}$ by means of the Wigner matrices $D (\vect{v})$ \citep{Edmonds1960}. From the standard definition, one has 
\begin{equation}
\ket{j,m,\alpha}_{\vect{v}} = U_{\vect{v}}^{\otimes n} \ket{j,m,\alpha} = \sum_{m'} \mathscr{D}_{m',m}^j (\vect{v}) \ket{j,m',\alpha} \,,
\end{equation}
where $U_{\vect{v}} \in {\rm SU}(2)$ is a rotation on a single copy, and the matrix elements are $\mathscr{D}_{m',m}^j = \bra{j,m',\alpha} D(\vect{v})\ket{j,m,\alpha}$. Hence $\rho_j^{(\alpha)}$ takes the general form
\begin{align}\label{ch3/rhoblock2}
\rho_j^{(\alpha)} = \left(\frac{1-r^2}{4}\right)^{N/2-j} \sum_{m=-j}^j & \left(\frac{1-r}{2}\right)^{j-m} \left(\frac{1+r}{2}\right)^{j+m} \nonumber\\
& \hspace{0.8cm}\otimes D(\vect{v})\ketbrad{j,m,\alpha}D^\dagger(\vect{v}) \,,
\end{align}
which is the same for all the equivalent irreducible representations (i.e., its coefficients do not depend on the label $\alpha$). Note that for pure states $\rho^{\otimes N}$ has projection only in the symmetric $(N+1)$-dimensional subspace $\rm\bf J=N/2$, whereas for mixed states it has components in all subspaces, including equivalent representations, $\rm\bf j^{(\alpha)}$.

The only thing left to do is to determine how many equivalent irreducible representations are for each $\rm\bf j$, that is the multiplicity $\nu_j$. It reads off from simple combinatorics. The value of $j$ associated to a subspace $\rm\bf j^{(\alpha)}$ is determined by the shape of its Young diagram, that is the particular partition of $N$ boxes in two rows such that the length of the second row is equal or shorter than that of the first. The different values that $\alpha$ can take correspond to all the possible \emph{standard} Young tableaux that can be built with that diagram. Given a diagram, a Young tableau is obtained by filling the boxes with integer numbers, from 0 to $N$; it is called standard if the following rules are fulfilled: (i) the entries in each row are in increasing order, from left to right, and (ii) the entries in each column are in increasing order, from top to bottom. For example, for $N=4$ the possible Young diagrams are\\

\begin{center}
\ytableausetup{boxsize=2em}
\ydiagram{4}
\qquad\qquad
\ydiagram{3,1}
\qquad\qquad
\ydiagram{2,2}
\end{center}
They are associated to the subspaces $\rm\bf j = 2$, $\rm\bf j = 1$, and $\rm\bf j = 0$, respectively.
With the first diagram only one standard Young tableaux can be constructed: 
\ytableausetup{smalltableaux}
\begin{ytableau}
1 & 2 & 3 & 4
\end{ytableau}\,.
With the second, 
\begin{ytableau}
1 & 2 & 3 \\
4
\end{ytableau}\,,
\begin{ytableau}
1 & 3 & 4 \\
2
\end{ytableau}\,, and
\begin{ytableau}
1 & 2 & 4 \\
3
\end{ytableau}\,.
Finally, for the third diagram one finds
\begin{ytableau}
1 & 2 \\
3 & 4
\end{ytableau}
and
\begin{ytableau}
1 & 3 \\
2 & 4
\end{ytableau}\,.
This means that, in the representation of invariant subspaces, the fully-symmetric subspace $\rm\bf 2$ occurs one time, the subspace $\rm\bf 1$ occurs three times and $\rm\bf 0$ occurs two times\footnote{Recalling that a subspace $\rm\bf j$ has dimension $2j+1$, one can check at this stage that the dimension of the total state $\rho^{\otimes 4}$ in this representation is indeed correct: $5\times1+3\times3+1\times2=2^4=16$.}. 
Following this reasoning one can see that the multiplicity $\nu_j$ of an arbitrary subspace $\rm\bf j$ is given by
\begin{equation}\label{ch3/rhoblock3}
\nu_j = \binom{N}{N/2-j} - \binom{N}{N/2-j-1} \,.
\end{equation}
\begin{figure}[t]
\begin{center}
\includegraphics[scale=.9]{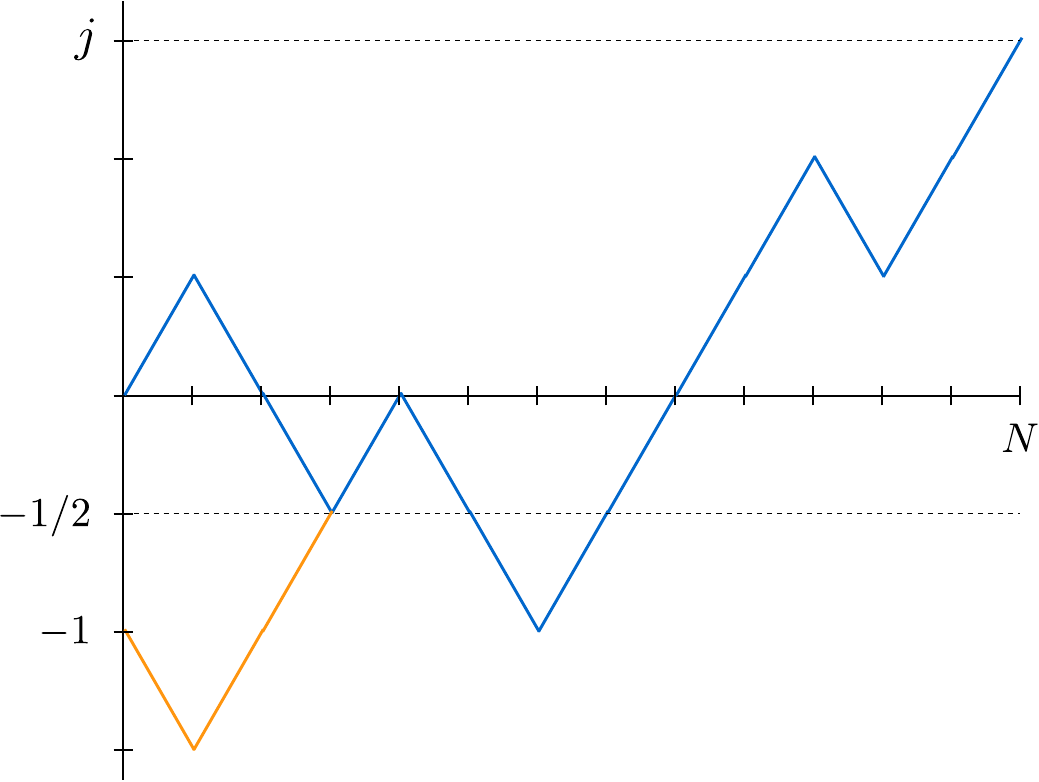}
\end{center}
\caption[Reflection theorem]{A ``bad'' path made of $N$ steps of length $1/2$ that ends at $j$ (blue). Reflecting vertically the part of the path at the left of the first step that crosses $-1/2$, one obtains a path that starts at $-1$ and ends at $j$ (orange). For every bad path of the blue type there exists one of the orange type.}\label{ch3/fig:fig7}
\end{figure}
\hspace{-1.1mm}However, there is a much simpler way to derive this formula: counting random walks. A certain $j$ can be thought of as the 
end point of a random walk of $N$ steps of length $1/2$, where each step can be taken either forward or backward, 
much as like $N$ spins $1/2$ are sequentially coupled to give a total angular momentum $j$.
A particular path ending in $j$ corresponds to a particular equivalent representation of the subspace $\rm\bf j$. Counting how many equivalent representations there are for a certain $j$ is thus the same as counting the number of (valid) paths that lead to the same $j$. 
If all the steps are made forward, the end point is the maximum value $N/2$ that $j$ can take, and, of course, that is the only path that reaches it. To get to the previous value $N/2-1$ one step shall be made backward at some point; all the different points at which this may happen account for all the paths ending at $N/2-1$. For an arbitrary $j$, the number of back steps needed is $N/2-j$, and the total number of paths with that many back steps is given by the first binomial in Eq.~\eqref{ch3/rhoblock3}.
Now, some of these paths go through negative values at some point. Let us call them ``bad'' paths. Clearly, bad paths do not correspond to valid coupling sequences of angular momenta, so they must be discarded. To see how many of these there are, note that each bad path must necessarily go through the value $-1/2$ at some point. Taking the reflection with respect to the value $-1/2$ of all the previous steps made up that point (see Fig.~\ref{ch3/fig:fig7}),
one obtains a new path that ends at $j$ but starts at $-1$, instead of at zero, and this is a one-to-one correspondence: each bad path can be associated with another path starting at $-1$. Hence the total number of bad paths is just all the possible paths that end at $j$ and start at $-1$, and this is the second binomial in Eq.~\eqref{ch3/rhoblock3}.
%

The block-decomposition of $\rho^{\otimes N}$, comprised by Eqs.~\eqref{ch3/rhoblock1}, \eqref{ch3/rhoblock2} and \eqref{ch3/rhoblock3}, turns out to be very useful in the computation of Eq.~\eqref{ch3/helstromN}. Since the trace norm operation is base independent, one can write the states $\rho_1^{\otimes N}$ and $\rho_2^{\otimes N}$ in the basis that block-diagonalizes them to split the trace norm over the global states into a sum of trace norms over each orthogonal subspace (hence reducing drastically the dimension of the matrices involved in the computation), i.e.,
\begin{equation}
\trnorm{\eta_1\rho_1^{\otimes N} - \eta_2\rho_2^{\otimes N}} = \sum_{j=0,1/2}^{N/2} \nu_j \trnorm{\eta_1\rho_{1,j}-\eta_2\rho_{2,j}} \,.
\end{equation}
Furthermore, for each $j$, the contribution of all the equivalent representations $\rm\bf j^{(\alpha)}$ boils down to a multiplicative factor (its multiplicity $\nu_j$), since $\rho_j^{(\alpha)}$ is the same matrix for all values of $\alpha$.
\\

\subsection{Infinitely many copies: the quantum Chernoff bound}\label{ch3/sec:quantumchernoff}

In the same spirit as Section~\ref{ch3/sec:classicalchernoff}, it is interesting to study the behaviour of the minimum-error probability in the asymptotic limit of infinite \mbox{copies. As it} happens with the minimum-error probability for distinguishing classical probability distributions, the trace norm, as a distance measure between quantum states, lacks monotonicity under the increase of the tensor powers of its arguments. That is to say, it is not difficult to find two pairs of states $\rho_1, \rho_2$ and $\sigma_1, \sigma_2$ for which $\trnorm{\rho_1-\rho_2} < \trnorm{\sigma_1-\sigma_2}$, but 
$\parallel\rho_1^{\otimes 2}-\rho_2^{\otimes 2}\parallel_1 \,>\, \parallel\sigma_1^{\otimes 2}-\sigma_2^{\otimes 2}\parallel_1$. 
It is thus desirable to count with a distance measure that does not explicitly depend on the provided number of copies $N$. In an analogous way to the Chernoff bound~\eqref{ch3/classicalchernoffbound}, the minimum-error probability for distinguishing two quantum states, defined in Eq.~\eqref{ch3/helstromN}, is upper-bounded by the \emph{quantum Chernoff bound} \citep{Audenaert2007}
\begin{equation}\label{ch3/quantumchernoffbound}
P_e(N) \leqslant \min_{s\in[0,1]} \eta_1^s \eta_2^{1-s} \,\tr \rho_1^s \rho_2^{1-s} \,,
\end{equation}
which is tight in the asymptotic limit $N\to\infty$\footnote{The upper bound is a direct application of the relation $\tr(A^s B^{1-s}) \geqslant \tr (A+B-|A-B|)/2$, that holds for any two positive operators $A$ and $B$ and for all $0\leqslant s\leqslant 1$. A lower bound for $P_e(N)$ was found in \citep{Nussbaum2009a} that coincides with the upper bound introduced in \citep{Audenaert2007} when $N\to\infty$, thus proving attainability.}. Furthermore, the error probability decreases exponentially with the number $N$ of copies as $N$ goes to infinity \citep{Cover2006}, and the rate exponent is determined by the quantum Chernoff bound. That is
\begin{equation}
P_e(N\to\infty) \sim e^{-N D(\rho_1,\rho_2)} \,,
\end{equation}
where
\begin{equation}
D(\rho_1,\rho_2) = -\min_{s\in[0,1]} \log \tr \rho_1^s \rho_2^{1-s}
\end{equation}
is known as the \emph{quantum Chernoff distance}.

As the classical Chernoff distance, defined in Eq.~\eqref{ch3/classicalchernoffdistance}, its quantum counterpart gives a proper measure of distinguishability between quantum states \citep{Calsamiglia2008}. Most importantly, although it is operationally based in a discrimination protocol (and consequently in a measurement procedure),  this measure defines the optimal error rate in a device-independent way. The quantity $D(\rho_1,\rho_2)$ thus provides a nice tool for benchmarking particular strategies. In contrast to the classical case, in quantum discrimination one has to optimize the strategy, and if there are restrictions over the available measurements this can be a 
rather involved
process. A quick test to see if a particular strategy is optimal is to compare the error rate that it gives with $D(\rho_1,\rho_2)$: if both match, then optimality is guaranteed.

An additional feature of the quantum Chernoff distance is that it induces a physically motivated metric to the space of quantum states, thus endowing it with a geometrical structure \citep{Petz1996a}. This enables a relation between geometrical concepts (e.g., distance, volume, curvature) to physical ones (e.g., state discrimination and estimation). The metric is obtained, roughly speaking, by defining a line element  between the infinitesimally close states $\rho$ and $\rho-d\rho$ through the distinguishability measure $D(\rho,\rho-d\rho)$. In particular, the so-called Chernoff metric \citep{Audenaert2007,Calsamiglia2008} provides an operationally defined volume element $d\rho^{\rm Ch}$, that for qubits with purity $r$ reads
\begin{equation}
d\rho^{\rm Ch} = \frac{1}{\pi-2} \frac{\left(\sqrt{1+r}-\sqrt{1-r}\right)^2}{\sqrt{1-r^2}} dr \frac{d\Omega}{4\pi} \,,
\end{equation}
where $d\Omega/4\pi$ is the invariant measure on the 2-sphere. Alternatively, there exist other metrics that are based on different criteria, such as the Bures metric, induced by the fidelity distance \citep{Zyczkowski2005} (see Section~\ref{ch4/sec:universal} for more details on different metrics for the qubits state space).\\

\section{Final comments}\label{ch3/sec:othertasks}

So far the standard problem of quantum state discrimination and the main approaches therein have been reviewed. In these, the answer sought is the identity of some unknown state---given an ensemble of known possibilities---and the figure of merit that benchmarks a particular strategy is the probability of a successful identification. However, as noted in Section~\ref{ch3/sec:theunknown}, state determination tasks encompass a broader variety of protocols. Different strategies serve the purpose in different situations, depending on the prior information one has and the questions one expects to answer about the unknown state.
Although a thorough review of these variations falls beyond the scope of this thesis, I would like to finish this Chapter by briefly going through the ones that, in a way or another, connect with the discrimination problems treated here.
\\

\subsection*{Discrimination with maximum confidence}

An alternative to the probability of success---or error---as the figure of merit for quantum state discrimination is the \emph{confidence} of the measurement outcomes, that is, the probability that the unknown state was indeed $\rho_i$ given that outcome $i$ was obtained. If the possible states are linearly independent, then the optimal measurement for unambiguous discrimination identifies the state with certainty, i.e., it provides outcomes with confidence one. If this is not the case, fully unambiguous answers are not achievable, but one can still try to find the measurement that allows to be as confident as possible that the state inferred from an outcome is the correct one. Hence, for linearly dependent states, measurement outcomes with maximum confidence is as unambiguous as it gets.
 
The discrimination of quantum states with maximum confidence was introduced in \citep{Croke2006}. The authors consider a system known to be prepared in one of $N$ possible states $\{\rho_i\}$, with associated \emph{a priori} probabilities $\{\eta_i\}$.
Say a measurement over the system provides the outcome $i$; consequently, one infers that the state of the system was $\rho_i$. The probability of this inference to be true, that is, the confidence of outcome $i$, is defined through Bayes' rule as
\begin{equation}\label{ch3/confidence_i}
C(i)\equiv p(\rho_i|i) = \frac{p(i|\rho_i)p(\rho_i)}{p(i)} = \frac{\eta_i \,\tr (\rho_i E_i)}{\tr (\rho E_i)} \,,
\end{equation}
where $\rho=\sum_{i=1}^N \eta_i \rho_i$, and $E_i$ is the measurement operator associated to outcome $i$. In \citep{Croke2006}, the optimality criterion is chosen to be the maximisation of the confidence $C(i)$ of \emph{every} possible outcome.
As each operator $E_i$ is optimised independently of the others, in general the set of operators $\{E_i\}_{i=1}^N$ will not describe a proper POVM. An inconclusive outcome that completes the identity will be required in most cases, with an associated operator $E_0=\id-\sum_{i=1}^N E_i$. Lastly, since multiplicative factors in the elements $E_i$ cancel out in Eq.~\eqref{ch3/confidence_i}, one can choose to give the largest factors to the elements for $i=1,\ldots,N$---and hence the smallest to the element $E_0$---compatible with the positivity condition $E_0\geqslant 0$, much as like it was done for deriving the optimal unambiguous strategy (see Section~\ref{ch3/sec:unambiguous}). One obtains in this way the maximum confidence strategy that minimises the probability of inconclusive results.

A very similar approach worth remarking is found in \citep{Fiurasek2003}, where the goal is to maximise the \emph{relative} probability of success, defined as the average success probability given that the measurement produces a conclusive answer. The relative success probability can be understood as a sort of overall confidence of the measurement apparatus, in contrast to the individual confidences $C(i)$. It is expressed as
\begin{equation}\label{ch3/confidence}
C = \frac{P_{\rm s}}{1-Q} \,.
\end{equation}
When the possible states $\rho_i$ are symmetrically distributed and have equal \emph{a priori} probabilities $\eta_i=1/N$, Eqs.~\eqref{ch3/confidence_i} and \eqref{ch3/confidence} coincide\footnote{Note that $\tr(\rho_i E_i)$ and $\tr(\rho E_i)$ cannot depend on the index $i$ in such symmetric problem. Then, it is enough to write $P_{\rm s} = \sum_{i=1}^N \eta_i \tr (\rho_i E_i) = \tr(\rho_1 E_1)$ and $1-Q=P_{\rm s} + P_{\rm e}=\sum_{i,j}\eta_i \tr(\rho_i E_j) = N \tr(\rho E_1)$.}, thus the two optimisation procedures are equivalent.

Now, recall that in the general scheme of discrimination with an error margin---or, equivalently, with a fixed rate of abstention---the optimal POVM is completely determined by the margin $r$---or the probability of abstention $Q$. The maximisation of $C$ yields the best ratio of correct identifications over conclusive answers, which is a scenario covered by these schemes. 
Maximum confidence discrimination thus corresponds to a particular instance of discrimination with an error margin. It yields the POVM associated with the margin $r$ above which the success probability increases linearly.
In the equivalent parametrization in terms of $Q$, the POVM for maximum confidence corresponds to the value of $Q$ above which the success probability decreases linearly.
\\

\subsection*{Comparison of quantum states}

Given two quantum systems prepared in two unknown states, in the absence of any other information one can still answer the question of whether the states of the systems are equal or different to each other. This is the objective of quantum state comparison \citep{Barnett2003}\footnote{Its extension to sets of multiple states, either all equal or at least one different, is analysed in \citep{Jex2004}.}. Since no information about the particular possible states is provided, the strategy relies on the symmetry of the collective state of the two systems. 
The total Hilbert space can be split into a symmetric subspace $\cal S$ and an antisymmetric subspace $\cal A$ such that $\cal S\oplus A = H\otimes H$, where $\cal H$ is the Hilbert space of each subsystem.
If the two states are equal, permuting the systems leaves the collective state invariant. This means that such collective state only has projection onto the symmetric subspace $\cal S$.  On the other hand, if the states are different, the collective state has projection onto $\cal S$ as well as $\cal A$. It follows that a measurement checking the presence of the collective state in these subspaces would be able with some probability to tell with certainty if the states are different, but not if they are equal. The measurement operators $E_{\rm diff} = \Pi_A$ and $E_0 = \Pi_S$ accomplish the task optimally, where $\Pi_X$ is a projector onto subspace $X$. As usual, the outcome $0$ is the inconclusive one.

The procedure above is the only resource one has for completely unknown pure states. When a known set of possible states for each system is provided, quantum state comparison becomes a special instance of standard quantum state discrimination. Having this extra knowledge allows, for instance, unambiguous answers for both possible cases, i.e., equal states and different states. To see this, take the possible states of each system to be $\ket{\psi_i}$, $i=1,2$, as defined in Eq.~\eqref{ch3/minerr_vecpsi}. The collective state $\ket{\psi_i}\otimes\ket{\psi_j}$ can be expressed as a combination of three parts: a symmetric entangled state, present only when $i=j$; an antisymmetric state, happening only when $i\neq j$; and some combination of the symmetric states $\ket{0}\otimes\ket{0}$ and $\ket{1}\otimes\ket{1}$, present in both situations. A standard unambiguous measurement made of projectors onto the subspaces spanned by these three parts is able to generate conclusive answers in the two possible scenarios, and it does so with a minimum probability of inconclusive results.
As for the minimum-error approach---now enabled since we know what the states $\ket{\psi_i}$ are---, it amounts to applying the optimal Helstrom measurement for standard discrimination between the global states
\begin{align}
\rho_{\rm eq} &= \left(\eta_1^2 \ketbrad{\psi_1}\otimes\ketbrad{\psi_1}+\eta_2^2 \ketbrad{\psi_2}\otimes\ketbrad{\psi_2}\right) \,,\\
\rho_{\rm diff} &= \eta_1 \eta_2\left(\ketbrad{\psi_1}\otimes\ketbrad{\psi_2}+ \ketbrad{\psi_2}\otimes\ketbrad{\psi_1}\right) \,,
\end{align}
where $\eta_i$ is the \emph{a priori} probability associated to $\ket{\psi_i}$. The minimum probability of error is just $P_{\rm e}=(1-\trnorm{\rho_{\rm eq}-\rho_{\rm diff}})/2$.
Of course, the same technique applies to the comparison of two mixed states with minimum error.
\\

\subsection*{Estimation of quantum states}

In contrast to the identification of an unknown quantum state by discriminating between a set of known hypotheses, 
when this set is also unknown the task becomes that of \emph{estimating} the state,
for which the quantum state discrimination toolbox reviewed so far is of no use.
Since this time one knows nothing, one shall assume that the state could be \emph{any} state of the Hilbert space of the system, that is, an infinite set of possible states. Any measurement we could perform over the system will have a finite number of outcomes, thus the association of one outcome with one possible state, as done in discrimination, is no longer 
feasible.
In fact, in the absence of any prior information, any measurement could hardly tell us much about the original state if we count with only one copy. 
A more realistic 
setting
is to have several copies of the unknown state, measure them, and then give an \emph{estimate}.
Such a process is called \emph{quantum state tomography}\footnote{A review on the broader field of quantum state estimation, which includes tomography, can be found in \citep{Paris2004}.}.

The scenario is usually described as follows. A source repeatedly produces identical copies of some quantum state $\rho$. An experimentalist, commissioned to characterize the source specifications, performs some measurement with $K$ outcomes over, say, $N$ copies of $\rho$ (in a collective state $\rho^{\otimes N}$). On the basis of the measurement outcome he will guess that the state was $\sigma_k$, where $k=1,\ldots,K$. In general the guess $\sigma_k$ will be wrong, hence the goal of the experimentalist is to design a measurement such that the produced guess is, on average, as close to $\rho$ as possible. The function that quantifies this closeness is the \emph{fidelity}, that for completely general (pure and mixed) quantum states reads \citep{Fuchs1996}
\begin{equation}
F(\sigma_k,\rho)=\tr \sqrt{\sigma_k^{1/2} \,\rho\, \sigma_k^{1/2}} \,,
\end{equation}
where, for some nonnegative operator $A$, $A^{1/2}$ is defined as the unique nonnegative operator such that $A^{1/2}A^{1/2}=A$ \footnote{Note that the fidelity is symmetric under the permutation of its arguments, i.e., $F(\sigma_k,\rho)=F(\rho,\sigma_k)$. Note also that, if the states are pure, i.e., $\rho=\ketbrad{\psi}$ and $\sigma_k=\ketbrad{\varphi_k}$, it reduces to the squared overlap $F(\varphi_k,\psi)=|\braket{\varphi_k}{\psi}|^2$.}. The figure of merit that the experimentalist shall try to maximise is the average fidelity over all possible guesses and all possible states, that is
\begin{equation}
\bar{F}_N = \sum_{k=1}^K \int P(\sigma_k|\rho) F(\sigma_k,\rho) d\rho \,,
\end{equation}
where $P(\sigma_k|\rho)$ is the probability that the guess is $\sigma_k$ given that the state is $\rho$---provided $N$ copies of $\rho$---, and $d\rho$ is some suitable probability density of the possible states produced by the source. All the properties of the estimation measurement, including the number of outcomes $K$, are to be determined from the maximisation of $\bar{F}_N$. Also, any prior information the experimentalist may have about the source is introduced through the computation of $P(\sigma_k|\rho)$.\\

Let me conclude by bringing up again the Bayesian interpretation of an unknown quantum state exposed in Section~\ref{ch3/sec:theunknown}, which, at first glance, might seem to loose consistency in the context of quantum state tomography\footnote{For more details in the Bayesian perspective of quantum state tomography, see~\citep{Fuchs2004}.}. As already anticipated, if quantum states are nothing more than states of belief of some agent rather than properties of nature, tomography is a clear an example where there is no other agent who might possess the state of knowledge $\rho$ that the experimentalist is trying to unravel. What is the unknown state $\rho$, then? The quick answer is that there is no need for a second agent, hence nor for the term ``unknown state'', in the Bayesian formulation of the problem.

The only assumption that the experimentalist needs to make is that the states produced by the source are indeed indistinguishable from each other, and nothing else. That is, if $\rho^{(N)}$ comprises his overall state of knowledge of the $N$ copies prior to any measurement, he will assign the same state $\rho^{(N)}$ to any permutation of the copies\footnote{Actually, the requirement that $\rho^{(N)}$ be derivable from $\rho^{(N+1)}$ for any $N$ is also necessary.}. The key point now is the \emph{quantum de Finetti representation theorem}, which states that, if $\rho^{(N)}$ corresponds to an exchangeable sequence of states---in the above sense---, it can be expressed as
\begin{equation}
\rho^{(N)} = \int P(\rho)\rho^{\otimes N} d\rho \,,
\end{equation}
where $P(\rho)$ is a prior probability distribution for $\rho$. Plainly put into words, the experimentalist can regard his prior state of knowledge $\rho^{(N)}$ \emph{as if} it were a probabilistic mixture of tensor product states $\rho^{\otimes N}$, where all he can tell about the unknown state is encapsulated in the probability distribution $P(\rho)$. 

Upon obtaining information from measuring some of the copies, $P(\rho)$ is shaped accordingly by means of Bayes' rule, until, when enough copies are measured, the state of knowledge of the experimentalist for the remaining copies resembles a product state. Furthermore, this updating process guarantees that two independent agents would come to agreement based on the same measurement outcomes,
regardless the prior probability with which each one starts with. Say the initial overall state of $N+M$ copies produced by the source is
\begin{equation}
\rho^{(N+M)} = \int P(\rho)\rho^{\otimes (N+M)} d\rho \,.
\end{equation}
The first $N$ copies are measured, and the obtained information is represented by the (multidimensional) random variable $k$. It can be shown that the remaining $M$ copies are left in the post-measurement state
\begin{equation}
\rho^{(M)}_k = \int P(\rho|k) \rho^{\otimes M} d\rho \,,
\end{equation}
where $P(\rho|k)$ is calculated through Bayes' rule. When $N$ is large enough, the probability $P(\rho|k)$ gets highly peaked on a certain state $\rho_k$ determined by the measurement outcomes, independently of the prior probability $P(\rho)$. If two agents start with different priors $P_i(\rho)$, $i=1,2$, both will adjust their state of knowledge after the measurement to the same product state $\rho_k^{\otimes M}$, as $\int P_i(\rho|k) \rho^{\otimes M} d\rho \rightarrow \rho_k^{\otimes M}$ for $N$ sufficiently large.

The Bayesian interpretation of the tomography process shifts in this way the focus from accessing the ``true'' state of the system to agents agreeing to a common state of knowledge in the light of evidence. Bayesian theory, as already said in Chapter~\ref{ch2_essentials}, does not describe how the physical world behaves, but rather how us, observers of that world, should act if we want to make rational assessments about it. 
The remaining chapters of this dissertation find strong support in these ideas.



\chapter{\chnamefour}

\label{ch4_pqsd}


The standard theory of quantum state discrimination, covered in Chapter~\ref{ch3_whatisknown}, is built on the premise of a measurer agent receiving both \emph{quantum} and \emph{classical} information, namely a quantum system in an unknown state, and a description of the possible states of the system and their \emph{a priori} probabilities. The agent uses all this available information to devise the discrimination machine that best determines the state of the system. As a consequence, the machine is 
tailored
to that particular discrimination instance: the given hypotheses are hard-coded into its design, and the machine 
becomes unreliable
in facing any other set of hypotheses.

It is then natural to wonder whether a device for discriminating arbitrary pairs of states---a universal (multipurpose) quantum-measurement apparatus so to say---, can be constructed. 
Such a ``quantum multimeter'' can be understood at an abstract level as a programmable quantum processor~\citep{Buzek2006}, that is, a device with a \emph{data} port and a \emph{program} port, where the input at the program port determines the operation 
to be performed on the input at the data port\footnote{Programmable quantum processors were first considered by Nielsen and Chuang as gate arrays~\citep{Nielsen1997}. They restricted their study to the case where a unitary operation, rather than a measurement or a more general completely positive linear map, is performed on the state in the data port.}.
The usual discrimination task between known states would correspond to a processor specifically programmed by a set of instructions---the \emph{classical} description of the possible states---to determine the state of a system in the data port, very much as programming a computer to perform a task by setting dials or switches to particular positions, each task requiring a different configuration.
Programmable quantum processors admit a much more general approach, that is to consider that the programming is carried out, not by a human agent manipulating switches, but directly by raw information in a \emph{quantum} form, i.e., information stored in the state of some quantum system.
In the state discrimination context, this means that the information about the possible states of the system at the data port is provided 
by quantum systems in particular states entering the program port of the processor.
A quantum processor programmed in this way would be able to read this quantum information by itself and adjust accordingly a discrimination measurement performed on the data system without human intervention. It could even take advantage of quantum correlated joint measurements over both the program and data systems to carry out the task more efficiently. In short, supplied with the correct programs, this machine would be capable of discriminating between any pair of quantum states.

Programmable quantum state discrimination machines have been extensively analysed in the literature. A programmable device that uses projective measurements to discriminate the state of a qubit, the basis of the projection being specified by the program, was discussed in \citep{Fiurasek2002,Fiurasek2004}. In \citep{Dusek2002} the case of distinguishing two equatorial qubits with generalized measurements was considered. The separation angle between the states, which specifies the POVM, was encoded in a single-qubit program, yielding a good but suboptimal performance. 
Later, Bergou and collaborators 
proposed 
a different encoding system: their machine has two program ports, each of them fed with a system in one of the possible states, and a data port, fed with the state to be identified.
The authors obtained the optimal solution in both the unambiguous and the minimum-error schemes for general pure qubit states~\citep{Bergou2005,Bergou2006b}, which works by exploiting the difference between the permutation symmetry of the global state of the three ports in the two alternatives.
This last approach benefits from not requiring beforehand any classical information about the hypotheses in order to prepare a specific encoding, as copies of the possible states---whatever they are---are just plugged into the program ports, perhaps coming out from some other quantum information processing device. Several other works, as well as the contents of this Chapter, extend further this idea\footnote{See, e.g., \citep{Hayashi2005,Hayashi2006,Bergou2006a,Zhang2006,He2007,Ishida2008,Herzog2008,Sedlak2007,Sedlak2009,Bartuskova2008,Zhou2011,Zhou2013,Colin2012}.}.

Interestingly, these devices can also be regarded as \emph{learning machines}: the device is instructed, or trained, through the program ports about different states, and, based on the acquired knowledge, it associates the state in the data port with one of the states belonging to the training set. This view implies that the discrimination task is carried out by two separate operations, an initial training step and a subsequent identification step, i.e., it considers a particular type of process happening inside the quantum processor (Chapter~\ref{ch5_learning} is devoted entirely to make clear this distinction). 
Furthermore, programmable discrimination machines are mathematically equivalent to a change-point problem~\citep{Akimoto2011}: a source produces states of one type and, either at time $t_1$ or at time $t_2$, it starts producing states of a different type; the change-point problem consists in identifying whether the time at which the change occurs is $t_1$ or $t_2$.

In this Chapter we consider the programmable discrimination of two general qubit states, although most of our results can be generalized to higher dimensional systems. For simplicity we assume that the prior occurrence probability of each state is identical and compute the unambiguous and minimum-error rates for optimal programmable devices when an arbitrary number of copies of the states is provided at every port.
We first study the performance of such devices for pure states. 
Some of these results are already available in the literature\footnote{See, e.g., \citep{He2007}, although no closed expressions for the error rates were given there.}, but the way we formalize the problem here is crucial to treat the more general mixed state case. In addition, we obtain analytical expressions that enable us to present the results and study limiting cases in a unified way. In particular, when the program ports are loaded with an infinitely large number of copies of the states  we recover the usual state discrimination problem for known states (see Section~\ref{ch3/sec:minimumerror})\footnote{An infinite number of copies of an unknown state permits perfect quantum state tomography. 
As a result, one has as much information as the classical description of the states entering the program ports.}.
On the other hand, when the number of copies at the data port is infinitely large, while the number of copies at the program ports are kept finite, we recover the state comparison problem (see Section~\ref{ch3/sec:othertasks}). 

We extend the previous pure state study to the case of mixed input states. In this scenario we only compute the minimum-error probability, as no unambiguous answers can be given if the states have the same support\footnote{See Section~\ref{ch3/sec:unambiguous} for details. As we will see, this is indeed the case here, since the global states entering the machine are full-rank matrices.}. The performance of the device for a given purity of the input states allows to quantify how the discrimination power is degraded in the presence of noise. The expressions here are much more involved, however one can still exploit the permutation symmetry of the input states to write the problem in a block-diagonal form, as shown in Section~\ref{ch3/sec:blockdecomposition}. We then obtain closed expressions for the  probability of error that can be computed analytically for small number of copies and numerically evaluated for a fairly large number of copies. 
We also obtain analytical expressions for some  asymptotic rates. Again, the leading term, as in the pure state case, is seen to coincide with the average minimum error for known states.


We also analyse the fully universal discrimination machine, i.e.,  a device that works optimally for completely unknown input states. In this case one has to
assume a uniform distribution for the purity. In contrast to the pure state distribution,  there is no unique choice~\citep{Petz1996}, and different reasonable assumptions
lead to different uniform priors. Here we consider the hard-sphere, Bures, and Chernoff priors.  
%
\\


\section{Pure states}\label{ch4/sec:pure}

Let us start by fixing the notation and conventions that we use. We label the two program ports by $A$ and $C$. 
These will be loaded with states $\ket{\psi_1}$ and $\ket{\psi_2}$, respectively. The data port, $B$, is the middle one
and will be loaded with the states we wish to identify as of type $1$ or type $2$. We also use the short hand notation $[\psi]$ to denote $\ketbrad{\psi}$, and similarly
$[\psi \phi\ldots] =[\psi]\otimes[\phi]\otimes\cdots=\ketbrad{\psi}\otimes \ketbrad{\phi}\otimes\cdots$.  We may also omit the subscripts  $A, B$ and $C$ when no confusion arises. We assume that the program ports are fed with $n$ copies of each state and the data port with $n'$ copies of the unknown state.
This is a rather general case for which closed expressions of the error probabilities can be given. The case with arbitrary $n_A, n_B,$ and $n_C$ copies at each port is discussed in Appendix~\ref{appA/sec:na-nb-nc}. The expressions are more involved but the techniques are a straightforward extension of the ones presented here.

When the state at the data port is $\ket{\psi_1}^{\otimes n'}$ or  $\ket{\psi_2}^{\otimes n'}$, the effective states entering the machine are given by the averages
\begin{eqnarray}\label{ch4/multi-copy-pure}
\sigma_1 &=& \int d\psi_1 d\psi_2 [\psi_1^{\otimes n}]_A [\psi_1^{\otimes n'}]_B [\psi_2^{\otimes n}]_C \,, \nonumber \\
\sigma_2 &=& \int d\psi_1 d\psi_2 [\psi_1^{\otimes n}]_A [\psi_2^{\otimes n'}]_B [\psi_2^{\otimes n}]_C \,, \label{ch4/sigmas}
\end{eqnarray}
respectively. Note that, by taking the average over all possible input states, $\sigma_1$ and $\sigma_2$ summarize our absolute lack of knowledge about $\ket{\psi_1}$ and $\ket{\psi_2}$ in a Bayesian way, very much as it was emphasized in Section~\ref{ch3/sec:theunknown}. Note also that this allows us to assess the performance of the machine in a state-independent way, in turn characterizing a machine that works for any $\ket{\psi_1}$ and $\ket{\psi_2}$.
The integrals in Eq.~\eqref{ch4/sigmas} can be easily computed using the Schur's lemma $\int d\phi [\phi]_X = \openone_X/d_X$, where $d_X$ is the dimension of the
Hilbert space spanned by $\{\ket{\phi}\}$ and $\openone_X$ is the projector onto this space. Hence
\begin{eqnarray}\label{ch4/rho1-rho2}
		\sigma_1 &=&    \frac{1}{d_{AB} d_C} \openone_{AB} \otimes \openone_{C} \,,\nonumber \\
		\sigma_2 &=&    \frac{1}{d_{A} d_{BC}} \openone_{A} \otimes \openone_{BC}\,,
\end{eqnarray}
where $ \openone_{XY} $ is the projector onto the completely symmetric subspace of $\mathcal{H}_X\otimes \mathcal{H}_Y$,
and $d_{XY}=\tr\openone_{XY}$ is its dimension. For qubits we have $d_A=d_C=n+1$ and $d_{AB}=d_{BC}=n+n'+1$.

\begin{figure}[t]
\begin{center}
\includegraphics[scale=0.8]{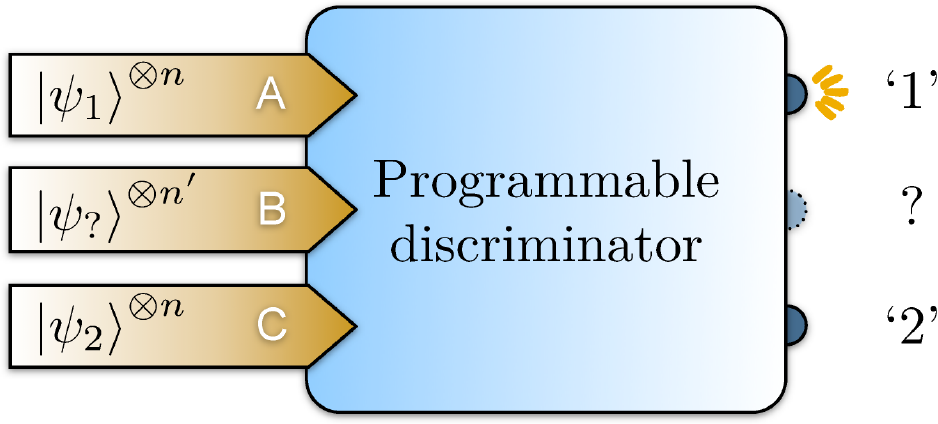}
\end{center}
\caption[A programmable discrimination machine]{A programmable discrimination machine with $n$ copies of the possible states entering the program ports $A$ and $C$, and $n'$ copies of the state to be identified entering the data port $B$. The machine has two possible outcomes if the discrimination is done within the minimum-error approach. If the unambiguous approach is used instead, a third (inconclusive) outcome has to be considered.\label{ch4/fig:fig0}}
\end{figure}

The structure of the states \eqref{ch4/rho1-rho2} suggests the use of the angular momentum basis: $\ket{j_A,j_B (j_{AB}), j_{C};J M}$ for
$\sigma_1$, and $\ket{j_{A},j_B,j_C (j_{BC});J M}$ for $\sigma_2$. The quantum numbers $j_{AB}=j_A+j_B$ and $j_{BC}=j_B+j_C$  recall the way
the three spins are coupled to give the total
angular momentum $J$. Here the angular momenta have a fixed value determined by the number
of copies at the ports, $j_A=j_C=n/2$ and $j_B=n'/2$, hence we can very much ease the notation by only writing explicitly the  labels $j_{AB}$ and $j_{BC}$.
We would like to stress, however, that, in general, one needs to keep track of all the quantum numbers, specially when dealing with mixed states as in Section~\ref{ch4/sec:mixed}.

In $\sigma_1$ the first $n+n'$ spins are coupled in a symmetric way, while in  $\sigma_2$  the symmetrized spins are the last $n+n'$, thus
$j_{AB}=(n+n')/2=j_{BC}$.  The states are diagonal in the angular momentum bases discussed previously, and we have
\begin{eqnarray}
\sigma_1 &=& \frac{1}{d_{AB} d_C}\sum_{J=0,1/2}^{n'/2+n} \sum_{M=-J}^{J} [j_{AB};J M]\,,\nonumber \\
\sigma_2 &=& \frac{1}{d_A d_{BC}}\sum_{J=0,1/2}^{n'/2+n} \sum_{M=-J}^{J} [j_{BC};J M]\,,
\end{eqnarray}
where the lower limit of the first summation takes the value 0 (1/2) for $n'$ even (odd).
Note that the spectrum of both matrices is identical and that the basis elements of their support
differ only in the way the three spins are coupled. Further, the key feature of the total angular momentum bases is the orthogonality relation
\begin{equation}\label{ch4/orthogonality}
\braket{j_{AB} ;J M}{j_{BC} ;J' M'} =0 \,,\quad \forall J \neq J' \;\mathrm{or}\;  M\neq M' \,.
\end{equation}
%
Bases obeying an orthogonality relation of the form \eqref{ch4/orthogonality} exist for any two subspaces, and are known as Jordan bases~\citep{Bergou2006a}. Since a state of the first basis has nonzero overlap with only one element of the second basis, the problem of discriminating $\sigma_1$ from $\sigma_2$ can be cast as pure state discrimination in each Jordan subspace, which we label by the quantum numbers $J$ and $M$ (although we will soon drop the label $M$). Then, the total error probability is simply the sum of all the contributions.

In the unambiguous approach, the minimum probability of an inconclusive result
for a pair of states $\ket{\phi_1},\ket{\phi_2}$  with equal priors is simply given by Eq.~\eqref{ch3/UA_p?} 
as $\PUA(\ket{\phi_1},\ket{\phi_2})=|\braket{\phi_1}{\phi_2}|$, hence
\begin{equation}\label{ch4/inconclusive}
\PUA=\frac{1}{d_{AB} d_{C}}\sum_{JM}|\braket{j_{AB} ;J M}{j_{BC} ;J M}| \,.
\end{equation}
These overlaps can be computed in terms of Wigner's 6$j$-symbols (see Appendix~\ref{appA/sec:wigner6j}):
\begin{align}\label{ch4/6j}
& \braket{j_{AB}; JM}{j_{BC}; J M} \nonumber\\
& \hspace{1cm}= (-1)^{j_A+j_B+j_C+J} \sqrt{(2j_{AB}+1)(2j_{BC}+1)}
\begin{Bmatrix} j_A & j_B & j_{AB} \\ j_C & J & j_{BC} \end{Bmatrix} \,.
\end{align}
Note that the 6$j$-symbols are independent of $M$, therefore in what follows we omit writing the quantum number $M$, and we perform the sum over $M$ in Eq.~\eqref{ch4/inconclusive} trivially by adding the multiplicative factor $2J+1$.  Substituting the value of the $6j$-symbols for $j_A=j_C=n/2$, $j_B=n'/2$, $j_{AB}=j_{BC}=(n+n')/2$, and setting $J=n'/2+k$, we obtain
\begin{equation}\label{ch4/overlap-nm}
\braket{j_{AB}; J}{j_{BC}; J}=
 \displaystyle \binom{n}{k} \binom{n+n'}{n-k}^{-1} \,,
\end{equation}
with $k=0,1,\ldots,n$ (observe that $J$ takes values from $J=n+n'/2$ of the totally symmetric space down to $J=n'/2$).

Plugging the overlaps in Eq.~\eqref{ch4/overlap-nm}  into Eq.~\eqref{ch4/inconclusive}, we obtain
\begin{equation}\label{ch4/ua-nm}
\PUA = \sum_{k=0}^{n} \frac{n'+2k+1}{(n+n'+1)(n+1)}\frac{(n'+k)!n!}{(n'+n)!k!} = 1- \frac{n n'}{(n+1)(n'+2)} \,,
\end{equation}
where the dimension of the subspace of total angular momentum $J$ is $n'+2k+1$, and in the second equality we have used the binomial sums
\begin{eqnarray}
\sum_{k=0}^n {n'+k \choose n'}&=&{n+n'+1 \choose n'+1} \,,\nonumber \\
\sum_{k=0}^n k {n'+k \choose n'}&=&{n+n'+1 \choose n'+1} \frac{n(n'+1)}{n'+2} \,.
\end{eqnarray}

In the minimum-error approach no inconclusive results are allowed, but the  machine  is permitted to give wrong answers with some probability
that one tries to minimize. This minimum-error probability can be computed along the same lines as in the previous case.
Recall that the error probability $\PME$ for two pure states $\ket{\phi_1},\ket{\phi_2}$ and equal \emph{a priori} probabilities is given by Eq~\eqref{ch3/helstrompure}, i.e.,
\begin{equation} \label{ch4/minerr}
\PME(\ket{\phi_1},\ket{\phi_2}) =  \frac{1}{2} \left( 1 - \sqrt{1-|\braket{\phi_1}{\phi_2}|^2} \right) \,.
\end{equation}
The total error probability is just the sum of the contribution of each pair of states  with the same quantum numbers $JM$, $\{\ket{j_{AB}; JM}, \ket{j_{BC}; JM}\}$,
\begin{equation}\label{ch4/min-nm}
\PME = \frac{1}{2} \left( 1-\sum_{k=0}^n \frac{n'+2k+1}{(n+1)(n+n'+1)} \sqrt{1-\left(\frac{(n'+k)!n!}{(n'+n)!k!}\right)^2} \right) \,.
\end{equation}
%

It is instructive to obtain the well-known results when the ports are loaded with just one copy of each state~\citep{Bergou2005} (i.e., $n=n'=1$).
The inconclusive probability in the unambiguous approach reads
\begin{equation}
\PUA = \frac{1}{6}\sum_{J=1/2}^{3/2} (2J+1) |\braket{j_{AB}=1 ;J }{j_{BC}=1; J }| = \frac{5}{6} \,;
\end{equation}
in average, five out of six times the machine gives an inconclusive result and only 1/6 of the times it identifies the state without error. Note that the overlaps for $J=3/2$ are one. This must be so since $J=3/2$ corresponds to the totally symmetric subspace, which is independent of the way the spins are coupled. That is, this subspace is identical for $\sigma_1$ and $\sigma_2$. This is the main contribution to $Q$ as it supplies $4/6=4/6\times 1$  out of the total $5/6$ probability of inconclusive results. The remaining $1/6=2/6\times 1/2$ is the  contribution of the $J=1/2$ subspace, where the $2/6$ is the probability of having an outcome on this subspace and $1/2$ is the overlap between the states [cf. Eq.~\eqref{ch4/overlap-nm}].

The minimum-error probability in the one copy case reads
\begin{equation} \label{ch4/minerr-tot}
\PME = \frac{1}{2} \left( 1 - \frac{1}{6} \sum_{J=1/2}^{3/2} (2J+1) \sqrt{1- |\braket{j_{AB}=1 ;J }{j_{BC}=1; J }|^2} \right) \,,
\end{equation}
which, by using either Eq.~\eqref{ch4/overlap-nm} or directly Eq.~\eqref{ch4/min-nm}, gives
\begin{equation} \label{ch4/minerr-tot-2}
\PME =  \frac{1}{2} \left( 1 - \frac{1}{2\sqrt{3}} \right)\simeq 0.356 \;.
\end{equation}
That is, approximately 1/3 of the times the outcome of the machine will be incorrect.

The error probability in both minimum-error and unambiguous approaches will, of course, decrease when using more copies of the states at the ports of the discrimination machine.
Equations \eqref{ch4/ua-nm} and \eqref{ch4/min-nm} give the unambiguous and minimum-error
probability for arbitrary values of $n$ and $n'$. They enable us to study  the behaviour of the machine for a large
number of copies in the program and the data ports, which is what we next discuss.
\\




\subsection{Asymptotic limits for pure states}\label{ch4/sec:pure-asymptotics}

Let us start by considering the case of an asymptotically large number of copies at the program ports ($n \to \infty $) while keeping finite the number of copies $n'$ at the data port.
For unambiguous discrimination, from Eq.~\eqref{ch4/ua-nm} one obtains
\begin{equation}\label{ch4/SUA-limit-n}
\lim_{n\to\infty} \PUA=\frac{ 2}{n'+2} \,.
\end{equation}
We wish to show that, in this limit, the programmable machine has a performance that
is equivalent to a protocol consisting in first estimating the states at the program ports and then performing a discrimination of \emph{known} states over the data port. The average of the  inconclusive  probability of such protocol over all input states should coincide with Eq.~\eqref{ch4/SUA-limit-n}.  Recall that, for known $\ket{\psi_1}$ and $\ket{\psi_2}$, when
a number $n'$ of copies of the unknown state is given, this probability reads
\begin{equation}\label{ch4/SUA-known}
\PUA (\psi_1,\psi_2)=\left|\braket{\psi_1}{\psi_2}\right |^{n'} \,.
\end{equation}
One can do an explicit calculation of the average 
\begin{equation}
\aver{\PUA (\psi_1,\psi_2)}=\frac{1}{2} \int_0^{\pi} \sin\theta \cos^{n'}\frac{\theta}{2} d\theta \,, 
\end{equation}
but it is interesting to obtain it in a very simple way from the Schur's lemma:
\begin{eqnarray}\label{ch4/SUA-known-average}
\int d\psi_2 \left(\left|\braket{\psi_1}{\psi_2}\right |^2\right)^{\frac{n'}{2}} &=&\bra{\psi_1}^{\otimes \frac{n'}{2}}\left(\int d\psi_2 [\psi_2]^{\otimes \frac{n'}{2}}\right) \ket{\psi_1}^{\otimes\frac{n'}{2}} \nonumber \\
	 &=& \frac{1}{d_{n'/2}}=\frac{1}{n'/2+1}\, ,
\end{eqnarray}
where $d_{n'/2}$ is the dimension of the symmetric space of $n'/2$ qubits (note that \emph{sensu stricto} this procedure is only
valid for $n'$ even). Plugging this average into Eq.~\eqref{ch4/SUA-known} one immediately recovers Eq.~\eqref{ch4/SUA-limit-n}.

Now we turn our attention to the minimum-error probability. 
The details of this computation are given in Appendix~\ref{appA/sec:minerr_n_to_infty}. 
In the limit $n\to\infty$, the leading term is found to be
\begin{eqnarray}\label{ch4/SME-sum}
\lim_{n\to\infty} \PME&=&\frac{ 1}{2}\left[ 1-2\int_0^1 dx \, x \sqrt{1-x^{2n'}}  \right] \nonumber \\
&=&\frac{1}{2}\left[  1-\frac{\sqrt{\pi}}{2}\frac{\Gamma(1+1/n')}{\Gamma(3/2+1/n')} \right]\, ,
\end{eqnarray}
where we have defined $x\equiv k/n$ and used the Euler-McLaurin summation formula at leading order to approximate the sum in Eq.~\eqref{ch4/min-nm}.
This result could be easily anticipated  from the minimum-error probability with classical knowledge of the  pure states.
Recall that  the minimum-error probability
given  $n'$ identical copies is
\begin{equation}
\PME(\psi_1,\psi_2)=\frac{1}{2} \left(1-\sqrt{1-|\braket{\psi_1}{\psi_2}|^{2n'}} \right) \,,
\end{equation}
so we just have to compute the average for all pairs of states of the above expression.  Using $|\braket{\psi_1}{\psi_2}|^2= \cos^2\theta/2$, where $\theta$ is the relative angle between the Bloch vectors of the two states, one has
\begin{equation}\label{ch4/PME-int}
\aver{\PME(\psi_1,\psi_2)}=\frac{1}{2}\left[ 1-\frac{1}{2} \int_0^\pi d\theta \sin\theta \sqrt{1-\cos^{2n'}(\theta/2)}\right] \,,
\end{equation}
and, performing the change of variables $x=\sin\theta/2$, this equation is cast exactly in the form of Eq.~\eqref{ch4/SME-sum}.

What cannot be anticipated is the next order $O(1/n)$, which  gives very relevant information on how fast the protocol reaches the asymptotic value~\eqref{ch4/SME-sum}. 
After some algebra we obtain that the coefficient of this subleading term coincides with the second term in Eq.~\eqref{ch4/SME-sum},
hence at this order we can write
\begin{equation}\label{ch4/SME-subleading}
\PME=\frac{1}{2}-\frac{\sqrt{\pi}}{4}\frac{\Gamma(1+1/n')}{\Gamma(3/2+1/n')} \left(1-\frac{1}{n}\right) \,.
\end{equation}

We now analyse the complementary case, that is,  when the number of copies at the data port is infinitely large ($n'\to \infty$) while the number $n$  of copies at the program ports is kept finite. In this limit we have perfect knowledge of the data state $\ket{\psi}$, but we do not know to which program port it should be associated. Observe that this situation is very much the same as state comparison (see Section~\ref{ch3/sec:othertasks}).

In this scenario, the inconclusive probability in the unambiguous approach reads  from Eq.~\eqref{ch4/ua-nm} as
\begin{equation}\label{ch4/SUA-limit-m}
\lim_{n'\to\infty} \PUA=\frac{1}{n+1} \,.
\end{equation}
Let us see that this agrees with the average performance of a standard state comparison protocol.  If the data state is the same as the program state in the  upper or lower port,
the effective states to be discriminated  are
\begin{eqnarray}\label{ch4/state-minfinity}
  \sigma_1&=&\frac{1}{d_n}[\psi^{\otimes n}]\otimes\openone_n \,,\nonumber \\
  \sigma_2&=&\frac{1}{d_n}\openone_n\otimes[\psi^{\otimes n}] \,,
\end{eqnarray}
respectively, where $d_n=n+1$ is the dimension of the symmetric space of $n$-qubits and $\openone_n$ is the projector onto this subspace. The minimal inconclusive probability for these two states can be obtained with a POVM with elements $E_1= [\psi^{\otimes n}]\otimes[\psi^{\otimes n}]^{\bot} ,E_2=[\psi^{\otimes n}]^{\bot}\otimes[\psi^{\otimes n}]$, both representing conclusive answers, and $E_0=\openone \otimes \openone -E_1 -E_2$, which represents the inconclusive one. In these expressions $[\psi^{\otimes n}]^{\bot}=\openone_n-[\psi^{\otimes n}]$. Note that this POVM checks whether the state in each register is  $\ket{\psi}$ or not. The probability of obtaining the inconclusive answer reads
\begin{equation}\label{ch4/eq:UA-averaged}
\PUA(\psi)=\frac{1}{2}\left(\tr E_0 \sigma_1+ \tr E_0 \sigma_2 \right)=\frac{1}{n+1}
\end{equation}
independently of the state $ \ket{\psi} $.

The minimum-error probability in this limit can be tackled in a similar fashion. The asymptotic expression of Eq.~\eqref{ch4/min-nm}, though not as direct as in the unambiguous case,  is rather straightforward to obtain. Note that the dominant factor in the  term containing factorials inside the square root is $n'\,^{-2(n-k)}$. Hence we can effectively replace the square root term by 1, for all $k<n$. Taking into account that for $k=n$ the square root vanishes, we have
\begin{equation}\label{ch4/SME-limit-m}
\lim_{n'\to\infty} \PME=\frac{ 1}{2}\left( 1- \frac{n}{n+1}\right)=\frac{1}{2(n+1)} \, .
\end{equation}

The minimum-error probability of a strategy that first estimates perfectly the data states and then tries to associate the correct label to them is given by the Helstrom formula~\eqref{ch3/helstrom} for $\sigma_1$ and $\sigma_2$, that is
\begin{equation}\label{ch4/Helstrom-m}
\PME  = \frac{1}{2} \left(  1-\frac{1}{2} \trnorm{\sigma_1-\sigma_2} \right) \,.
\end{equation}
%
Substituting the expression of the states \eqref{ch4/state-minfinity} we obtain
\begin{eqnarray}
\PME &=&\frac{1}{2}\left( 1-\frac{1}{2(n+1)} \trnorm{ [\psi^{\otimes n}] \otimes [\psi^{\otimes n}]^{\bot}  -  [\psi^{\otimes n}]^{\bot}\otimes[\psi^{\otimes n}] } \right) \nonumber \\
         &=&\frac{1}{2}\left(  1-\frac{2}{2(n+1)} \trnorm{ [\psi^{\otimes n}]\otimes [\psi^{\otimes n}]^{\bot}}\right)\nonumber \\
          &=&\frac{1}{2}\left(  1-\frac{n}{n+1} \right)=\frac{1}{2(n+1)} \,, \label{ch4/eq:ME-averaged}
\end{eqnarray}
where in the first equality we have subtracted the common term $[\psi^{\otimes n}]\otimes[\psi^{\otimes n}]$ from both states, in the second we have used the orthogonality of the operators and in the last equality we have taken into account that $\tr [\psi^{\otimes n}]^{\bot}=\tr (\openone_n-[\psi^{\otimes n}])= n$ (i.e., one unit less than the dimension of the corresponding symmetric space). As expected, the result is again independent of $\ket{\psi}$.
It is worth noting that this minimum-error probability is achieved by a strategy that uses the optimal POVM for unambiguous discrimination above, which returns an inconclusive outcome with probability $1/(n+1)$, and, whenever this outcome is obtained, guesses randomly for either $\sigma_1$ or $\sigma_2$. This means that, in the limit $n'\to\infty$, the only difference between the unambiguous and the minimum-error approaches is in the post-processing of the outcomes, not in the physical measurement operation.

To end this section we compute the asymptotic error probabilities for the symmetric case, that is, when all the ports are loaded with the same $n'=n$ (and large) number of copies.

In the unambiguous approach, when $n=n'\to \infty$ the first nonvanishing order of~\eqref{ch4/ua-nm} reads
\begin{equation}
\PUA=\frac{3}{n}+\ldots
\end{equation}

To compute the minimum-error probability, it is convenient to write Eq.~\eqref{ch4/min-nm} for $n=n'$ as
\begin{equation}
\PME=\frac{1}{2}\sum_{k=0}^n p_k \left(1- \sqrt{1-c_k^2}\right)\, ,
\end{equation}
where
\begin{equation}\label{ch4/pk}
p_k=\frac{n+1+2k}{(2n+1)(n+1)}\,,
\end{equation}
and
\begin{equation}
c_k = \binom{n+k}{n} {\binom{2n}{n}}^{-1} \, .
\end{equation}
We first observe that $c_k$ is a monotonically increasing function and hence it takes its maximum value  at $k=n$. Second, we note that around this point
\begin{eqnarray}
\binom{n+k}{n}&\simeq& 2^{(n+k) H(\frac{n}{n+k})}\nonumber \\
&\simeq& 2^{(n+k) H(1/2)}=2^{n+k}\, ,
\end{eqnarray}
where $H(x)=-x\ln x-(1-x)\ln(1-x)$ is the Shannon entropy of a binary random variable, and we have used that $k\approx n$ and $H(1/2)=1$.
Similarly, one has
\begin{equation}
\binom{2n}{n} \simeq 2^{2n H(1/2)}=2^{2n} \,,
\end{equation}
and hence $ c_k\simeq 2^{-(n-k)} $. With this, the probability of error in this limit reads
\begin{equation}
\PME=\frac{1}{2}\sum_{k=0}^\infty p_k \left(1- \sqrt{1-\left(\frac{1}{4}\right)^{n-k}}\right)\,.
\end{equation}
Finally, we perform the change of variables $k\to n-k$ and use that in Eq.~\eqref{ch4/pk} $p_{n-k}\simeq 3/(2 n)$ for $k\simeq 0$ to obtain
\begin{equation}\label{ch4/pure-assymptotic}
\PME=\frac{3}{4n}\zeta(1/4)\approx \frac{0.882}{n}\, ,
\end{equation}
where we have defined the function
\begin{equation}
\zeta(x)=\sum_{k=0}^\infty\left(1-\sqrt{1-x^k}\right)\, ,
\end{equation}
which converges very quickly to its exact value (the first  four terms already give a value that differ in less than $10^{-3}$ from the exact value).
\\


\section{Mixed states}\label{ch4/sec:mixed}

We now move to the case when the program and data ports are loaded with mixed states. This situation arises for instance when there are imperfections in the preparation or noise in the transmission of the states.  It is reasonable to suppose that these  imperfections have the same effect on all states (i.e. to consider that the states have all the same purity $r$). The input states are then tensor products of
\begin{equation}\label{ch4/mixed-qubit}
\rho_i=\frac{\openone + r \,\vec{n}_i \,\vec{\sigma}}{2} \,,
\end{equation}
where $\vec{n}_i$ is a unitary vector and $\vec{\sigma}=(\sigma_x,\sigma_y,\sigma_z)$ are the usual Pauli matrices.  In what follows we assume that only the purity is known, i.e., one knows the characteristics of  the noise affecting the states, but nothing else. This  means that the averages will be performed over the isotropic Haar measure of the $\mathbb{S}^2$ sphere, in the same manner as for pure states. At the end of this section we also analyse the performance of a fully universal discrimination machine, that is, when not even the purity is considered to be known.

Note that mixed states can only be unambiguously discriminated if they have different supports (see Section~\ref{ch3/sec:unambiguous}), 
which is not the case when the ports are loaded with copies of the states \eqref{ch4/mixed-qubit}  as they are  full-rank matrices.
Therefore, only the minimum-error discrimination approach will be analysed here. It is worth stressing that the computation of the optimal error probability in the multicopy case is highly nontrivial, even for known qubit mixed states. Only recently have feasible methods for computing the minimum-error probability for a rather large number of copies been developed~\citep{Calsamiglia2010}, and the asymptotic expression of such probability obtained\footnote{This is achieved via attainability of the quantum Chernoff bound. See Section~\ref{ch3/sec:quantumchernoff} for details.}. The main difficulty  can be traced back to the computation of the trace norm [see Eq.\eqref{ch4/Helstrom-m}] of large matrices. The dimension of the matrices grows exponentially with the total number of copies entering the machine, and for a relative small number of them the problem becomes unmanageable. However, as it will be clear, it is possible to exploit the permutation symmetry of the input states to write them in the block-diagonal form given in Eq.~\eqref{ch3/rhoblock1}, crucially reducing the complexity of the problem.

The two effective states we have to discriminate are
\begin{eqnarray}\label{ch4/s1-s2-int}
\sigma_1 &= & \int dn_1 dn_2 \rho^{\otimes n}_{1\,  A} \otimes \rho_{1\,  B}^{\otimes n'} \otimes \rho^{\otimes n}_{2\,  C}  \,,\nonumber \\
\sigma_2 &=&  \int dn_1 dn_2 \rho^{\otimes n}_{1\,  A} \otimes \rho_{2\,  B}^{\otimes n'} \otimes \rho^{\otimes n}_{2\,  C} \,,
\end{eqnarray}
where $dn_i= d\Omega_i/(4\pi)$ is the invariant measure on the 2-sphere. Recall that, as discussed in Section~\ref{ch3/sec:blockdecomposition}, any state having permutation invariance (e.g.,  $ \rho^{\otimes n} $)  can be written in a
block-diagonal form using the irreducible representations of the symmetric group $S_n$. Each block is specified by the total angular momentum $j$ and a label $\alpha$  that distinguishes the different equivalent representations for a given $j$
\begin{equation}
\rho^{\otimes n}=\bigoplus_{j,\alpha} \rho_j^{(\alpha)} \,.
\end{equation}
The angular momentum takes values \mbox{$j=n/2, n/2-1, \ldots, 1/2\,(0)$}  for odd (even) $n$, and the number of equivalent representations for each $j$
is [cf. Eq.~\eqref{ch3/rhoblock3}]
\begin{equation}\label{ch4/repeated}
\nu_j^n={n \choose n/2-j} - {n \choose n/2-j-1}\,,
\end{equation}
that is $ \alpha=1,\ldots,\nu_j^n $. For each block we have
%
\begin{equation}\label{ch4/deconst}
\tr \rho_j^{(\alpha)} = \left(\frac{1-r^2}{4}\right)^{n/2-j} \sum_{k=-j}^j \left(\frac{1-r}{2}\right)^{j-k} \left(\frac{1+r}{2}\right)^{j+k} \equiv (2j+1) C^n_j \,,
\end{equation}
which, of course, is the same for all equivalent irreducible representations (i.e., independent on the label $\alpha$).
The origin of the factors appearing in Eq.~\eqref{ch4/deconst} was outlined in deducing Eq.~\eqref{ch3/rhoblock_diagform}, but let us briefly remember it here\footnote{Also, full details can be found in~\citep{Bagan2006}.}.
The first factor comes from the contribution from the $n/2-j$ singlets present in a representation $j$ made up of $n$ spin-1/2 states. The summation term is the trace of the projection of the remaining states in the  symmetric subspace with total angular momentum $j$, where we can use the rotational invariance of the trace to write each state in the form ${\rm diag}\left(\frac{1+r}{2} , \frac{1-r}{2} \right)$. This
term simply reads
\begin{equation}
t_j = \sum_{k=-j}^j \left(\frac{1-r}{2}\right)^{j-k} \left(\frac{1+r}{2}\right)^{j+k} = \frac{1}{r}\left[ \left( \frac{1+r}{2} \right)^{2j+1} - \left( \frac{1-r}{2} \right)^{2j+1} \right]\,,
\end{equation}
and hence
\begin{equation}\label{ch4/deconstt}
C_j^n=\frac{1}{2j+1}\left(\frac{1-r^2}{4}\right)^{n/2-j} t_j \; .
\end{equation}

Very much in the same way as it happened  in previous sections, the only difference between the diagonal basis of $\sigma_1$ and $\sigma_2$ is the ordering of the  angular momenta couplings. In $ \sigma_1 $ we first couple subspaces $A$ and $B$ and obtain
\begin{equation}\label{ch4/rhoAB}
\rho_{AB} = \int dn_1  \rho^{\otimes n}_{1\,  A} \otimes \rho_{1\,  B}^{\otimes n'}
= \sum_{\xi_{AB}} C_{j_{AB}}^{n+n'} \openone_{\xi_{AB}}\; ,
\end{equation}
where
\begin{equation}
\openone_{\xi_{AB}}=\sum_{M_{AB}}\ketbrad{\xi_{AB} M_{AB}}
\end{equation}
is the projector onto the subspace with associated quantum numbers $\xi_{AB} = \{j_A,\alpha_A,j_B,\alpha_B,j_{AB}\}$, and $C_{j_{AB}}^{n+n'} $ is defined in
Eq.~\eqref{ch4/deconst}.
Note that $C_{j_{AB}}^{n+n'}$  depends only on the purity of the state and on the total angular momentum $j_{AB}$.  Note also that the tensor product of a mixed
state has projections in all subspaces and the blocks are not uniquely determined by the value of  $j_{AB}$, i.e.,  one has to keep track of the labels $j_A$ and $j_B$ as
well. Of course, subspaces with different quantum numbers
$ \xi_{AB} $ are orthogonal, i.e.,  $\tr [\openone_{\xi}\openone_{\xi'}]=\delta_{\xi \xi'}\tr \openone_{\xi} $. When coupling the third
system one plainly adds  the quantum numbers
$\xi_C=\{j_C, \alpha_C\}$.

The diagonal bases of $\sigma_1$ and $\sigma_2$  are written as $\mathcal{B}_1=\{\ket{\xi_{AB}\xi_C; JM}\}$ and
$\mathcal{B}_2=\{\ket{\xi_A\xi_{BC}; JM}\}$, respectively. Obviously, each set contains  $2^{2n+n'}$  orthonormal states and Eq.~\eqref{ch4/s1-s2-int} reads
\begin{eqnarray}\label{ch4/s1-s2-int-2}
\sigma_1 &=&  \sum_{\xi_{AB}\xi_C}\sum_{JM} C_{j_{AB}}^{n+n'} C_{j_C}^n [\xi_{AB} \xi_C; J  M] \,,\nonumber\\
\sigma_2 &=&  \sum_{\xi_A\xi_{BC}}\sum_{JM} C_{j_A}^n C_{j_{BC}}^{n+n'} [\xi_A  \xi_{BC} ; J  M] \,.
\end{eqnarray}
We just have to compute the minimum-error probability from the Helstrom formula \eqref{ch4/Helstrom-m} for these two states. It is convenient to define the trace norm term
\begin{equation}\label{ch4/tracenorm-mixed}
T=\trnorm{\sigma_1-\sigma_2} \,,
\end{equation}
so that
\begin{equation}
\PME=\frac{1}{2}\left( 1-\frac{T}{2} \right) \,.
\end{equation}
To compute $T$ we need to know the unitary matrix $\Lambda$  that transforms $\mathcal{B}_2$ into $\mathcal{B}_1$ or vice versa.
The elements of this unitary are given by the overlaps between the elements of both bases\footnote{Note that these are just the elements of the permutation operation that interchanges subsystems $A$ and $C$.}
$\braket{\xi_{AB}\xi_{C}; JM}{\xi'_{A}\xi'_{BC};J'M'}$. We observe that these overlaps are nonvanishing only if  $j_X=j'_X $ ,  $\alpha_X=\alpha'_X $ ($X=A,B,C$) and
$J=J', M=M'$. Furthermore, as mentioned previously,  their value does not depend on $M$ or $\alpha_X$, thus sums over these quantum numbers simply amount to
introduce the corresponding multiplicative factors. Therefore, it is useful to introduce a label containing the quantum numbers  that determine the orthogonal
blocks in $\mathcal{B}_1$ and $\mathcal{B}_2$  that may have nonvanishing overlaps, $\xi=\{j_A, j_B,j_C, J\}$, and the corresponding multiplicative factor
\begin{equation}\label{ch4/multiplicative-factor}
\gamma_{\xi}=\nu_{j_A}^{n}\nu_{j_B}^{n'}\nu_{j_C}^{n}(2J+1) \,,
\end{equation}
where  $\nu^n_j$ is given in Eq.~\eqref{ch4/repeated}. Eq.~\eqref{ch4/tracenorm-mixed} then reads
\begin{equation}\label{ch4/T}
T = \sum_\xi \gamma_\xi T^{\xi}= \sum_\xi \gamma_\xi \trnorm{\sigma^{(\xi)}_1-
\Lambda^{(\xi)} \sigma^{(\xi)}_2 {\Lambda^{(\xi)}}^{T} } \,,
\end{equation}
where the explicit expressions of the matrix elements are
\begin{eqnarray}\label{ch4/sigma-xi}
\left[\sigma_1^{(\xi)}\right]_{j_{AB}j'_{AB}}&=&\delta_{j_{AB}j'_{AB}} C_{j_{AB}}^{n+n'} C_{j_C}^n \,,\nonumber \\
\left[\sigma_2^{(\xi)}\right]_{j_{BC}j'_{BC}}&=&\delta_{j_{BC}j'_{BC}} C_{j_{A}}^{n} C_{j_{BC}}^{n+n'} \,,
\end{eqnarray}
and
\begin{equation}\label{ch4/lambda}
\Lambda^{(\xi)}_{j_{AB},j_{BC}}=\braket{\xi, j_{AB}}{\xi, j_{BC}} \,.
\end{equation}
Recall that the overlap \eqref{ch4/lambda} is independent of the quantum number labelling the equivalent representations (recall also that it is independent of $M$), and therefore is given by Eq.~\eqref{ch4/6j}\footnote{Note that Eq.~\eqref{ch4/lambda} is a generalization of Eq.~\eqref{ch4/orthogonality}, i.e., when the dimension of the Jordan subspaces is greater than one.}.

The computation of the minimum-error probability  reduces  to a sum of trace norms of small-size Helstrom matrices that have dimensions of the allowed values of  $j_{AB}$ and $j_{BC}$ for given $\xi=\{j_A,j_B, j_C, J\}$. Hence
\begin{equation}\label{ch4/pme-general}
\PME = \frac{1}{2} \left( 1 - \frac{1}{2}\sum_{\xi} \gamma_\xi T^{\xi}\right) \,,
\end{equation}
and this computation can be done very efficiently.

We would like to show the analytical results for  the simplest case of having just one state at each port, i.e., when $n=n'=1$. In this situation we have fixed values $ j_A=j_B=j_C=1/2 $, so the total angular momentum can be $ J=3/2,1/2 $, and $ j_{AB}=1,0 $ (and similarly for $ j_{BC} $). Here there is no degeneracy,
the number of equivalent representations defined in Eq.~\eqref{ch4/repeated}  is 1,
and, therefore, the multiplicative factor \eqref{ch4/multiplicative-factor}  simply reads $\gamma_\xi=2J+1$.
The only relevant quantum number in this case is $\xi=J$,  as all the others are fixed, and we do not need to write them explicitly.  The minimum-error probability is then
\begin{equation}\label{ch4/mixedn1m1}
\PME=\frac{1}{2} \left[ 1-\frac{1}{2}\sum_{J=1/2}^{3/2}(2J+1)\trnorm{\sigma_1^{(J)}-\Lambda^{(J)}\sigma_2^{(J)}{\Lambda^{(J)}}^T} \right] \,.
\end{equation}
The term of the sum corresponding to $ J=3/2 $ vanishes since it corresponds to the projection of $ \sigma_{1,2} $ onto the completely symmetric
subspace, which is identical for both states.
Indeed, in  this subspace $\sigma_1^{(3/2)}=\sigma_2^{(3/2)}=C^2_1 C^1_{1/2}=(3+r^2)/24$, where we have used Eq.~\eqref{ch4/deconstt},
and from Eq.~\eqref{ch4/lambda} we obtain $\Lambda^{(3/2)}=1$. In the subspace $J=1/2$ we have
\begin{equation}
		 \sigma_1^{(1/2)} =  \sigma_2^{(1/2)}  =
		\begin{pmatrix}
			C^2_1 C^1_{1/2} & 0 \\
			0 & C^2_0 C^1_{1/2}
		\end{pmatrix}
		= 
		\begin{pmatrix}
			\frac{1}{24}\left(3+r^2\right) & 0 \\
			0 & \frac{1}{8}\left(1-r^2\right)
		\end{pmatrix} \,,
\end{equation}
and
\begin{equation}\label{ch4/ss3}
	\Lambda^{(1/2)} =
		\begin{pmatrix}
			\frac{1}{2} & \frac{\sqrt{3}}{2} \\
			\frac{\sqrt{3}}{2} & -\frac{1}{2}
		\end{pmatrix} \, .
\end{equation}
Plugging these expressions into Eq.~\eqref{ch4/mixedn1m1} we obtain the minimum-error probability for the one-copy state case
\begin{equation}\label{ch4/1x1x1}
\PME=\frac{1}{2}\left(1-\frac{r^2}{2\sqrt{3}}\right) \,.
\end{equation}
As expected, when $r\to 1$ we recover the pure state value~\eqref{ch4/minerr-tot-2}.
\begin{figure}[t]
\begin{center}
\includegraphics[scale=1.5]{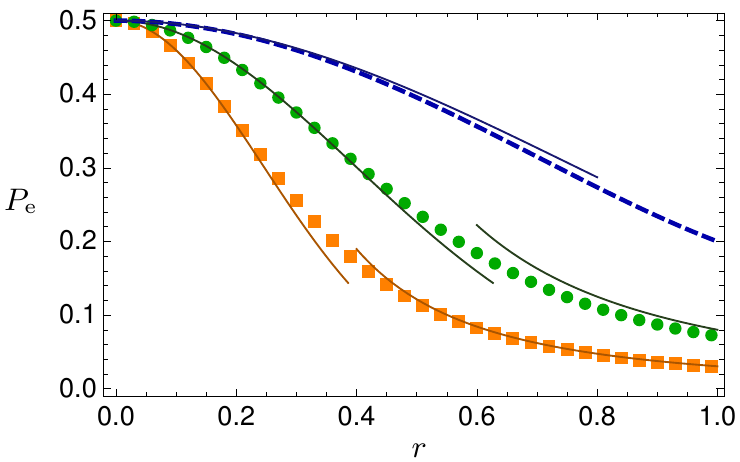}
\end{center}
\caption[Error probability of a programmable device with mixed qubit states, as a function of the purity]{Error probability $ \PME $ for $ n=n'=3 $ (blue dashed line), $11$ (green circles) and $ 29 $ (yellow squares) versus purity. The fit $ \PME \simeq 0.882/(n r^2) $ in the regime of high purities for $ n=11 $ and $ n=29 $ and the Gaussian approximation $\PME\simeq 1/2 \exp[-n r^2/(2\sqrt{3})]$ in the regime of low purities for all cases is represented (solid lines).\label{ch4/fig:fig1}}
\end{figure}
\begin{figure}[t]
\begin{center}
\includegraphics[scale=1.5]{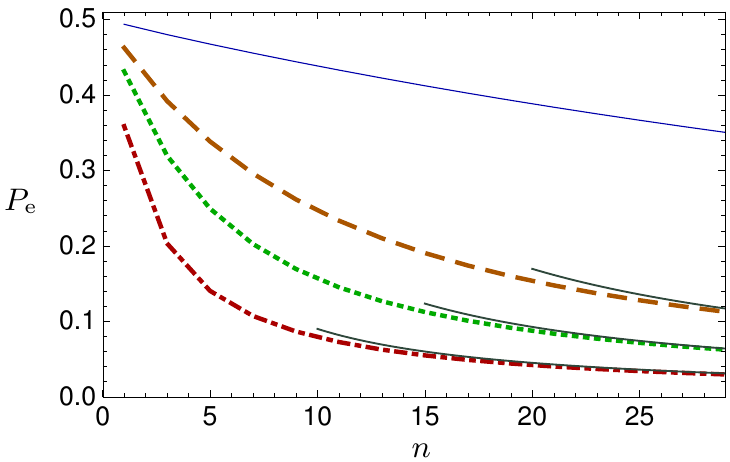}
\end{center}
\caption[Error probability of a programmable device with mixed qubit states, as a function of the number $n$ of copies]{Error probability $ \PME $ for $ r=0.2 $ (blue thin solid line), $ r=0.5 $ (brown dashed line), $ r=0.7 $ (green dotted line), and $ r=1 $ (red dot-dashed line) versus $ n $ ($ n=n' $ is assumed). Numerical points have been joined for an easier visualization.  The approximation $0.882/(nr^2)$  is represented (thin solid lines).\label{ch4/fig:fig2}}
\end{figure}

Numerical results  of the minimum-error probability as a function of the purity of the input states for the symmetric case $n=n'$ are depicted in  Fig.~\ref{ch4/fig:fig1}. One sees that, for low values of $n$ ($n\lesssim 3$), the dependence on the purity is not very marked: the curves are concave almost in the whole range of the purity. For larger $n$, however, there is an interval of purities where the behaviour changes quite significantly.  For instance, for $n=29$, the inflection point occurs at $r\approx 0.3$.  At very large values of $n$ one expects a step-like shape with an inflection point approaching $r=0$ because the probability of error remains very small for $r\neq0$ and is strictly 1/2 at $r=0$.  The shape of the curves is explained by the existence of two distinct  regimes. For high purities the probability of error is well fitted by a linear function in the inverse of the number of copies. We get  $\PME\simeq 0.88/(nr^2)$, where the value $0.88$ coincides with the analytical value computed for pure states in Eq.~\eqref{ch4/pure-assymptotic}.  Of course,  this  approximation cannot be valid for low purities.  In the low-purity regime, the minimum-error probability is very well approximated by the Gaussian function $\PME\simeq 1/2 \exp[-n r^2/(2\sqrt{3})]$, where we have taken the argument of the exponential  from the exponentiation of the error probability for the exact $1\times 1\times 1$ case, given in Eq.~\eqref{ch4/1x1x1}.  This approximation works for purities in the interval of  the width of the Gaussian, i.e., up to  $\sim1/\sqrt{n}$.  Therefore, as $n$ increases  the asymptotic approximation  $\PME\propto 1/(nr^2)$ extends its validity to almost the whole range of purities, and the expected jump discontinuity develops in $r=0$ as $n\to \infty$.
Similar information is depicted in Fig.~\ref{ch4/fig:fig2}, where the error probability is plotted as function of the number of copies $n$ for different purities. We have superimposed the asymptotic result, which  is seen to yield a very good approximation to the exact error probability already for $n\gtrsim 20$.
\\


\subsection{\boldmath Asymptotic $n\times 1\times n$}\label{ch4/sec:mixed-limits}

As in previous sections, it is interesting to study the performance of the machine in the asymptotic regimes. A particularly important instance where it is possible
to obtain closed expressions is the case when the number of copies at the program ports is asymptotically large and there is one state at the data port.
This regime will also be of interest for Chapter~\ref{ch5_learning}.
We show how to  compute the leading order and  sketch the generalizations needed to obtain the subleading term.

Observe first that $j_{AB}$  can only take the values $j_{AB}=j_A\pm1/2$, and similarly for $j_{BC}$. Therefore $\sigma_{1,2}^{(\xi)}$ are $2\times2$ matrices (except in the extremal case of $J=j_A+j_C+1/2$, in which are one-dimensional).  It is useful to write
\begin{equation}\label{ch4/sigma-j}
\sigma(j)=C^{n}_{j_{A}}C^{n}_{j_{C}} 
\begin{pmatrix}  
R_{+}(j) & 0 \\
0  & R_{-}(j)
\end{pmatrix} \,,
\end{equation}
with
\begin{equation}
R_{\pm}(j)=\frac{C_{j\pm1/2}^{n+1}}{C_j^n} \,.
\end{equation}
With this definition one simply has [see Eq.\eqref{ch4/sigma-xi}]
\begin{equation}
\sigma_1^{(\xi)}=\sigma(j_A) \ \ \ \ \ \mathrm{and} \ \ \ \ \ \sigma_2^{(\xi)}=\sigma(j_C).
\end{equation}
We further note that for large $n$
\begin{equation}\label{ch4/gauss-approx}
\nu_j^n C_j^n \approx \frac{1}{n/2+j+1} \frac{1+r}{2r}  \sqrt{\frac{2}{n \pi (1-r^2)}}
\exp \left[-n\frac{(2j/n-r)^2}{2(1-r^2)}\right] \,.
\end{equation}
Defining $y=2j/n$  and using the  Euler-Maclaurin summation
formula~\eqref{appA/euler_maclaurin},
we have for a generic function $f(j)$
\begin{equation}\label{ch4/gauss-approx-2}
\sum_j \nu_j^n C_j^n f(j)   \approx \frac{1+r}{2r} \int_{-\infty}^{\infty}\frac{dy  \, G_n(y) }{n/2+n y/2+1}
f\left(\frac{ny}{2}\right) \,,
\end{equation}
where we have extended  limits of integration from (0,1)  to $(-\infty,\infty)$, which is legitimate for large $n$, and defined
\begin{equation}
G_n (y)=\sqrt{\frac{n}{2 \pi (1-r^2)}}\exp \left[-n\frac{(y-r)^2}{2(1-r^2)}\right] \,,
\end{equation}
i.e., a Gaussian distribution centred at $y=r$  with variance $\sigma^2=(1-r^2)/n$.
Note that, at leading order in the limit $n\to \infty$,  $G_{\infty}\approx \delta(y-r)$, and hence
\begin{equation}
\sum_j \nu_j^n C_j^n f(j)   \approx \frac{1}{n r}  f\left(\frac{nr}{2}\right) \,.
\end{equation}
Note also that at this order
\begin{equation}\label{ch4/rpm-leading}
R_{\pm}(j)\approx R_{\pm}\left(\frac{nr}{2}\right)=\frac{1\pm r}{2} \,.
\end{equation}
There only remains  to compute the unitary matrix Eq.~\eqref{ch4/lambda}. Observe that the  total angular momentum takes values $J=|j_A -j_C|+1/2+k$, with $k=0,1,\ldots,2 \min\{j_A,j_C\}$.  The leading order is rather easy to write (the subleading term, although  straightforward, is far more involved and we will not show it here).  At this order we have $J=1/2+k$ and  $k=0,1,\ldots, n r$, and the matrix elements computed from Eq.~\eqref{ch4/6j}  yield
\begin{equation}\label{ch4/lambda-leading}
\Lambda^{(\xi)}=\frac{1}{nr}
\begin{pmatrix} 
k & \sqrt{(nr)^2-k^2} \\
\sqrt{(nr)^2-k^2} & -k
\end{pmatrix} \,.
\end{equation}
Plugging  Eqs.~(\ref{ch4/sigma-j}-\ref{ch4/lambda-leading}) into Eq.~\eqref{ch4/T} one gets
\begin{equation}
T\simeq \sum_{k=0}^{nr} 2k \frac{2}{n^3 r^2} \sqrt{(nr)^2-k^2} \,,
\end{equation}
where the sum over $j_A$ and $j_C$ has been trivially performed by substituting  their central value $n r/2$ in the summand, and the only remaining multiplicative of $\gamma_\xi$ [cf. Eq.~\eqref{ch4/multiplicative-factor}]  is $2J+1\simeq 2k$.  Finally, defining $x\equiv k/nr$ and using the Euler-Maclaurin approximation \eqref{appA/euler_maclaurin} 
we obtain
\begin{equation}
T\simeq 4 r\int_0^1 dx \, x\sqrt{1-x^2}=\frac{4r}{3} \,,
\end{equation}
and hence
\begin{equation}\label{ch4/mixed-leading}
\PME\simeq \frac{1}{2}-\frac{r}{3} \,,
\end{equation}
which obviously coincides with the pure state result Eq.~\eqref{ch4/SME-sum} for $n'=1$ and $r\to 1$.

As for the computation of the next-to-leading order, the integrals approximating the sums over $j_A$ and $j_C$  have to incorporate the fluctuations around the central value, that is, one defines  $j_A=\frac{n}{2}(r+\eta_A)$ and $j_C=\frac{n}{2}(r+\eta_C)$, where the variables $\eta_X$ have effective dimension $n^{-1/2}$.
Then one can expand the matrix elements of $\sigma_{1,2}$, $\Lambda$, and the terms of $\nu_j^{n}$ present in Eq.~\eqref{ch4/gauss-approx-2}, taking into account the effective dimensionality of all the terms [notice that $k\to n(r+\eta) x$, where the integration range of  $x$ is $(0,1)$]. One then performs the sum in $k$ by means of the Euler-Maclaurin summation formula as before. Finally one computes the  integration in $j_{A/B}$ taking into account that the range of the variables $\eta_{A/B}$ can be taken to be $(-\infty,\infty)$.
After a somewhat lengthy calculation we obtain
\begin{equation}\label{ch4/mixed-subleading}
\PME\simeq \frac{1}{2}-\frac{r}{3}+\frac{1}{3n r} \,.
\end{equation}
Note that the limit $r=0$ is singular and not surprisingly the  expansion breaks down for purities of order $1/n$. As it should, the error probability \eqref{ch4/mixed-subleading}
increases monotonically with the purity.
\begin{figure}[t]
\begin{center}
\includegraphics[scale=1.5]{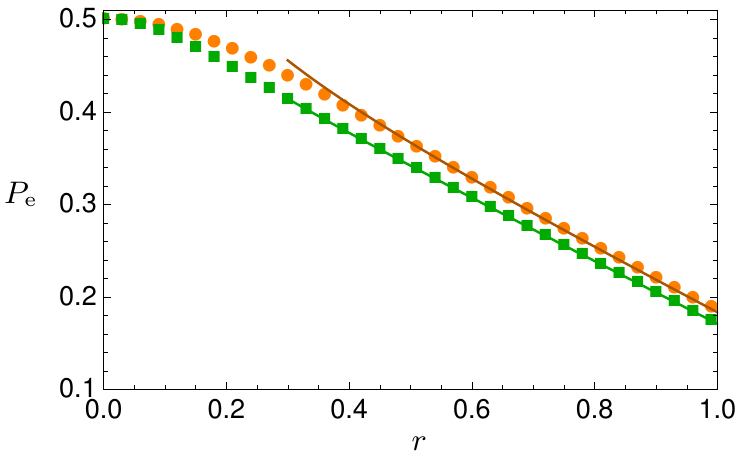}
\end{center}
\caption[Asymptotic behaviour of the error probability versus purity]{Error probability $ \PME $ for $ n=20 $ (yellow circles) and $ n=79 $ (green squares) versus purity. The asymptotic behaviour given by Eq.~\eqref{ch4/mixed-subleading} is represented for both cases.\label{ch4/fig:fig3}}
\end{figure}

In Fig.~\ref{ch4/fig:fig3} we plot the error probability as a function of the purity for $n=20$ and $n=79$. One sees that the asymptotic expression \eqref{ch4/mixed-subleading} approximates very well the minimum-error probability even for a small number of copies. For larger $n$ (e.g., for $n=79$) the approximation works extremely well down to values below $r=0.3$.

We finish this section by showing that the leading term \eqref{ch4/mixed-leading}  coincides with the average error of a device that first estimates the mixed states at the program ports and afterwards does the usual minimum-error discrimination of the data state. From the Helstrom formula~\eqref{ch4/Helstrom-m} particularized for mixed qubit states one has
\begin{equation}
\PME = \Bigr\langle \frac{1}{2}\left( 1-\frac{1}{2}|\vec{r}_1-\vec{r}_2| \right) \Bigr\rangle \,,
\end{equation}
where the average is taken over all possible orientations of the Bloch vectors $\vec{r_1}$ and $\vec{r}_2$. For equal purity states it simply reads
\begin{equation}
\PME =\frac{1}{2}\left(1-\frac{r}{2}\int_0^\pi d\theta  \sin\theta \sin\theta/2\right) =\frac{1}{2}-\frac{r}{3} \,.
\end{equation}
\\


\section{Universal discrimination}\label{ch4/sec:universal}

Let us finally address the  fully universal discrimination machine, that is a machine that distinguishes states from which nothing is assumed to be known, not even its purity.  For this type of machine, we  need to specify
a prior distribution for the purity.
While the isotropy of the angular variables yields a unique uniform distribution for the angular variables, the Haar measure on the 2-sphere used in previous sections, the corresponding expression for a fully unbiased distribution of the purity $w(r)$ is not uniquely determined.  This is a longstanding issue, and several priors haven been suggested depending on the assumptions made~\citep{Petz1996,Bengtsson2006}. Here we will  not stick to a particular distribution, rather we will show results for three reasonable distributions. The actual values of the probability of error may depend on the chosen prior, but the overall performance is seen to be very similar.

\begin{figure}[t]
\begin{center}
\includegraphics[scale=1.5]{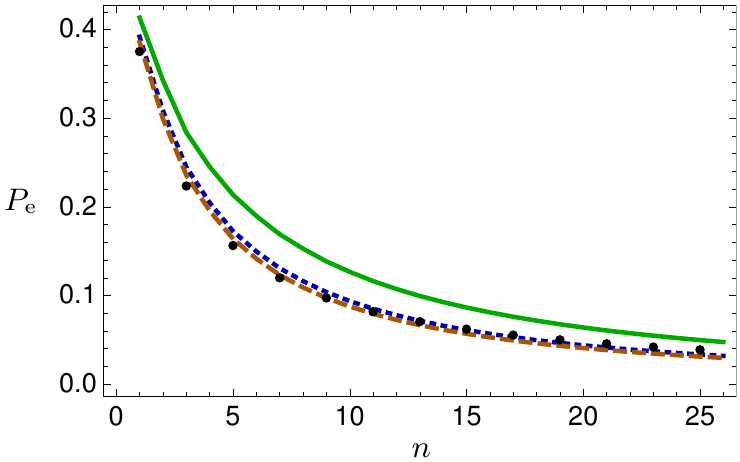}
\end{center}
\caption[Error probability of a fully universal discrimination machine]{Error probability $ \PME $ for hard-sphere (green solid line), Bures (blue dotted line) and Chernoff (red dashed line) priors versus $ n $ ($ n=n' $ is assumed). The points correspond to the error probability for a fixed $ r=0.9 $; its proximity to the Chernoff curve exposes the fact that this prior gives larger weights to states of high purity.\label{ch4/fig:fig4}}
\end{figure}

The most straightforward, but perhaps not very well grounded, choice is that of the distribution of a hard-sphere $w(r)\propto r^2$, that is, a normalized integration measure given by
\begin{equation}\label{ch4/hard}
d\rho^\mathrm{HS} = 3 r^2 dr \frac{d\Omega}{4\pi} \,.
\end{equation}

The Bures distribution is far better motivated. It corresponds to the volume element induced by the fidelity distance~\citep{Zyczkowski2005}. It is monotonically decreasing under coarse graining~\citep{Petz1996} and it has been argued that it corresponds to maximal randomness of the signal states~\citep{Hall1998}.  In this case one has $w(r)\propto r^2/\sqrt{1-r^2}$. Note that this distribution assigns larger weights to pure states, as their distinguishability  in terms of the fidelity is larger than that of mixed states.
The integration measure reads
\begin{equation}\label{ch4/bures}
d\rho^{\mathrm{Bu} }= \frac{4}{\pi}\frac{r^2}{\sqrt{1-r^2}} dr \frac{d\Omega}{4\pi} \,.
\end{equation}

Lastly, we also consider the 
Chernoff distribution~\citep{Audenaert2007,Calsamiglia2008}. It is the prior induced by the Chernoff distance, which has a clear operational meaning in terms of the distinguishability between states (see Section~\ref{ch3/sec:quantumchernoff}). By construction it is  monotonically decreasing under coarse graining. This measure assigns even larger weights to states of high purity and lower to the very mixed ones. This assignment is, again, based on distinguishability properties, but in terms of the asymptotic behaviour of the error probability.  The measure can be written as~\citep{Audenaert2007,Calsamiglia2008}
\begin{equation}\label{ch4/chernoff}
d\rho^{\mathrm{Ch} }= \frac{1}{\pi-2}\frac{\left(\sqrt{1+r}-\sqrt{1-r}\right)^2}{\sqrt{1-r^2}} dr \frac{d\Omega}{4\pi} \,.
\end{equation}

The effective states we have to discriminate are
\begin{equation}\label{ch4/s-universal}
\Sigma_i = \int d\rho_1 d\rho_2 \rho^{\otimes n}_{1\,  A} \otimes \rho_{i\,  B}^{\otimes n'} \otimes \rho^{\otimes n}_{2\,  C} \,, \qquad i=1,2 \,,
\end{equation}
where $d\rho_i$ takes the expressions of the measures \eqref{ch4/hard} through \eqref{ch4/chernoff}. Note that the block structure of the states is preserved, as it only depends on the permutation invariance of the input states,  which remains untouched. Further, we can use rotational invariance in the same fashion as in Eqs.~\eqref{ch4/rhoAB} and~\eqref{ch4/s1-s2-int-2}.  Therefore,  here it is only required  to compute the average of the coefficients $C_j^{n}$ in Eq.~\eqref{ch4/deconst} according to priors \eqref{ch4/hard} through \eqref{ch4/chernoff}. To calculate the minimum-error probability of this fully universal machine one simply uses Eq.~\eqref{ch4/pme-general}  for the states~\eqref{ch4/s1-s2-int-2} with the averaged coefficients $\aver{C_j^n}$ computed in
Appendix~\ref{app/ch4/averages}.

In Fig.~\ref{ch4/fig:fig4} we present the minimum-error probability of the fully universal machine for the three priors discussed for an equal number of program and data states up to $n=n'=26$.
As anticipated, the smaller average error corresponds to the  Chernoff distance, because states with higher purity are assigned a larger weight, and  these are easier to discriminate.  The probability of error, as somehow expected, is inversely proportional to the number of copies, and attains very similar values than for the discrimination of states with fixed known purity of the order of $r\sim 0.9$.
\\


\section{Programmable discrimination with an error margin}\label{ch4/sec:margins}

In this Section we analyse the paradigm of quantum state discrimination with an error margin, presented in Section~\ref{ch3/sec:errormargin} for known states, in the context of programmable discrimination machines for pure qubit states, when $n$ copies of the program states and $n'$ copies of the data state are provided. By doing so we connect the results for unambiguous and minimum-error discrimination derived in Section~\ref{ch4/sec:pure}. We will show that, by relaxing the zero-error condition slightly, the resulting scheme provides an important enhancement in performance over the widely used unambiguous scheme for programmable machines. We discuss the two ways of imposing an error margin to the error probability, i.e., via a \emph{weak} condition and a \emph{strong} condition.

Although so far not much attention has been paid to the POVM that represents the machine, for this Section it is convenient to explicitly refer to it.
A programmable discriminator is generically defined by a POVM with three elements $\mathcal{E}=\{E_1,E_2,E_0\}$. Recall that, as a consequence of the orthogonality relation of the Jordan bases~\eqref{ch4/orthogonality}, the averaged global states $\sigma_1$ and $\sigma_2$ have a block-diagonal structure in the angular momentum basis, each block corresponding to a Jordan subspace with an associated total angular momentum $J$. Hence the total Hilbert space of the states is of the form $\mathcal{H}=\bigoplus_J \mathcal{H}_J$, and, consequently, the optimal POVM can also be chosen to be of the form $\mathcal{E}=\bigoplus_J \mathcal{E}_J$, where, clearly, $\mathcal{E}_J$ acts on $\mathcal{H}_J$.
To ease the notation, rather than labelling the various subspaces ${\cal H}_J$ by their total angular momentum~$J$, let us simply enumerate them hereafter by natural numbers\footnote{No confusion should arise with the labels $\alpha$ of the equivalent representations of a given subspace, as for pure states these do not play any role.}, $\alpha=1,2,\dots,n+1$, and sort them by increasing value of~$J$. Hence $J=\alpha+n'/2-1$. 
With a slight abuse of notation, we will accordingly write ${\cal H}_\alpha$ and enumerate the corresponding
POVMs and overlaps as ${\cal E}_\alpha$ and $c_{\alpha}$, respectively,  where one has [cf. Eq.~\eqref{ch4/overlap-nm}]
\begin{equation}
c_\alpha=\begin{pmatrix} n'+\alpha-1\\ n' \end{pmatrix} \begin{pmatrix}n+n'\\ n'\end{pmatrix}^{-1}
\label{ch4/the_c_alpha} .
\end{equation}
%

A direct consequence of the block structure of the averaged states and $\mathcal{E}$ is that the overall success probability of a programmable discriminator can be expressed as
\begin{eqnarray}
P_{\rm s} &=& \sum_{\alpha=1}^{n+1} p_\alpha P_{{\rm s},\alpha} \, , \label{ch4/weakj} \\
p_\alpha &=& \tr(\sigma_i \id_\alpha) = \frac{2\alpha+n'-1}{(n+1)(n+n'+1)}\,,\quad i=1,2\,,
\end{eqnarray}
where $P_{{\rm s},\alpha}$ is the success probability of discrimination in the subspace ${\cal H}_\alpha$, and $p_\alpha$ is the probability of $\sigma_1$ and~$\sigma_2$ projecting onto that subspace 
upon performing the measurement $\{\id_\alpha\}$. Likewise, $P_{\rm e}$ and $Q$ can be expressed as a convex combination of the form~(\ref{ch4/weakj}). 
\\




\subsection{Weak error margin}

Let us start by considering the weak condition. If we denote the error margin by $R$, the weak condition reads~$P_{\rm e}\leqslant R$. According to the previous paragraph, the optimal strategy and the corresponding success probability $P_{\rm s}$ are defined through the maximization problem
\begin{equation}\label{ch4/max_probl}
P_{\rm s} = \max_{\cal{E}} \sum_{\alpha=1}^{n+1} p_\alpha P_{{\rm s},\alpha}
\quad
\mbox{subject to}
\quad
\sum_{\alpha=1}^{n+1} p_\alpha P_{{\rm e},\alpha}\leqslant R.
\end{equation}
Recall now that the POVMs ${\cal E}_\alpha$ are independent,
and that each of them is parametrized through Eq.~(\ref{ch3/optimal-phi}) by a margin~\mbox{$r=r_\alpha$} which, moreover, satisfies the constraint~$P_{{\rm e},\alpha}\leqslant r_\alpha$.  Therefore, Eq.~(\ref{ch4/max_probl}) can be cast as
\begin{equation}\label{ch4/max-r-j}
P_{\rm s} = \max_{\{r_\alpha\}} \sum_{\alpha=1}^{n+1} p_\alpha P_{{\rm s},\alpha}^{W}(r_\alpha) 
\quad 
\mbox{subject to}
\quad \sum_{\alpha=1}^{n+1} p_\alpha r_\alpha = R \,, 
\end{equation}
where the functions $P_{{\rm s},\alpha}^W$ are defined as in Eq.~(\ref{ch3/weak}) with~\mbox{$c=c_\alpha$}. In other words, these functions give the success probability of discrimination in the subspaces~${\cal H}_\alpha$ with {\em weak} error margins $r_\alpha$.
%
The maximization of  the success probability 
translates into finding the optimal set of weak margins $\{r_\alpha\}_{\alpha=1}^{n+1}$ 
which average, $\sum_{\alpha=1}^{n+1} p_\alpha r_\alpha$, equals a (global) margin~$R$.

Let us start by discussing the extreme cases of this scheme. On the unambiguous side, $R=0$, the only possible choice is $r_\alpha=0$ for all values of~$\alpha$, and the success probability is hence $P_{\rm s}^{\rm UA} = 1 - \PUA$, where $\PUA$ is given by Eq.~\eqref{ch4/ua-nm}. At the other end point, if~$R\geqslant R_c=\sum_{\alpha=1}^{n+1} p_\alpha r_{c,\alpha}$, where $r_{c,\alpha}$ is the critical margin in the subspace~${\cal H}_\alpha$, given by Eq.~\eqref{ch3/critical-r} with $c=c_\alpha$, we immediately recover the minimum-error result $P_{\rm s}^{\rm ME} = 1-\PME$, with $\PME$ given by Eq.~\eqref{ch4/min-nm}. We will refer to $R_c$ as the global critical margin.

An explicit expression for $P_{\rm s}$ if $0 < R < R_c$ is most easily derived by starting at the unambiguous end and progressively increasing the margin $R$. For a very small error margin,
%
the Lagrange multiplier method provides the maximum. It occurs at $r_\alpha=r^{(1)}_\alpha$, where
%
%
\begin{equation}\label{ch4/lagrange}
r^{(1)}_\alpha=\frac{1-c_\alpha}{\sum_{\alpha=1}^{n+1} p_\alpha (1-c_\alpha)} R \, .
\end{equation}
%
This solution is valid only when all (partial) error margins are below their critical values,
$ r^{(1)}_\alpha\leqslant r_{c,\alpha}$. If this inequality holds, the maximum success probability is~$P_{\rm s}=\sum_\alpha p_\alpha P_{{\rm s},\alpha}^W(r^{(1)}_\alpha)$. 
The use of the superscript ``$(1)$'' will become clear shortly.

If we keep on increasing the global margin~$R$, it will eventually reach a value
$R=R_1$ at which the error margin of the first subspace ${\cal H}_1$ is saturated, namely,  where $r^{(1)}_1=r_{c,1}$. 
This is so because the overlaps, given in Eq.~\eqref{ch4/the_c_alpha}, satisfy $c_1 < c_2 < \hdots < c_{n+1}=1$.
Hence we have
$r_1^{(1)} > r_2^{(1)} > \hdots > r_{n+1}^{(1)}$ 
and
$r_{c,1} < r_{c,2} <\dots < r_{c,n+1}$, according to Eqs.~\eqref{ch4/lagrange} and \eqref{ch3/critical-r}, respectively.
The expression for $R_1$ can be read off from Eq.~\eqref{ch4/lagrange}:
\begin{equation}
R_1=\frac{r_{c,1}}{1-c_1} \sum_{\alpha=1}^{n+1} p_\alpha (1-c_\alpha) \, .
\end{equation}
For $R>R_1$, the optimal value of the margin of subspace~${\cal H}_1$ is then frozen at the value $r_{1}=r_{c,1}$, and the remaining margins are obtained by excluding the fixed contribution of the subspace~${\cal H}_1$, i.e., by computing the maximum on the right-hand side of
%
%
\begin{eqnarray}
&\displaystyle 
P_{\rm s}-p_1 P_{{\rm s},1}^W(r_{c,1})=
\max_{\{r_\alpha\}
} \;\sum_{\alpha=2}^{n+1} p_\alpha P_{{\rm s},\alpha}^{W}(r_\alpha)
&\nonumber
\\[.5em]
&\text{subject to}& \label{ch4/r-1}
\\[.5em]
&\displaystyle
\sum_{\alpha=2}^{n+1} p_\alpha r_\alpha = R-p_1 r_{c,1}\, .
&\nonumber
\end{eqnarray}
%
The location of this maximum, which we denote by $\{r_\alpha^{(2)}\}_{\alpha=2}^{n+1}$, is formally given by Eq.~\eqref{ch4/lagrange} with $R$ replaced by $R-p_{1} r_{c,1}$ and the sum in the denominator running from~$\alpha=2$ to~$n+1$. 
In this case, we have
\begin{equation}\label{ch4/P_s at 1}
P_{\rm s}=p_1 P_{{\rm s},1}^W(r_{c,1})+\sum_{\alpha=2}^{n+1}p_\alpha
P_{{\rm s},\alpha}^W(r_{\alpha}^{(2)}) .
\end{equation}
Again, this is valid only until $R$ reaches a second saturation point $R_2$, i.e., provided $R_1<R<R_2$, and so on. Clearly, the margins $r_\alpha$ saturate in an orderly fashion as we increase $R$.

Iterating the procedure described above, the optimal error margins in the interval $R_{\beta-1}\leqslant R\leqslant R_{\beta}$ (throughout the remaining of the Chapter, Greek indexes run from $1$ to $n+1$), where $R_0\equiv0$ and $R_{n+1}\equiv R_c$, are found to be
%
%
\begin{equation}\label{ch4/lagrange k}
r^{(\beta)}_\alpha=\frac{1-c_\alpha}{\chi_\beta} \left(R-\xi_\beta \right) \, ,
\end{equation}
where
%
%
\begin{equation}\label{ch4/R-k}
R_\beta = \displaystyle\frac{r_{c,\beta} }{1-c_\beta} \chi_{\beta}+\xi_{\beta} \, , 
\end{equation}
and
\begin{equation}
\xi_\beta = \sum_{\alpha=1}^{\beta-1} p_\alpha r_{c,\alpha} \, ,
\qquad  \chi_\beta = \sum_{\alpha=\beta}^{n+1} p_\alpha (1-c_\alpha) \, .
\end{equation}
The success probability in this interval [analogous to Eq.~(\ref{ch4/P_s at 1})] is
\begin{equation}
P_{\rm s}=P^{\rm sat}_{{\rm s},\beta}+\sum_{\alpha=\beta}^{n+1}p_\alpha P_{{\rm s},\alpha}^W(r_\alpha^{(\beta)}),
\end{equation}
where
%
%
\begin{equation}\label{ch4/ps-appendix}
P^{\rm sat}_{{\rm s},\beta} = \sum_{\alpha=1}^{\beta-1} p_\alpha P_{{\rm s},\alpha}(r_{c,\alpha}) = \frac{1}{2}\sum_{\alpha=1}^{\beta-1} p_\alpha \left( 1+\sqrt{1-c_\alpha^2} \right)
\end{equation}
is the contribution to the success probability of the subspaces where the error margins are frozen at their critical values.
After some algebra, we find that the success  probability can be written in a quite compact form as
%
%
\begin{equation}\label{ch4/ps-general}
P_{\rm s} = P^{\rm sat}_{{\rm s},\beta} +\left(\sqrt{R - \xi_\beta} + \sqrt{\chi_\beta} \right)^{2},\quad
R_{\beta-1}\leqslant R\leqslant R_{\beta}.
\end{equation}
%
Eqs.~\eqref{ch4/lagrange k} through \eqref{ch4/ps-general} comprise our main result.
\\


\subsection{Strong error margin}

The concept of a strong margin for programmable machines requires a more careful formulation than that of a weak margin since, in principle, there are different conditions one can impose on the various probabilities involved. For instance, one could require the strong conditions (\ref{ch3/strong1}) and (\ref{ch3/strong2})  for \textit{every} possible pair of states fed into  the machine, that is, for every given $\{\rho_1=[\psi_1],\rho_2=[\psi_2]\}$. This approach is quickly seen to be trivial since the machine, which performance is independent of the states, is required to satisfy the condition in a worst case scenario, in which $\ket{\psi_1}$ and $\ket{\psi_2}$ are arbitrarily close to each other. For any value of the error margin less than~$1/2$ the inconclusive probability must then approach unity, i.e., $Q \to1$. This implies that both $P_{\mathrm{s}}$ and 
$P_{\mathrm{e}}$ vanish. A similar argument leads to the trivial solution~$P_{\rm s}=P_{\rm e}=1/2$ if the margin is larger than or equal to~$1/2$.

The task performed by a programmable discriminator can be most naturally viewed as state labelling:  
the machine attaches the label $1$ ($2$) to the data if its state is identified, by a ``clicking'' of the operator~$E_1$~($E_2$), to be that of the qubits loaded through  program port $A$ ($C$); i.e., the state of the ports has the pattern $[\psi_1^{\otimes n}][\psi_1^{\otimes n'}][\psi_2^{\otimes n}]$ ($[\psi_1^{\otimes n}][\psi_2^{\otimes n'}][\psi_2^{\otimes n}]$).
For this task, the relevant error probabilities are $p(2|E_1)$ and $p(1|E_2)$, namely, the probability of wrongly assigning the labels 1 and 2, respectively.
It seems, therefore, more suitable for programmable discrimination to impose the strong margin conditions $p(2|E_1)\leqslant R$ and $p(1|E_2)\leqslant R$.
In terms of the average states $\sigma_1$ and $\sigma_2$ in Eq.~\eqref{ch4/multi-copy-pure} these conditions are
%
%
%
\begin{equation}\label{ch4/strongprog}
p(2|E_1)=\frac{\tr E_1 \sigma_2}{\tr E_1 \sigma_1 + \tr E_1 \sigma_2} \leqslant R \, ,
\end{equation}
and likewise for $p(1|E_2)$.

Note that, in contrast to the weak case, here the conditional probabilities are nonlinear functions of the POVM elements, thus the maximization of the success probability under these conditions is {\em a priori} more involved. To circumvent this problem, we can use the relation~(\ref{ch3/mwms1}), which 
for programmable discrimination also holds, and reads
\begin{equation}\label{ch4/MWMS}
R^S=\frac{R^W}{P_{\rm s}(R^W)+R^W}
\end{equation}
to express the (global) weak  error margin $R^W$ in terms of the strong one $R^S$. 
%
Then, one simply uses Eqs.~\eqref{ch4/lagrange k} through \eqref{ch4/ps-general} to obtain the maximum success probability. 
The inversion of Eq.~(\ref{ch4/MWMS}) is somewhat lengthy but straightforward.  The difficulty arises from the fact that the success probability, Eq.~(\ref{ch4/ps-general}), is a piecewise function which expression depends specifically on how many 
margins $r_\alpha$ have reached their critical value $r_{c,\alpha}$ for a given~$R^S$. 
Thus we need to compute the strong saturation points~$R_\beta^S$, analogous to~\eqref{ch4/R-k}, through the relation~\eqref{ch4/MWMS}.
\\

\subsection{Analysis of the results}

\begin{figure}[t]
\begin{center}
\includegraphics[scale=1.5]{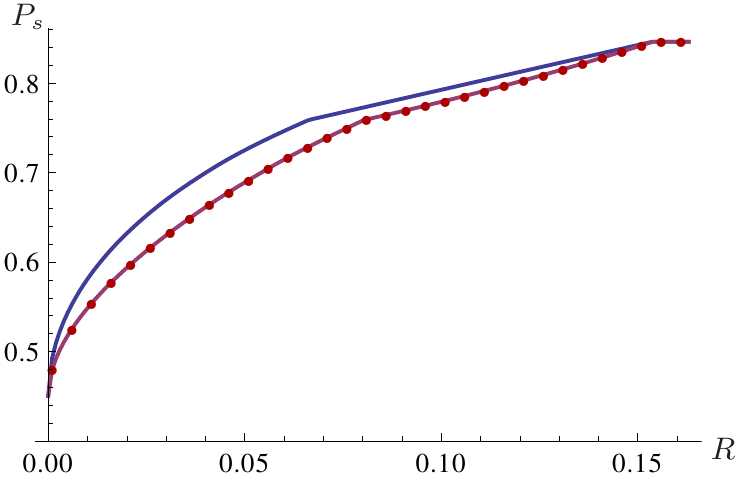}
\end{center}
\caption[$P_{\rm s}$ versus $R$ for a weak and a strong margin condition]{$P_{\rm s}$ versus $R$ for a weak (upper line) and a strong (lower line) condition, for $n=9$ and $n'=2$. The global critical margin is $R_c \simeq 0.154$. A numerical maximization of the success probability under the strong condition~(\ref{ch4/strongprog})  (points) is seen to agree with our analytical solution.}\label{ch4/fig:fig5}
\end{figure}

\begin{figure}[t]
\begin{center}
\includegraphics[scale=1.2]{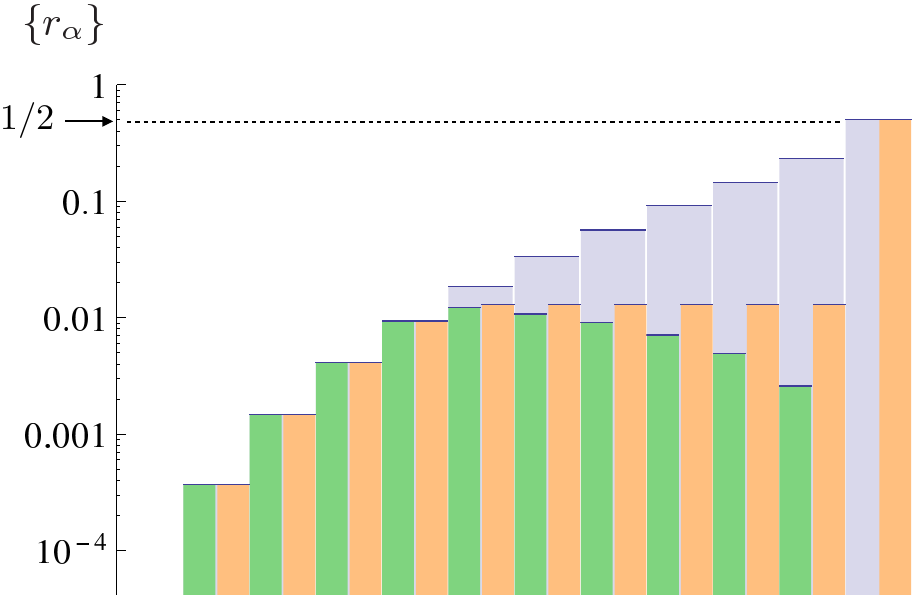}
\end{center}
\caption[Error margins for each subspace $\alpha$]{The various error margins for $n=11$, \mbox{$n'=2$} and a (global) margin~$R=0.0055$. The full heights of the wide bars in the background  (blue)  represent the values of the critical margins $r_{c,\alpha}$, 
starting from $\alpha=1$ (leftmost) up to~$\alpha=12$ (rightmost). 
For the same values of $\alpha$, each pair of narrow bars represents 
the weak margin $r_\alpha^W$ [left (green)] and the strong margin~$r_\alpha^S$ [right (orange)]. 
We note that the first five error margins have reached their critical value. The values  
for $\alpha=1$ are very small, which explains why the corresponding bars do not show up in the chart.}\label{ch4/fig:fig6}
\end{figure}

In Fig.~\ref{ch4/fig:fig5} we plot the maximum success probabilities for both the weak and the strong conditions
as a function of a common (global) margin $R$, for nine program and two data copies.
We also show in Fig.~\ref{ch4/fig:fig5} the results of a numerical optimization with the strong condition (dots), which exhibit perfect agreement with our analytical solution.
We observe that by allowing just a 5\% error margin, the success probability increases by more than~50\%.
This is just an example of a general feature of programmable discrimination with an error margin: the success probability increases sharply for small values of the error margin.

A comment about the effect of the subspace ${\cal H}_{n+1}$ 
on the shape of the plots is in order. 
This subspace contains the completely symmetric states of the whole system~$ABC$ and, hence, it is impossible to tell if the state of the data~($B$) coincides with that of one program~($A$) or that of the other~($C$);
more succinctly, $c_{n+1}=1$.
Therefore, half the number of conclusive answers will be correct and half of them will be wrong, and~$P^W_{{\rm s},n+1}=r_{n+1}$, provided~$r_{n+1}\leqslant r_{c,n+1}=1/2$. 
Increasing the error margin simply allows for an equal increase in the success probability. 
This is reflected in the linear stretch in the upper curve 
in Fig.~\ref{ch4/fig:fig5},
right before the (rightmost) flat plateau. For the strong condition, 
the same situation arises in the interval $R_n^S\leqslant R\leqslant R_c$, but the plot of the success probability is {\em not} a straight line due to the nonlinear relation~\eqref{ch4/MWMS} between the weak and the strong margin.

An alternative (though completely equivalent) way to compute the maximum success probability with a strong margin is based on the observation that the POVMs ${\cal E}_\alpha$ are also fully determined by strong margins, $r_\alpha^S$, through Eq.~(\ref{ch3/phi-strong}), with the exception of~${\cal E}_{n+1}$, for which~$c=c_{n+1}=1$ [giving rise to an ambiguity, as discussed after Eq.~(\ref{ch3/phi-strong})].
In this approach, the success probability becomes a convex combination  of $P^S_{{\rm s},\alpha}(r_\alpha^S)$, as in Eq.~\eqref{ch4/weakj}, where these functions are given in Eq.~\eqref{ch3/strong} with~$c=c_\alpha$.
%
The optimal set~$\{r^{S\,(\beta)}_{\alpha}\}$ can be readily obtained from the weak margins in Eq.~\eqref{ch4/lagrange k} using the relation \eqref{ch3/mwms1}. 
%
The strategy in the last subspace~${\cal H}_{n+1}$ can be easily seen to consist~in abstention with a certain probability, and a random choice of the labels~1 \mbox{and~2~otherwise.}

The bar chart in Fig.~\ref{ch4/fig:fig6} represents an optimal strategy in terms of the corresponding weak and strong error margins. For this example, we have chosen~$11$ program and two data copies. For illustration purposes, the (global) margin is set to a low value of~$0.0055$. The wide vertical bars in the background depict the critical margins $r_{c,\alpha}$. There are~12 of them, displayed in increasing order of~$\alpha$ (the first one is not visible because of the small value of~$r_{c,1}$). On their left (right) halves, a narrow green (orange) bar depicts the optimal weak (strong) margin $r_\alpha^W$ ($r_\alpha^S$) (we attach the subscripts~$W$ and~$S$ through the rest of the Section to avoid confusion). We note that the first~5 margins ($\alpha\leqslant5$) have reached their critical value. For $\alpha>5$, the weak margins decrease monotonically according to Eq.~\eqref{ch4/lagrange k}. For the last one, we have~$r^W_{n+1}=r^W_{12}=0$, which holds for any value of $R$, provided $R\leqslant R_n$. This must be so, since we recall that the projections of $\sigma_1$ and $\sigma_2$ onto the subspace with maximum angular momentum are indistinguishable. Clearly,  allowing  for $r_{n+1}^W>0$ while there is still room for the other margins to increase cannot be optimal.

Also noticeable in Fig.~\ref{ch4/fig:fig6} is that the set of strong margins that have not reached their critical  value $r_{c,\alpha}$  has a flat profile (this does not apply to $r^S_{n+1}$ that is always frozen to its critical value of~$1/2$). 
To provide an explanation for this, we write the equality in~Eq.~\eqref{ch4/strongprog}, which is attained if $R\leqslant R_c$,  as~$R P_{\rm s} - (1-R) P_{\rm e}=0$, using once again the symmetry of the problem. We next write the success and error probabilities as a convex sum over~$\alpha$ and use the equality in the strong conditions~(\ref{ch3/strong1}) and~(\ref{ch3/strong2}) for each subspace ${\cal H}_\alpha$ to express~$P_{{\rm e},\alpha}^S$ in terms of~$P_{{\rm s},\alpha}^S$. We obtain the strong condition 
%
\begin{equation}\label{ch4/lagrange-strong}
\sum_\alpha p_\alpha P_{{\rm s},\alpha}^S (r_\alpha^S) \left[ R  - (1-R) \frac{r_\alpha^S}{1-r_\alpha^S} \right] =0 .
\end{equation}
The terms in square brackets can be positive or negative depending on $r^S_\alpha$ being smaller or larger than $R$, both of which are possible.
So, at face value, this equation cannot explain the flat profile of $r_\alpha^S$ and more work is needed.
Next, we use the Lagrange multiplier method to maximize $P_{\rm s}=\sum_\alpha p_\alpha P_{{\rm s},\alpha}^S (r_\alpha^S)$ and note that the dependence of~$P_{{\rm s},\alpha}^S$ on~$\alpha$  (i.e., the term $1-c_\alpha$) factorizes, as can be checked from Eq.~(\ref{ch3/strong}). Without further calculation, we can anticipate that the optimal margins will be determined by $n+1$ equations of the form $p_\alpha(1-c_\alpha)f(r^S_\alpha)=0$, where $f$ can be a function only of $R$, the Lagrange multiplier and the number of margins below their critical value. Hence, all the (unfrozen) margins will have the same optimal value.
%
%
For $\beta=1$ (no frozen margins) we have the simple solution $r_\alpha^{S,(1)}=R$ for all~$\alpha$, and the corresponding success probability is
\begin{equation}
P_{{\rm s}}=\left(\frac{\sqrt{1-R}}{\sqrt{R}-\sqrt{1-R}}\right)^{\!\!2}  \frac{n n'}{(n+1)(n'+2)} 
\end{equation}
for a sufficiently small strong margin $R$.
\\



\section{Discussion}

In the first part of the Chapter, we have analysed the problem of programmable discrimination of two unknown general qubit states when multiple copies of the states are provided.
For pure states we have obtained the  optimal unambiguous discrimination and minimum-error probabilities (Section~\ref{ch4/sec:pure}). 
Knowing the error in the asymptotic regimes is very relevant information, as it allows to assess and compare the performance of devices in a way that is independent on the number of copies. 
We have obtained analytical expressions for the leading and subleading terms in several cases of interest. As could be anticipated, when the number of copies at the program ports is asymptotically large, at leading order we recover the average of the usual discrimination problem of known states in both unambiguous and minimum-error approaches. When the data port is loaded with an asymptotically large number of copies, we recover the state comparison averaged errors. These cases correspond to estimate-and-discriminate protocols, where the estimation unveils the classical information about the states.

We have also addressed, for the first time, the programmable discrimination of copies of mixed states (Section~\ref{ch4/sec:mixed}). By taking advantage of the block decomposition of permutationally invariant states to crucially reduce the computational complexity of the problem, we have obtained the minimum-error probability when the ports are loaded with copies of qubits of known purity.
We have assumed that all states have the same purity. This would correspond to a scenario where all the initially pure data and program states are subject to the same depolarizing noise before entering the machine.
Closed analytical results for a small number of copies can be obtained and efficiently computable expressions for a fairly large number of copies are given. The asymptotic analytical results show very good agreement with the numerics.
The latter show a characteristic $1/N$ dependence with the number $N$ of available copies---in contrast to the  usual exponential decay found in standard (nonuniversal) state discrimination---and provide a very good approximation already for a relatively low number of copies when the states have high purity.
For very mixed states the error probability has a drastically different behaviour.  Logically, in both cases the error probability monotonically decreases with increasing purity $r$, but in the low-purity regime the dependence is much less pronounced. The range of purities exhibiting this behaviour shrinks as the number of copies increases, and the characteristic $1/N$ behaviour of the asymptotic regime extends its validity over almost the whole range of purities.



We have analysed next the fully universal discrimination machine, a device that takes in states of which nothing is known, not even their purity (Section~\ref{ch4/sec:universal}). We have computed the minimum-error probability for three reasonable prior distributions of the purity: the hard-sphere, Bures, and Chernoff. The latter is seen to give the lowest error probability.
This comes as no surprise, since the Chernoff distribution assigns larger weights to pure states (because they are better distinguished). Our results also indicate that the fully universal discrimination machine yields an error probability comparable to the discrimination of states of known purity, being that remarkably large ($r\sim 0.9$).

Finally, we have provided two generalizations of programmable state discrimination that enable control on the rate with which errors inevitably arise because of the very principles of quantum mechanics (Section~\ref{ch4/sec:margins}). In the first, a margin is set on the average error probability of mislabelling the input data states (weak condition). In the second, a more stringent condition is required that, for each label, the probability of it being wrongly assigned is within a given margin (strong condition). Generically, in both cases, the discrimination protocol may result sometimes in an inconclusive  outcome (i.e., in being unable to assign a label to the data). We have shown that there is a one-to-one correspondence between these two margins, so that weak and strong conditions turn out to be the same if their margins are related by a simple equation.
These generalizations extend the range of applicability of programmable discriminators  to scenarios where some rate of errors and some rate of inconclusive outcomes are both affordable; or, more specifically, to situations where a trade-off between these two rates is acceptable, which depart from the standard unambiguous (zero error) and  minimum-error (zero abstention) discrimination scenarios. 

Our results include the analytical expression of the success probability for the optimal programmable device as a function of both weak and strong error margins, as well as the characterization of the POVM that specifies such optimal device.
From the analysis of these results, we conclude that small error margins can significantly boost the success probability; i.e., a small departure from the unambiguous scheme can translate into an important increase of the success rate while still having very reliable results (very low error rate). We provide an example of this, where a mere error margin value of~$5\%$ adds about $50\%$ to the success probability.\\

Throughout this Chapter we have considered programmable discriminators to be black boxes, as we optimized always over completely general POVMs. It is very relevant to examine restricted measurement schemes compatible with a machine learning scenario, in which the machine first ``learns'' about the states at the program ports and then assigns a label to the states at the data port, in that particular order. In Chapter~\ref{ch5_learning} we consider this scenario in detail, and we contrast the results with the ones obtained here. 

\chapter{\chnamefive}

\label{ch5_learning}

Programmable processors, as pointed out in Chapter \ref{ch4_pqsd}, 
are expected to automate information processing tasks, lessening human intervention by adapting their functioning according to some input program. This adjustment, that is, the process of extraction and assimilation of information relevant to perform efficiently some task, is often called \emph{learning}, borrowing a word most naturally linked to living beings. 
\emph{Machine learning} is a broad research field that seeks to endow machines with this sort of ability, 
so that they
can ``learn'' from past experience, perform ``pattern recognition'' or ``discover patterns in scrambled data'' \citep{MacKay2003,Bishop2006}. Algorithms featuring learning capabilities have numerous practical applications, including speech and text recognition, image analysis, and data mining. In \emph{supervised} machine learning, a machine is trained using a learning algorithm that takes a dataset as input, namely a \emph{training set} (TS), consisting in some observations on the characteristics of certain objects. Once trained, the machine is expected to recognize these (\emph{classification}) or other (\emph{regression}) characteristics in upcoming new objects. On the other hand, \emph{unsupervised} learning machines try to find structure hidden in unlabelled data.

Whereas conventional machine learning theory implicitly assumes the TS to be fundamentally classical---a set of classical features of classical objects, an array of symbols and numbers\mbox{---,} its quantum variant 
explores training with quantum objects, and, in doing so, it links the notion of learning in the real---quantum---world with the underlying physical theory on which it is grounded.
\emph{Quantum learning}~\citep{Aimeur2006} has recently raised great attention. Particularly, the use of programmable quantum processors has been investigated to address machine learning tasks such as pattern matching~\citep{Sasaki2002}, binary classification~\citep{Guta2010a,Neven2009,Pudenz2013}, feedback-adaptive quantum measurements~\citep{Hentschel2010}, learning of unitary transformations~\citep{Bisio2010}, Probably Approximately Correct learning~\citep{Servedio2004}, and unsupervised clustering~\citep{Lloyd2013}.
Quantum learning algorithms not only provide improvements over some classical learning problems, but also have a wider range of applicability. Quantum learning has also strong links with quantum control theory, and is becoming a significant element of the quantum information processing toolbox. 

This Chapter is concerned with a simple, yet fundamental instance of quantum state identification, which finds its motivation in learning theory. A source produces two unknown pure qubit states with equal probability. A human expert (who knows the source specifications, for instance) classifies a number of $2n$ states produced by this source into two sets of size roughly $n$ (statistical fluctuations of order $\sqrt n$ should be expected) and attaches the labels $0$ and~$1$ to them. We view these~$2n$ states as a training sample, and we set ourselves to find a universal machine that uses this sample to assign the right label to a new unknown state produced by the same source with the smallest error rate. We refer to this task as quantum classification for short. Clearly, quantum classification can be understood as a supervised quantum learning problem, as has been noticed by Guta and Kotlowski in their recent work~\citep{Guta2010a} (though they use a slightly different setting).

It is worth mentioning that a very similar problem was proposed in~\citep{Sasaki2002} under the name of ``universal quantum matching machine''. The task of this machine differs from that of ours in that, rather than identifying the unknown qubit as one of the states in the TS, it determines to which of them is \emph{closest}, thus a fidelity-related figure of merit is used instead of the error probability. The work of Sasaki and Carlini pioneered the view on the quantum classification problem as a learning protocol, and set an inspiration for later works on---the more general---programmable discrimination machines.

Of course, an absolute limit on the minimum error in quantum classification is provided by the optimal programmable discrimination machine (see Chapter~\ref{ch4_pqsd}). 
In that context, to ensure optimality one assumes that a fully general two-outcome joint measurement is performed on {\em both} the~$2n$ training qubits and the qubit we~wish to classify, where the observed outcome determines which of the two labels,~$0$ or~$1$, is assigned to the latter qubit.
Thus, in principle, this assumption implies that, in a learning scenario, a quantum memory is needed to store the training sample till the very moment we wish to classify the unknown qubit.
The issue of whether or not the joint measurement assumption can be relaxed has not yet been addressed.
Nor has the issue of how the information left after the joint measurement can be used to classify a second unknown qubit produced by the same source, unless a fresh new TS is provided (which may seem unnatural in a learning context).

The main objective of this Chapter is to show that, for a sizable TS (asymptotically large $n$), the absolute lower bound on the probability of misclassifying the unknown qubit, set by programmable discrimination, can be attained by first performing a suitable measurement on the TS followed by a Stern-Gerlach type of measurement on the unknown qubit, where forward classical communication is used to control the parameters of the second measurement\footnote{Interestingly, this result is the opposite to the one found by Sasaki and Carlini for their universal quantum matching machine, where any strategy of two separate measurements is suboptimal~\citep{Sasaki2002}. Again, their protocol is slightly different to ours.}.
The whole protocol can thus be undersood as a learning machine (LM), which requires much less demanding assumptions while still having the same accuracy as the optimal programmable discrimination machine. All the relevant information about the TS needed to control the Stern-Gerlach measurement is kept in a {\em classical} memory, thus classification can be executed any time after the learning process is completed. Once trained, this machine can be subsequently used an arbitrary number of times to classify states produced by the same source. Moreover,~this optimal LM is robust under noise, i.e., it still attains optimal performance if the states produced by the source undergo depolarization to any degree.
Interestingly enough, in the ideal scenario where the qubit states are pure and~the TS consists in exactly the same number of copies of each of the two types 0/1  (no statistical fluctuations are allowed) this LM attains the optimal programmable discrimination bound for {\em any} size $2n$ of the TS, not necessarily asymptotically large.


%

At this point it should be noted that LMs without~quantum memory can be naturally assembled from~two quantum information primitives: state estimation and state discrimination. We will refer to these specific constructions as ``estimate-and-discriminate'' (E\&D) machines. The protocol they execute is as follows: by performing, e.g., an optimal covariant measurement on the~$n$ qubits in the TS labelled $0$, their state~$|\psi_0\rangle$ is estimated with some accuracy, and likewise the state~$|\psi_1\rangle$ of the other~$n$ qubits that carry the label~$1$ is characterized. 
This classical information is stored and subsequently used to discriminate an unknown qubit state. 
It will be shown that the excess risk (i.e., excess average error over classification when the states $|\psi_0\rangle$ and $|\psi_1\rangle$ are perfectly known) of this protocol is twice  that of the optimal LM.
The fact that the E\&D machine is suboptimal means that the kind of information retrieved from the TS and stored in the classical memory of the optimal LM is specific to the classification problem at hand, and that the machine itself is more than the mere assemblage of well known protocols.

We will first present our results for the ideal scenario where states are pure and no statistical fluctuation in the number of copies of each type of state is allowed. The effect of these fluctuations and the robustness of the LM optimality against noise will be postponed to the end of the Chapter. 
\\

\section{The learning machine}

In this Chapter we use the notation and conventions of Chapter~\ref{ch4_pqsd}. Before presenting our results, let us briefly recall the setting of the problem for programmable machines and its optimal solution.
Neglecting statistical fluctuations, the TS of size $2n$ is given by a state pattern of the form $[\psi_0^{\otimes n}]\otimes[\psi_1^{\otimes n}]$, 
where no knowledge about the actual states $|\psi_0\rangle$ and $|\psi_1\rangle$ is assumed (the figure of merit will be an average over {\em all} states of this form). The qubit state that we wish to label (the \emph{data} qubit) belongs either to the first group (it is $[\psi_0]$)  or to the second one (it is $[\psi_1]$). Thus the optimal machine must discriminate between the two possible states: either $\varrho^n_0=[\psi_0^{\otimes (n+1)}]_{AB}\otimes[\psi_1^{\otimes n}]_{C}$, in which case it should output the label $0$, or $\varrho^n_1=[\psi_0^{\otimes n}]_{A}\otimes[\psi_1^{\otimes (n+1)}]_{BC}$, in which case the machine should output the label $1$. Here and when needed for clarity, we name the three subsystems involved in this problem~$A$,~$B$ and~$C$, where $AC$ is the TS and~$B$ is the data qubit. In order to discriminate~$\varrho^n_0$ from~$\varrho^n_1$, a joined two-outcome measurement, independent of the actual states $|\psi_0\rangle$ and $|\psi_1\rangle$, is performed on all~$2n+1$ qubits. Mathematically, it is  represented by a two-outcome POVM ${\mathscr E}=\{E_0,E_1=\openone-E_0\}$. 
The minimum average error probability of the quantum classification process is [cf. Eq.~\eqref{ch3/helstrom}]
%
%
%
$P_{\rm e} = (1-\Delta/2)/2$, where
\begin{equation}
\Delta = 2\max_{E_0}\tr\left[\left(\sigma^n_0-\sigma^n_1\right)E_0\right]= \trnorm{\sigma^n_0-\sigma^n_1} \,,
\label{ch5/Delta}
\end{equation}
%
and $\sigma^n_{0/1}$ are average states analogous to the states \eqref{ch4/rho1-rho2}, defined in this case as
\begin{eqnarray}\label{ch5/sigma states}
\sigma^n_0&=&\frac{\id_{n+1}\otimes\id_n}{d_{n+1}d_{n}}=\frac{\id_{AB}\otimes\id_C}{d_{AB}d_{C}}  \,, \nonumber \\
\sigma^n_1&=&\frac{\id_{n}\otimes\id_{n+1} }{d_{n}d_{n+1}} =\frac{\id_{A}\otimes\id_{BC}}{d_{A}d_{BC}} \,,
\end{eqnarray}
where $\id_m$ stands for the projector onto the fully symmetric invariant subspace of $m$ qubits, which has dimension $d_{m}=m+1$. Sometimes, it turns out to be more convenient to use the subsystem labels, as on the right of~Eq.~\eqref{ch5/sigma states}. 

Recall that the trace norm in Eq.~\eqref{ch5/Delta} can be computed by switching to the total angular momentum basis, $\{\ket{J,M}\}$, and splitting it in the different contributions of the orthogonal Jordan subspaces (see Section~\ref{ch4/sec:pure} for details). The final answer is given by Eq.~\eqref{ch4/min-nm}, with $n'=1$. It takes the simple form
%
%
%
%
\begin{equation}
P^{\rm opt}_{\rm e}={1\over2}-\frac{1}{d_n^2 d_{n+1}} \sum_{k=0}^n k \sqrt{d_n^2-k^2} \,,
 \label{ch5/OptDisc}
\end{equation}
where we have written the various values of the total angular momentum as $J=k+1/2$.
%
%
The formula~\eqref{ch4/SME-subleading} gives the asymptotic expression of $P^{\rm opt}_{\rm e}$ for large $n$, which in this case simply reads
\begin{equation}\label{ch5/progr asymp}
P^{\rm opt}_{\rm e}\simeq {1\over 6}+{1\over3n} .
\end{equation}
The leading order ($1/6$) coincides with the average error probability for \emph{known} states $\int d\psi_0\,d\psi_1 \,p^{\rm opt}_{\rm e}(\psi_0,\psi_1)$, where $p^{\rm opt}_{\rm e}(\psi_0,\psi_1)$ is the minimum error in discrimination between two given states $\ket{\psi_0}$ and $\ket{\psi_1}$.

\begin{figure}[t]
\begin{center}
\includegraphics[scale=0.5]{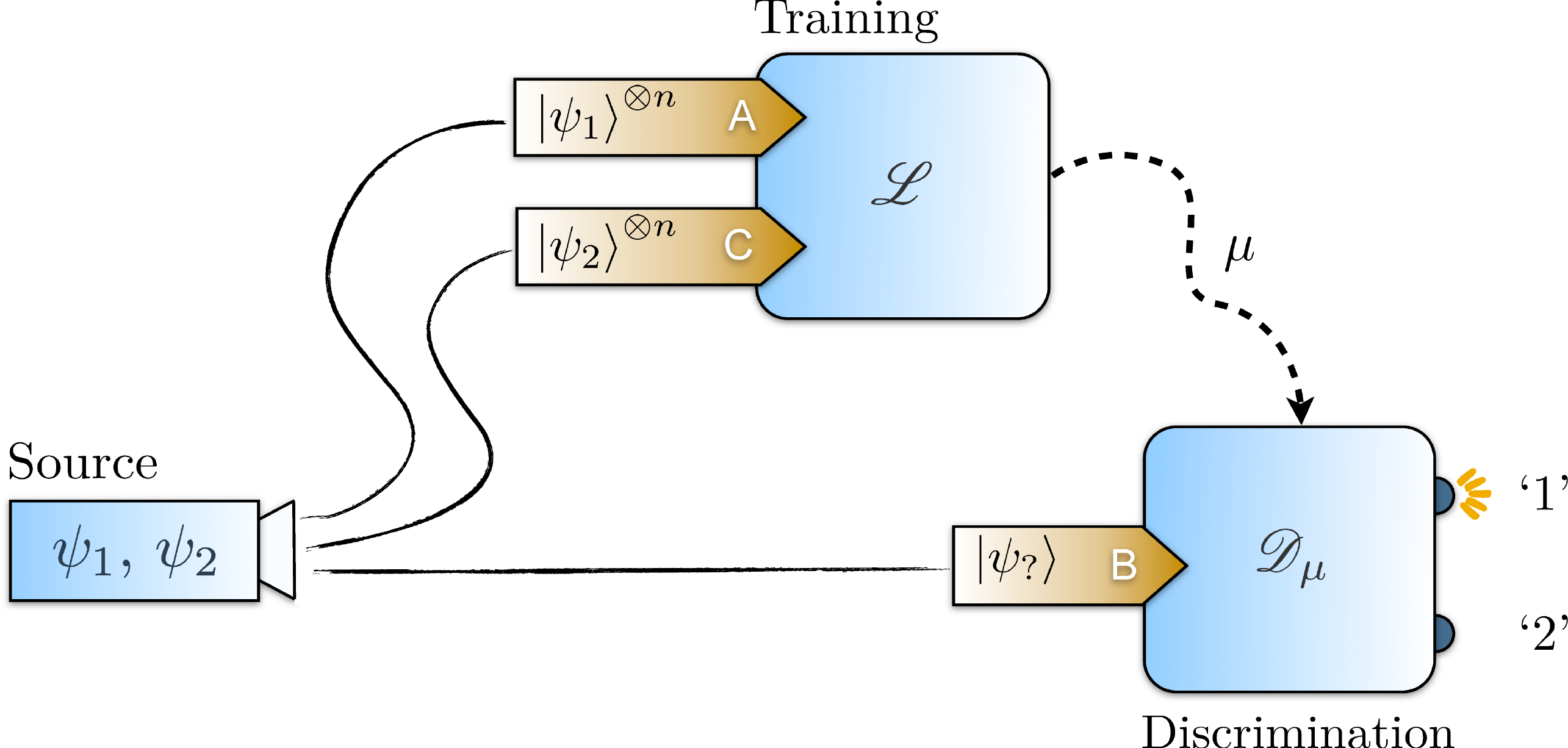}
\end{center}
\caption[Learning machine for qubit classification]{A learning protocol for qubit classification. First, a measurement $\mathscr{L}$ is performed over the TS (system $AC$), from which the information $\mu$ is extracted. Then, $\mu$ is used for defining a two-outcome measurement $\mathscr{D}_\mu$ that classifies the data qubit (system $B$) with some probability of success.}\label{ch5/fig:fig0}
\end{figure}

The formulas above give an absolute lower bound to the error probability  that can be physically attainable. We wish to show that this bound can actually be attained by a learning machine that uses a classical register to store all the relevant information obtained in the  learning process regardless the size, $2n$, of the TS.
A first hint that this may be possible is that the optimal measurement~${\mathscr E}$ can be shown to have positive partial transposition with respect to the partition TS/data qubit. Indeed this is a necessary condition for any measurement that consists of a local POVM on the TS which outcome is fed-forward to a second POVM on the data qubit. This class of one-way adaptive measurement can be characterized as
\begin{equation}
E_0=\sum_\mu L_\mu\otimes D_\mu,\quad
E_1=\sum_\mu L_\mu\otimes (\id_1-D_\mu),
\label{ch5/LM POVM}
\end{equation}
where the positive operators  $L_\mu$ ($D_\mu$) act on the Hilbert space of the TS (data qubit we wish to classify), and $\sum_\mu L_\mu=\id_n\otimes\id_n$.  The POVM ${\mathscr L}=\{L_\mu\}$ represents the learning process, and the parameter $\mu$, which \emph{a priori} may be discrete or continuous, encodes the information gathered in the measurement and required at the classification stage. For each possible value of~$\mu$, ${\mathscr D}_\mu=\{D_\mu,\id_1-D_\mu\}$ defines the measurement on the data qubit, which two outcomes represent the classification decision (see Fig.~\ref{ch5/fig:fig0}).  Clearly, the size of the required classical memory will be determined by the information content of the random variable~$\mu$.
\\

\section{Covariance and structure of $\mathscr{L}$}

We will next prove that the POVM $\mathscr L$, which extracts the relevant information from the TS, can be chosen to be covariant. This will also shed some light on the physical interpretation of the classical variable $\mu$.
The states \eqref{ch5/sigma states} are by definition invariant under a rigid rotation acting on subsystems $AC$ and $B$ of the form $U=U_{AC} \otimes u$, where, throughout this Chapter, $U$ stands for an element of the appropriate representation of SU(2), which should be obvious by context (in this case $U_{AC}=u^{\otimes 2n}$, where $u$ is in the fundamental representation). Since
$
\tr ( E_0 \sigma^n_{0/1} ) = \tr (E_0 U^{\dagger} \sigma^n_{0/1} U ) = \tr (U E_0 U^{\dagger} \sigma^n_{0/1} )
$, the positive operator $UE_0U^\dagger$ gives the same error probability as $E_0$ for {\em any} choice of $U$ [as can be seen from, e.g., Eq.~\eqref{ch5/Delta}]. The same property thus holds for their average over the whole SU(2) group 
$
\bar E_0 = \int du \, U E_0 U^{\dagger}
$, 
which is invariant under rotations, and where $du$ denotes the SU(2) Haar measure. By further exploiting rotation invariance (see Appendix~\ref{appB/sec:covariance} for full details), $\bar E_0$ can be written as
 %
\begin{equation}\label{ch5/LM covariant POVM}
\bar E_0=\int du \, \left(U_{AC}\,{\Omega}\,U^\dagger_{AC}\right)\otimes \left(u [\,\uparrow\,] u^{\dagger}\right)
\end{equation}
%
%
for some positive operator $\Omega$, where we use the shorthand notation $[\,\uparrow\,]\equiv\ketbrad{\half,\half}$. Similarly, the second POVM element can be chosen to be an average,~$\bar E_1$, of the form~\eqref{ch5/LM covariant POVM}, with $[\,\downarrow\,]\equiv\ketbrad{\half,-\half} $ instead of~\mbox{$[\,\uparrow\,]$}. We immediately recognize $\bar{\mathscr E}=\{\bar E_0,\bar E_1\}$ to be of the form~\eqref{ch5/LM POVM}, where $u$,  $L_u\equiv U_{AC}\,{\Omega}\,{U}_{AC}^{\dagger}$ and $D_u\equiv u[\,\uparrow\,]u^\dagger$ play the role of~$\mu$, $L_\mu$ and~$D_\mu$, respectively. 
Hence, without loss of generality we can choose ${\mathscr L}=\{U_{AC}\,{\Omega}\,{U}_{AC}^{\dagger}\}_{\rm SU(2)}$, which is a covariant POVM with seed~$\Omega$.
Note that $u$ entirely defines the Stern-Gerlach measurement, ${\mathscr D}_u=\{u[\,\uparrow\,]u^\dagger, u[\,\downarrow\,]u^\dagger \}$, 
i.e., $u$ specifies the direction along which the Stern-Gerlach has to be oriented. This is the relevant information that has to be retrieved from the TS and kept in the classical memory of the LM.
%
%

Covariance has also implications on the structure of~${\Omega}$. 
In Appendix~\ref{appB/sec:covariance}, we show that this seed can always be written as
\begin{equation}\label{ch5/omega}
{\Omega} = \sum_{m=-n}^n {\Omega}_m \, ; \quad {\Omega}_m \geqslant 0 \, ,
\end{equation}
where
\begin{equation}\label{ch5/omega_m}
\sum_{m=-j}^j \langle j,m|\Omega_m| j,m\rangle= 2j+1 ,\quad 0\leqslant j\leqslant n,
\end{equation}
%
%
and $j$ ($m$) stands for the total angular momentum~$j_{AC}$ (magnetic number $m_{AC}$) of the qubits in the~TS. In other words,~the seed is a direct sum of operators with a well defined magnetic number. As a result, we can interpret that ${\Omega}$ points along the $z$-axis. The constraint~\eqref{ch5/omega_m} ensures~that $\mathscr L$ is a resolution of the identity.

To gain more insight into the structure of~$\Omega$, we trace subsystems~$B$ in the definition of $\Delta$, given by the first equality in Eq.~\eqref{ch5/Delta}. For the covariant POVM~\eqref{ch5/LM covariant POVM},  rotational invariance enables us to express this quantity as 
%
\begin{equation}
\Delta^{\mathrm{LM}} = 2\max_\Omega \tr \left\{(\sigma^n_0-\sigma^n_1) \Omega\otimes[\,\uparrow\,]\right\}= 2\max_\Omega \tr (\Gamma_\uparrow \Omega),
\label{ch5/Delta SDP}
\end{equation}
where we have defined 
\begin{equation}\label{ch5/def GammaUp}
\Gamma_\uparrow=\tr_{\!B}\{[\,\uparrow\,] (\sigma^n_0-\sigma^n_1)\}
\end{equation}
(the two resulting terms in the right-hand side are the post-measurement states of~$AC$ conditioned to the outcome~$\uparrow$ after the Stern-Gerlach measurement ${\mathscr D}_{\scrvec z}$ is performed on~$B$), and the maximization is over valid seeds (i.e., over positive operators~$\Omega$ such that $\int du\, U_{AC}\,\Omega\, U^\dagger\kern-.3em{}_{AC}=\id_{AC}$). We calculate~$\Gamma_\uparrow$ in Appendix~\ref{appB/sec:Gamma_up}. The resulting expression can be cast in the simple and transparent form 
%
%
\begin{equation}
\Gamma_\uparrow={\hat J^{A}_z-\hat J^{C}_z\over d_n^2 d_{n+1}},
\label{ch5/JA-JC}
\end{equation}
%
where $\hat J^{A/C}_z$ is the $z$ component of the total angular momentum operator acting on subsystem $A/C$,
i.e., on the training qubits to which the human expert assigned the label~0/1.
Eq.~\eqref{ch5/JA-JC} suggests that the optimal $\Omega$ should project on the subspace of $A$ ($C$) with maximum (minimum) magnetic number, which implies that $m_{AC}=0$. An obvious candidate is 
\begin{equation}\label{ch5/seed}
\Omega=[\phi^0]\,, \quad
\ket{\phi^0} = \sum_{j=0}^n \sqrt{2j+1} \ket{j,0} .
\end{equation}
Below we prove that indeed this seed generates the optimal~LM~POVM. 
\\

\section{Optimality of the LM}

We now prove our main result: the POVM $\bar{\mathscr E}=\{\bar{E}_0,\bar{E}_1\}$, generated from the seed state in Eq.~\eqref{ch5/seed}, gives an error probability $P_{\rm e}^{\rm LM}=(1-\Delta^{\rm LM}/2)/2$  equal to the minimum-error probability~$P^{\rm opt}_{\rm e}$ of the optimal programmable discriminator,~Eq.~\eqref{ch5/OptDisc}. It is, therefore, optimal and, moreover, it attains the absolute minimum allowed by quantum physics. 

The proof goes as follows. From the very definition of error probability,
\begin{equation}
P_{\rm e}^{\rm LM} = \frac{1}{2} \left(\tr \sigma^n_1\bar E_0+\tr \sigma^n_0\bar E_1\right),
\end{equation}
we have
\begin{equation}
P_{\rm e}^{\rm LM} =
{\tr\left(\id_{A}\otimes\id_{BC} [\phi^0]\otimes{[\uparrow]}\right)+
\tr\left(\id_{AB}\otimes\id_{C} [\phi^0]\otimes[\downarrow]\right)\over 2 d_n d_{n+1}} ,
\end{equation}
%
%
where we have used rotational invariance. We can further simplify this expression by writing it as 
\begin{equation}
P_{\rm e}^{\rm LM} ={
\norm{\id_A \otimes \id_{BC} \ket{\phi_0}\ket{\uparrow}}^{\,2}
 + \norm{\id_{AB} \otimes \id_C \ket{\phi_0}\ket{\downarrow}}^{\,2} 
\over 2 d_n d_{n+1}} .
\label{ch5/P^LM norm}
\end{equation}
%
%
%
%
To compute the projections inside the norm signs
we first write $\ket{\phi^0}\ket{\uparrow}$ ($\ket{\phi^0}\ket{\downarrow}$ will be considered below) in the total angular momentum basis $|J,M\rangle_{\xyZ ACB}\,$, where the attached subscripts remind us how subsystems $A$, $B$ and~$C$ are  both \emph{ordered} and {\em coupled} to give the total angular momentum~$J$ (note that a permutation of subsystems, prior to fixing the coupling, can only give rise to a global phase, thus not affecting the value of the norm we wish to compute). This is a trivial task since $\ket{\phi^0}\ket{\uparrow}\equiv\ket{\phi^0}_{AC}\ket{\uparrow}_B$, i.e., subsystems are ordered and coupled as the subscript $(AC)B$ specifies, so we just need the Clebsch-Gordan coefficients 
\begin{equation}
\braket{j\pm\half,\half}{j,0;\half,\half}=\pm\sqrt{\frac{j+\frac{1}{2}\pm\frac{1}{2}}{2j+1}} .
\end{equation}

The projector $\id_{A}\otimes\id_{BC}$, however, is naturally written as
$
\id_{A}\otimes\id_{BC}=\sum_{J,M}\ket{J,M}_{\Xyz ACB}\!\bra{J,M}
$.
This basis differs from that above in the coupling of the subsystems. To compute the projection $\id_A\otimes\id_{BC} \ket{\phi^0}\ket{\uparrow}$ we only need to know the overlaps between the two bases ${{}_{\Xyz ACB}\kern-.1em\braket{J,M}{J,M}_{\xyZ ACB}}$ . Wigner's 6$j$-symbols provide this information as a function of the angular momenta of the various subsystems (the overlaps are computed explicitly in Appendix~\ref{appB/sec:overlaps}).
 
Using the Clebsch-Gordan coefficients and the overlaps between the two bases, it is not difficult to obtain
\begin{equation}\label{ch5/projection}
 \id_{A}\otimes\id_{BC} |\phi^0\rangle\ket{\uparrow} =
 \sum_{j=1}^{n+1}
\sqrt{j} {\sqrt{d_n+j}-\sqrt{d_n-j}\over\sqrt2 d_n}\ket{j-\half,\half}_{\Xyz ACB}\,,
 \end{equation}
 %
An identical expression can be obtained for $\id_{AB}\otimes\id_C \ket{\phi^0}\ket{\downarrow}$ in the basis $\ket{J,M}_{\xyZ BAC}\,$. To finish the proof, we compute the norm squared of Eq.~\eqref{ch5/projection} and 
substitute in Eq.~\eqref{ch5/P^LM norm}.
It is easy to check that this gives the expression of the error probability~\eqref{ch5/OptDisc}, i.e.,~$P_{\rm e}^{\rm LM}=P^{\rm opt}_{\rm e}$. 
\\

\section{Memory of the LM}

Let us go back to the POVM condition, specifically to the minimum number of unitary transformations needed to ensure that, given a suitable discretization $\int du \to \sum_\mu p_\mu$ of Eq.~\eqref{ch5/LM covariant POVM}, $\{p_\mu U_\mu[\,\phi^0\,]U^\dagger_\mu\}$ is a resolution of the identity for arbitrary~$n$. 
This issue is addressed in~\citep{Bagan2001}, where an explicit algorithm for constructing finite POVMs, including the ones we need here, is given.  From the results there, we can bound the minimum number of outcomes of $\mathscr L$ by $2(n+1)(2n+1)$.
This figure is important because its binary logarithm gives an upper bound to the minimum memory required. 
We see that it  grows at most logarithmically with the size of the TS.\\

\section{E\&D machines}

E\&D machines can be discussed within this very framework, as they are particular instances of LMs. In this case the POVM~$\mathscr L$ has the form
$L_{\al i} =M_\al\otimes M'_i$, where ${\mathscr M}=\{M_\al\}$ and ${\mathscr M}'=\{M'_i\}$ are themselves POVMs on the TS subsystems $A$ and~$C$, respectively. The role of~${\mathscr M}$ and~${\mathscr M}'$ is to estimate (optimally) the qubit states in these subsystems~\citep{Holevo1982}. The measurement on~$B$ (the data qubit) now depends on the pair of outcomes of~${\mathscr M}$ and~${\mathscr M}'$: ${\mathscr D}_{\al i}=\{D_{\al i},\id_1-D_{\al i}\}$. It performs standard one-qubit discrimination according to the two pure-state specifications, say, the unit Bloch vectors $\vec s_0^\al$ and $\vec s_1^i$, estimated with ${\mathscr M}$ and~${\mathscr M}'$. 
In this Section, we wish to show that E\&D machines perform worse than the optimal LM. 

We start by tracing subsystems~$AC$ in Eq.~\eqref{ch5/Delta}, which for E\&D reads
\begin{equation}
\Delta^{\rm E\&D}=2\max_{{\mathscr M}, {\mathscr M}'}\tr_{\!B} \max_{\{{\mathscr D}_{\alpha i}\}}\tr_{\!AC}[(\sigma^n_0-\sigma^n_1)E_0].
\end{equation}
If we write $\Delta^{\rm E\&D}=\max_{\mathscr {\mathscr M},{\mathscr M}'}\Delta_{{\mathscr M},{\mathscr M}'}$, we have
\begin{equation}
\Delta_{{\mathscr M},{\mathscr M}'}=\sum_{\al i}p_\al p'_i |\vec r_0^\al-\vec r_1^i| ,
\label{ch5/E+D}
\end{equation} 
where $\vec r_0^\al$ and $\vec r_1^i$ are the Bloch vectors of the data qubit states 
\begin{equation}\label{ch5/conditioned}
\rho^\alpha_0={1\over p_\alpha}\tr_{\!A}\left({\id^{AB}_{n+1}\over d_{n+1}}M_\alpha\right),
\quad
\rho^i_1={1\over p'_i}\tr_{\!C}\left({\id^{BC}_{n+1}\over d_{n+1}}M'_i\right),
\end{equation}
conditioned to the outcomes~$\alpha$ and~$i$ respectively,  
and~$p_\alpha=d^{-1}_n\tr M_\alpha$, $p'_i=d^{-1}_n\tr M'_i$ are their probabilities. 
We now recall that optimal estimation necessarily requires that all elements of $\mathscr M$ must be of the form $M_\alpha=c_\alpha U_\alpha[\psi^0] U^\dagger_\alpha$, where~$|\psi^0\rangle=|\mbox{\footnotesize${n\over2}$},\mbox{\footnotesize${n\over2}$}\rangle$, $c_\alpha>0$, and~$\{U_\alpha\}$ are appropriate SU(2) rotations  (analogous necessary conditions are required for~${\mathscr M}'$)~\citep{Derka1998}. Substituting in Eq.~\eqref{ch5/conditioned} we obtain 
$
p_\alpha={c_\alpha/d_n}$,  %
and
\begin{equation}
u^\dagger_\alpha\rho^\alpha_{0} u_\alpha={1\over d_{n+1}}\left(d_n[\,\uparrow\,]+[\,\downarrow\,]\right) 
\end{equation}
(a similar expression holds for $\rho^i_1$).
This means that the Bloch vector of the data qubit conditioned to outcome~$\alpha$ is proportional to $\vec s^\alpha_0$ (the Bloch vector of the corresponding estimate) and is shrunk by a factor $n/d_{n+1}=n/(n+2)\equiv\eta$. Note in passing that the shrinking factor~$\eta$ is independent of the measurements, provided it is optimal. 

Surprisingly at first sight, POVMs that are optimal, and thus equivalent, for estimation may lead to different minimum-error probabilities. In particular, the continuous covariant POVM is outperformed  in the problem at hand by those with a finite number of outcomes. 
Optimal POVMs with few outcomes enforce large angles between the estimates~$\vec s_0^\al$ and $\vec s_1^i$, and thus between $\vec r_0^\al$ and $\vec r_1^i$ ($\pi/2$ in the $n=1$ example below). This translates into increased discrimination efficiency, as shown by Eq.~\eqref{ch5/E+D}, without compromising the quality of the estimation itself. Hence the orientation of  ${\mathscr M}$ relative to ${\mathscr M}'$ (which for two continuous POVMs does not even make sense) plays an important role, as it does the actual number of outcomes. 
With an increasing size of the TS, the optimal estimation POVMs require also a larger number of outcomes and the angle between the estimates decreases in average, since they tend to fill the 2-sphere isotropically. Hence the minimum-error probability is expected to approach that of two continuous POVMs. This is supported by numerical calculations. 
The problem of finding the optimal E\&D machine for arbitrary $n$ appears to be a hard one.
Here we will give the absolute optimal E\&D machine for $n=1$ and, also, we will compute the minimum-error probability for both~$\mathscr M$ and~${\mathscr M}'$ being the continuous POVM that is optimal for estimation. The later, as mentioned,  is expected to attain the optimal E\&D error probability asymptotically.

We can obtain an upper bound on Eq.~\eqref{ch5/E+D} by applying the Schwarz inequality.
We~readily find that 
%
\begin{eqnarray}
\Delta_{{\mathscr M},{\mathscr M}'} &\leqslant& \sqrt{\sum_{\al i} p_\al p'_i|\vec r_0^\alpha-\vec r_1^i|^2} \nonumber \\
&=&\sqrt{\sum_{\al} p_\al|\vec r_0^\alpha|^2+\sum_{i}p'_i |\vec r_1^i|^2} \,,
\end{eqnarray}
where we have used that $\sum_\al p_\alpha \vec r_0^\alpha=\sum_i p'_i \vec r_1^i=0$, as follows from the POVM condition on $\mathscr M$ and ${\mathscr M}'$. The maximum norm of $\vec r^\alpha_0$ and $\vec r^i_1$ 
is bounded by $1/3$ (the shrinking factor $\eta$ for $n=1$). 
Thus 
\begin{equation}
\Delta_{{\mathscr M},{\mathscr M}'} \leqslant \sqrt{2}/3<1/\sqrt3=\Delta^{\rm LM}\,,
\label{ch5/Delta n=1}
\end{equation}
where
the value of $\Delta^{\rm LM}$ can be read off from Eq.~\eqref{ch5/OptDisc}.
The E\&D bound $\sqrt2/3$ is attained by the choices \mbox{$M_{\uparrow/\downarrow}=[\,\uparrow\!/\!\downarrow\,]$} and $M'_{+/-}=[+\!/\!-]$, where we have used the definition~$\ket{\pm}=(\ket{\uparrow}\pm\ket{\downarrow})/\sqrt2$.

For arbitrary $n$, a simple expression for the error probability can be derived in the continuous POVM case, ${\mathscr M}={\mathscr M}'=\{{d_n} U_{\scrvec s}[\,\psi^0\,]U^\dagger_{\scrvec s}\}_{\scrvec s\in{\mathbb S}^2}$, where 
$\vec s$ is a unit vector (a point on the 2-sphere ${\mathbb  S}^2$) and~$U_{\scrvec s}$ is the representation of the rotation that takes the unit vector along the $z$-axis,~$\vec z$, into~$\vec s$. Here $\vec s$ labels the outcomes of the measurement and thus plays the role of~$\alpha$ and $i$. 
The continuous version of Eq.~\eqref{ch5/E+D} can be easily computed to be
\begin{equation}
\Delta^{\rm E\&D}=\eta  \int d\vec s\,|{\vec z}-\vec s|=\frac{4n}{3(n+2)}\,.
\end{equation}
Asymptotically, we have $P_{\rm e}^{\rm E\&D}=1/6+2/(3n)+\dots$. Therefore, the excess risk, which we recall is the difference between the average error probability of the machine under consideration and that of the optimal discrimination protocol for {\em known} qubit states 
($1/6$), is $R^{\rm E\&D}=2/(3n)+\dots$. This is twice the excess risk of the optimal programmable machine and the optimal LM, which can be read off from Eq.~\eqref{ch5/progr asymp}: 
\begin{equation}\label{ch5/ex risk}
R^{\rm LM}=R^{\rm opt}={1\over 3n}+\dots\,.
\end{equation}
For $n=1$, Eq.~\eqref{ch5/Delta n=1} leads to $R^{\rm E\&D}=(4-\sqrt2)/12$.
This value is already~$15\%$ larger than excess risk of the optimal LM:~$R^{\rm LM}=(4-\sqrt3)/12$.
\\



\section{Robustness of LMs}

So far we have adhered to the simplifying assumptions that the two types of states produced by the source are pure and, moreover, exactly equal in number. Neither of these two assumptions is likely to hold in practice, as both, interaction with the environment, i.e., decoherence and noise, and statistical fluctuations in the numbers of states of each type, will certainly take place. Here we prove that the performance of the optimal LM  is not altered by these effects in the asymptotic limit of large TS. More precisely, the excess risk of the optimal~LM remains equal to that of the optimal programmable discriminator to leading order in $1/n$ when noise and statistical fluctuations are taken into account.

Let us first consider the impact of noise,  which we will assume isotropic and uncorrelated. Hence, instead of producing $[\psi_{0/1}]$, the source produces copies of
%
%
%
\begin{equation}
\rho_{0/1} = r [\psi_{0/1}] + (1-r) \frac{\id}{2}\,,\quad 0<r\leqslant 1 \,.
\label{ch5/rho mixed}
\end{equation}
%
In contrast to the pure qubits case, where $[\psi^{\otimes n}_{0/1}]$ belongs to the fully symmetric invariant subspace of maximum angular momentum $j=n/2$, the state of $A/C$ is now a full-rank matrix of the form $\rho^{\otimes n}_{0/1}$. Hence, as showed in Section~\ref{ch3/sec:blockdecomposition}, it has projections on all the orthogonal subspaces
$\mathscr{S}_{j}\otimes{\mathbb C}^{\nu^n_j}$, where~$\mathscr{S}_{j}={\rm span}(\{\ket{j,m}\}_{m=-j}^j)$, ${\mathbb C}^{\nu^n_j}$ is the $\nu^n_j$-dimensional multiplicity space of the representation with total angular momentum $j$,
and~$j$ is in the range from~$0$~($1/2$) to $n/2$ if $n$ is even (odd). Therefore $\rho^{\otimes n}_{0/1}$ is block-diagonal in the total angular momentum eigenbasis. The multiplicity space~${\mathbb C}^{\nu^n_j}$ carries the label of the $\nu^n_j$ different equivalent representations of given~$j$, which arise from the various ways the individual qubits can couple to produce total angular momentum~$j$. 
For permutation invariant states (such as~$\rho^{\otimes n}_{0/1}$), this has no physical relevance and the only effect of ${\mathbb C}^{\nu^n_j}$ in calculations is through its dimension $\nu^n_j$, given by Eq.~\eqref{ch3/rhoblock3}. The multiplicity space will hence be dropped throughout the rest of the Chapter. 

The average states now become a direct sum of the form
\begin{eqnarray}
 \int d\psi_0\,d\psi_1\, \rho_0^{\otimes (n+1)} \otimes \rho_1^{\otimes n} &=& \sum_\xi p^n_\xi \sigma^n_{0,\xi} \label{ch5/sigmaxi1} , \\
 \int d\psi_0\,d\psi_1\, \rho_0^{\otimes n} \otimes \rho_1^{\otimes (n+1)} &=&\sum_{\xi} p^n_{\xi}\sigma^n_{1,\xi}   \label{ch5/sigmaxi2} ,
\end{eqnarray}
where we use the shorthand notation $\xi=\{ j_A, j_C\}$ 
[each angular momentum ranges from $0$ ($1/2$) to $n/2$ for $n$ even (odd)], and
$p^n_\xi=p^n_{j_A}p^n_{j_C}$ is the probability of any of the two average states projecting on the block labelled~$\xi$. 
%
%
%
%
%
Hence 
\begin{equation}\label{ch5/delta LM sum xi}
\Delta^{\rm LM}= \sum_\xi p^n_\xi \trnorm{ \sigma^n_{0,\xi} -\sigma^n_{1,\xi} } \,.
\end{equation}
%
%
%
The number of terms in Eq.~\eqref{ch5/delta LM sum xi} is $[(2n+3\pm 1)/4]^2$ for even/odd $n$.
It grows quadratically with $n$, in contrast to the pure state  case for which there is a single contribution corresponding to $j_A=j_C=n/2$. In the asymptotic limit of large $n$, however, a big simplification arises because of the following two results\footnote{Here we just state the results. We derive them in detail in Appendices~\ref{appB/sec:block} and \ref{appB/sec:robust}.}. The first result is that, for each $\xi$ of the form $\xi=\{j,j\}$ \mbox{($j_A=j_C=j$)}, the relation
\begin{equation}\label{ch5/ja=jc}
{\sigma^n_{0,\xi} -\sigma^n_{1,\xi} } ={r\langle\hat{J}_z\rangle_j\over j} \left({\sigma^{2j}_{0}-\sigma^{2j}_{1}}\right) 
\end{equation}
holds, where $\sigma^{2j}_{0/1}$ are the average states~\eqref{ch5/sigma states} for a number of~$2j$ \emph{pure} qubits. Here~$\langle\hat{J}_z\rangle_j$ is the expectation value restricted to~$\mathscr{S}_{j}$ of the \mbox{$z$-component} of the angular momentum in the state $\rho^{\otimes n}$, where $\rho$ has Bloch vector $r\vec{z}$. Eq.~\eqref{ch5/ja=jc} is an exact algebraic identity that holds for any value of $j$, $n$ and $r$ (it bears no relation whatsoever to measurements of any kind).  The second result is that, for large~$n$, both~$p^n_{j_A}$ and~$p^n_{j_C}$ become continuous probability distributions, $p_n(x_A)$ and $p_n(x_C)$, where $x_{A/C}=2 j_{A/C}/n\in[0,1]$. Asymptotically, they approach Dirac delta functions peaked at~\mbox{$x_A=x_C=r$}. Hence the only relevant  contribution to~$\Delta^{\rm LM}$ comes from~$\xi=\{rn/2,rn/2\}$. It then follows that in the asymptotic limit
\begin{equation}\label{ch5/s-s asymp}
\sum_\xi p^n_\xi\left(\sigma^n_{0,\xi}-\sigma^n_{1,\xi}\right)\simeq {2\langle\hat{J}_z\rangle_{rn/2}\over n}
\left(\sigma^{rn}_{0}-\sigma^{rn}_{1}\right)\,.
\end{equation}
%
%
This last equation tells us that mixed-state quantum classification using a TS of~size $2n$ is equivalent to its {\em pure}-state  version for a TS of size $2nr$, provided $n$ is asymptotically large. In particular, our proof of optimality above also holds for {\em arbitrary}~$r\in(0,1]$ if the TS is sizable enough, and~\mbox{$R^{\rm LM}\simeq R^{\rm opt}$}.
%
This result is much stronger than robustness against decoherence, which only would require optimality for values of $r$ close to unity. 

From Eqs.~\eqref{ch5/delta LM sum xi} and~\eqref{ch5/s-s asymp} one can easily compute~$\Delta^{\rm LM}$ for arbitrary $r$ using that~\citep{Gendra2012} $\langle\hat{J}_z\rangle_j\simeq j-(1-r)/(2r)$ up to exponentially vanishing terms. The trace norm of~$\sigma^{rn}_{0}-\sigma^{rn}_{1}$ can be retrieved from, e.g., Eq.~\eqref{ch5/ex risk}. For~$rn$ {\em pure} qubits one has $\trnorm{\sigma^{rn}_{0}-\sigma^{rn}_{1}}\simeq (4/3)[1-1/(rn)]$. After some trivial algebra we obtain 
\begin{equation}
P^{\rm LM}_{\rm e}={1\over2}-{r\over3}+{1\over3rn}+O(n^{-1})
\end{equation}
for the error probability, in agreement with the optimal programmable machine value given by Eq.~\eqref{ch4/mixed-subleading}, as claimed above. This corresponds to an 
excess risk of
\begin{equation}\label{ch5/rlm=ropt}
R^{\rm LM}={1\over3rn}+O(n^{-1}) = R^{\rm opt}\,.
\end{equation}

In the nonasymptotic case, the sum in Eq.~\eqref{ch5/delta LM sum xi} is not restricted to $\xi=\{j,j\}$ and the calculation of the excess risk becomes very involved. Rather than attempting to obtain an analytical result, for small training samples we have resorted to a numerical optimization.
%
We first note that
Eqs.~\eqref{ch5/omega} through \eqref{ch5/JA-JC} define a \emph{semidefinite programming} optimization problem (SDP), for which very efficient numerical algorithms have been developed~\citep{Vandenberghe1996}.  In this framework, one maximizes the objective function~$\Delta^{\rm LM}$ [second equality in Eq.~\eqref{ch5/Delta SDP}] of the SDP variables~${\Omega}_m\geqslant 0$, subject to the linear condition \eqref{ch5/omega_m}.
We use this approach to  compute the error probability, or equivalently, the excess risk of a LM for mixed-state quantum classification of small samples ($n\leqslant 5$), where no analytical expression of the optimal seed is known. 
For mixed states the expression of $\Gamma_\uparrow$ and $\Omega_m$ can be found in the Appendix, Eqs.~\eqref{appB/jA - jC 2} through~\eqref{appB/SDP methods 2}.

\begin{figure}[t]
\begin{center}
\includegraphics[scale=1.2]{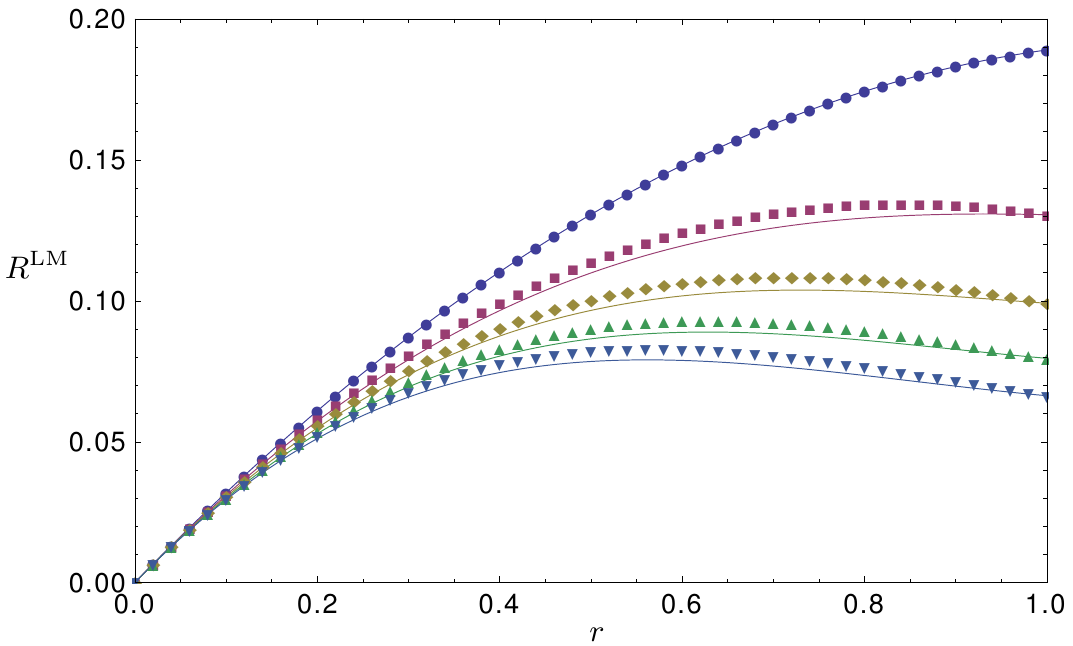}
\end{center}
\caption[Excess risks for optimal learning and programmable machines, as a function of the purity]{\label{ch5/fig:fig1}Excess risk $R^{\rm LM}$ (points) and its corresponding lower bound $R^{\rm opt}$ (lines), both as a function of the purity $r$, and for values of $n$ ranging from 1 to 5 (from top to bottom).}
\end{figure}

%
%
%
Our results are shown in Fig.~\ref{ch5/fig:fig1}, where we plot $R^{\rm LM}$ (shaped dots) and the lower bounds given by $R^{\rm opt}$ (solid lines) as a function of the purity $r$ for up to~$n=5$. We note that the excess risk of the optimal LM is always remarkably close to the absolute minimum provided by the optimal programmable machine, and in the worst case ($n=2$) it is only $0.4\%$ larger.
For $n=1$ we see that $R^{\rm LM}=R^{\rm opt}$ for any value of~$r$. This must be the case since for a single qubit in $A$ and $C$ one has $j_A=j_C=1/2$, and Eq.~\eqref{ch5/ja=jc} holds.
%
%
%


We now turn to robustness against statistical fluctuations in the number of states of each type produced by the source. In a real scenario one has to expect that~$j_A=n_A/2\not=n_C/2=j_C$, $n_A+n_B=2n$. Hence $\Gamma_\uparrow$ has the general form~\eqref{appB/jA - jC 2}, which gives us a hint that our choice $\Omega=\Omega_{m=0}$ may not be optimal for finite~$n$. This has been confirmed by numerical analysis using the same SDP approach discussed above. Here, we show that the asymptotic performance (for large training samples) of the optimal LM, however, is still the same as that of the optimal programmable discriminator running under the same conditions (mixed states and statistical fluctuations in $n_{A/C}$).  

Asymptotically, a real source for the problem at hand will typically produce~$n_{A/C}=n\pm \delta\sqrt{n}$ mixed copies of each type. 
In Appendix~\ref{appB/sec:robust}, it is shown that the relation~\eqref{ch5/s-s asymp} still holds in this case if $n$ is large. It reads
\begin{equation}\label{ch5/ja=jc asymp}
\sigma^n_{0,\xi}-\sigma^n_{1,\xi}  \simeq r \left(1-{1-r\over n r^2}\right)  \left(\sigma^{rn}_{0}-\sigma^{rn}_{1}\right)
\end{equation}
($\delta$ first appears at order $n^{-3/2}$). Hence the effect of both statistical fluctuations in $n_{A/C}$ and noise (already considered above) is independent of the machine used for quantum classification (i.e., it is the same for LM, programmable machines, E\&D, \dots). In particular, the relation~\eqref{ch5/rlm=ropt}, $R^{\rm LM}=R^{\rm opt}$, between the excess rate of the optimal LM and its absolute limit given by the optimal programmable discriminator still holds asymptotically, which proves robustness.

%
%
%
%
%

To illustrate this, let us consider the effect of statistical fluctuations in~$n_{A/C}$ for pure states.
The optimal programmable machine for arbitrary $n_A$, $n_B$ and $n_C$ is discussed in Appendix~\ref{appA/sec:na-nb-nc}. The error probability for the case at hand ($n_B=1$)
can be read off directly from Eq.~\eqref{ch4/measym}, and its asymptotic form when $n_A$ and $n_C$ are both very large can be easily derived using Euler-Maclaurin's summation formula. The result up to subleading order is
\begin{equation*}
P_{\rm e}^{\rm opt} \simeq \frac{1}{6} \left(1+\frac{1}{n_A}+\frac{1}{n_C}\right),
\end{equation*}
which leads to 
%
%
\begin{equation}\label{ch5/opt asym}
R^{\rm opt} = \frac{1}{6} \left(\frac{1}{n_A}+\frac{1}{n_C}\right)+\dots\,.
\end{equation}
We see that when~$n_{A/C}=n\pm \delta\sqrt{n}$ (i.e., when statistical fluctuations in $n_{A/C}$ are taken into account) one still has~$R^{\rm opt}\simeq1/(3n)\simeq R^{\rm LM}$. 
\\

\section{Discussion}


We have presented  a \emph{supervised} quantum learning machine that classifies a single qubit prepared in a pure but otherwise unknown state after it has been trained with a number of already classified qubits.
Its performance attains the absolute bound given by the optimal programmable discrimination machine. 
This learning machine does not require quantum memory and can also be reused without retraining, 
which may save a lot of resources.
The machine has been shown to be robust against noise and statistical fluctuations in the number of states of each type produced by the source.
For small sized training sets the machine is very close to optimal, attaining an excess risk that is larger than the absolute lower limit by at most $0.4\%$. In the absence of noise and statistical fluctuations, the machine attains optimality for {\em any} size of the training set.


One may rise the question of whether or not the separated measurements on the training set and data qubit can be reversed in time; in a classical scenario where, e.g., one has  to identify one of two faces based on a stack of training portraits, it is obvious that, without memory limitations, the order of training and data observation can be reversed (in both cases the final decision is taken based on the very same information).
We will briefly show that this is not so in the quantum world.
In the reversed setting, the machine first performs a measurement~${\mathscr D}$, with each element of rank one, \mbox{$u_\mu[\,\uparrow\,] u^\dagger_\mu$}, and stores the information (which of the possible outcomes is obtained) in the classical memory to control the measurement to be performed on the training set in a later time. 
The probability of error conditioned to one of the outcomes, say $\uparrow$, is given by the Helstrom formula $P_{\rm e}^{\uparrow}=(1-\trnorm{\Gamma_\uparrow}/2)/2$, where $\Gamma_\uparrow$ is defined in Eq.~\eqref{ch5/def GammaUp}. Using Eq.~\eqref{ch5/JA-JC} one has $\trnorm{\Gamma_\uparrow}=d^{-2}_n d^{-1}_{n+1}\sum_{m,m'}|m-m'|=n/[3(n+1)]$. The averaged error probability is then
\begin{equation}\label{ch5/PeLMback}
P^{\stackrel{{\rm LM}}{\mbox{\tiny$\leftarrow$}}}_{\rm e}=\frac{1}{2} \left(1-\frac{1}{6}\frac{n}{n+1}\right).
\end{equation}
In the limit of infinite copies we obtain $P^{\stackrel{{\rm LM}}{\mbox{\tiny$\leftarrow$}}}_{\rm e}\simeq 5/12$, which is way larger than $P^{\rm LM}_{\rm e}\simeq 1/6$. The same minimum-error probability of~Eq.~\eqref{ch5/PeLMback} can be attained by performing a Stern-Gerlach measurement on the data qubit, which requires just one bit of classical memory. This is all the classical information that we can hope to retrieve from the data qubit, in agreement with Holevo's bound~\citep{Holevo1973}. This clearly limits the possibilities of a correct classification---very much in the same way as in face identification with limited memory size.
In contrast, the amount of classical information ``sent forward'' in the optimal learning machine goes as the logarithm of the size of the training sample.  This asymmetry also shows that, despite the separability of the measurements, nonclassical correlations between the training set and the data qubit play an important role in quantum learning. 



\chapter{\chnamesix}

\label{ch6_learningcv}

This Chapter analyses the effect of uncertainty in discriminating between two coherent states in a learning context, following the scheme for qubits presented in the previous chapter. Coherent states are the states produced by an ideal laser, and they comprise a very specific class among the states of continuous-variables (CV) systems, i.e., quantum systems with Hilbert spaces of infinite dimension like, for instance, the bosonic modes of an electromagnetic field. States of this type have been absent up to this point in the dissertation (only finite-dimensional systems have been considered so far), hence a few words about them are in order. Also, the mathematical toolbox required to deal with CV systems is quite different. For a technical overview on the basic tools needed for this Chapter, refer to Appendix~\ref{appC}.

The quantum information research field divides itself in two branches, depending on the subject of study: finite dimensional systems, and CV systems. While traditionally the biggest efforts were put into the former type of systems, the study of CV systems as resources for quantum information processing has gradually become a matter of paramount importance. 
CV states have displayed great versatility within the field,
from the ease in their preparation and control in the experimental ground to their utility as subjects of genuinely quantum information processing tasks, such as 
quantum teleportation, quantum cloning, quantum key distribution, and quantum dense coding~\citep{Braunstein2005,Eisert2003,Cerf2007}. 
Most of the attention in the field of quantum information with CV systems is focused on Gaussian states, that is, the class of CV states that follow Gaussian statistics \mbox{\citep{Weedbrook2012}}. 
%
This is mainly so for two reasons: first, Gaussian states have a very simple mathematical characterization 
and, second, they describe appropriately the most common states of light that are realized with current technology.

The discrimination of Gaussian states plays a central role in the CV framework and, among all Gaussian states, coherent states stand out for its relevance in quantum optical communication theory. Lasers are widely used in current telecommunication systems, and the transmission of information can be theoretically modelled by bits encoded in the amplitude or phase modulation of a laser beam. The basic task of distinguishing two coherent states in an optimal way is thus of great interest, since lower chances of misidentification translate into higher transfer rates between the sender and the receiver.

The discrimination of coherent states has been considered within the two main approaches, namely minimum-error (Section~\ref{ch3/sec:minimumerror}) and unambiguous discrimination (Section~\ref{ch3/sec:unambiguous}), although the former is more developed. Generically, a logical bit can be encoded in two possible coherent states $\ket{\alpha}$ and $\ket{-\alpha}$, via a phase shift, or in the states $\ket{0}$ and $\ket{2\alpha}$, via amplitude modulation. Both encoding schemes are equivalent, since one can move from one to the other by applying a displacement operator $\hat{D}(\alpha)$ [cf. Eq.~\eqref{appC/weyl}] to both states. In the minimum-error approach, the theoretical minimum for the probability of error is simply given by the Helstrom formula for pure states \eqref{ch3/helstrompure}, as
\begin{equation}
P_{\rm e} = \frac{1}{2} \left(1-\sqrt{1-e^{-4|\alpha|^2}}\right) \,,
\end{equation}
where the overlap $|\braket{\alpha}{\beta}|^2=e^{-|\alpha-\beta|^2}$ has been used, and the probabilities of occurrence of each possible state have been taken to be equal for simplicity. A variety of implementations have been devised to achieve 
this discrimination task,
e.g., the Kennedy receiver \citep{Kennedy1973}, based on photon counting; the Dolinar receiver \citep{Dolinar1973}, a modification of the Kennedy receiver with real-time quantum feedback; and the homodyne receiver (see Section~\ref{appC/sec:homodyne})\footnote{While the latter is the simplest procedure, it does not achieve optimality. However, for weak coherent states ($|\alpha|^2<0.4$), it yields an error probability very close to the optimal value $P_{\rm e}$, and it is optimal among all Gaussian measurements \citep{Takeoka2008}. In fact, just one of the three mentioned, the Dolinar receiver, is optimal.}. Concerning the unambiguous approach to the discrimination problem, results include the unambiguous discrimination between two known coherent states~\citep{Chefles1998a,Banaszek1999}, and its \emph{programmable} version (see Chapter~\ref{ch4_pqsd}), i.e., when the value of the amplitude $\alpha$ is completely unknown~\citep{Sedlak2007,Sedlak2009,Bartuskova2008}.

The purpose of this Chapter is to explore the fundamental task of discriminating between two coherent states with minimum error, when the available information about their amplitudes is incomplete. The simplest instance of such problem is a partial knowledge situation: the discrimination between the vacuum state, $\ket{0}$, and some coherent state, $\ket{\alpha}$, where the value of $\alpha$ is not provided beforehand in the classical sense, but instead embedded in a number $n$ of auxiliary modes in the state $\ket{\alpha}^{\otimes n}$. 
Again, such discrimination scheme can be cast as a learning protocol, thus extending the concepts established in Chapters \ref{ch4_pqsd} and \ref{ch5_learning} to the CV realm, and we face the question of whether this learning form matches the performance of the most general quantum protocol.
%

Before starting with our results and to motivate the problem investigated in this Chapter, let me define the specifics of the setting in the context of a quantum-enhanced readout of classically-stored information.
\\

\section{Quantum reading of classical information}

Imagine a classical memory register modelled by an array of cells, where each cell contains a reflective medium with two possible reflectivities $r_0$ and $r_1$. To read the information stored in the register, one shines light into one of the cells and analyses its reflection. The task essentially consists in discriminating the two possible states of the reflected signal, which depend on the reflectivity of the medium and thus encode the logical bit stored in the cell. In the seminal paper of \emph{quantum reading} \citep{Pirandola2011}, the author takes advantage of ancillary modes to prepare an initial entangled state between those and the signal. The reflected signal is sent together with the ancillas to a detector, where a joint discrimination measurement is performed. A purely quantum resource---entanglement---is thus introduced, enhancing the probability of a successful identification of the encoded bit\footnote{In particular, Pirandola shows that a two-mode squeezed vacuum state outperforms any classical light, in the regime of few photons and high reflectivity memories.}.
The idea of using nonclassical light to retrieve classical information can be traced back to the precursory work of \emph{quantum illumination} \citep{Lloyd2008a,Tan2008}, where the presence of a low-reflectivity object in a bright thermal-noise bath is detected with higher accuracy when entangled light is sent to illuminate the target region.


%
\begin{figure}[t]
\begin{center}
\includegraphics[scale=.6]{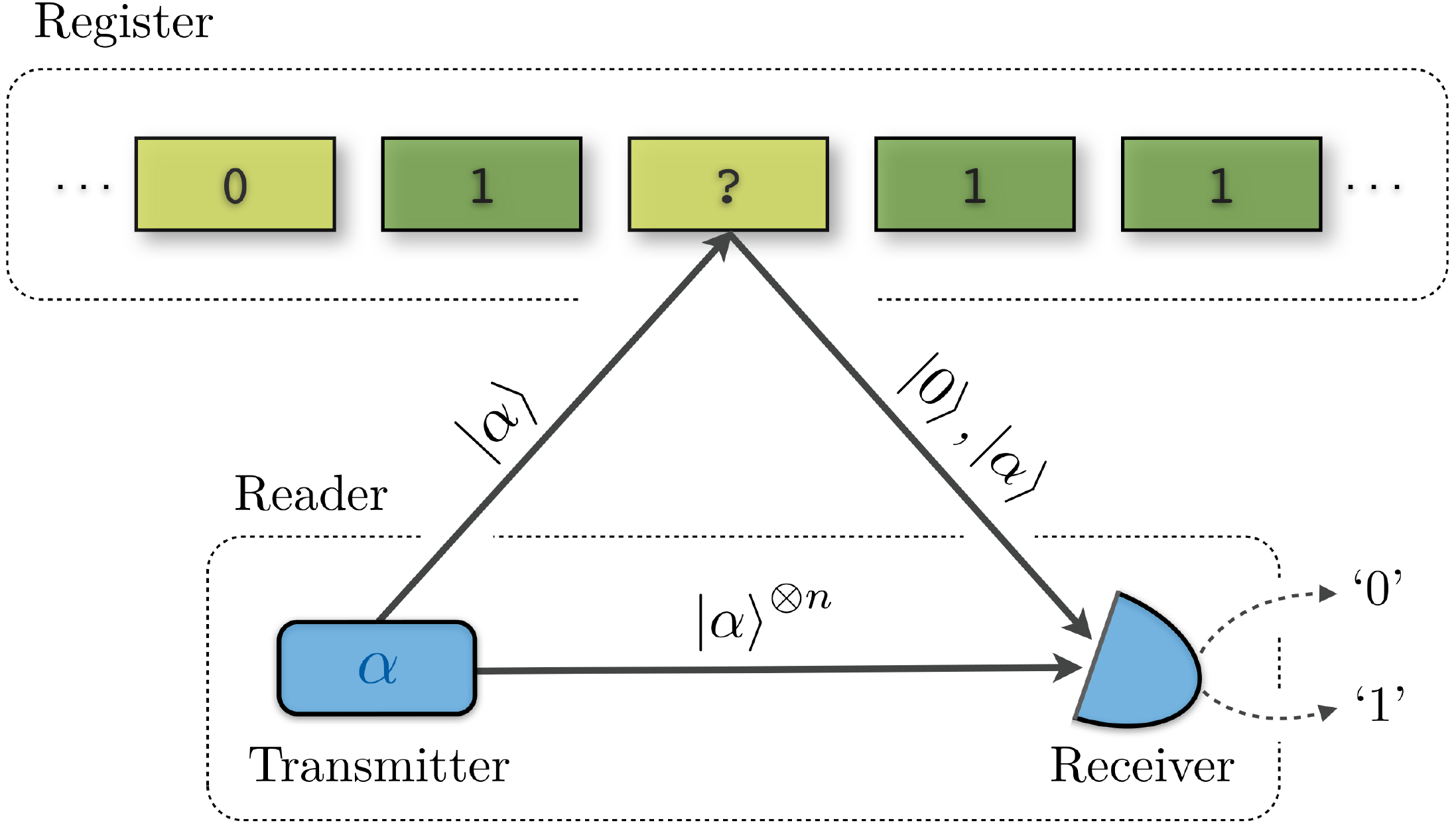}
\end{center}
\caption[Scheme of quantum reading]{A quantum reading scheme that uses a coherent signal $\ket{\alpha}$, produced by a transmitter, to illuminate a cell of a register that stores a bit of information. A receiver extracts this bit by distinguishing between the two possible states of the reflected signal, $\ket{0}$ and $\ket{\alpha}$, assisted by $n$ auxiliary modes sent directly by the transmitter.\label{ch6/fig:fig1}}
\end{figure}

In this Chapter we consider a reading scenario with an imperfect coherent light source and no initial entanglement involved. The proposed scheme is as follows (see Fig.~\ref{ch6/fig:fig1}). We model an ideal classical memory by a \emph{register} made of cells that contain either a transparent medium ($r_0=0$) or a highly reflective one ($r_1=1$). A \emph{reader}, comprised by a \emph{transmitter} and a \emph{receiver}, extracts the information of each cell. The transmitter is a source that produces coherent states of a certain amplitude $\alpha$. The value of $\alpha$ is not known with certainty due, for instance, to imperfections in the source, but it can be statistically localised in a Gaussian distribution around some (known) $\alpha_0$.
A signal state $\ket{\alpha}$ is sent toward a cell of the register and, if it contains the transparent medium, it goes through; if it hits the highly reflective medium, it is reflected back to the receiver in an unperturbed form. This means that we have two possibilities at the entrance of the receiver upon arrival of the signal: either nothing arrives, and we represent this situation as the vacuum state $\ket{0}$, or it is the reflected signal, which is represented by the same signal state $\ket{\alpha}$. To aid in the discrimination of the signal, we alleviate the effects of the uncertainty in $\alpha$ by considering that $n$ auxiliary modes are produced by the transmitter in the global state $\ket{\alpha}^{\otimes n}$ and sent directly to the receiver. The receiver then performs measurements over the signal and the auxiliary modes and outputs a binary result, corresponding with some probability to the bit stored in the irradiated cell.

We now set ourselves to answer the following questions: (i) which is the optimal (unrestricted) measurement, in terms of the error probability, that the receiver can perform? and (ii) is a \emph{joint} measurement, performed over the signal together with the auxiliary modes, necessary to achieve optimality? To do so, we first obtain the optimal minimum-error probability considering collective measurements (Section~\ref{ch6/sec:collective}). 
Then, we contrast the result with that of the obvious estimate-and-discriminate (E\&D) strategy,
consisting in first estimating $\alpha$ by measuring the auxiliary modes, and then using the acquired information to determine the signal state by a discrimination measurement tuned to distinguish the vacuum state $\ket{0}$ from a coherent state with the estimated amplitude (Section~\ref{ch6/sec:local}). 
In order to compare the performance of the two strategies we focus on the asymptotic limit of large $n$.
We show that a collective measurement provides a lower excess risk\footnote{Recall, from Chapter~\ref{ch5_learning}, that the excess risk is defined as the excess asymptotic average error over discrimination when the states ($\alpha$ in this case) are known.} than any Gaussian E\&D strategy, and we conjecture (and provide strong evidence) that this is the case for all local strategies.
\\

\section{Collective strategy}\label{ch6/sec:collective}

The global state that arrives at the receiver can be expressed as either \mbox{$\left[ \alpha \right]^{\otimes n} \otimes [0]$} or $\left[ \alpha \right]^{\otimes n} \otimes [\alpha]$, where recall the shorthand notation $[\,\cdot\,]\equiv\ketbrad{\,\cdot\,}$. 
For simplicity, we take equal \emph{a priori} probabilities of occurrence of each state. We will always consider the signal state to be that of the last mode, and all the previous modes will be the auxiliary ones.
First of all, note that the information carried by the auxiliary modes can be conveniently ``concentrated'' into a single mode by means of a sequence of unbalanced beam splitters\footnote{See, e.g., Section III A in \citep{Sedlak2008} for details.}. The action of a beam splitter over a pair of coherent states $\ket{\alpha}\otimes\ket{\beta}$ yields
\begin{equation}
\ket{\alpha}\otimes\ket{\beta} \,\longrightarrow\, |\sqrt{T}\alpha+\sqrt{R}\beta\rangle\otimes|-\sqrt{R}\alpha+\sqrt{T}\beta\rangle \,,
\end{equation}
where $T$ is the transmissivity of the beam splitter, $R$ is its reflectivity, and $T+R=1$. A balanced beam splitter ($T=R=1/2$) acting on the first two auxiliary modes thus returns $\ket{\alpha}\otimes\ket{\alpha}\longrightarrow|\sqrt{2}\alpha\rangle\otimes\ket{0}$.
Since the beam splitter preserves the tensor product structure of the two modes, one can treat separately the first output mode and use it as input in a second beam splitter, together with the next auxiliary mode. By choosing appropriately the values of $T$ and $R$, the transformation $|\sqrt{2}\alpha\rangle\otimes\ket{\alpha}\longrightarrow|\sqrt{3}\alpha\rangle\otimes\ket{0}$ can be achieved. Applying this process sequentially over the $n$ auxiliary modes, we perform the transformation
\begin{equation}
\ket{\alpha}^{\otimes n} \,\longrightarrow\, |\sqrt{n}\alpha\rangle\otimes\ket{0}^{\otimes n-1}\,.
\end{equation}
Note that this is a deterministic process, and that no information is lost, for it is contained completely in the complex parameter $\alpha$. This operation allows us to effectively deal with only two modes. The two possible global states entering the receiver hence become $[\sqrt{n}\alpha]\otimes[0]$ and $[\sqrt{n}\alpha]\otimes[\alpha]$.

The parameter $\alpha$ is not known with certainty. Building on the Bayesian ideas used in Chapter~\ref{ch4_pqsd} to embed this lack of information into \emph{average} global states, we immediately see that a flat prior distribution for $\alpha$, as we considered for qubits, is not reasonable in this case. On the one hand, such prior would yield divergent average states of infinite energy, since the phase space is infinite. On the other hand, in a real situation it is not reasonable at all to assume that \emph{all} amplitudes $\alpha$ are equally probable. 
The usual procedure in these cases is to consider that a small number of auxiliary modes is used to make a rough estimation of $\alpha$, such that our prior becomes a Gaussian probability distribution centred at $\alpha_0$, which width goes as $\sim 1/\sqrt{n}$ \footnote{Since we are interested in comparing the asymptotic performance of discrimination strategies in the limit of large $n$, the number of modes used for the rough estimation is negligible, i.e., $\tilde{n}=n^{1-\epsilon}$. Then, it can be shown that $\alpha$ belongs to a neighbourhood of size $n^{-1/2+\epsilon}$ centred at $\alpha_0$, with probability converging to one (this is shown, though in a classical statistical context, in \citep{Gill1995}). Moreover, this happens to be true for any model of i.i.d. quantum states $\rho$ (regardless their dimensionality), hence the analysis of the asymptotic behaviour of any estimation model of this sort can be restricted to a local Gaussian model, centred at a fixed state $\rho_0$. This is known as \emph{local asymptotic normality} \citep{Gill2013}.}. 
Under these considerations, we express the true amplitude $\alpha$ as
\begin{equation}\label{ch6/alphaisalpha0}
\alpha \approx \alpha_0 + u/\sqrt{n} \, , \quad u \in \mathbb{C} \,,
\end{equation}
where the parameter $u$ follows the Gaussian distribution
\begin{equation}\label{ch6/gaussian}
G(u) = \frac{1}{\pi \mu^2} e^{-u^2/\mu^2} \,.
\end{equation}
%
To avoid divergences, we have introduced the free parameter $\mu$ as a temporal energy cut-off that defines the width of $G(u)$. Once we have expressions for the excess risks in the asymptotic regime of large $n$, we will remove the cut-off dependence by taking the limit $\mu\to\infty$.

Using the prior information acquired through the rough estimation, that is Eqs.~\eqref{ch6/alphaisalpha0} and \eqref{ch6/gaussian}, we compute the average global states arriving at the receiver
\begin{eqnarray}
\sigma_1 &=& \int G(u) \, [ \sqrt{n} \alpha_0 + u ] \otimes [0] \,d^2u \, ,\\
\sigma_2 &=& \int G(u) \, [ \sqrt{n} \alpha_0 + u ] \otimes [\alpha_0 + u/\sqrt{n}\,]\, d^2u \, .
\end{eqnarray}
The optimal measurement to determine the state of the signal is the Helstrom measurement for the discrimination of the states $\sigma_1$ and $\sigma_2$, that yields the average minimum-error probability [cf. Eq.~\eqref{ch3/helstrom}]
\begin{equation}\label{ch6/perror col}
P_{\rm e}^{\rm opt}(n)= \frac{1}{2} \left(1-\frac{1}{2}\trnorm{\sigma_1-\sigma_2} \right) \,.
\end{equation}
\footnote{Note that, \emph{sensu stricto}, the dependence of $P_{\rm e}^{\rm opt}(n)$ on the localisation parameter $\alpha_0$ should be made explicit. Keep in mind that, in general, all quantities computed in this Chapter will depend on $\alpha_0$. Thus for the sake of notation clarity, we omit it hereafter when no confusion arises.}The technical difficulty in computing $P_{\rm e}^{\rm opt}(n)$ resides in that $\sigma_1-\sigma_2$ is an infinite-dimensional full-rank matrix, hence its trace norm does not have a computable analytic expression for arbitrary finite $n$. Despite this, one can still resort to analytical methods in the asymptotic regime $n\to\infty$ by treating the states perturbatively. 

To ease this calculation, we first apply the displacement operator 
\begin{equation}\label{ch6/displacement}
\hat{D}(\alpha_0) = \hat{D}_1 (-\sqrt{n}\alpha_0) \otimes \hat{D}_2(-\alpha_0)
\end{equation}
to the states $\sigma_1$ and $\sigma_2$, where $\hat{D}_1$ ($\hat{D}_2$) acts on the first (second) mode, and we obtain the displaced global states
\begin{eqnarray}
\bar{\sigma}_1 &=& \hat{D}(\alpha_0)\sigma_1\hat{D}^\dagger(\alpha_0) = \int G(u) \left[  u \right] \otimes \left[-\alpha_0 \right] d^2u \,,\label{ch6/barsigma1}\\
\bar{\sigma}_2 &=& \hat{D}(\alpha_0)\sigma_2\hat{D}^\dagger(\alpha_0) = \int G(u) \left[  u \right] \otimes [ u/\sqrt{n}\,]\, d^2u \,.\label{ch6/barsigma2}
\end{eqnarray}
Since both states have been displaced the same amount, the trace norm does not change, i.e., $\trnorm{\sigma_0-\sigma_1}=\trnorm{\bar{\sigma}_0-\bar{\sigma}_1}$. Eq.~\eqref{ch6/barsigma1} directly yields
\begin{equation}\label{ch6/barsigma1exp}
\bar{\sigma}_1 = \sum_{k=0}^\infty c_k [k] \otimes [-\alpha_0] \, ,
\end{equation}
where $c_k = \mu^{2k}/[(\mu^2+1)^{k+1}]$. Note that, as a result of the average, the first mode in Eq.~\eqref{ch6/barsigma1exp} corresponds to a thermal state with average photon number $\mu^2$. Note also that the $n$-dependence is entirely in $\bar{\sigma}_2$. In the limit $n\to\infty$, we can expand the second mode of $\bar{\sigma}_2$ by expressing it in the Fock basis as
\begin{equation}
|u/\sqrt{n}\,\rangle = e^{-\frac{|u|^2}{2n}} \sum_k \frac{(u/\sqrt{n})^k}{\sqrt{k!}} \ket{k} \,.
\end{equation}
Then, up to order $1/n$ its asymptotic expansion gives
\begin{align}\label{ch6/series_u}
[u/\sqrt{n}\,] &\sim \ketbrad{0} + \frac{1}{\sqrt{n}} \left( u \ketbra{1}{0} + u^* \ketbra{0}{1} \right) \nonumber\\
&+ \frac{1}{n} \left\{ |u|^2 \left( \ketbrad{1}-\ketbrad{0}\right) + \frac{1}{\sqrt{2}} \left[ u^2 \ketbra{2}{0} + \left(u^*\right)^2 \ketbra{0}{2} \right] \right\} \, .
\end{align}
%
%
Inserting Eq.~\eqref{ch6/series_u} into Eq.~\eqref{ch6/barsigma2} and computing the corresponding averages of each term in the expansion, we obtain a state of the form 
\begin{equation}\label{ch6/barsigma2exp}
\bar{\sigma}_2 \sim \bar{\sigma}_2^{(0)} + \frac{1}{\sqrt{n}}\bar{\sigma}_2^{(1)} + \frac{1}{n}\bar{\sigma}_2^{(2)} \,.
\end{equation}
We can now use Eqs.~\eqref{ch6/barsigma1exp} and \eqref{ch6/barsigma2exp} to compute the trace norm $\trnorm{\bar{\sigma}_1-\bar{\sigma}_2}$ in the asymptotic regime of large $n$, up to order $1/n$, by applying perturbation theory. The explicit form of the terms in Eq.~\eqref{ch6/barsigma2exp}, as well as the details of the computation of the trace norm, are given in Appendix~\ref{appD/sec:tracenormcol}.
Here we just show the result: the average minimum-error probability $P_{\rm e}^{\rm opt}(n)$, defined in Eq.~\eqref{ch6/perror col}, can be written in the asymptotic limit as
\begin{equation}\label{ch6/perror col2}
P_{\rm e}^{\rm opt} \equiv P_{\rm e}^{\rm opt}(n\to\infty) \sim \frac{1}{2} \left[1-\sqrt{1-e^{-|\alpha_0|^2}} -\frac{1}{2n}\left(\Lambda_+^{(2)}-\Lambda_-^{(2)} \right) \right] \,,
\end{equation}
where $\Lambda_\pm^{(2)}$ is given by Eq.~\eqref{appD/Lambda2pm}.
\\

\subsection*{Excess risk}

The figure of merit that we use to assess the performance of our protocol is the \textit{excess risk}, defined as the difference between the asymptotic average error probability $P_{\rm e}^{\rm opt}$ and the average error probability for the optimal strategy when $\alpha$ is perfectly known. As we said at the beginning of the section, the true value of $\alpha$ is $\alpha_0+u/\sqrt{n}$ for a particular realization, thus knowing $u$ equates knowing $\alpha$. The minimum-error probability for the discrimination between the \emph{known} states $\ket{0}$ and $\ket{\alpha_0+u/\sqrt{n}}$, $P_{\rm e}^*(u,n)$, averaged over the Gaussian distribution $G(u)$, takes the form
\begin{align}
P_{\rm e}^*(n) &= \int G(u) \,P_{\rm e}^* (u,n) \,d^2u \nonumber\\
&= \int G(u) \,\frac{1}{2} \left(1-\sqrt{1-|\!\braket{0}{\alpha_0+u/\sqrt{n}}\!|^2} \right) d^2u \,. \label{ch6/perror known int}
\end{align}
To compute this integral we do a series expansion of the overlap in the limit $n\rightarrow \infty$ and we use Eqs.~\eqref{ch6/int u1}, \eqref{ch6/int u2}, and \eqref{ch6/int u3}. After some algebra we obtain
\begin{equation}\label{ch6/perror known}
P_{\rm e}^* \equiv P_{\rm e}^*(n\to\infty) \sim \frac{1}{2} \left(1-\sqrt{1-e^{-|\alpha_0|^2}} + \frac{1}{n} \Lambda^* \right)\,,
\end{equation}
where
\begin{equation}\label{ch6/perror known delta}
\Lambda^* = \frac{\mu^2 \left[ 2\left(e^{-|\alpha_0|^2}-1\right) +|\alpha_0|^2 \left(2-e^{-|\alpha_0|^2}\right) \right]}{4 \left(e^{|\alpha_0|^2}-1\right) \sqrt{1-e^{-|\alpha_0|^2}}} \,.
\end{equation}
The excess risk is then given by Eqs.~\eqref{ch6/perror col2} and \eqref{ch6/perror known} as
\begin{equation}
R^{\rm opt}_\mu = n \left(P_{\rm e}^{\rm opt} - P_{\rm e}^*\right) \,.
\end{equation}
Finally, we remove the cut-off imposed at the beginning by taking the limit $\mu \rightarrow \infty$ and we obtain
\begin{equation}\label{ch6/excess_risk_opt}
R^{\rm opt} =\lim_{\mu\to\infty} R^{\rm opt}_\mu = \frac{|\alpha_0|^2 e^{-|\alpha_0|^2/2}\left(2e^{|\alpha_0|^2}-1\right)}{16 \left( e^{|\alpha_0|^2}-1 \right)^{3/2}} \,.
\end{equation}
Note that the excess risk only depends on the module of $\alpha_0$, i.e., on the average distance between $\ket{\alpha}$ and $\ket{0}$. The excess risk is thus phase-invariant, as it should.

Eq.~\eqref{ch6/excess_risk_opt} is the first piece of information we need for addressing the main question posed at the beginning, namely whether the optimal performance of the collective strategy 
is achievable by an estimate-and-discriminate (E\&D) strategy.
We now move on for the second piece.
\\

\section{E\&D strategy}\label{ch6/sec:local}

An alternative---and more restrictive---strategy to determine the state of the signal consists in the natural combination of two fundamental tasks: state estimation, and state discrimination of known states. In such an 
E\&D strategy, \emph{all} auxiliary modes are used to better estimate the unknown amplitude $\alpha$. Then, the obtained information is used to tune a discrimination measurement over the signal that distinguishes the vacuum state from a coherent state with the estimated amplitude. In this Section we find the optimal E\&D strategy based on Gaussian measurements and compute its excess risk $R^{\rm E\&D}$. Then, we compare the result with that of the optimal collective strategy $R^{\rm opt}$.

The most general Gaussian measurement that one can use to estimate the state of the auxiliary mode $|\sqrt{n}\alpha\rangle$ is a \emph{generalized heterodyne measurement} (see Appendix \ref{appC/sec:heterodyne}), represented by a POVM with elements
\begin{equation}\label{ch6/localpovm}
E_{\bar{\beta}} = \frac{1}{\pi} \,|\bar{\beta},r,\phi\rangle\!\langle\bar{\beta},r,\phi| \,,
\end{equation}
i.e., projectors onto pure Gaussian states with amplitude $\bar{\beta}$ and squeezing $r$ along the direction $\phi$. The outcome of such heterodyne measurement $\bar{\beta}=\sqrt{n}\beta$ produces an estimate for $\sqrt{n}\alpha$, hence $\beta$ stands for an estimate of $\alpha$ \footnote{In our notation, the outcome of the measurement also labels the estimate, so $\beta$ stands for both indistinctly. This should generate no confusion, since the trivial guess function that uses outcome $\bar{\beta}$ to produce the estimate $\beta$ does not vary throughout the chapter.}. Upon obtaining $\bar{\beta}$, the prior information that we have about $\alpha$ gets updated according to Bayes' rule, so that now the signal state can be either $\ketbrad{0}$ or some state $\rho(\beta)$. The form of this second hypothesis is given by
\begin{equation}
\rho(\beta) = \int p(\alpha|\beta) \ketbrad{\alpha} d^2\alpha \,,
\end{equation}
where $p(\alpha|\beta)$ encodes the \emph{posterior} information that we have acquired via the heterodyne measurement. It represents the conditional probability of the state of the auxiliary mode being $\ket{\sqrt{n}\alpha}$, given that we obtained the 
outcome $\bar{\beta}$. Bayes' rule dictates
\begin{equation}
p(\alpha|\beta) = \frac{p(\beta|\alpha) p(\alpha)}{p(\beta)} \,,
\end{equation}
where $p(\beta|\alpha)$ is given by (see Appendix \ref{appD/sec:heterodyne_prob})
\begin{equation}\label{ch6/heterodyne_prob_ab}
p(\beta|\alpha) = \frac{1}{\pi \cosh r} e^{-|\sqrt{n} \alpha - \bar{\beta}|^2-{\rm Re}[(\sqrt{n} \alpha-\bar{\beta})^2 e^{-i 2 \phi}] \tanh r} \,,
\end{equation}
$p(\alpha)$ is the prior information of $\alpha$ before the heterodyne measurement, and 
\begin{equation}
p(\beta) = \int p(\alpha) p(\beta|\alpha) d^2\alpha
\end{equation}
is the total probability of giving the estimate $\beta$.

The error probability of the E\&D strategy, averaged over all possible estimates $\beta$, is then
\begin{equation}\label{ch6/perror_eyd}
P_{\rm e}^{\rm E\&D}(n) = \frac{1}{2} \left(1-\frac{1}{2} \int p(\beta) \trnorm{\ketbrad{0}-\rho(\beta)} d^2\beta \right) \,.
\end{equation}
Note that the estimate $\beta$ depends ultimately on the number $n$ of auxiliary modes, hence the explicit dependence in the left-hand side of Eq.~\eqref{ch6/perror_eyd}.

We are interested in the asymptotic expression of Eq.~\eqref{ch6/perror_eyd}, so let us now move to the $n\to\infty$ scenario. Recall that an initial rough estimation of $\alpha$ permits the localisation of the prior $p(\alpha)$ around a central point $\alpha_0$, such that $\alpha \approx \alpha_0 + u/\sqrt{n}$, where $u$ is distributed according to $G(u)$, defined in Eq.~\eqref{ch6/gaussian}. Consequently, the estimate $\beta$ will also be localised around the same point, i.e., $\beta\approx\alpha_0+v/\sqrt{n}$, $v\in\mathbb{C}$. As a result, we can effectively shift from amplitudes $\alpha$ and $\beta$ to a local Gaussian model around $\alpha_0$, parametrized by $u$ and $v$. According to this new model, we make the following transformations:
\begin{eqnarray}
p(\alpha) &\rightarrow& G(u) \,,\\
p(\beta|\alpha) &\rightarrow& p(v|u) = \frac{1}{\pi \cosh r} e^{-|u-v|^2-{\rm Re}[(u-v)^2] \tanh r} \,,\\
p(\beta) &\rightarrow& p(v) = \int p(v|u) G(u) du = \frac{1}{\pi \cosh r} \frac{1}{\sqrt{1+\mu^2\left(2+\frac{\mu^2}{\cosh^2 r}\right)}} \nonumber\\
&\,&\times\; {\rm exp}\left(\frac{|v|^2\left(1+\frac{\mu^2}{\cosh^2 r}\right)+{\rm Re}[v^2]\tanh r}{\mu^4 \tanh^2 r-\left(\mu^2+1\right)^2}\right) \,,\label{ch6/pv}\\
p(\alpha|\beta) &\rightarrow& p(u|v)=\frac{p(v|u) G(u)}{p(v)} \,,\label{ch6/puv}
\end{eqnarray}
where, for simplicity, we have assumed $\alpha_0$ to be real. Note that this can be done without loss of generality. Note also that, by the symmetry of the problem, this assumption implies $\phi=0$.

The shifting to the local model 
transforms the trace norm in Eq.~\eqref{ch6/perror_eyd} as
\begin{equation}\label{ch6/tracenorm_eyd}
\trnorm{\ketbrad{0}-\rho(\beta)} \quad\rightarrow\quad \trnorm{\ketbrad{-\alpha_0}-\rho(v)} \,,
\end{equation}
where
\begin{equation}
\rho(v)=\int p(u|v) \, |u/\sqrt{n}\rangle\!\langle u/\sqrt{n}| \, d^2u \,.
\end{equation}
To compute the explicit expression of $\rho(v)$ we proceed as in the collective strategy. That is, we expand $\ketbrad{u/\sqrt{n}}$ in the limit \mbox{$n\rightarrow\infty$} up to order $1/n$, as in Eq.~\eqref{ch6/series_u}, and we compute the trace norm using perturbation theory (see Appendix~\ref{ch6/sec:eyd_tracenorm} for details).
The result allows us to express the asymptotic average error probability of the E\&D strategy as
\begin{equation}\label{ch6/perror_eyd_asymp}
P_{\rm e}^{\rm E\&D} \equiv P_{\rm e}^{\rm E\&D}(n\to\infty) \sim \frac{1}{2}\left(1-\sqrt{1-e^{-\alpha_0^2}} + \frac{1}{n} \Delta^{\rm E\&D} \right)\,,
\end{equation}
where $\Delta^{\rm E\&D}$ is given by Eq.~\eqref{appD/DeltaEyD}.
\\

\subsection*{Excess risk}

The excess risk associated to the E\&D strategy is generically expressed as
\begin{equation}\label{ch6/excess_risk_eyd}
R^{\rm E\&D}(r) = n \lim_{\mu\to\infty} \left(P_{\rm e}^{\rm E\&D}-P_{\rm e}^*\right) \,,
\end{equation}
where $P_{\rm e}^*$ is the error probability for known $\alpha$, given in Eq.~\eqref{ch6/perror known}, and $P_{\rm e}^{\rm E\&D}$ is the result from the previous section, i.e., Eq.~\eqref{ch6/perror_eyd_asymp}. 
The full analytical expression for $R^{\rm E\&D}(r)$ is given in Eq.~\eqref{appD/excessrisk_eyd_r}.
Note that we have to take the limit $\mu\to\infty$ in the excess risk, as we did for the collective case. Note also that all the expressions calculated so far explicitly depend on the squeezing parameter $r$ (apart from $\alpha_0$). This parameter stands for the squeezing of the generalized heterodyne measurement in Eq.~\eqref{ch6/localpovm}, which we have left unfixed on purpose. As a result, we now define, through the squeezing $r$, the optimal heterodyne measurement over the auxiliary mode to be that which yields the lowest excess risk \eqref{ch6/excess_risk_eyd}, i.e.,
\begin{equation}\label{ch6/excess_risk_eyd_2}
R^{\rm E\&D} = \min_{r} R^{\rm E\&D} (r) \,.
\end{equation}

To find the optimal $r$, we look at the parameter estimation theory of Gaussian models (see, e.g., \citep{Gill2013}). In a generic two-dimensional Gaussian shift model, the optimal measurement for the estimation of a parameter $\theta = (q,p)$ is a generalized heterodyne measurement\footnote{This is the case whenever the covariance of the Gaussian model is known, and the mean is a linear transformation of the unknown parameter.} of the type \eqref{ch6/localpovm}. Such measurement yields a quadratic risk of the form
\begin{equation}
R_{\hat{\theta}}=\int p(\theta) ((\hat{\theta}-\theta)^T G (\hat{\theta}-\theta)) d^2\theta \,,
\end{equation}
where $p(\theta)$ is some probability distribution, $\hat{\theta}$ is an estimator of $\theta$, and $G$ is a two-dimensional matrix. One can always switch to the coordinates system in which $G$ is diagonal, $G={\rm diag}(g_q,g_p)$, to write
\begin{equation}\label{ch6/quadratic_risk}
R_{\hat{\theta}}=g_q \int p(\theta) (\hat{q}-q)^2 d^2\theta + g_p \int p(\theta) (\hat{p}-p)^2 d^2\theta \,.
\end{equation}
It can be shown \citep{Gill2013} that the optimal squeezing of the estimation measurement, i.e., that for which the quadratic risk $R_{\hat{\theta}}$ is minimal, is given by
\begin{equation}
r=\frac{1}{4}\ln \left(\frac{g_q}{g_p}\right) \,.
\end{equation}
We can then simply compare Eq.~\eqref{ch6/quadratic_risk} with Eq.~\eqref{ch6/excess_risk_eyd} to deduce the values of $g_q$ and $g_p$ for our case. By doing so, we obtain that the optimal squeezing reads
\begin{figure}[t]
\begin{center}
\includegraphics[scale=1.4]{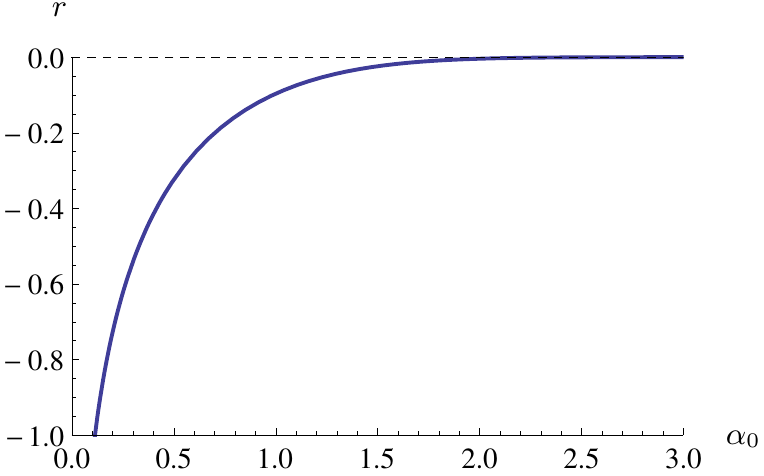}
\end{center}
\caption[Optimal squeezing for the generalized heterodyne measurement in a E\&D strategy]{Optimal squeezing $r$ for the generalized heterodyne measurement in a E\&D strategy, as a function of $\alpha_0$.\label{ch6/fig:fig2}}
\end{figure}
\begin{equation}\label{ch6/optimalsqueezing}
r = \frac{1}{4} \ln \left(\frac{f(\alpha_0)+\alpha_0^2}{f(\alpha_0)-\alpha_0^2}\right) \,,
\end{equation}
where
\begin{equation}
f(\alpha_0) = 2 e^{\alpha_0^2} \left(e^{\alpha_0^2}-1\right)\left(\sqrt{1-e^{-\alpha_0^2}}-1\right) 
+ \alpha_0^2\left(1-2e^{\alpha_0^2}\sqrt{1-e^{-\alpha_0^2}}\right) \,.
\end{equation}

Eq.~\eqref{ch6/optimalsqueezing} tells us that the optimal squeezing $r$ is a function of $\alpha_0$ that takes negative values, and asymptotically approaches zero when $\alpha_0$ is large (see Fig.~\ref{ch6/fig:fig2}). This means that the optimal estimation measurement over the auxiliary mode is comprised by projectors onto coherent states antisqueezed along the line between $\alpha_0$ and the origin (which represents the vacuum) in phase space. In other words, the estimation is tailored to have better resolution along that axis because of the subsequent discrimination of the signal state. This makes sense: since the error probability in the discrimination depends primarily on the distance between the hypotheses, it is more important to estimate this distance more accurately rather than along the orthogonal direction.
For large amplitudes, the estimation converges to a (standard) heterodyne measurement with no squeezing. As $\alpha_0$ approaches 0 the states of the signal become more and more indistinguishable, and the projectors of the heterodyne measurement approach infinitely squeezed coherent states, thus converging to a homodyne measurement.

\begin{figure}[t]
\begin{center}
\includegraphics[scale=1.4]{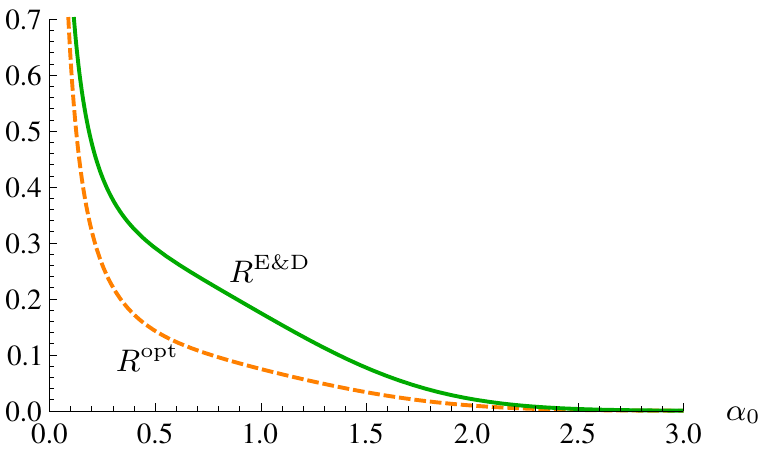}
\end{center}
\caption[Excess risk for the collective strategy, $R^{\rm opt}$, and for the E\&D strategy, $R^{\rm E\&D}$]{Excess risk for the collective strategy, $R^{\rm opt}$, and for the E\&D strategy, $R^{\rm E\&D}$, as a function of $\alpha_0$.\label{ch6/fig:fig3}}
\end{figure}

Inserting Eq.~\eqref{ch6/optimalsqueezing} into Eq.~\eqref{ch6/excess_risk_eyd_2} we finally obtain the expression of $R^{\rm E\&D}$ as a function of $\alpha_0$, which we can now compare with the excess risk for the collective strategy $R^{\rm opt}$, given in Eq.~\eqref{ch6/excess_risk_opt}. We plot both functions in Fig.~\ref{ch6/fig:fig3}. For small amplitudes in the range $\alpha_0\sim (0.3-1.5)$ there is a noticeable difference in the performance of the two strategies, reaching more than a factor two at some points. We also observe that the gap closes for large amplitudes. This behaviour is expected, since the problem becomes classical when the energy of the signal is sufficiently large. Very weak energies also render the strategies almost equivalent.\\

\section{Completely general estimation measurements}

We have showed that a local strategy based on the estimation of the auxiliary state via a generalized heterodyne measurement, followed by the corresponding discrimination measurement on the signal mode, performs worse than the most general (collective) strategy. However, this procedure does not encompass \emph{all} local strategies. The heterodyne measurement, although with some nonzero squeezing, still 
detects the phase space around $\alpha_0$ in a Gaussian way. A more general measurement that produces a non-Gaussian probability distribution for the estimate $\beta$ might perform better in terms of the excess risk and even match optimal performance, closing the gap between the curves in Fig.~\ref{ch6/fig:fig3}.
Here we show that the observed difference in performance between the collective and the local strategy is not due to lack of generality of the latter. We do so by considering a simplified although nontrivial version of the problem that allows us to obtain a fully general solution.

One could think, at first, that a non-Gaussian probability distribution for $\beta$ might give an advantage for the following reason.
Imagine that we restrict $\alpha$ further to be on the positive real axis. Then, the true $\alpha$ is either to the left of $\alpha_0$ or to the right, depending on the sign of the local parameter $u$. In the former case, $\alpha$ is closer to the vacuum, so the error in discriminating between them is larger than for the states on the other side. One would then expect that it is desirable to estimate better the negative parameters $u$, compared to the positive ones. Gaussian measurements like the heterodyne do not contemplate this situation, as they are translationally invariant, and that might be the reason behind the gap in Fig.~\ref{ch6/fig:fig3}.

To test this, we design the following simple example. Since the required methods are a straightforward extension of the ones used in the previous sections, we only sketch the procedure without showing any explicit calculation.
Imagine now that the true value of $\alpha$ is not Gaussian distributed around $\alpha_0$, but it can only take the values $\alpha=\alpha_0 \pm 1/\sqrt{n}$, representing the states that are closer to the vacuum and further away. Having only two possibilities for $\alpha$ allows us to solve analytically the most general local strategy, since estimating the auxiliary state becomes a discrimination problem between the states $\ket{\sqrt{n}\alpha_0+1}$ and $\ket{\sqrt{n}\alpha_0-1}$. The measurement that distinguishes the two possibilities is a two-outcome POVM $\mathcal{E}=\{\ketbrad{e_+},\ketbrad{e_-}\}$\footnote{Note that we have chosen the POVM elements to be rank-1 projectors. This is no loss of generality. Due to the convexity properties of the trace norm, POVMs with higher-rank elements cannot be optimal.}. We use the displacement operator~\eqref{ch6/displacement} to shift to the local model around $\alpha_0$, such that the state of the auxiliary mode is now either $\ket{1}$ or $\ket{-1}$. Then, the probabilities of correctly identifying each state are
\begin{equation}
p_+ = |\!\braket{e_+}{1}\!|^2\equiv c^2 \qquad {\rm and} \quad p_- = |\!\braket{e_-}{-1}\!|^2 = 1-c^2 \,.
\end{equation}
Since the vectors $\ket{e_+}$ and $\ket{e_-}$ are orthogonal by definition, the only freedom in choosing the POVM $\mathcal{E}$ is its relative orientation with respect to the pair of vectors $\ket{1}$ and $\ket{-1}$, which is parametrized by the overlap $c$. 
If the optimal estimation measurement is indeed asymmetric,
it should happen that \mbox{$c<1/2$}, i.e., that the probability of a correct identification is greater for the state $\ket{-1}$ than for $\ket{1}$.

From now on we proceed as for the E\&D strategy. We first compute the posterior state of the signal mode according to Bayes' rule. Then, we compute the optimal error probability in the discrimination of $\ketbrad{-\alpha_0}$ and the posterior state, which is a combination of $\ketbrad{1/\sqrt{n}}$ and $\ketbrad{-1/\sqrt{n}}$, weighted by the corresponding posterior probabilities. The $c$-dependence is carried by these probabilities. Going to the asymptotic limit $n\to\infty$, applying perturbation theory for computing the trace norm, and averaging the result over the two possible outcomes in the discrimination of the signal state, we finally obtain the asymptotic average error probability for the local strategy as a function of $c$. 
The asymptotic average error probability for the optimal collective strategy in this simple case is obtained exactly along the same lines as shown in Section~\ref{ch6/sec:collective}, and the one for \emph{known} states is given by the asymptotic expansion of Eq.~\eqref{ch6/perror known int}, substituting the average over $G(u)$ appropriately. 

Now we can compute the excess risk for the local and collective strategy, and optimize the local one over $c$. As already advanced at the beginning, the optimal solution yields $c=1/2$, i.e., the POVM $\mathcal{E}$ is symmetric with respect to the vectors $\ket{1}$ and $\ket{-1}$, hence both hypotheses receive the same treatment by the measurement in charge of determining the state of the auxiliary mode. Moreover, the gap between the excess risk of both strategies remains. This result leads us to conjecture that the optimal collective strategy performs better than \emph{any} local strategy.
\\

\section{Discussion}

In this Chapter we have proposed a learning scheme for coherent states of light, similar to the one proposed for qubits in Chapter~\ref{ch5_learning}. 
We have presented it in the context of a quantum-enhanced readout of classically-stored binary information, following a recent research line initiated in \citep{Pirandola2011}.
The reading of information, encoded in the state of a signal that comes reflected by a memory cell, is achieved by measuring the signal and deciding its state to be either the vacuum state or some coherent state of \emph{unknown} amplitude. The effect of this uncertainty is palliated by supplying a large number of auxiliary modes in the same coherent state. We have presented two strategies that make different uses of this (quantum) side information to determine the state of the signal: a collective strategy, consisting in measuring all modes at once and making the binary decision, and a local (E\&D) strategy, based on first estimating---\emph{learning}---the unknown amplitude, then using the acquired knowledge to tune a discrimination measurement over the signal. We have showed that the former outperforms any E\&D strategy that uses a Gaussian estimation measurement over the auxiliary modes. Furthermore, we conjecture that this is indeed the case for \emph{any} local strategy, on the light of a simplification of the original setting that allows us to consider completely general measurements.

Previous works on quantum reading rely on the use of specific preparations of nonclassical---entangled---states of light to improve the reading performance of a classical memory \citep{Pirandola2011,Nair2011,Spedalieri2012,Tej2013}. Our results indicate that, when there exists some uncertainty in the states produced by the source (and, consequently, the possibility of preparing a specific entangled signal state is highly diminished), quantum resources (collective measurements) still enhance the reading of classical information using classicaly correlated light.
It is worth mentioning that there are precedents of classically correlated coherent states exhibiting quantum phenomena of this sort. As an example, in the context of estimation of product coherent states, the optimal measure-and-prepare strategy on identical copies of $\ket{\alpha}$ can be achieved by LOCC (according to the fidelity criterion), but bipartite product states $\ket{\alpha}\!\ket{\alpha^*}$ require entangled measures~\citep{Niset2007}.

On a final note, the quantum enhancement found here is relevant on the regime of low energy signals\footnote{Note that here we have only considered sending a single-mode signal. However, in what coherent states are concerned, increasing the number of modes of the signal and increasing the energy of a single mode are equivalent situations.} (small amplitudes). This is in accordance to the advantage regime provided by nonclassical light sources, as discussed in other works. A low energy readout of memories is, in fact, of very practical interest. While---mathematically---the success probability of any readout protocol could be arbitrarily increased by sending signals with infinite energy, there are many situations where this is highly discouraged. For instance, the readout of photosensitive organic memories requires a high level of control over the amount of energy irradiated per cell. In those situations, the use of signals with very low energy benefits from quantum-enhanced performance, whereas highly energetic classical light could easily damage the memory.

\chapter{\chnameseven}

\label{ch7_povms}

The growth of quantum information theory and, in particular, the development of a vast variety of quantum processing techniques in the past few decades has drawn major attention towards the measurement process in quantum mechanics. Because no complete knowledge of the state of a quantum system can be retrieved from a single measurement, in general there are different incompatible measurement strategies that may yield very different results when applied to the same scenario. Hence, most often the design of a quantum processing technique involves finding which measurement best accomplishes a specific task, or which sequence of measurements is statistically optimal. These problems are the keystone of quantum estimation theory \citep{Helstrom1976}, and its solutions stand as a characteristic feature of many quantum processing tasks. 

Recent advances in experimental techniques 
have rendered many of these tasks realizable in a laboratory,
where a minimum resource perspective prevails. The sought for the minimum resources needed to implement a certain task has a paradigmatic example in quantum state preparation: to prepare all pure states of a bipartite system, it is enough to prepare only one maximally entangled pure state; then, by means of local operations and classical communication, one can obtain any bipartite pure state \citep{Nielsen2000}. The mathematical object that represents a general quantum measurement is a 
POVM (see Section~\ref{ch2/sec:measurement}), and therefore these kind of questions concern to the mathematical structure of POVMs. The aim of this Chapter is to address the following minimum resource problem: 
given a certain POVM, what are the simplest resources needed, and how one can implement it in terms of them?

POVMs form a convex set. This means that, given two known POVMs, any randomized implementation of them is also a POVM: just as mixed states are probabilistic mixtures of pure states, one can talk about measurements that can be regarded as probabilistic mixtures of POVMs. Those that cannot be expressed as combinations of other measurements are called extremal POVMs. Since many measurement optimization problems consist in maximizing a convex figure of merit, which leads to an extremal solution, this type of POVM appears quite frequently. It is no wonder then that the characterization of extremal POVMs has been extensively addressed in the literature\footnote{See, e.g., \citep{D'Ariano2005,Chiribella2010,Pellonpaa2011,Heinosaari2011}.}.
%

It is clear that the set of all extremal POVMs comprise the toolbox needed to effectively implement any measurement, as an appropriate convex combination of extremal POVMs will reproduce its statistics. A number of works have been devoted to prove the existence of such decompositions of measurements into extremals for finite \citep{D'Ariano2005,Haapasalo2011} as well as infinite dimensional systems \citep{Chiribella2007b}. 
However, the question of which are the minimal resources needed to implement a given POVM remains unclear from an operational point of view.
In this Chapter we provide a clear answer to this question by designing a constructive and efficient algorithm that takes as input any POVM with an arbitrary (but finite) number of outcomes and gives as output a convex combination of extremal POVMs that reproduces its statistics. We show that only rank-1 extremal POVMs are needed if one allows for a classical post-processing of the outcomes (in agreement to a similar result shown in \citep{Haapasalo2011}). 
The number of extremals that this algorithm produces is upper bounded by $(N-1)d+1$, where $N$ is the number of outcomes of the input POVM and $d$ is the dimension of its associated Hilbert space. This bound is significantly lower than the best previously known upper bound \citep{D'Ariano2005}, which scaled as $d^2$.
As a byproduct of our analysis, we obtain a simple geometrical characterization of extremal POVMs in terms of the generalized Bloch vectors associated to their elements.

In Section \ref{ch7/sec:simplecases} we fix the notation and illustrate how the algorithm works in a few simple cases. In Section \ref{ch7/sec:geometric} we set the mathematical tools we rely on and we derive from them a geometrical characterization of extremal POVMs. Section \ref{ch7/sec:algorithm} is devoted to the full description of the algorithm, and Section \ref{ch7/sec:ordereddecomp} to the discussion of further improvements. We finally summarize our results.
\\

\section{Simple cases}\label{ch7/sec:simplecases}


Let us start by fixing the notation and conventions used throughout this Chapter. A POVM is a set $\mathbb{P}=\{E_i\}$ of positive semidefinite operators acting on a Hilbert space $\mathcal{H}$ of dimension $d$, which satisfy the normalization condition $\sum_i E_i =\mathbb{I}$. The operator $E_i$ is called a \emph{POVM element}, and it is associated to the outcome $i$ of the POVM.
In this Chapter we focus on POVMs with a finite number of outcomes.
The elements $E_i$ might be zero for some $i$, meaning that the corresponding outcomes have zero probability of occurrence. Two POVMs that differ only in the number or position of their zero elements are considered to be physically equivalent. When characterizing a POVM by its number of outcomes we will refer only to those with physical meaning, that is to the outcomes with a  nonzero operator associated. In this spirit, we denote by $\mathbb{P}_N$ a POVM $\mathbb{P}$ with $N$ nonzero elements, and we will refer to it as a $N$-outcome POVM.

A convex combination of two POVMs is also a POVM: suppose that $\mathbb{P}_3^{(1)}=\left\{E_1,E_2,E_3,0,0\right\}$ and $\mathbb{P}_3^{(2)}=\left\{0,0,E_3,E_4,E_5\right\}$ are two 3-outcome POVMs, then
$\mathbb{P}_5 \equiv p_1\mathbb{P}_3^{(1)} + p_2\mathbb{P}_3^{(2)} = \left\{p_1E_1,p_1E_2,(p_1+p_2)E_3,p_2E_4,p_2E_5\right\}$ is also a POVM, where $p_1+p_2=1$. The convex combination $\mathbb{P}_5$ is the weighted sum element-by-element of $\mathbb{P}_3^{(1)}$ and $\mathbb{P}_3^{(2)}$.

\begin{figure}[t]
\begin{center}
\includegraphics[scale=1.3]{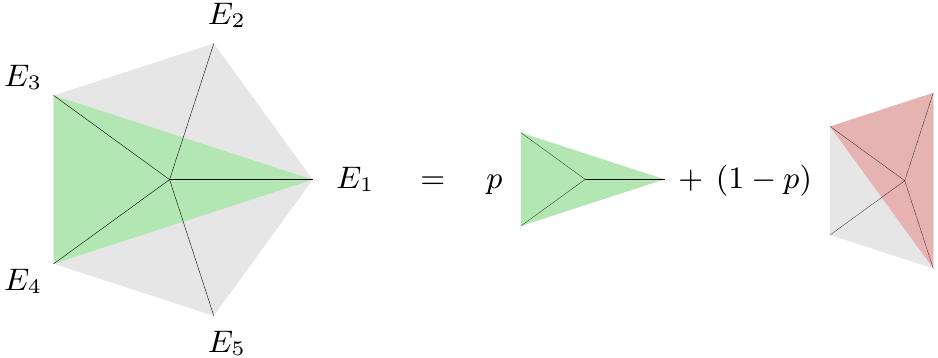}
\end{center}
\caption[Decomposition of a pentagon POVM]{First step of the decomposition of $\mathbb{P}_5$. The selection of elements (green) form the trine $\mathbb{P}_3^{(1)}$ which appears in the decomposition with associated probability $p$. After extracting it, we are left with $\mathbb{P}_4^{({\rm aux})}$ with associated probability $(1-p)$. In the second step we select another trine (red) from $\mathbb{P}_4^{({\rm aux})}$.}\label{ch7/fig:fig1}
\end{figure}

In this Chapter we are faced with the reverse situation: given a POVM, we want to find a decomposition into a convex combination of smaller (i.e. with less outcomes) POVMs. As a simple example of this type of decomposition, consider the POVM needed in the eavesdropping of the ``BB84'' protocol \citep{Nielsen2000} 
\begin{equation}
  \mathbb{P}_4 =
  \left\{{\footnotesize{1\over2}} \ketbrad{0},{\footnotesize{1\over2}} \ketbrad{1},
  {\footnotesize{1\over2}} \ketbrad{+}, {\footnotesize{1\over2}}\ketbrad{-}
  \right\}\,.
\end{equation}
Note that $\mathbb{P}_4$ can be expressed as
\begin{equation}
  \mathbb{P}_4 = {\footnotesize{1\over2}}\mathbb{P}_2^{(z)} + {\footnotesize{1\over2}}\mathbb{P}_2^{(x)} \,,
\end{equation}
where
\begin{eqnarray}
  \mathbb{P}_2^{(z)} &= \left\{\ketbrad{0},\ketbrad{1},0,0\right\} \\
  \mathbb{P}_2^{(x)} &= \left\{0,0,\ketbrad{+},\ketbrad{-}\right\}\,.
\end{eqnarray}
Thus, the POVM $\mathbb{P}_4$ can be effectively implemented by tossing an
unbiased coin, and then performing either $\mathbb{P}_2^{(x)}$ or
$\mathbb{P}_2^{(z)}$ based on the outcome of this toss.
In this case it is trivial to identify at sight the two pairs of orthogonal operators and their weights in the decomposition. This will not be so for an arbitrary measurement. The next example is presented to gain insight on how this operation can be performed algorithmically. Consider the POVM with five outcomes
\begin{equation}
\mathbb{P}_5=\left\{{\footnotesize{2\over 5}}E_1,{\footnotesize{2\over 5}}E_2,{\footnotesize{2\over 5}}E_3,{\footnotesize{2\over 5}}E_4,{\footnotesize{2\over 5}}E_5\right\}\,,
\end{equation}
where $E_i$ are rank-1 projectors lying on the equator of the Bloch sphere and aligned on the directions shown in Fig~\ref{ch7/fig:fig1}. To carry out its decomposition, one first notices that some subsets of $\{E_i\}$ may form a smaller POVM by themselves with appropriate weights. Then, by selecting one of these subsets (for instance the trine formed by elements 1, 3 and 4), one can rewrite the original POVM as
\begin{equation}
\mathbb{P}_5=p \mathbb{P}_3^{(1)} + (1-p) \mathbb{P}_4^{({\rm aux})}\,,
\end{equation}
where $p=1/\sqrt{5}$ and
\begin{eqnarray}
\mathbb{P}_3^{(1)} \!\!&\;\;=\; \left\{ {\footnotesize{2\over \sqrt{5}}} E_1,0, \left(1-{\footnotesize{1\over \sqrt{5}}}\right) E_3, \left(1-{\footnotesize{1\over \sqrt{5}}}\right) E_4,0\right\} \,, \\ \nonumber\\
\mathbb{P}_4^{({\rm aux})} \!\!&=\; \left\{ 0,{\footnotesize{2\over 5-\sqrt{5}}} E_2, {\footnotesize{3-\sqrt{5}\over 5-\sqrt{5}}} E_3, {\footnotesize{3-\sqrt{5}\over 5-\sqrt{5}}} E_4, {\footnotesize{2\over 5-\sqrt{5}}} E_5\right\}\,.
\end{eqnarray}
Note that both $\mathbb{P}_3^{(1)}$ and $\mathbb{P}_4^{({\rm aux})}$ are strictly smaller POVMs than $\mathbb{P}_5$. The operation just performed consists in algebraically extracting $\mathbb{P}_3^{(1)}$, in such a way that the remaining elements form a POVM with at least one less outcome (in the following section we prove that such an operation is always possible). Note also that $\mathbb{P}_4^{({\rm aux})}$ is further decomposable. Proceeding iteratively, one can select this time the elements 2, 3 and 5 and write the convex combination of trines 
\begin{equation}
\mathbb{P}_4^{({\rm aux})}=p' \mathbb{P}_3^{(2)}+(1-p')\mathbb{P}_3^{(3)}\,,
\end{equation}
where $p'=1/2$ and
\begin{eqnarray}
\mathbb{P}_3^{(2)} \!\!\!&\;\;=\; \left\{ 0,\left(1-{\footnotesize{1\over \sqrt{5}}}\right) E_2, \left(1-{\footnotesize{1\over \sqrt{5}}}\right) E_3, 0,{\footnotesize{2\over \sqrt{5}}} E_5\right\} \,, \\ \nonumber\\
\mathbb{P}_3^{(3)} \!\!\!&\;\;=\; \left\{ 0,{\footnotesize{2\over \sqrt{5}}} E_2, 0, \left(1-{\footnotesize{1\over \sqrt{5}}}\right) E_4,\left(1-{\footnotesize{1\over \sqrt{5}}}\right) E_5\right\}\,.
\end{eqnarray}
Finally, the original 5-outcome POVM can be expressed as a convex combination of 3-outcome POVMs as
\begin{equation}
\mathbb{P}_5=p_1\mathbb{P}_3^{(1)}+p_2\mathbb{P}_3^{(2)}+p_3\mathbb{P}_3^{(3)}\,
\end{equation}
where $p_1=p$, $p_2=(1-p)p'$ and $p_3=(1-p)(1-p')$.

Note that both $\mathbb{P}_5$ and $\mathbb{P}_4$ in the previous examples are rank-1 POVMs\footnote{A POVM is called rank-1 iff all its nonzero elements $E_i$ are rank-1 operators, i.e. they can be written as $E_i=e_i P_i$, where $0< e_i\leqslant 1$ and $P_i$ is a normalized one-dimensional projector.}, and hence we need no more than convex combinations of rank-1 POVMs to implement them. However, consider the full-rank 2-outcome POVM
\begin{equation}
\mathbb{P}_2=\left\{{\footnotesize{1\over 2}} \ketbrad{0}, {\footnotesize{1\over 2}} \ketbrad{0}+\ketbrad{1}\right\} \, .
\end{equation}
In this case it will be enough to measure $\mathbb{P}_2^{(z)}=\{\ketbrad{0},\ketbrad{1}\}$ and, if its first outcome is obtained, then toss an unbiased coin to decide between the two outcomes of $\mathbb{P}_2$. The projector $\ketbrad{0}$, an element of $\mathbb{P}_2^{(z)}$, is associated with more than one element of $\mathbb{P}_2$, thus the association of the obtained outcome with an original outcome is not immediate. This situation requires an additional step: classical post-processing of the outcomes. This kind of operation has been previously introduced in the literature under the name of \emph{relabelling} \citep{Haapasalo2011}. In general, the post-processing step will be necessary whenever $\rank{\mathbb{P}_N}>1$. For any original element $E_i$ such that $\rank{E_i}>1$, we will split it into a combination of rank-1 operators (by writing it in its eigenbasis) and consider such operators as additional outcomes, thus obtaining a rank-1 POVM that is statistically equivalent to the original one. Of course, to reproduce the statistics accordingly, a map from such new outcomes to the original ones is needed.
We address in full detail the case of POVMs of higher rank and the inclusion of a post-processing step in Section \ref{ch7/sec:algorithm}.


We have seen in this Section examples of measurements that are mixtures of other measurements. The mathematical structure of POVMs is convex: any inner point of the set of POVMs corresponds to a mixed measurement, i.e. it can be expressed as a convex combination of two different POVMs. We denote by $\mathcal{P}_N$ the convex set of POVMs with $N$ elements on $\mathcal{H}$. Note that for any $\mathbb{P} \in \mathcal{P}_N$ we can construct a physically equivalent POVM $\tilde{\mathbb{P}} \in \mathcal{P}_M$, with $M \geqslant N$, just by adding zero-elements to $\mathbb{P}$. The limit of infinite elements yields the convex set of all POVMs $\mathcal{P}$.

An \emph{extremal} POVM is a measurement that cannot be expressed as a mixture of two other POVMs. The 2- and 3-outcome POVMs obtained in the examples above are extremal. If a POVM with $N$ elements $\mathbb{P}$ is extremal in the convex set $\mathcal{P}_N$, then any physically equivalent POVM with $M$ elements $\tilde{\mathbb{P}}$, $M\geqslant N$, is also extremal in $\mathcal{P}_M$. Ultimately, $\mathbb{P}$ will be associated with a set of extremal points of $\mathcal{P}$. 
So far we have used an apparently more restricted definition of extremality. From the logic of the decompositions presented, it follows that we are considering a rank-1 POVM $\mathbb{P}_N=\{E_i\}$ to be extremal iff there does not exist any subset $\{E_k\}\subset\mathbb{P}_N$, $k=1,\ldots,M<N$ such that $\mathbb{P}_M=\{a_k E_k\}$ is itself a POVM for a suitable set of positive coefficients $\{a_k\}$. We have seen that if such a subset exists, then $\mathbb{P}_N$ can be split in $\mathbb{P}_M$ plus another POVM. We are therefore considering only decompositions into extremals formed by a subset of elements of the original $\mathbb{P}_N$. However, we prove in Section \ref{ch7/sec:geometric} that looking for such subsets is sufficient to check for extremality of a given POVM.
\\

\section{Selection of extremal POVMs and geometric characterization}\label{ch7/sec:geometric}

The decomposition of the POVMs presented as examples above is achieved through the selection of subsets of their elements capable of forming a POVM by themselves. In order to give some insight on how to perform this selection for a general POVM $\mathbb{P}$ with $N$ outcomes, we now examine the conditions under which a set of $n$ arbitrary rank-1 operators $\{E_i\}$ can comprise a POVM, that is, there is a set of positive coefficients $\{a_i\}$ such that $\sum_{i=1}^{n} a_i E_i=\id$. For simplicity and w.l.o.g. we will assume the operators $E_i$ to be normalized (i.e., $\tr E_i=1$). Recall that, for a $d$-dimensional Hilbert space, we can express $E_i$ in a generalized Bloch-like representation as
\begin{equation}\label{ch7/elements}
E_i=\left(\frac{1}{d} \id + \frac{1}{2} \sum_j \mean{\hat{\lambda}_j}_i \hat{\lambda}_j \right) \, ,
\end{equation}
where the operators $\hat{\lambda}_j$, $j=1,\dots,d^2-1$ are an orthogonal basis of generators of SU$(d)$ and the generalized Bloch vector $\vec{v}_i$ is defined with their expectation values: \mbox{$\vec{v}_i \equiv (\mean{\hat{\lambda}_1}_i,\dots,\mean{\hat{\lambda}_{d^2-1}}_i)$}. In this representation, pure states have associated a generalized Bloch vector of fixed length $|\vec{v}|=\sqrt{2(d-1)/d}$. Then, the POVM condition may be equivalently written as

\begin{eqnarray}
\sum_i a_i = d \label{ch7/cond3} \, ,\\
\sum_i a_i \vec{v}_i = \vec{0} \label{ch7/cond4} \, ,
\end{eqnarray}
that is a system of $d^2$ linear equations. At this point we are only interested in checking the consistency of \eqref{ch7/cond3} and \eqref{ch7/cond4}. Therefore, the existence of the set $\{a_i\}$ can be cast as a \emph{linear programming feasibility problem}. 

Before proceeding further, let us briefly overview the standard linear programming formalism (for an extensive review on the topic see e.g. \citep{Optimization2004,Todd2002}). A general \emph{linear program} (LP) has the standard form
\begin{eqnarray}\label{ch7/LP}
\min &\quad& c^T x \nonumber \\
\mbox{subject to} &\quad& Ax=b \nonumber \\
&\quad& x\geqslant 0 \, ,
\end{eqnarray}
where $A\in \mathbb{R}^{p \times q}$, $b\in \mathbb{R}^p$ and $c\in \mathbb{R}^q$ are the given data, and the vector $x\in \mathbb{R}^q$ is the variable to optimize. We call \eqref{ch7/LP} \emph{feasible} if there exists $x\in \mathbb{R}^q$ such that $Ax=b$, $x\geqslant0$. Any LP of the standard form above has a \emph{dual problem} of the form
%
\begin{eqnarray}\label{ch7/dual}
\max &\quad& -b^T \nu \nonumber \\
\mbox{subject to} &\quad& A^T \nu +c \geqslant 0 \, ,
\end{eqnarray}
where $\nu \in \mathbb{R}^p$. Let us assume that both LPs~\eqref{ch7/LP} and \eqref{ch7/dual} are feasible. Then, we may write
\begin{equation}\label{ch7/dualcheck}
c^T x + b^T \nu = x^T c + x^T A^T \nu = x^T (c+A^T \nu) \geqslant 0 \, .
\end{equation}
In order to obtain feasibility conditions of the LP \eqref{ch7/LP}, we now set $c=0$ and solve it. The existence of a solution implies that \eqref{ch7/LP} is feasible and, from \eqref{ch7/dual} and \eqref{ch7/dualcheck}, 
that for all vectors $\nu$, $A^T \nu \geqslant 0$ implies $b^T \nu \geqslant 0$.
If the dual problem does not have a solution, then its corresponding LP neither has one. Conversely, the existence of a vector $\nu$ that verifies the conditions
\begin{eqnarray}
A^T \nu &\leqslant& 0 \label{ch7/dualcond1} \, , \\
b^T \nu &>& 0 \label{ch7/dualcond2} \, ,
\end{eqnarray}
implies the infeasibility of \eqref{ch7/LP}. Notice that finding a $\nu$ subject to $A^T \nu \geqslant 0$, $b^T \nu <0$ is an equivalent problem.

We are now in the position to reinterpret the problem of finding the set of coefficients $\{a_i\}$ within the general linear program scheme presented above. The components of the vector $x$ are the coefficients we want to determine, that is
$
x=\{a_1,a_2,\dots,a_n\} .
$
Conditions \eqref{ch7/cond3} and \eqref{ch7/cond4} can be cast together in the \mbox{$Ax=b$} equation: $A$ is a matrix whose columns are given by vectors $v_i=(\vec{v}_i,1)$, and $b=(\vec{0},d)$. Therefore, the dimensions of this linear program are given by $p\equiv d^2, q\equiv n$. In the dual problem the vector $\nu$ has dimension $d^2$ and is unrestricted. However, for later convenience and w.l.o.g. let us choose the specific form
$
\nu=(\beta \vec{\nu}, \alpha) \, ,
$
where $\alpha \in \mathbb{R}, \beta \in \mathbb{R}^+$ are arbitrary constants and $|\vec{\nu}|=\sqrt{2(d-1)/d}$. From Eqs.~\eqref{ch7/dualcond1} and \eqref{ch7/dualcond2} we have
\begin{eqnarray}
\beta \vec{v}_i \cdot \vec{\nu} + \alpha \leqslant 0 \, , \\
\alpha > 0 \, .
\end{eqnarray}
A vector $\nu$ will simultaneously satisfy these conditions if and only if $\vec{v}_i \cdot \vec{\nu} < -\alpha/\beta$. We can always choose $\beta$ sufficiently large such that $-\alpha/\beta \rightarrow 0$, so the least restrictive condition has the form
\begin{equation}\label{ch7/hemisphere}
\vec{v}_i \cdot \vec{\nu} < 0 
\end{equation}
[taking the complementary equations to \eqref{ch7/dualcond1} and \eqref{ch7/dualcond2} would have led to the equivalent condition $\vec{v}_i \cdot \vec{\nu} > 0$]. To summarize, as long as there exists a vector $\vec{\nu}$ whose scalar product with every other generalized Bloch vector $\vec{v}_i$ is negative, we can always choose two positive constants $\alpha, \beta$ such that $\nu=\left( \beta \vec{\nu},\alpha \right)$ satisfies Eqs.~\eqref{ch7/dualcond1} and \eqref{ch7/dualcond2}. Hence, the LP \eqref{ch7/LP} is infeasible and the set of operators $\{E_i\}$ cannot form a POVM. 

Condition \eqref{ch7/hemisphere} has a clear geometrical interpretation: $\vec{\nu}$ defines a hyperplane in $\mathbb{R}^{d^2-1}$ which includes the $\vec{0}$ point and splits a $(d^2-2)$-sphere such that all $\vec{v}_i$ points are situated at one side of the hyperplane. Obviously, if the vectors $\vec{v}_i$ do not span $\mathbb{R}^{d^2-1}$ but a subspace of smaller dimension $d'$, it will suffice to consider hyperplanes of dimension $d'-1$. This hyperplane condition is equivalent to stating that the convex hull of the $\vec{v}_i$ points does not contain the $\vec{0}$ point.

We now state and prove next that, given a POVM with $n>d^2$ nonzero elements, it is always possible to select a subset of at most $d^2$ which is also a POVM, up to a suitable redistribution of weights. This is easily derived from the LP feasibility formulation: Eqs.~\eqref{ch7/cond3} and \eqref{ch7/cond4} represent a system of $d^2$ equality conditions and $n$ variables; if such a system is feasible, it would have a single solution for some value of $n\leqslant d^2$. For $n>d^2$ its solution will have $n-d^2$ extra degrees of freedom, and hence we will always be able to fix $n-d^2$ variables to zero. Since this statement is not valid when $n\leqslant d^2$ (except for the case in which vectors $\vec{v}_i$ span a smaller subspace of $\mathbb{R}^{d^2-1}$), it follows that an extremal POVM will have at most $d^2$ nonzero elements, as it has been noted in previous works \citep{D'Ariano2005,Haapasalo2011}.

The geometrical interpretation of the POVM condition provides a clear and useful picture of the results in the previous paragraph in terms of the distribution of vectors $\vec{v}_i$. Note that the number of vectors needed to subtend a solid angle in $\mathbb{R}^{d^2-1}$ is $d^2-1$. 
The conical hull defined by such vectors contains a portion of a hypersphere $S^{d^2-2}$.
It is then easy to convince oneself that the minimum number of vectors required to cover the whole $S^{d^2-2}$ as a union of conical hulls is $d^2$ [note that such a distribution necessarily implies the violation of condition \eqref{ch7/hemisphere} and, therefore, the fulfilment of \eqref{ch7/cond4}].
This means that, given such a set of $d^2$ vectors, if we add an extra vector, it will necessarily fall in a conical hull defined by a certain subset of $d^2-1$ vectors of the original set and thus it could be expressed as a conical combination of those (i.e. as a linear combination with nonnegative coefficients).
Hence, given $d^2+1$ POVM elements whose Bloch vectors satisfy condition \eqref{ch7/cond4}, one can always choose one of the vectors and 
replace it by a conical combination of $d^2-1$ other vectors:
the remaining set of $d^2$ vectors still satisfies condition \eqref{ch7/cond4}.

In general, Bloch vectors $\vec{v}_i$ will be contained in $\mathbb{R}^{d^2-1}$. When $n<d^2$, additional restrictions over vectors $\vec{v}_i$ derive from \eqref{ch7/hemisphere}. If $n=2$ then the generalized Bloch vectors $\vec{v}_1$ and $\vec{v}_2$ should span a 1-dimensional space in order to be able to violate condition \eqref{ch7/hemisphere}. In fact, the condition is violated only if $\vec{v}_1=-\vec{v}_2$. If $n=3$, vectors $\vec{v}_1, \vec{v}_2$ and $\vec{v}_3$ should lie on a plane and not belong to the same semicircle (defined by a line). For any $n$ we should have 
\begin{equation}
\{\vec{v}_1,\vec{v}_2,\dots,\vec{v}_n\} \in S^{n-2} \subset \mathbb{R}^{n-1} \, ,
\end{equation}
where vectors $\vec{v}_i$ do not belong to any hemisphere of $S^{n-2}$. Note that the extremality statement in the previous paragraph extends to $\mathbb{R}^{n-1}$: if we have $n' \geqslant n+1$ vectors (whose associated operators form a POVM) that span $\mathbb{R}^{n-1}$, then we can always find subsets of at most $n$ vectors which violate condition \eqref{ch7/hemisphere}, and thus are able to form an extremal POVM.

To finish this section and for clarity purposes, note that it has been assumed that the solutions of the LP feasibility problem correspond to extremal POVMs, i.e. extremal points not only of the set of feasible points but also of the set of all POVMs. This is indeed the case: on one hand, such a solution corresponds to a set of linearly independent POVM elements $\{E_i\}$; on the other hand, any POVM with at most $d^2$ rank-1 linearly independent elements is extremal (see, e.g., Proposition 3 in \citep{Haapasalo2011}).
\\

\section{The algorithm}\label{ch7/sec:algorithm}

In this section, we present our constructive algorithm for decomposing a POVM into extremals. We first address the case of rank-1 POVMs, and then we extend the algorithm to higher-rank cases.
We are given a rank-1 POVM $\mathbb{P}_N=\{a_i E_i\}$, $i=1,\ldots,N$, where $E_i$ are normalized operators given by \eqref{ch7/elements} and $a_i>0$. Our aim is to express it as
\begin{equation}\label{ch7/decomp}
\mathbb{P}_N=\sum_k p_k \mathbb{P}^{(k)}_n ,
\end{equation}
where $\mathbb{P}^{(k)}_n$ is an extremal rank-1 POVM with $n \leqslant d^2$ outcomes. 
This means that in order to implement $\mathbb{P}_N$ it will suffice to randomly select a value of $k$ from the probability distribution $p_k$, and then perform $\mathbb{P}^{(k)}_n$.
The algorithm we propose to carry out such a decomposition works as follows:

We first define the LP feasibility problem
\begin{eqnarray}\label{ch7/LP2}
\mbox{find} &\quad& x \nonumber \\
\mbox{subject to} &\quad& Ax=b \nonumber \\
&\quad& x\geqslant 0 \, ,
\end{eqnarray}
where $x$ is a vector of $N$ variables, $A$ is a matrix whose columns are given by vectors $v_i=(\vec{v}_i,1)$, and $b=(\vec{0},d)$. The set of feasible points of this LP, i.e. the values of $x$ compatible with the conditions of the LP, define a convex polytope $K$ in the space of coefficients:
\begin{equation}
K = \{x \,/\; Ax=b, x\geqslant 0\} \subset \mathbb{R}^N .
\end{equation}
The vertices of $K$ are its extremal points, and the region of $\mathbb{R}^N$ defined by the convex hull of all the vertices contains all the points that can be expressed as convex combinations of these extremal points. Dantzig's \emph{simplex method} for solving LPs \citep{Todd2002} starts at a vertex of $K$, and it moves from vertex to vertex minimizing a cost function, until there is no preferred direction of minimization; then, the optimal solution has been found. Since there is no cost function in a feasibility problem, the simplex method applied to \eqref{ch7/LP2} terminates at its first step: when it finds the first vertex. 
The convex polytope $K$ is isomorphic to a subset of $\mathcal{P}_N$, i.e. there is a one-to-one correspondence between all their elements, and they behave equivalently.
Therefore, such a vertex $x^{(1)}=\{x^{(1)}_i\}$ found as the solution of the LP corresponds to the set of coefficients of an extremal POVM, and as such $x^{(1)}$ will have at most $d^2$ and at least $d$ nonzero elements. The vertices of the polytope $K$ correspond to all the extremal POVMs that one can comprise using only the original elements $\{E_i\}$, and its interior region contains all the possible POVMs generated by these extremals.

Once we have found $x^{(1)}$, we algebraically subtract it from the original set of coefficients $\{a_i\}$. To illustrate this operation, let us assume $d=2$ and $x^{(1)}=\{x^{(1)}_1,x^{(1)}_2,0,\ldots,0\}$. Then, $\{a_i\}$ is rewritten as
\begin{eqnarray}\label{ch7/1step}
& \{a_1,a_2,a_3,\ldots,a_N\} = p\,x^{(1)} + (1-p) x^{\rm (aux)}\,, \\ \nonumber\\
& x^{\rm (aux)}=\left\{\frac{a_1-p\,x^{(1)}_1}{1-p},\frac{a_2-p\,x^{(1)}_2}{1-p},\frac{a_3}{1-p},\ldots,\frac{a_N}{1-p}\right\}\,.
\end{eqnarray}
For $x^{\rm (aux)}$ to be an element of $K$, the inequality 
\begin{equation}\label{ch7/pcond}
p \leqslant a_i/x^{(1)}_i \leqslant 1
\end{equation}
has to hold for all $i$ such that $x^{(1)}_i>0$. To guarantee the left-hand side of \eqref{ch7/pcond}, we take
\begin{equation}
p=\min_i \frac{a_i}{x^{(1)}_i}\,.
\end{equation}
Let us reorder the coefficients $\{a_i\}$ and $x^{(1)}$ such that $p=a_1/x^{(1)}_1$. This choice of $p$ makes the first coefficient of $x^{\rm (aux)}$ to be zero (it could happen that more than one element turns to be zero, thus accelerating the algorithm, but we consider from now on the worst case scenario in which one element is eliminated at a time). 
Also, the right-hand side of \eqref{ch7/pcond} is immediately satisfied since $a_1<x^{(1)}_1$.
Note that $p\in\left[0,1\right]$, thus it is a probability.
Now, \eqref{ch7/1step} can be understood as a probabilistic (convex) combination of $x^{(1)}$ and $x^{\rm (aux)}$, both set of coefficients corresponding to an extremal POVM $\mathbb{P}_2^{(1)}$ and a POVM with $N-1$ outcomes $\mathbb{P}^{\rm (aux)}_{N-1}$. Hence, as a result of the first step of the algorithm, we can write
\begin{equation}
\mathbb{P}_N=p\,\mathbb{P}_2^{(1)}+(1-p)\,\mathbb{P}_{N-1}^{\rm (aux)}\,.
\end{equation}
We then repeat this process redefining the LP with $\mathbb{P}_{N-1}^{\rm (aux)}$ as the initial POVM, which gives us another vertex $x^{(2)}$ associated to an extremal POVM with $n$ outcomes $\mathbb{P}^{(2)}_n$, a remainder $\mathbb{P}^{\rm (aux)}_{N-2}$ and its corresponding probabilities. Of course, in general $d\leqslant n\leqslant d^2$. We iterate this process $N-n_L$ times, where $n_L$ is the number of outcomes of the last extremal POVM obtained. At the last step the simplex algorithm will identify a unique solution with probability 1, corresponding to the input set $x^{\rm (aux)}=x^{(N-n_L)}$.

It is important to stress that the polytopes of the LPs at each step of the algorithm, $K^k$, are subsequent subsets of each other, that is 
\begin{equation}
K\supset K^1\supset \ldots \supset K^{N-n_L+1}.
\end{equation}
The result of each step is the elimination of one of the original elements $\{E_i\}$, and with it all the vertices that required that element. 
Thus, each step projects the polytope onto a subspace of the space of coefficients by reducing its dimension by one.
As a consequence, in the end all the vertices selected by the simplex algorithm were vertices of the original $K$.
\\

When the rank of $\mathbb{P}_N$ is higher than 1 we can still apply the same algorithm, just adding two extra steps: one preparation step and one post-processing step. The preparation step works as follows: for every $i$ such that \mbox{$\rank{E_i}>1$}, express $E_i$ in its eigenbasis $\{\ket{v_{ij}}\}$ as
\begin{equation}
E_i=\sum_j \lambda_j \ketbrad{v_{ij}}=\sum_j E_{ij}.
\end{equation}
Consider each rank-1 operator $E_{ij}$ as a new outcome and denote the new (rank-1) POVM by $\mathbb{P}_{\bar{N}}=\{\bar{E}_l\}_{l=1}^{\bar{N}}$, where $\bar{N}=\sum_i \rank{E_i}>N$. The label $l(i,j)$ carries the information contained in labels $i$ and $j$. Now, the algorithm described above can be applied directly over $\mathbb{P}_{\bar{N}}$. The post-processing step is needed for associating the outcomes of the measure finally performed ($l$) to the outcomes of the original $\mathbb{P}_N$ ($i$). 
\\

A generic algorithm for decomposing a point in a convex set into a combination of extremal points of that set can be found in \citep{D'Ariano2005}. Although in this paper D'Ariano \emph{et al.} specialize it for a general $\mathbb{P} \in \mathcal{P}_N$, we would like to remark that significant differences stand between our algorithm and the one presented there. The algorithm of \citep{D'Ariano2005} consists in a recursive splitting of an inner point of the convex set into a convex combination of two points that lie on a facet of the convex set (and thus a subset of a strictly smaller dimension). After enough steps it yields a number of extremal points along with some weights in a tree-like form, thus statistically reproducing the original point as a mixture of extremal points. The direction in which the splitting is done at each step is determined through an eigenvalue evaluation.
The particular decomposition we have presented in this Chapter may be considered within this general scheme (we also do binary partitions at each step), however two main differences arise. On one hand, the process of obtaining extremal points (i.e. the direction of splitting) is radically different. We associate a polytope $K$ to a subset of the convex set $\mathcal{P}_N$ via an isomorphism, and then we move efficiently along the directions marked by the vertices of $K$. Thus, there is no need to analyse the whole convex set $\mathcal{P}_N$ (which is strongly convex, i.e. its extremal points are not isolated but lie on a continuum) for a given $\mathbb{P}$: our algorithm does not optimize a direction among a continuum of possibilities at each step but selects any direction of a given finite set. On the other hand, the authors in \citep{D'Ariano2005} state that their algorithm provides a minimal decomposition, with a number of extremals upperbounded by $(N-1) d^2 +1$. We have found that our algorithm yields the tighter bound $(N-1)d+1$.
\\

\section{Ordered decompositions}\label{ch7/sec:ordereddecomp}

The algorithm described in Section \ref{ch7/sec:algorithm} will produce one of many possible decompositions of the initial POVM into at most $N-n_L+1$ extremals (recall that $n_L$ ranges from $d$ to $d^2$), even if we only consider extremals made of original elements. Because at each step any of the vertices of the polytope could be identified and extracted, the final decomposition obtained is not unique and depends on the particular implementation of the simplex method for solving the LP. That being said, one could be interested in a particular decomposition that exhibits certain properties.
We observe that there is room in our algorithm for these extra requirements while maintaining its structure, that is to efficiently produce decompositions into at most $N-n_L+1$ extremals obtained through a LP solved by the simplex method.
To obtain a particular decomposition with this structure that verifies a certain desired property we will simply have to establish some ranking among the vertices of the polytope in agreement to that property or associated criterion, and tweak the algorithm to choose first the ones at the top of the ranking. This is what we call an \emph{ordered} decomposition.

A desirable ordering from the point of view of an experimental realization may be, for instance, to prioritize the vertices with more zero elements, if there is any. Those vertices would correspond to extremals with less outcomes. In the case of $d=2$, for instance, extremal POVMs can have 2, 3 or 4 outcomes. Such a decomposition would seek first for 2-outcome (Stern-Gerlach measurements), then 3-outcome and finally 4-outcome POVMs.

The simplex method is an efficient way of finding the optimal vertex of a polytope according to some criterion, which is 
implemented as
a cost function. This is done by minimizing or maximizing such a cost function. In the description of the algorithm we chose this function to be independent of the variables, because we were only interested in finding a feasible point. The choice of the cost function will vary the direction taken by the simplex algorithm when it moves from one vertex to another, and it is therefore a way to establish a ranking among the vertices. Consider for instance the cost function
\begin{equation}\label{ch7/costQ}
Q_n = \sum_{i=1}^n x_i^2 \, .
\end{equation}
The maximization of $Q_n$ on its own could in principle work for finding the vertices with more zeros: if we would have no other constraint but a fixed quantity $d$ to distribute among the $n$ parties $x_i$, the strategy that maximizes $Q_n$ is to give all to one party and zero to the others. But we have more constraints in \eqref{ch7/LP2}. Let us take a look on the minimum and maximum values of $Q_4$, that is for extremals with 4 outcomes. The value of $Q_4$ will only depend on the geometric distribution of the outcomes of the extremal. On one hand, $Q_4$ takes its minimum value when $d=\sum_i x_i$ is equally distributed among the variables $x_i$, that is when the 4 associated Bloch vectors $\vec{v}_i$ are orthogonal in pairs (i.e. the POVM is a combination of two Stern-Gerlachs). This value is $Q_4^{\rm min}=(d/4)^2 \times 4=d^2/4$. On the other hand, $Q_4$ reaches its maximum value if three of the vectors are parallel and the fourth is orthogonal to all the others (this is the way to put a maximum weight on one of the $x_i$), that is $Q_4^{\rm max}=(d/2)^2+(d/6)^2\times 3=d^2/3$. Applying the same reasoning for 3-outcome extremals we have $Q_3^{\rm min}=d^2/3$ and $Q_3^{\rm max}=3d^2/8$, and 2-outcomes can only give $Q_2=d^2/2$. Since
\begin{equation}\label{ch7/Qd2}
Q_2>Q_3^{\rm max} > Q_3^{\rm min} = Q_4^{\rm max} > Q_4^{\rm min} \, ,
\end{equation}
the maximization of function $Q_n$ prioritizes the extremals with fewer outcomes at least for $d=2$, when the maximum number of nonzero elements in a vertex is $n=4$.
This, unfortunately, stops being valid for $n>4$, which in general happens if $d>2$. 

The general problem of maximizing a convex function over a convex set of feasible points is called \emph{convex maximization}. The problem at hand belongs to this category. While the more standard class of \emph{convex minimization} problems (i.e. minimizing a convex function over a convex polytope) count on efficient solving algorithms, this is not the case for convex maximization, except for very special cases. The efficiency of the convex minimization relies on the uniqueness of the convex function's minimum, which is an inner point of the polytope. Conversely, its maxima are located on the vertices of the polytope and all but one are \emph{local} maxima. This fact makes the convex maximization problems intractable in general, and so it is the maximization of \eqref{ch7/costQ}. The difficulty lies on the fact that an algorithm might find a local maximum (a vertex), but there is no way to certificate its global optimality (although there are algorithms that, despite no proof certificate, provide good guesses \citep{Fortin2010}). 

Any global search algorithm (able to guarantee global optimality) for convex maximization somehow \emph{enumerates} all the vertices, and thus its efficiency highly depends on the number of those. Of course, the ordered decomposition we are looking for is immediately obtained if one enumerates all the vertices of $K$. With such a list, we would just have to pick up first those vertices with more zero elements, corresponding to the extremals with fewer outcomes (or according to any other criterion we may wish). Furthermore, no additional optimization is required since we can extract from the same list the vertex required at each step, thus keeping us from solving a LP for doing so.
The problem of enumerating the vertices of a bounded polyhedron is NP hard in the general case \citep{Khachiyan2008}, but has efficient algorithms able to generate all vertices in polynomial time (typically linear in the number of vertices) for several special cases. For instance, in \citep{Avis1992} there is an algorithm that enumerates the $v$ vertices of a convex polyhedron in $\mathbb{R}^m$ defined by a system of $D$ linear inequalities in time $O(mDv)$. Our polytope $K$ is of this type, and hence we could use the algorithm for our purpose. Note however that $v$ has a direct dependence on $m$ and $D$. The problem of computing $v$ for a given polytope is NP-hard, but a bound can be provided \citep{Barvinok2011}: the number of vertices of our polytope $K\subset\mathbb{R}^m$ is at least exponential in $m$.

In summary, an ordered decomposition of a POVM can be carried out in two ways. On one hand, nonlinear programming techniques can be used to maximize a cost function subject to the constraints of \eqref{ch7/LP2}, but none of them will perform with perfect accuracy. We have found a cost function that prioritizes the extremals with less outcomes for $d=2$, but not for greater dimensions. Finding a cost function is problem-specific, and it seems to be highly nontrivial: its maximization should lead first to a vertex of the polytope, and secondly it should move from one to another maximizing the desired property. On the other hand, an alternative method is to enumerate all the vertices of the polytope $K$ defined by the constraints of \eqref{ch7/LP2}, but the number of vertices and thus the time required to carry out the enumeration grows exponentially with the number of elements of the original POVM.
\\

\section{Discussion}

We have presented an efficient algorithm to decompose any POVM \mbox{$\mathbb{P} \in \mathcal{P}_N$} into extremal ones. The decomposition achieved consists of a convex combination of at least $N-n_L+1$ (if $\mathbb{P}$ is rank-1) and at most $Nd-n_L+1$ (if $\mathbb{P}$ is full-rank) extremal measurements, where $n_L$ ranges from $d$ to $d^2$ and its value is determined by each particular $\mathbb{P}$. In the case in which $\mathbb{P}$ presents some symmetry (as the BB84 POVM shown as an example in Section~\ref{ch7/sec:simplecases}), more than one element may be eliminated in one step of the algorithm and thus the number of extremals would be even less. We have shown that only extremal rank-1 POVMs are required to effectively implement $\mathbb{P}$ by introducing a classical post-processing of the outcomes. The decomposition is efficiently carried out by an algorithm based on resolutions of LPs using the simplex method, within polynomial time in $N$ and $d$. The efficiency is achieved by restricting the analysis to a polytope-shaped subset of $\mathcal{P}_N$ for a given $\mathbb{P}$, and thus by taking into consideration only a finite number of extremals (the vertices of the polytope), in contrast to what other authors have considered so far (see, e.g., \citep{D'Ariano2005}). Furthermore, in \citep{D'Ariano2005}, a generic decomposition algorithm that yields a certain maximum number of extremals is provided. We have found that our algorithm beats this performance in a worst case scenario.

Since a given POVM admits many decompositions, we also explore the possibility of obtaining a particular decomposition that exhibits a certain desired property, introduced in the algorithm as an input. We call these decompositions \emph{ordered}, and they are based on prioritizations of extremals that can be made out of subsets of the elements of $\mathbb{P}$. As an example we give a method to prioritize extremal POVMs with less outcomes in the case of $d=2$, and show that either efficiency or accuracy necessarily get compromised.

\chapter*{\conclusions}
\addcontentsline{toc}{chapter}{\conclusions}
\chaptermark{\conclusions}

The specific conclusions of the research projects addressed in this thesis have already been discussed at the end of each corresponding chapter.
Here, I would like to finish by giving a brief outlook on future research lines and open problems that naturally arise from within the covered topics.

The group-theoretic concepts used in Chapter~\ref{ch4_pqsd} to compute the optimal programmable discrimination machine for qubits can also be applied to higher-dimensional systems. 
In fact, some results are already available in the literature for pure states of arbitrary dimension \citep{Hayashi2005,Hayashi2006,Akimoto2011}, but the mixed states case remains an open problem, and so does the fully universal discrimination machine, for states of more than two dimensions. In this line of generalizations, the extreme case of infinite dimensions, i.e., programmable discrimination of continuous-variables systems, has only been discussed before for coherent states and unambiguous discrimination \citep{Sedlak2007,Sedlak2009}. Although Chapter~\ref{ch6_learningcv} provides an instance of programmable minimum-error discrimination with coherent states, there is much work to be done. Extending the applicability of programmable discrimination protocols to general Gaussian states, or even more complex cases such as multimode entangled states, would be of great fundamental and practical interest.

In Chapter~\ref{ch5_learning}, I analysed the classification of qubit states in a \emph{supervised} learning scenario. The most obvious generalization, and the most promising one, is to consider \emph{unsupervised} scenarios, where no human expert classifies the training sample. This is a challenging problem with direct practical applications in quantum control and information processing. Although this topic is fairly new, it is beginning to raise much attention (see, e.g., \citep{Lloyd2013}).

Another generalization of both programmable and learning machines is to consider more than two possible states, although the scarcity of results in general multihypothesis quantum state discrimination is somewhat discouraging $\!\!\!$\mbox{---it} is expected that only very special cases will be analytically tractable.
A more promising extension
is to analyse the behaviour of the proposed programmable and learning machines under the more general scheme of discrimination with an error margin. On the one hand, programmable discrimination of mixed states has yet to be considered when a limiting margin is imposed on the rate of errors. On the other hand, a very interesting question that remains unanswered to date is whether the optimality of the learning protocol proposed in Chapter~\ref{ch5_learning} is compromised---and if so, to which extent---when one allows for some proportion of inconclusive answers.



As for the decomposition of quantum measurements examined in Chapter~\ref{ch7_povms}, there are at least two directions worth exploring further. 
The first goes along the idea of ordered decompositions, that is, the search of convex combinations of extremal POVMs satisfying a particular criterion. Apart from the proposed pursuing of extremal POVMs with fewer elements, finding efficient search algorithms tuned to look for other potentially desirable properties of measurements is work to be done. The second direction prompts upon relaxing the requisite that the decompositions shall reproduce \emph{exactly} the statistics of the original POVM. Looking for convex combinations of extremal POVMs that only \emph{approximate} it gives one more freedom to search for ``convenient'' decompositions that might not be possible to assemble using the original POVM elements\footnote{These could be, for instance, decompositions made of symmetric informationally complete measurements (SIC-POVMs), covariant measurements, measurements with a fixed number of outcomes, etc.}. 

Then, of course, one would need to consider what a ``good enough approximation'' means. A definition can be found in Winter's measurement compression theorem~\citep{Winter2001}, which gives a decomposition of any quantum measurement into an ``intrinsic'' part (information) and an ``extrinsic'' part (noise). The theorem considers approximate POVM simulations for an asymptotically large number of realizations, and puts them in a communication context: a sender implements many measurement instances and sends the outcomes to a receiver, using as little communication as possible, and counting with some amount of shared randomness as a resource. The achievability of a ``faithful'' simulation depends on the amount of this randomness and the classical communication rate between the two parties. The algorithm proposed in Chapter~\ref{ch7_povms} provides decompositions with a fewer number of extremal POVMs than its predecessors, and POVMs with less outcomes can be prioritized using a cost function; these two features can in principle be directly related to shared randomness and communication rates in the above context, respectively. The details of this relation remain a stimulating open question.

\appendix

\chapter{Technical details of Chapter~\ref{ch4_pqsd}}

\section{Wigner's 6$j$-symbols}\label{appA/sec:wigner6j}

Let us consider three angular momenta $j_1, j_2, j_3$ that couple to give a total $J$. Note that there is no unique way~to carry out this coupling; we might first couple $j_1$ and~$j_2$ to give a resultant $j_{12}$, and couple this to $j_3$ to give~$J$, or alternatively, we may couple $j_1$ to the resultant~$j_{23}$ of coupling $j_2$ and $j_3$. Moreover, the intermediate couplings can give in principle different values of~$j_{12}$ or~$j_{23}$ which, when coupled to $j_3$ or $j_1$, end up giving the same value of $J$. All these possibilities lead to linearly independent states with the same $J$ and $M$, thus they must be distinguished by specifying the intermediate angular momentum and the order of coupling. There exists a unitary transformation that maps the states obtained from the two possible orderings of the coupling; Wigner's 6j-symbols~\citep{Edmonds1960}, denoted in the next equation by $\{ \, {}^{\cdots}_{\cdots} \, \}$, provide the coefficients of this transformation:
\begin{align}\label{appA/wigner6j}
& \braket{(j_1\,j_2)j_{12},j_3;J,M}{j_1,(j_2\,j_3)j_{23};J,M} \nonumber\\[.5em]
&\hspace{1cm} = (-1)^{j_1+j_2+j_3+J} \sqrt{(2j_{12}+1)(2j_{23}+1)}
\begin{Bmatrix}
j_1 & j_2 & j_{12} \\
j_3 & J & j_{23}
\end{Bmatrix} \,.
\end{align}
Note that this overlap is independent of $M$.
\\

\section{Arbitrary number of copies}\label{appA/sec:na-nb-nc}

In this Section we present the probabilities for unambiguous and minimum-error discrimination when the number of copies $n_A, n_B, n_C$ loaded at the machine ports is completely  arbitrary. Note that, in this case, the global states $ \sigma_1 $ and $ \sigma_2 $ [cf. Eq.~\eqref{ch4/rho1-rho2}] may have different dimensions, for  $ d_1 = (\na+\nb+1)(\nc+1) $ is in general not equal to  $ d_2 = (\na+1)(\nb+\nc+1) $.  One can easily convince oneself that the support  of the state with smallest dimension is always contained in the support of the other, and hence the problem can be solved in very much the same way as in the main text as far as the intersection of the supports is concerned. The remaining of the state with higher dimension yields a trivial contribution to the error probabilities.
Without loss of generality we can assume from now on that $\na\geqslant \nc$.
As discussed in the main text, the error probabilities are computed by adding the pairwise contributions of the state bases in the common support, the main difference being that $ \sigma_1 $ and $ \sigma_2 $ do not have equal coefficients in front of the projectors and hence the prior probabilities of each pair of states are different. Also, the overlaps in Eq.~\eqref{ch4/6j}  will have a slightly more complicated expression. Here we have $j_A=\na/2 $, $ j_B=\nb/2 $, $ j_C=\nc/2 $, $ j_{AB}=(\na+\nb)/2 $ and $ j_{BC}=(\nb+\nc)/2$. The minimum $ J $ available for $ \sigma_1 $ is $ j_B+j_A-j_C \equiv J_{\rm min}^1$, and   $ |j_B+j_C-j_A| \equiv J_{\rm min}^2 $ for $ \sigma_2 $. The maximum angular momentum $ j_A+j_B+j_C \equiv J_{\rm max} $ is reachable for both states. 
For equal prior probabilities for $ \sigma_1 $ and $ \sigma_2 $, we can write
\begin{eqnarray*}
\frac{1}{2} \sigma_1 &=& \sum_{J=J_{\rm min}^1}^{J_{\rm max}}\sum_{M=-J}^J p_J\, \pi^{1}_J { [j_{AB} ; J M] } \,, \\
\frac{1}{2} \sigma_2 &=& \sum_{J=J_{\rm min}^2}^{J_{\rm max}}\sum_{M=-J}^J p_J\, \pi^{2}_J { [j_{BC} ; J M] } \,,
\end{eqnarray*}
where  
$
p_J={1\over 2}\left({1\over d_1}+{1\over d_2}\right)$, $\pi^{1}_J={1\over 2p_J\,d_1}$, $\pi^{2}_J={1\over 2p_J\,d_2}
$
for $J_{\rm min}^1\le J\le J_{\rm max}$, whereas
$
p_J={1\over 2d_2}$, $\pi^{1}_J=0$, $\pi^{2}_J=1
$
for $J_{\rm min}^2\le J<J_{\rm min}^1$. We view $ p_J $ as the probability of obtaining the outcome ($M$) $J$ in a measurement of the ($z$ component of the) total angular momentum on the unknown state.
Likewise, we view $\pi^{1}_J$, $\pi^{2}_J = 1-\pi^1_J$ as the probabilities that the unknown state be $[j_{AB} ; J M]$ or $[j_{BC} ; J M]$ for that specific pair of outcomes $J$ and $M$ (note that these probabilities are actually independent of $M$).
If the condition
\begin{equation}\label{ch4/uaineq}
{c^2_{J} \over 1+c^2_{J}}\le \pi_J^{AB}\le{1 \over 1+c^2_{J}} \,,
\end{equation}
where $c_J=|\langle j_{AB} ; J M | j_{BC} ; J M \rangle|$ is given by Eq.~\eqref{ch4/6j}, holds, then the probability of obtaining an inconclusive answer when we finally discriminate between $[j_{AB} ; J M]$ and $[j_{BC} ; J M]$ is $\PUA_J=2\sqrt{\pi_J^1 \pi_J^2} c_J $ [cf. Eq.~\eqref{ch3/UA_eta1eta2}].
If Eq.~\eqref{ch4/uaineq} is satisfied for $ {\hat J}=J_{\rm max}-1 $, then it will be satisfied all over this range of $ J $, since $ c_J $ is a monotonically increasing function of $ J $. The overlap $ c_{\hat J} $ has the very simple form
\begin{equation*}
c^2_{\hat J}={n_A n_C\over (n_A+n_B)(n_B+n_C)} \,.
\end{equation*}
Thus Eq.~\eqref{ch4/uaineq} is equivalent to
\begin{eqnarray*}
{n_A n_C\over (n_A+n_B)(n_B+n_C)}
&\le &
{(n_A+n_B+1)(n_C+1)\over(n_B+n_C+1)(n_A+1)} \nonumber \\
&\le & {(n_A+n_B)(n_B+n_C)\over n_A n_C} \,,
\end{eqnarray*}
which is clearly true. Eq.~\eqref{ch4/uaineq} does not hold if $ J=J_{\rm max} $, for which we have $\PUA_{J_{\rm max}}=1$. 
Note that since no error is made for $J_{\rm min}^2\le J<J_{\rm min}^1$, for $\pi^{1}_J=0$, the total
inconclusive probability  reads $\PUA=\sum_{J=J_{\rm min}^1}^{J_{\rm max}}
p_J\, (2J+1)\PUA_J$, which has the explicit expression
\begin{align*}
\PUA &= \frac{1}{2} \left(\frac{1}{\sqrt{d_1}}-\frac{1}{\sqrt{d_2}}\right)^2 d_{ABC}  + \frac{1}{\sqrt{d_1d_2}} \sum_{k=0}^{n_C} (n_A+n_B-n_C+2k+1) \nonumber \\
& \hspace{5cm}  \times
\sqrt{
{
\scriptsize
\begin{pmatrix} n_A+n_B-n_C+k \\  n_B \end{pmatrix}
\begin{pmatrix} n_B+k \\  n_B \end{pmatrix}
              \over
\begin{pmatrix} n_A+n_B \\ n_B \end{pmatrix}
\begin{pmatrix} n_C+n_B \\ n_B \end{pmatrix}
}
} \,,
\end{align*}
where $ d_{ABC}=n_A+n_B+n_C+1 $. Note also that, when $\na = \nc$, the term proportional to $ d_{ABC} $ vanishes and the square root term simplifies, so we recover the closed form given in the main text [cf. Eq.~\eqref{ch4/ua-nm}].

The minimum-error probability can be computed entirely along the same lines. For a pair of states we have $P_{{\rm e},J}=\frac{1}{2}\left(1-\sqrt{1-4\pi^{1}_J\pi^{2}_Jc_J^2} \right)$ [cf. Eq.~\eqref{ch3/helstrompure}], and the total error probability reads
\begin{align}\label{ch4/measym}
\PME &\;=\; {1\over4}\left\{1+{d_1\over d_2}-{d_1+d_2\over d_1d_2}\sum_{k=0}^{n_C}
(n_A+n_B-n_C+2k+1) \right. \nonumber\\
&\hspace{2cm} \times\left.
\sqrt{
1-4{d_1d_2\over(d_1+d_2)^2}
{
\scriptsize
\begin{pmatrix} n_A+n_B-n_C+k \\  n_B \end{pmatrix}
\begin{pmatrix} n_B+k \\  n_B \end{pmatrix}
              \over
\begin{pmatrix} n_A+n_B \\ n_B \end{pmatrix}
\begin{pmatrix} n_C+n_B \\ n_B \end{pmatrix}
}
}
\right\} \,.
\end{align}
%
%
This expression coincides with Eq. (31) of~\citep{Akimoto2011}.
\\

\section{\boldmath Limit $n\to\infty$ for minimum error and pure states}\label{appA/sec:minerr_n_to_infty}

In this Section we determine the asymptotic form of the minimum-error probability $\PME$, given by Eq.~\eqref{ch4/min-nm}, in the limit of large $n$.
We first define $x=k/n$ and approximate the factorials in $\PME$ using the Stirling approximation $z!\approx z^z \mathrm{e}^{-z}\sqrt{2 \pi z}$. Expanding up to order $1/n$ in the limit $n\to\infty$, we can write
\begin{equation*}
\log \frac{(n'+k)!n!}{(n'+n)!k!} = n' \log x +\frac{n' (n'+1)(1-x)}{2xn} + O(n^{-2}) \,,
\end{equation*}
or, equivalently,
\begin{equation*}
\frac{(n'+k)!n!}{(n'+n)!k!} = x^{n'} +\frac{n'(n'+1)(1-x)x^{n'-1}}{2n} + O(n^{-2}) \,.
\end{equation*}
Hence, up to order $1/n$, the square root in the formula of the error probability is
\begin{equation*}
\sqrt{1-x^{2n'}} - \frac{n'(n'+1)(1-x)x^{2n'-1}}{2n\sqrt{1-x^{2n'}}} \,.
\end{equation*}
Also, note that
\begin{equation*}
\frac{n'+2k+1}{(n+1)(n+n'+1)} = \frac{2x}{n} + \frac{n'+1-2x(n'+2)}{n^2} + O(n^{-3}) \,.
\end{equation*}
Combining the two last equations we can write the nontrivial factor in $\PME$ as
\begin{equation*}
\frac{2x\sqrt{1-x^{2n'}}}{n} + \frac{[n'+1-2(n'+2)x](1-x^{2n'})-n'(n'+1)(1-x)x^{2n'}}{n^2\sqrt{1-x^{2n'}}}+O(n^{-2})
\end{equation*}
We next use the Euler-MacLaurin formula 
\begin{equation}\label{appA/euler_maclaurin}
\sum_{k=0}^n f(k) \approx n \int_0^1 dx f(x) + {f(1)+f(0)\over 2}
\end{equation}
to express the leading term in $\PME$ as
\begin{equation*}
\PME = {1\over2}\left\{1-2\int_0^1 dx\,x \sqrt{1-x^{2n'}}\right\} \,.
\end{equation*}
The change of variables $x=t^{1/2n'}$ leads to
\begin{equation*}
\PME = {1\over2}\left\{1-{1\over n'}\int_0^1dt \, t^{{1\over n'}-1}(1-t)^{1/2}\right\}
={1\over2}-{B({3\over2},{1\over n'})\over2 n'} \,,
\end{equation*}
where $B(a,b)$ is the standard Beta Function. Finally, we obtain
\begin{equation*}
\PME = {1\over2}-{\sqrt\pi\,\Gamma(1+{1\over n'})\over4 \Gamma({3\over2}+{1\over n'})} \,.
\end{equation*}

A lengthy, but rather straightforward, calculation yields the remarkable result that the subleading term has a coefficient which coincides with the value of the integral $\int_0^1 dx \, x \sqrt{1-x^{2n'}}$. At this order we therefore can write
\begin{equation*}
\PME=\frac{1}{2}-\frac{\sqrt{\pi}}{4}\frac{\Gamma(1+1/n')}{\Gamma(3/2+1/n')} \left(1-\frac{1}{n}\right) \,.
\end{equation*}
\\

\section{\boldmath Averaged $C_j^n$ coefficients}\label{app/ch4/averages}

Here we compute the average of the coefficients [see Eq.~\eqref{ch4/deconst}]
\begin{equation*}
C^n_j = \frac{1}{2j+1}\left(\frac{1-r^2}{4}\right)^{n/2-j} \sum_{k=-j}^j \left(\frac{1-r}{2}\right)^{j-k} \left(\frac{1+r}{2}\right)^{j+k}
\end{equation*}
for the hard-sphere, Bures and Chernoff priors, given by Eqs.~\eqref{ch4/hard} through \eqref{ch4/chernoff}, considered in the fully universal discrimination machine.

For the hard-sphere prior we have
\begin{equation*}
\aver{C_j^n}_\mathrm{HS}  = 3 \int C_j^n r^2 dr = 6 \,\frac{\Gamma(n/2+j+2) \Gamma(n/2-j+1)}{\Gamma(n+4)} \,.
\end{equation*}

The Bures distribution yields
\begin{equation*}
\aver{C_j^n}_{\mathrm{Bu}} = \frac{4}{\pi} \int C_j^n \frac{r^2}{\sqrt{1-r^2}} dr = \frac{4}{\pi}\frac{\Gamma(n/2+j+3/2) \Gamma(n/2-j+1/2)}{\Gamma(n+3)} \,.
\end{equation*}

The averages for the  Chernoff prior are a bit more involved, but  still can be given in a closed form as
\begin{eqnarray*}
\aver{C_j^n}_{\mathrm{Ch}} &=& \frac{1}{\pi-2} \int C_j^n \frac{\left(\sqrt{1+r}-\sqrt{1-r}\right)^2}{\sqrt{1-r^2}} dr \nonumber \\
&=&  \frac{2}{(\pi-2)(2j+1)} \sum_{m=-j}^j \left[ B_{1/2} \left( \tfrac{n+1-2m}{2},\tfrac{n+1+2m}{2}\right) \right.\nonumber \\
&&\phantom{xxxxxxxxxxx}\left. - 2 B_{1/2} \left( \tfrac{n-2m+2}{2},\tfrac{n+2m+2}{2}\right)\right]\,,
\end{eqnarray*}
where $ B_x(a,b)=\int_0^x t^{a-1} (1-t)^{b-1} dt $ is the incomplete beta function~\citep{Abramowitz1972}.

\chapter{Technical details of Chapter~\ref{ch5_learning}}

\section{\boldmath Covariance and structure of $\mathscr{L}$}\label{appB/sec:covariance}

We start with a POVM element of the form $\bar{E}_0=\int du\,U\, E_0\, U^{\dagger}$. Since $D_\mu$ must be a rank-one projector, it can always be written as $D_\mu = u_\mu\, [\,\uparrow\,] \,u_\mu^\dagger$ for a suitable SU(2) rotation $u_{\mu}$. Thus,
\begin{equation*}
{\bar E}_0 = \sum_\mu\int du \left(U_{AC} L_\mu U_{AC}^\dagger\right)\otimes   \left(u u_\mu[\,\uparrow\,]u_\mu^\dagger u^\dagger \right)\,.
\end{equation*}
We next use the invariance of the Haar measure $du$ to make the change of variable $u\,u_{\mu}\rightarrow u'$ and, accordingly,  $U_{AC} \rightarrow U'_{AC} U^{\dagger}_{\mu \, {AC}}$. After regrouping terms we have
\begin{eqnarray}\label{appB/E0bar}
{\bar E}_0 &=&  \sum_\mu  \int   du' \left( U'_{AC} U^\dagger_{\mu\, {AC}}  L_\mu U_{\mu\, {AC}}  U'{}_{AC}^\dagger\right)  \otimes   \left( u'[\,\uparrow\,]u'{}^\dagger\right)  \nonumber \\
&=&   \int   du'  \left[U'_{AC} \left(  \sum_\mu U^\dagger_{\mu\, {AC}}  L_\mu U_{\mu\, {AC}}  \right)  U'{}_{AC}^\dagger\right] \otimes   \left(u'[\,\uparrow\,]u'{}^\dagger \right)\nonumber \\
&=& \int   du \left(U_{AC}\, \Omega\, U^\dagger_{AC}\right)\otimes \left(u[\,\uparrow\,] u^{\dagger}\right) \,, 
\end{eqnarray}
where we have defined 
\begin{equation*}
{\Omega} =\sum_\mu U^\dagger_{\mu\,AC}L_\mu U_{\mu\,AC}\geqslant 0 \,.
\end{equation*}
The POVM element $\bar{E}_1$ is obtained by replacing~\mbox{$[\,\uparrow\,]$} by~\mbox{$[\,\downarrow\,]$} in the expressions above. From the POVM~condition $\sum_\mu L_\mu=\id_{AC}$ it immediately follows that
\begin{equation*}
\int du  U_{AC} \,{\Omega}\, U^\dagger\kern-.3em{}_{AC}= \id_{AC} \,,
\end{equation*}
where $\id_{AC}$ is the identity on the Hilbert space of the TS, i.e., $ \id_{AC}=\id_A \otimes \id_C$. Therefore ${\mathscr L}=\{U_{AC} \,{\Omega} \,U^\dagger\kern-.3em{}_{AC}\}_{\rm SU(2)}$ is a covariant POVM. The positive operator $\Omega$ is called the seed of the covariant POVM~$\mathscr L$. 

Now, let $u_z(\varphi)$ be a rotation about the $z$-axis, which leaves $[\,\uparrow\,]$ invariant. By performing the change of variables $u\rightarrow u'u_z(\varphi)$ [and $U_{\,AC}\rightarrow U'_{AC}U_{\,zAC}(\varphi)$] in Eq.~\eqref{appB/E0bar}, we readily see that $\Omega$ and $U_{\,zAC}(\varphi)\,\Omega\,U^\dagger_{\,zAC}(\varphi)$ both give the same average operator $\bar E_0$ for any $\varphi\in[0,4\pi)$. So, its average over $\varphi$,
\begin{equation*}
\int_0^{4\pi} {d\varphi\over4\pi} U_z(\varphi)\,\Omega\, U_z^\dagger(\varphi) \,,
\end{equation*}
can be used as a seed without loss of generality, where we have dropped the subscript $AC$ to simplify the notation. Such a seed is by construction invariant under the group of rotations about the $z$-axis (just like $[\,\uparrow\,]$) and, by Schur's lemma, a direct sum of operators with well defined magnetic number. Therefore, in the total angular momentum basis for~$AC$, we can always choose the seed of $\mathscr L$ as
%
\begin{equation*}
 \Omega=\sum_{m=-n}^n\Omega_m \,; 
\quad \Omega_m\geqslant0 \,.
\end{equation*}
%
%
The constraint \eqref{ch5/omega_m} follows from the POVM condition ${\id}_{AC}=\int du\, U\,\Omega\, U^\dagger$ and Schur's lemma.
The result also holds if~$A$ and $C$ have different number of copies (provided they add up to $2n$). 
It also holds for mixed states.
\\

\section{Overlaps}\label{appB/sec:overlaps}

For the proof of optimality of the LM, we couple subsystems~$A$,~$B$ and~$C$ in two ways: $A(CB)$ and $(AC)B$ to produce the states $\ket{j_A,(j_C\,j_B)j_{CB};J,M}$ and $\ket{(j_A\,j_C)j_{AC},j_B;J,M}$, which we denote by $|J,M\rangle_{\Xyz ACB}$ and $|J,M\rangle_{\xyZ ACB}$ respectively for short. The various angular momenta involved are fixed to \mbox{$j_A=j_C=\mbox{\footnotesize${n\over2}$}$}, $j_B=\half$, $j_{AC}=j$, $j_{CB}=\mbox{\footnotesize${n\over2}$}+\half$, whereas $J=j\pm\half$. With these values, the general expression~\eqref{appA/wigner6j} gives us the overlaps that we need:
\begin{equation*}
  {{}_{\Xyz ACB}\kern-.1em\langle j\pm\half,\half|j\pm\half,\half\rangle_{\xyZ ACB}}=
  \sqrt{\frac{
  {n+\mbox{\footnotesize${3\over2}$}\pm(j+\half)}}{2(n+1)}}\,.
\end{equation*}
%
\\

\section{\boldmath Measurement of a block-diagonal $\rho^{\otimes n}$}\label{appB/sec:block}

The state $\rho^{\otimes n}$ of $n$ identical copies of a general qubit state $\rho$ with purity~$r$ and Bloch vector $r\vec{s}$, has a block diagonal form in the basis of the total angular momentum (see Section~\ref{ch3/sec:blockdecomposition}) given by
\begin{equation*}
\rho^{\otimes n} = \sum_{j} p^n_j  \rho_{j} \otimes {\id_j\over \nu^n_j} \,.
\end{equation*}
Here $j=0\,(1/2),\hdots,n/2$ if $n$ is even (odd), $\idd_j$ is the identity in the multiplicity space ${\mathbb C}^{\nu^n_j}$, of dimension $\nu^n_j$ (the multiplicity of the representation with total angular momentum $j$), where
\begin{equation*}
\nu^n_j= {n\choose n/2-j} - {n \choose n/2-j-1} \,.
\end{equation*}
[cf. Eq.~\eqref{ch3/rhoblock3}]. The normalized state $\rho_{j}$, which is supported on the representation  subspace $\mathscr{S}_{j}={\rm span}\{\ket{j,m}\}$ of dimension~$2j+1=d_{2j}$,~is
\begin{equation*}
\rho_{j} = U_{\scrvec  s}\left( \sum_{m=-j}^{j} a^j_m\; [j,m]\right)U_{\scrvec  s}^{\dagger}  \,,
\end{equation*}
where
\begin{equation}
a^j_m=\frac{1}{c_j} \left(\frac{1-r}{2}\right)^{j-m} \left(\frac{1+r}{2}\right)^{j+m} \,,
\label{appB/a_m}
\end{equation}
and
\begin{equation*}
c_j = \frac{1}{r} \left\{\left(\frac{1+r}{2}\right)^{2j+1}\!\!\!-\left(\frac{1-r}{2}\right)^{2j+1}\right\} \,,
\end{equation*}
so that $\sum_{m=-j}^j a^j_m=1$,
and we stick to our shorthand notation $[\,\cdot\,]\equiv|\,\cdot\,\rangle\langle\,\cdot\,|$, i.e., $[j,m]\equiv \ketbrad{j,m}$.
The measurement on $\rho^{\otimes n}$ defined by the set of projectors on the various subspaces~$\mathscr{S}_{j}$ will produce $\rho_{j}$ as a posterior state with probability
\begin{equation*}
p^n_j = \nu^n_j c_j \left(\frac{1-r^2}{4}\right)^{n/2-j} \,.
\end{equation*}
One can easily check that $\sum_j p^n_j=1$. 

In the large $n$ limit, we can replace $p^n_j$ for a continuous probability distribution $p_n(x)$ in $[0,1]$, where $x=2j/n$. Applying Stirling approximation to $p_j$ one obtains
\begin{equation*}
p_n(x)\simeq \sqrt{\frac{n}{2 \pi}}\frac{1}{\sqrt{1-x^2}} {x(1+r)\over r(1+x)}\;\mathrm{e}^{- n H(\frac{1+x}{2}\parallel\frac{1+r}{2})} \,,
\end{equation*}
where $H(s\parallel t)$ is the (binary) relative entropy
\begin{equation*}
H(s\parallel t)=s \log\frac{s}{t}+ (1-s)\log\frac{1-s}{1-t} \,.
\end{equation*}
The approximation is valid for $x$ and $r$ both  in the open unit interval~$(0,1)$. For nonvanishing $r$, $p_n(x)$ becomes a Dirac delta function peaked at $x=r$, $p_\infty(x)=\delta(x-r)$, which corresponds to~$j=nr/2$.
\\

\section{Derivation of Eqs.~\eqref{ch5/ja=jc} and~\eqref{ch5/ja=jc asymp}}\label{appB/sec:robust}

Let us start with the general case where $\xi=\{j,j'\}$. To obtain $\sigma^n_{0,\xi}$ we first write Eqs.~\eqref{ch5/sigmaxi1} and~\eqref{ch5/sigmaxi2} as the SU(2) group integrals 
%
\begin{eqnarray*}
\sigma^n_{0,\xi} &=& \int du \, U_{AB} \left( \sum_{m=-j}^j a^j_m [j,m]_A \otimes \rho^B_0 \right)  U_{AB}^{\dagger} \\
&&\hspace{.6cm} \otimes  \int du' \, U'_C \left(\sum_{m=-j'}^{j'} a^{j'}_m [j',m]_C \right) U'^{\dagger}_C  \,,
\end{eqnarray*}
where $a^j_m$ is given in Eq.~\eqref{appB/a_m},
%
%
and $\rho^B_0$ is the mixed state~$\rho_0$, Eq.~\eqref{ch5/rho mixed}, of the qubit~$B$. We next couple $A$ with $B$ (more precisely, their subspaces of angular momentum~$j$) using the Clebsch-Gordan coefficients 
\begin{eqnarray*}
|\!\braket{j+\half,m+\half}{j,m;\half,\half}\!|^2 &=&\frac{j+ m+1}{2j+1} \, , \\
|\!\braket{j-\half,m+\half}{j,m;\half,\half}\!|^2 &=&\frac{j- m}{2j+1} \, .
\end{eqnarray*}
The resulting expressions can be easily integrated using Schur's lemma. Note that the integrals of crossed terms of the form $\ketbra{j,m}{j',m}$ will vanish for all $j\neq j'$. We~readily obtain
\begin{equation*}
\sigma^n_{0,\xi}= \sum_{m=-j}^j a^j_m  \left( \frac{j+1+m r}{d_{2j}}\, {\id_{2j+1}^{AB} \over  d_{2j+1}}+ \frac{j-m r}{d_{2j}}\,{ \id_{2j-1}^{AB}\over  d_{2j-1} } \right) \otimes \frac{ \id_{2j'}^C}{d_{2j'}} \,,
\end{equation*}
%
%
%
where 
$\id_{2j}$ is the projector on~$\mathscr{S}_{j}$ and $d_{2j}=2j+1=\dim\mathscr{S}_{j}$. The superscripts attached to the various projectors specify the subsystems to which they refer. These projectors are formally equal to those used in Eq.~\eqref{ch5/sigma states} (i.e., $\id_{2j}$ projects onto the fully symmetric subspace of $2j$ qubits), hence we stick to the same notation.   Note that $\tr\sigma^n_{0,\xi}=1$, as it should be. 

We can further simplify this expression by introducing $\langle\hat{J}_z\rangle_j=\sum_m m\, a^j_m$, i.e.,  the expectation value of the $z$-component of the total angular momentum in the state~$\rho_{j}$ (i.e., of $\id_{2j}\hat{J}_z\id_{2j}$ in the state~$\rho^{\otimes n}_{0/1}$)  for a Bloch vector~$r {\vec{z}} $:
\begin{equation*}
\sigma^n_{0,\xi}=  \left( \frac{j+1+ r\langle\hat{J}_z\rangle_{j}}{d_{2j}}\, {\id_{2j+1}^{AB} \over  d_{2j+1}}+ \frac{j-r\langle\hat{J}_z\rangle_{j}}{d_{2j}}\,{ \id_{2j-1}^{AB}\over  d_{2j-1} } \right) \otimes \frac{ \id_{2j'}^C}{d_{2j'}} \,.
\end{equation*}
%
%
Using the relation
\begin{equation*}
\id_{2j-1}^{AB}=\id_{2j}^{A}\otimes \id^B_{1}-\id_{2j+1}^{AB}\,,
\end{equation*}
%
and
$(j+1)/d_{2j+1}=j/d_{2j-1}=1/2$, we can write
\begin{equation}
\sigma^n_{0,\xi}=  \left( {r\langle\hat{J}_{z}\rangle_{j}\over j}{\id_{2j+1}^{AB} \over  d_{2j+1}}
+ \frac{j-r\langle\hat{J}_{z}\rangle_{j}}{j}\,{\id_{2j}^{A}\over d_{2j}}\otimes {\id^B_{1}\over2}\right) \otimes \frac{ \id_{2j'}^C}{d_{2j'}}\,.\label{appB/new sigma0xi}
\end{equation}
Similarly, we can show that
\begin{equation}
\sigma^n_{1,\xi}= \frac{ \id_{2j}^A}{d_{2j}} \otimes \left( \frac{r\langle\hat{J}_{z}\rangle_{j'}}{j'}{\id_{2j'+1}^{BC} \over  d_{2j'+1}}+ \frac{j'-r\langle\hat{J}_{z}\rangle_{j'}}{j'}\, {\id^B_{1}\over2}\otimes {\id^{C}_{2j'}\over d_{2j'}} \right) \,.
\label{appB/new sigma1xi}
\end{equation}
Therefore, if $j'=j$,
\begin{equation*}
\sigma^n_{0,\xi}-\sigma^n_{1,\xi}={r\langle\hat{J}_z\rangle_{j}\over j}\left({\id^{AB}_{2j+1}\over d_{2j+1}}\otimes{\id^{C}_{2j}\over d_{2j}}-{\id^{A}_{2j}\over d_{2j}}
\otimes {\id^{BC}_{2j+1}\over d_{2j+1}}\right) \,.
\end{equation*}
Comparing with Eq.~\eqref{ch5/sigma states}, 
the two terms in the second line can be understood as 
the average states for a number of $2j$ pure qubits, i.e., as $\sigma^{2j}_0$ and  $\sigma^{2j}_1$~respectively. Hence if $\xi=\{j,j\}$ we have the relation
\begin{equation*}
\sigma^n_{0,\xi}-\sigma^n_{1,\xi}  = \frac{r \mean{\hat{J}_z}_j}{j}   \left(\sigma^{2j}_{0}-\sigma^{2j}_{1}\right) \,,
\end{equation*}
which is Eq.~\eqref{ch5/ja=jc}. 
It is important to emphasize that this equation is exact (i.e.,~it holds for any value of $j$, $n$ and~$r$) and bears no relation whatsoever to measurements, for it is just an algebraic identity between the various operators involved.

In the asymptotic limit, for $n_A$ and $n_C$ of the form $n_{A/C}\simeq n\pm b n^{a}$,  $n\gg 1$, $a<1$, the probabilities $p^n_j$ and $p^n_{j'}$ are peaked at $j\simeq r n_A/2$ and~$j'\simeq r n_C/2$, as was explained in Section~\ref{appB/sec:block}. Hence only the average state components $\sigma^n_{0/1,\xi}$ with $\xi=\{j,j'\} $ such that $j \simeq (r/2) n(1+b n^{a-1})$ and $j' \simeq (r/2) n(1-b n^{a-1})$ are important. From~Eqs.~\eqref{appB/new sigma0xi} and~\eqref{appB/new sigma1xi} it is straightforward to obtain

\begin{equation*}
\sigma^n_{0,\xi}-\sigma^n_{1,\xi}  \simeq r \left(1-{1-r\over n r^2}\right)  \left(\sigma^{rn}_{0}-\sigma^{rn}_{1}\right) +o(n^{-1}) \,,
\end{equation*}
where we have used that~\citep{Gendra2012} $\langle\hat{J}_z\rangle_j\simeq j-(1-r)/(2r)$ up to exponentially vanishing terms. This relation, for the particular value of $a=1/2$, is used in the proof of robustness, Eq.~\eqref{ch5/ja=jc asymp}.
\\

\section{\boldmath Calculation of $\Gamma_\uparrow$}\label{appB/sec:Gamma_up}

Here we calculate $\Gamma_{\uparrow,\xi}=\tr_{\!B}\{[\,\uparrow\,] (\sigma^n_{0,\xi}-\sigma^n_{1,\xi})\}$, where the average states are defined in Eqs.~\eqref{ch5/sigmaxi1} and \eqref{ch5/sigmaxi2}, and explicitly given in Eqs.~\eqref{appB/new sigma0xi} and \eqref{appB/new sigma1xi}  for~$\xi=\{j,j'\}$. Let us first calculate the conditional state $\tr_{\!B}([\,\uparrow\,] \sigma^n_{0,\xi})$. For that, we need to express $\id^{AB}_{2j+1}=\sum_m [j+\half,m]$ in the original product basis $\{\ket{j_A,m_A}\otimes\ket{\uparrow/\downarrow}\}$.
Recalling the Clebsch-Gordan coefficients $|\braket{\half,\half; j, m}{j+\half,m+\half}|^2=(j+m+1)/(2j+1)$, one readily obtains
\begin{equation*}
\tr_{\!B}\left([\,\uparrow\,] {\id^{AB}_{2j+1}\over d_{2j+1}}\right)=\sum_{m=-j}^{j}\frac{j+1+m}{2(j+1)d_{2j}}[j,m]_A \,,
\end{equation*}
which can be written as
\begin{equation*}
\tr_{\!B}\left([\,\uparrow\,] {\id^{AB}_{2j+1}\over d_{2j+1}}\right)=
{1\over2}\left({\id^A_{2j}\over d_{2j}}+{1\over d_{2j}}{\hat{J}^A_z\over j+1}\right) \,,
\end{equation*}
where $\hat J^{A}_z$ is the $z$ component of the total angular momentum operator acting on subsystem $A$.
An analogous expression is obtained for  $\tr_{\!B}\left([\,\uparrow\,] \id^{BC}_{2j'+1}\right)$. Substituting in Eqs.~\eqref{appB/new sigma0xi} and~\eqref{appB/new sigma1xi} and subtracting the resulting expressions, one has $\Gamma_\uparrow=\sum_{\xi}p^n_\xi \Gamma_{\uparrow,\xi}$, with
\begin{equation}\label{appB/jA - jC 2}
\Gamma_{\uparrow,\xi}={1\over2d_{2j_A}d_{2j_C}}
\left(
{r\langle\hat{J}_z\rangle_{j_A}\over j_A}{\hat{J}^A_z\over j_A+1}
-
{r\langle\hat{J}_z\rangle_{j_C}\over j_C}{\hat{J}^C_z\over j_C+1}
\right) \,,
\end{equation}
where we have written $\xi=\{j_A,j_C\}$, instead of $\xi=\{j,j'\}$ used in the derivation. For pure states,
$r=1$, 
$j_A=j_C=n/2$, $\langle\hat{J}_z\rangle_{n/2}=n/2$, and we recover Eq.~\eqref{ch5/JA-JC}.

In order to minimize the excess risk using SDP, we find it convenient to write Eq.~\eqref{ch5/Delta SDP}  in the form
\begin{equation}\label{appB/SDP methods 1}
\Delta^{\rm LM}=2\max_{\{\Omega_{m,\xi}\}} \sum_\xi p^n_\xi \tr(\Gamma_{\uparrow,\xi}\Omega_{m,\xi}) \,,
\end{equation}
where we recall that $m=m_{AC}=m_A+m_C$, and  we assumed w.l.o.g. that the seed of the optimal POVM has the block form~$\Omega_{m}=\sum_\xi\Omega_{m,\xi}$. The POVM condition, Eq.~\eqref{ch5/omega_m} must now hold on each block, thus for~\mbox{$\xi=\{j_A,j_C\}$}, we must impose that
\begin{equation}\label{appB/SDP methods 2}
\sum_{m=-j}^j\langle j,m|\Omega_{m,\xi}|j,m\rangle=2j+1,\,|j_A-j_C|\leqslant j\leqslant j_A+j_C \,.
\end{equation}

\chapter{Continuous-variables systems}

\label{appC}

A continuous-variables (CV) system is a bosonic system described by a Hilbert space of infinite dimension. CV systems provide the appropriate description of the states of light, and they have earned an outstanding role in quantum information and communication, as quantum optical settings allow to successfully implement, with current technology, quantum processing tasks such as quantum teleportation \citep{Furusawa1998}, quantum key distribution \citep{Grosshans2003}, 
and quantum dense coding \citep{Li2002}.
Special tools are required for describing this type of systems. The purpose of this Section is to give an overview on the formalism of CV systems that underlies in Chapter~\ref{ch6_learningcv}. For more complete reviews on the topic, see \citep{Braunstein2005,Eisert2003,Cerf2007}.

A CV system of $N$ canonical bosonic modes is described by a Hilbert space $\mathcal{H}=\bigotimes_{i=1}^N \mathcal{H}_i$, resulting from the tensor product structure of infinite dimensional spaces $\mathcal{H}_i$, each of them associated to a single mode.
Each mode is described by a pair of canonical conjugate operators $\hat{q}_i$ and $\hat{p}_i$, acting on $\mathcal{H}_i$. These operators may correspond, for instance, to position and momentum operators associated to a second quantized electromagnetic field, which Hamiltonian 
\begin{equation}\label{appC/H}
\hat{H} = \sum_{i=1}^N \hslash \omega_i \left( \hat{a}_i^\dagger \hat{a}_i + \frac{1}{2} \right)
\end{equation}
describes a system of $N$ noninteracting harmonic oscillators with different frequencies $\omega_i$, the \emph{modes} of the field. Another example susceptible of a canonical description is the collective spin of a polarized ensemble of atoms \citep{Julsgaard2001}. The ladder operators $\hat{a}_k$ and $\hat{a}_k^\dagger$ relate to the quadrature phase operators (position and momentum) according to
\begin{equation}\label{appC/qp_a}
\hat{q}_k = \frac{\hat{a}_k+\hat{a}_k^\dagger}{\sqrt{2}} \,,\quad \hat{p}_k = \frac{\hat{a}_k-\hat{a}_k^\dagger}{i\sqrt{2}} \,,
\end{equation}
and they obey the canonical commutation relation (CCR)
\begin{equation*}
[\hat{a}_k,\hat{a}_l^\dagger] = \delta_{kl} \,,\quad [\hat{a}_k,\hat{a}_l]=[\hat{a}_k^\dagger,\hat{a}_l^\dagger]=0 \,,
\end{equation*}
which, in terms of $\hat{q}_k$ and $\hat{p}_k$, reads\footnote{The canonical operators are chosen to be adimensional, hence $\hslash$ does not appear explicitly in any of the equations.}
\begin{equation}\label{appC/CCR}
[\hat{q}_k,\hat{p}_k] = i \openone_k\,,
\end{equation}
where $\openone_k$ is the identity operator on mode $k$. The canonical operators of all modes of the system can be grouped in the vector $\hat{R}=(\hat{q}_1,\hat{p}_1,\ldots,\hat{q}_N,\hat{p}_N)^T$. In this notation, the CCR \eqref{appC/CCR} reads
\begin{equation*}
[\hat{R}_k,\hat{R}_l]=i\Omega_{kl}\,,
\end{equation*}
where $k,l=1,2,\ldots,2N$, and $\Omega$ is the symplectic matrix
\begin{equation*}
\Omega = \bigoplus_{i=1}^N \omega \,,\quad \omega = \begin{pmatrix} 0 & 1 \\ -1 & 0 \end{pmatrix} \,.
\end{equation*}
\\

\section{The phase-space picture}

The states of a CV system are the set of positive trace-class operators $\{\rho\}$ on the Hilbert space $\mathcal{H}=\bigotimes_{i=1}^N \mathcal{H}_i$. The complete description of any state $\rho$ of such an infinite-dimensional system can be conveniently provided by the ($0$-ordered) \emph{characteristic function}
\begin{equation}\label{appC/characteristicfunction}
\chi (\xi) = \tr(\rho \hat{D}_\xi) \,,
\end{equation}
where $\xi\in\mathbb{R}^{2N}$, and $\hat{D}_\xi$ is a Weyl operator (see below). The vector $\xi$ belongs to the $2N$-dimensional real vector space $\Gamma(\mathbb{R}^{2N},\Omega)$ called \emph{phase space}. From the form of Eq.\eqref{appC/characteristicfunction} one can readily see that the tensor-product structure of the Hilbert space is replaced by a direct sum structure in the phase space, such that $\Gamma = \bigoplus_{i=1}^N \Gamma_i$, where $\Gamma_i(\mathbb{R}^2,\omega)$ is the local phase space of mode $i$. The Weyl operator $\hat{D}_\xi$ acts in the states as a translation in the phase space. It is defined as
\begin{equation}\label{appC/weylN}
\hat{D}_\xi = e^{-i\hat{R}^T \Omega \xi} \,,
\end{equation}
and its action over an arbitrary vector of canonical operators $\hat{R}$ yields
\begin{equation*}
\hat{D}_\xi^\dagger \hat{R}_i \hat{D}_\xi = \hat{R}_i - \xi_i \openone\,.
\end{equation*}

The characteristic function $\chi(\xi)$ is related, via a Fourier transform, to the so-called \emph{Wigner function}
\begin{equation}\label{appC/wignergeneral}
W(\xi) = \frac{1}{(2\pi)^{2N}} \int_{\mathbb{R}^{2N}} d^{2N}\kappa \chi(\kappa) e^{i \kappa^T \Omega \xi} \,,
\end{equation}
that constitutes an alternative complete description of quantum states for CV systems. The Wigner function is a real-valued \emph{quasi-probability distribution}\footnote{There exist alternative ways of defining quasi-probability distributions for CV states for which the Wigner function is not an appropriate description. These variations are derived from alternative definitions of the characteristic function \citep{Leonhardt1997}.}. This denomination is motivated from the fact that the function $W(\xi)$ might be negative or ill-behaved in certain regions of the phase space, and nevertheless it quantifies the probability with which one might expect to obtain the values $\xi$ upon measuring simultaneously the canonical operators $\hat{R}$. The following properties are worth remarking:
\begin{enumerate}
\item $W(\xi)$ is normalized, i.e.,
\begin{equation*}
\int_{\mathbb{R}^{2N}} d^{2N}\kappa W(\kappa) = \tr \rho = \chi(0) = 1 \,.
\end{equation*}
\item In terms of $W(\xi)$, the purity of a state $\rho$ is expressed as
\begin{equation*}
\int_{\mathbb{R}^{2N}} d^{2N}\kappa W^2(\kappa) = \int_{\mathbb{R}^{2N}} d^{2N}\xi |\chi(\xi)|^2 = \tr \rho^2 = \mu \,.
\end{equation*}
\item The overlap between two states $\rho_1$ and $\rho_2$ corresponds to 
\begin{equation*}
\tr(\rho_1 \rho_2) = 2\pi \int_{\mathbb{R}^{2N}} d^{2N}\kappa W_1(\kappa)W_2(\kappa) \,.
\end{equation*}
\end{enumerate}

The phase-space formulation offers the theoretical tools to map states and operations of infinite-dimensional CV systems into relations in finite real spaces. Both the density matrix and the Wigner function provide a complete description of the state of a CV system, hence a one-to-one correspondence between them exists. For a single mode state, i.e., $\xi=(q,p)$, it is of the form
\begin{equation*}
W(q,p) = \frac{1}{\pi} \int_{-\infty}^{\infty} dx \expect{q+x}{\rho}{q-x} e^{-2ipx} \,,
\end{equation*}
which is Wigner's legendary formula \citep{Wigner1932}.
\\

\section{The states of light}

As stated above, the ($N$-mode) electromagnetic field, the paradigm of CV systems, can be modelled by the Hamiltonian of $N$ noninteracting harmonic oscillators given in Eq.~\eqref{appC/H}. The states of the harmonic oscillator associated to the $i$th mode belong to the Hilbert space $\mathcal{H}_i$, and this space is spanned by the eigenstates of the number operator $\hat{n}_i=\hat{a}_i^\dagger \hat{a}_i$ that represents the corresponding Hamiltonian. These states form the so-called Fock basis $\{\ket{n}_i\}$, verifying
\begin{equation*}
\hat{n}_i \ket{n}_i = n_i \ket{n}_i \,,
\end{equation*}
where $n_i=0,\ldots,\infty$ gives the quanta of excitations of mode $i$. The Hamiltonian of each mode is bounded from below, thus ensuring the stability of the system. For the $i$th mode, the ground state of the oscillator or \emph{vacuum state} of the field is that which is annihilated by the operator $\hat{a}_i$, i.e., $\hat{a}_i \ket{0}_i = 0$. The vacuum state of the global Hilbert space is just $\ket{0}=\bigotimes_i \ket{0}_i$. The Fock state $\ket{n}_i$ can be regarded as the $n$th excitation (photon) of the vacuum of mode $i$, obtained by the action of the annihilation ($\hat{a}_i$) and creation ($\hat{a}_i^\dagger$) operators (recall that $\hat{a}^\dagger\ket{n}=\sqrt{n+1}\ket{n+1}$ and $\hat{a}\ket{n}=\sqrt{n}\ket{n-1}$), i.e.,
\begin{equation*}
\ket{n}_i = \frac{(\hat{a}_i^\dagger)^n}{\sqrt{n_i!}}\ket{0}_i
\end{equation*}

The Fock states, with the exception of the vacuum, belong to the broader class of non-Gaussian states. In general, non-Gaussian states are difficult to handle, both mathematically and experimentally. By contrast, Gaussian states exhibit much nicer properties and comprise an extremely relevant class of CV states, since the vast majority of the states prepared in quantum optics laboratories are of this type\footnote{For a review on the uses of Gaussian states in quantum information applications, see~\citep{Weedbrook2012}.}.

The set of \emph{Gaussian states} is, by definition, the set of states with Gaussian characteristic functions and quasi-probability distributions on the multimode quantum phase space. Gaussian states include, among others, coherent, squeezed, and thermal states. 
From its very definition, it follows that a Gaussian state $\rho$ is completely characterized by the first and second statistical moments of the quadrature field operators, embodied in the vector of first moments $\bar{R}$ and the \emph{covariance matrix} (CM) $\bm\sigma$, respectively, which elements are
\begin{eqnarray*}
\bar{R}_i &=& \mean{\hat{R}_i} \,,\\
\sigma_{ij} &=& \mean{\hat{R}_i\hat{R}_j+\hat{R}_j\hat{R}_i} - 2\mean{\hat{R}}_i\mean{\hat{R}_j} \,,
\end{eqnarray*}
and where $i,j=1,\ldots,2N$. The Wigner function of a Gaussian state $\rho$ has the form
\begin{equation}\label{appC/wignergaussian}
W(X)=\frac{1}{\pi^N \sqrt{\det{\bm\sigma}}} e^{-(X-\bar{R}){\bm\sigma}^{-1}(X-\bar{R})^T} \,,
\end{equation}
where $X$ stands for the real phase-space vector $(q_1,p_1,\ldots,q_n,p_n)\in\Gamma$.

The vector of first moments $\bar{R}$ can be arbitrarily adjusted by local unitary operations, namely displacements in phase space by means of Weyl operators \eqref{appC/weylN}. Since the reduced state resulting from a partial trace operation over a subset of modes of a Gaussian state is still Gaussian, one can apply single-mode Weyl operators to locally re-center each such reduced Gaussian.
Such operations leave all the informationally relevant properties of the state invariant, hence in general the first moments can be adjusted to $0$ without loss of generality. It follows that, despite the infinite dimension of the associated Hilbert space, the complete description of an arbitrary Gaussian state (up to local unitary operations) is given by its $2N\times 2N$ CM $\bm\sigma$.
For a CM to describe a proper physical state, it must verify the condition
\begin{equation*}
{\bm\sigma}+i\Omega \geqslant 0 \,,
\end{equation*}
analogous to the semidefinite-positive condition for the density matrix $\rho\geqslant 0$.

Generically, a $N$-mode Gaussian state has a CM $\bm\sigma$ that can be written in terms of $2\times 2$ submatrices as
\begin{equation*}
{\bm\sigma}=
\begin{pmatrix}
{\bm\sigma}_1 & {\bm\epsilon}_{1,2} & \cdots & {\bm\epsilon}_{1,N} \\
{\bm\epsilon}_{1,2}^T & \ddots & \ddots & \vdots \\
\vdots & \ddots & \ddots & {\bm\epsilon}_{N-1,N} \\
{\bm\epsilon}_{1,N}^T & \cdots & {\bm\epsilon}_{N-1,N}^T & {\bm\sigma}_{N}
\end{pmatrix} \,.
\end{equation*}
The diagonal block $\bm\sigma_i$ is the local CM of the corresponding reduced state of mode $i$. On the other hand, the off-diagonal matrices $\bm\epsilon_{i,j}$ encode the intermodal correlations (both classical and quantum) between modes $i$ and $j$. A product state has no off-diagonal terms, hence its CM is simply the direct sum of the local CMs. Properties like the entanglement of a state and its purity, and linear transformations of first moments in phase space (symplectic transformations), can all be described within the CM formalism.

The three most important types of single-mode Gaussian states are coherent, squeezed, and thermal states.
\\

\subsection{Coherent states}

Coherent states are the states produced by an ideal laser. They are ubiquitous in CV quantum information, and, among all CV states, their dynamics is the one that most resembles the behaviour of a classical electromagnetic field. Coherent states have minimal quantum uncertainty, which means that fluctuations are symmetrically distributed between its quadratures. 

\begin{figure}[t]
\begin{center}
\includegraphics[scale=.55]{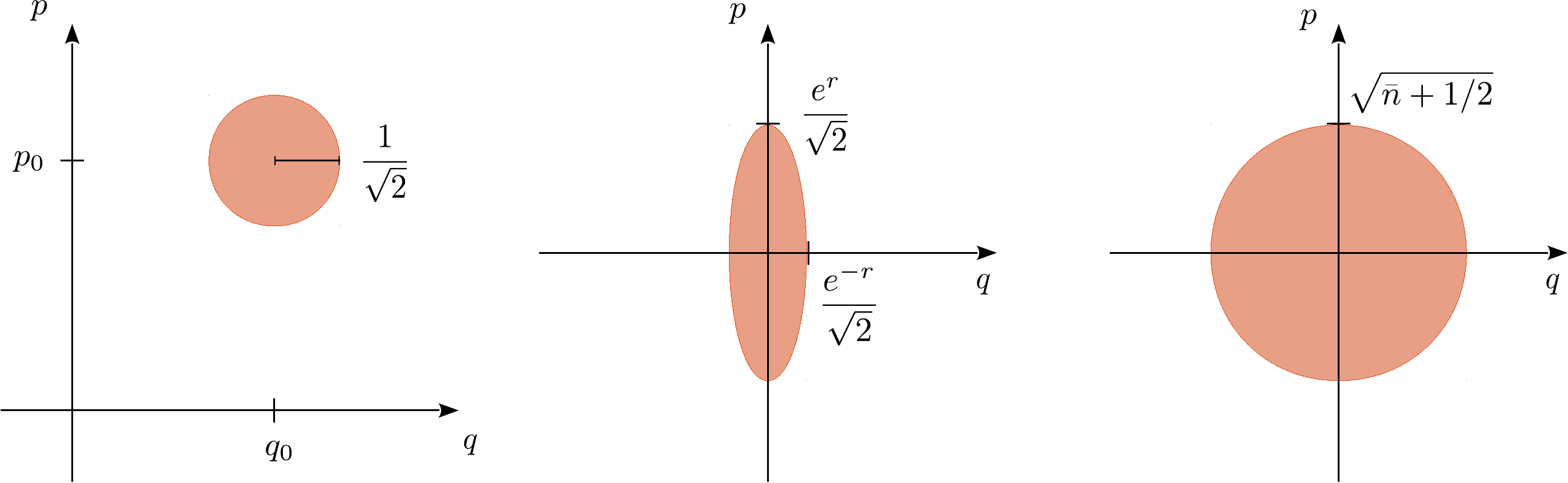}
\end{center}
\caption[Coherent, squeezed and thermal states in phase space]{From left to right: a coherent state of amplitude $\alpha=(q_0+ip_0)/\sqrt{2}$, a squeezed vacuum state with a squeezing parameter $r$, and a thermal state with average photon number $\bar{n}$.}\label{appC/fig:fig1}
\end{figure}

Coherent states can be defined as the eigenstates of the annihilation operator $\hat{a}$
\begin{equation*}
\hat{a}\ket{\alpha}=\alpha\ket{\alpha} \,,
\end{equation*}
where the eigenvalue $\alpha$, in general complex, is the \emph{amplitude} of the state $\ket{\alpha}$, and it is related to the quadratures through
\begin{equation*}
\alpha = \frac{q+ip}{\sqrt{2}} \,,
\end{equation*}
i.e., $q=\sqrt{2}{\rm Re}(\alpha)$ and $p=\sqrt{2}{\rm Im}(\alpha)$. The state $\ket{\alpha}$ results from applying the single-mode displacement operator $\hat{D}(\alpha)$ to the vacuum, that is 
\begin{equation*}
\ket{\alpha}=\hat{D}(\alpha)\ket{0} \,, 
\end{equation*}
where
\begin{equation}\label{appC/weyl}
\hat{D}(\alpha)=e^{\alpha \hat{a}^\dagger - \alpha^* \hat{a}} = e^{-|\alpha|^2/2}e^{\alpha\hat{a}^\dagger}e^{-\alpha^*\hat{a}}\,.
\end{equation}
The displacement operator can be identified with the single-mode Weyl operator~\eqref{appC/weylN} by using the relations in Eq.~\eqref{appC/qp_a}. In the Heisenberg picture, the action of $\hat{D}(\alpha)$ over the operator $\hat{a}$ yields the displacement
\begin{equation*}
\hat{D}^\dagger(\alpha)\,\hat{a}\,\hat{D}(\alpha) = \hat{a} + \alpha \,.
\end{equation*}
Another useful property is
\begin{equation}\label{appC/Dproperty}
\hat{D}^\dagger(\alpha)\hat{D}(\beta)=e^{-\frac{1}{2} \left(\alpha\beta^*-\beta\alpha^*\right)} \hat{D}(\beta-\alpha) \,.
\end{equation}

One can use the definition of the displacement operator, that is Eq.~\eqref{appC/weyl}, to express a coherent state in terms of Fock states:
\begin{eqnarray}
\ket{\alpha} &=& \hat{D}(\alpha)\ket{0} = e^{-|\alpha|^2/2}e^{\alpha\hat{a}^\dagger}e^{-\alpha^*\hat{a}} \ket{0} \nonumber\\
&=& e^{-|\alpha|^2/2} \sum_{n=0}^\infty \frac{(\alpha\hat{a}^\dagger)^n}{n!} \ket{0} \nonumber\\
&=& e^{-|\alpha|^2/2} \sum_{n=0}^\infty \frac{\alpha^n}{n!} \ket{n} \,.\label{appC/coherentFock}
\end{eqnarray}
The Fock representation~\eqref{appC/coherentFock} shows that a coherent state has the Poissonian photon statistics
\begin{equation*}
P(n)=|\braket{n}{\alpha}|^2 = \frac{|\alpha|^{2n}}{n!} e^{-|\alpha|^2} \,.
\end{equation*}
Note that the average photon number, or \emph{intensity}, of a coherent state is $\mean{\hat{n}}=\mean{\hat{a}^\dagger\hat{a}}=|\alpha|^2$.

The overlap between two coherent states $\ket{\alpha}$ and $\ket{\beta}$ can be readily seen to give, by means of Eq.~\eqref{appC/Dproperty},
\begin{eqnarray*}
\braket{\alpha}{\beta}&=&\expect{0}{\hat{D}^\dagger(\alpha)\hat{D}(\beta)}{0}\nonumber\\
&=&e^{-\frac{1}{2} \left(\alpha\beta^*-\beta\alpha^*\right)}\expect{0}{\hat{D}(\beta-\alpha)}{0} \nonumber\\
&=&e^{-\frac{1}{2} \left(\alpha\beta^*-\beta\alpha^*\right)}\braket{0}{\beta-\alpha} \nonumber\\
&=&e^{-|\alpha|^2/2-|\beta|^2/2+\alpha^*\beta}\,,
\end{eqnarray*}
hence
\begin{equation}\label{appC/aboverlap}
|\!\braket{\alpha}{\beta}\!|^2=e^{-|\alpha-\beta|^2}\,.
\end{equation}
Eq.~\eqref{appC/aboverlap} shows that two coherent states approach orthogonality only when their amplitude difference is large. Despite being nonorthogonal, coherent states form a basis in $\mathcal{H}$ (which is an \emph{overcomplete} basis, for this very reason), and fulfil the completeness relation
\begin{equation*}
\frac{1}{\pi} \int d^2\alpha \ketbrad{\alpha} = \openone \,.
\end{equation*}

The characteristic function of a coherent state can be straightforwardly obtained through Eqs.~\eqref{appC/characteristicfunction} and \eqref{appC/weyl}. One can then compute the corresponding Wigner function using Eq.~\eqref{appC/wignergeneral}, and compare the result with the Wigner function of a general Gaussian state, given by Eq.~\eqref{appC/wignergaussian}. This analysis shows that a coherent state $\ket{\alpha}$ has a displacement vector $\bar{R}=(q,p)$, and a CM $\bm\sigma=\openone$. This means that it has the same minimal fluctuations as the vacuum state, but displaced in phase space. Thus a coherent state can be depicted as a displaced circle of radius $1/\sqrt{2}$ in phase space (see Fig.~\ref{appC/fig:fig1}).
\\

\subsection{Squeezed states}

Squeezed states are states that have an asymmetrical distribution of fluctuations among their quadratures. That means, it is possible to reduce the uncertainty in one of the quadratures of a state, but this comes always at the expense of an increase in the noise of its conjugate variable, in accordance to Heisenberg's uncertainty principle. The preparation procedure of a squeezed state, that is the squeezing transformation, uses nonlinear optic elements and does not conserve the total photon number.

The single-mode squeezing operator is described by
\begin{equation}\label{appC/squeezedef}
\hat{S}(r,\phi) = e^{\frac{r}{2} \left(\hat{a}^2 e^{-2i\phi} - \hat{a}^{\dagger2} e^{2i\phi}\right)}\,,
\end{equation}
where $r\geqslant0$ is the \emph{squeezing parameter}. Its effect over the operators $\hat{a}$ and $\hat{a}^\dagger$ is
\begin{eqnarray*}
\hat{S}^\dagger(r,\phi) \hat{a} \hat{S}(r,\phi) &=& \hat{a} \cosh r-\hat{a}^\dagger e^{i\phi}\sinh r \,\\
\hat{S}^\dagger(r,\phi) \hat{a}^\dagger \hat{S}(r,\phi) &=& \hat{a}^\dagger \cosh r-\hat{a} e^{-i\phi}\sinh r \,.
\end{eqnarray*}
Applied instead to the rotated quadrature operators $\hat{q}_\phi=\hat{q}\cos\phi+\hat{p}\sin\phi$ and $\hat{p}_\phi=-\hat{q}\sin\phi+\hat{p}\cos\phi$, it yields
%
%
\begin{eqnarray*}
\hat{S}^\dagger(r,\phi)\hat{q}_\phi\hat{S}(r,\phi)&=&\hat{q}_\phi e^{-r} \,,\\
\hat{S}^\dagger(r,\phi)\hat{p}_\phi\hat{S}(r,\phi)&=&\hat{p}_\phi e^{r} \,.
\end{eqnarray*}

In the Fock representation, using Eq.~\eqref{appC/squeezedef} and taking $\phi=0$ for simplicity, the action of $\hat{S}$ over a vacuum state $\ket{0}$ results in the squeezed vacuum state
%
%
\begin{equation*}
\ket{0,r}=\hat{S}(r,0)\ket{0} = \frac{1}{\sqrt{\cosh r}}\sum_{n=0}^\infty \tanh^n r \frac{\sqrt{(2n)!}}{2^n n!} \ket{2n} \,.
\end{equation*}
The CM of the state $\ket{0,r}$ takes the simple form $\bm\sigma = {\rm diag} (e^{-2r},e^{2r})$, which accounts for the difference in the quadrature variances $\Delta^2q = \mean{\hat{q}^2}-\mean{\hat{q}}^2 = e^{-2r}/2$ and $\Delta^2p = \mean{\hat{p}^2}-\mean{\hat{p}}^2 = e^{2r}/2$. A squeezed vacuum state is thus depicted in phase space as an ellipse with an area equal to that of a minimal uncertainty state, i.e., $\pi/2$ (see Fig.~\ref{appC/fig:fig1}).
\\

\subsection{Thermal states}

The state of a single-mode field in thermal equilibrium with its environment is a thermal state, with density operator
\begin{equation}\label{appC/thermal}
\rho_{\rm th} = (1-e^{-\beta}) \sum_{n=0}^\infty e^{-\beta n} \ketbrad{n} \,,
\end{equation}
where $\beta=\omega/k_B T$ denotes the ratio between the energy $\omega$ and the temperature $T$ ($k_B$ stands for the Boltzmann's constant). To justify Eq.~\eqref{appC/thermal}, recall that in thermal equilibrium the density operator must be diagonal in the energy representation and that photons obey the Bose-Einstein statistics. The average photon number for the thermal state $\rho_{\rm th}$ is
\begin{equation*}
\bar{n}=\tr(\rho_{\rm th}\hat{n})=(1-e^{-\beta})\sum_{n=0}^\infty n e^{-\beta n} = \frac{1}{e^\beta-1} \,,
\end{equation*}
thus $\rho_{\rm th}$ can be expressed in terms of $\bar{n}$ as
\begin{equation*}
\rho_{\rm th}=\frac{1}{1+\bar{n}} \sum_{n=0}^\infty \left(\frac{\bar{n}}{1+\bar{n}}\right)^n \ketbrad{n}\,.
\end{equation*}

One can easily check that the Wigner function associated to the state $\rho_{\rm th}$ is Gaussian. The first moments vanish, hence its displacement is $\bar{R}=(0,0)$. Its CM is the diagonal matrix $\bm\sigma= (2\bar{n}+1)\openone$. The form of $\bm\sigma$ tells us that a thermal state has symmetric variances of its quadratures, and that these are proportional to $\bar{n}$ and, in turn, dependent on the temperature $T$. Thus a thermal state is a symmetric state of greater than minimal uncertainty. It can be depicted in phase space as a circle of radius $\sqrt{\bar{n}+1/2}$ (see Fig.~\ref{appC/fig:fig1}).\\

The most general \emph{mixed} Gaussian state is obtained by the sequential action of the squeezing~\eqref{appC/squeezedef} and displacement~\eqref{appC/weyl} operators on a thermal state~\eqref{appC/thermal}:
\begin{equation*}
\rho(\alpha,r,\phi) = \hat{D}(\alpha)\hat{S}(r,\phi)\rho_{\rm th}\hat{S}^\dagger(r,\phi)\hat{D}^\dagger(\alpha) \,.
\end{equation*}
The most general \emph{pure} Gaussian state is achieved by setting $\bar{n}=0$. This corresponds to a rotated, squeezed and displaced state $\ket{\alpha,r,\phi}=\hat{D}(\alpha)\hat{S}(r,\phi)\ket{0}$.
\\

\section{The measurements of light}

Quantum measurements of CV systems can be theoretically described by the POVM formalism. This is to say, one can describe a measurement by a set of positive-semidefinite operators $\{E_i\}$ such that $\sum_i E_i = \openone$. In contrast to the case of finite-dimensional systems, the set of outcomes of a measurement performed over a CV state $\rho$ is often continuous ($i\in\mathbb{R}$), so that $p(i)=\tr(E_i \rho)$ is a probability density function. A measurement is said to be Gaussian if, when applied to a Gaussian state, it yields outcomes that are Gaussian distributed. A property of such measurements is the following: given a $(N+M)$-mode Gaussian state, a Gaussian measurement of $N$ modes gives a Gaussian probability density function for the outcomes, and the remaining $M$ modes are left in a Gaussian state. From a practical point of view, any Gaussian measurement can be accomplished by homodyne detection, linear optics and Gaussian ancillary modes.
\\

\subsection{Homodyne detection}\label{appC/sec:homodyne}

The most common Gaussian measurement used in CV quantum information is homodyne detection. It consists in measuring one of the quadratures of a mode. Mathematically, this is done by projecting over the quadrature basis, i.e., if $\hat{q}$ ($\hat{p}$) is the quadrature to be measured, the POVM elements are $E_q=\ketbrad{q}$ ($E_p=\ketbrad{p}$), that is they are projectors onto infinitely squeezed states. Experimentally, the homodyne detection is implemented by combining the target quantum mode with a local oscillator (LO) in a balanced beam splitter and measuring the intensity of the two output modes with two photodetectors. The subtraction of the signal of both photodetectors gives a signal proportional to $\hat{q}$ ($\hat{p}$). 

The LO provides the phase reference $\phi$ for the quadrature measurement, thus by shifting the phase to $\phi\to\phi+\pi/2$ the other quadrature can be measured. For an arbitrary phase $\phi$, the POVM elements associated to the homodyne detection are
\begin{equation*}
E_{x_\phi} = \ketbrad{x_\phi} \,,
\end{equation*}
where $\hat{x}_\phi=\hat{q}\cos\phi + \hat{p}\sin\phi$.
\\

\subsection{Heterodyne detection}\label{appC/sec:heterodyne}

The heterodyne detection consists in, roughly speaking, measuring simultaneously \emph{both} quadratures. The target mode is mixed with the vacuum by means of a balanced beam splitter, then homodyne detection of the conjugate quadratures is performed over the outgoing signals. Note that, in this case, quantum mechanics does not raise any objections to the simultaneous measurement of conjugate quadratures. This can be understood by taking into account that the fluctuations of the vacuum field introduce extra noise in the signal, and, as a consequence, the precision in the measurement of each quadrature is diminished so that the Heisenberg's uncertainty principle is preserved.

The heterodyne measurement can be viewed as a POVM which elements are projectors onto coherent states, i.e., $E_\alpha = (1/\pi) \ketbrad{\alpha}$ \citep{Leonhardt1997}. This procedure can be generalized to any POVM composed of projectors over pure Gaussian states \citep{Giedke2002}. This means that the most general pure Gaussian measurement that yields information about both quadratures of a state, which may be called a \emph{generalized heterodyne measurement}, is achieved by a POVM with elements
\begin{equation*}
E_{\alpha,r,\phi} = \frac{1}{\pi} \ketbrad{\alpha,r,\phi} \,,
\end{equation*}
Moreover, such POVMs can be decomposed into a Gaussian unitary operation applied to the target mode and the ancillary modes (vacuum), the action of linear optical elements (beam splitters) and homodyne measurements on all output modes.
\\

\subsection{Photon counting and photodetection}

Despite being non-Gaussian measurements, photon counting and photodetection play an important role in certain quantum information tasks, such as discrimination of Gaussian states and entanglement distillation. The photon counting measurement consists in projecting onto the number-state basis, i.e.,
\begin{equation*}
E_n = \ketbrad{n} \,.
\end{equation*}
The measurement device is simply an optical receiver that converts light into electric current. When a single mode is excited, the receiver measures the intensity of the generated current, which is proportional to the photon number.

The photodetection measurement is a variant that serves to discriminate between two possible states: the vacuum, and one or more photons. The associated POVM elements are thus $E_0=\ketbrad{0}$ and $E_1=\openone-\ketbrad{0}$. In practice, photodetectors typically have a small efficiency, i.e., only a small fraction of photons is detected. Real photodetectors can be modelled by adding a beam splitter before an ideal photodetector, which transmissivity relates to the efficiency of the detector.

\chapter{Technical details of Chapter \ref{ch6_learningcv}}

\section{Trace norm for the collective strategy}\label{appD/sec:tracenormcol}

The global states that need to be discriminated in the collective strategy are $\bar{\sigma}_1$ and $\bar{\sigma}_2$. As shown in the main text, the first can be expressed as [cf. Eq.~\eqref{ch6/barsigma1exp}]
\begin{equation}\label{appD/barsigma1exp_bis}
\bar{\sigma}_1 = \sum_{k=0}^\infty c_k \ketbrad{k} \otimes \ketbrad{-\alpha_0} \,,
\end{equation}
whereas the second admits an asymptotic expansion [cf. Eq.~\eqref{ch6/barsigma2exp}]
\begin{equation}\label{appD/barsigma2exp_bis}
\bar{\sigma}_2 \sim \bar{\sigma}_2^{(0)} + \frac{1}{\sqrt{n}}\bar{\sigma}_2^{(1)} + \frac{1}{n}\bar{\sigma}_2^{(2)}
\end{equation}
as the result of taking the limit $n\to\infty$ up to order $1/n$ in Eq.~\eqref{ch6/barsigma2}. Computing the arising averages (see Appendix \ref{ch6/sec:gaussian integrals}), the terms in Eq.~\eqref{appD/barsigma2exp_bis} take the explicit form
%
%
\begin{eqnarray}
\bar{\sigma}_2^{(0)} &=& \sum_{k=0}^\infty   c_k  \ketbrad{k} \otimes \ketbrad{0}  \,, \nonumber \\
\bar{\sigma}_2^{(1)} &=& \sum_{k=0}^\infty  d_{k+1} \ketbra{k}{k+1} \otimes \ketbra{1}{0} + \tilde{d}_{k-1} \ketbra{k}{k-1} \otimes \ketbra{0}{1} \,, \label{ch6/sigma11} \\
\bar{\sigma}_2^{(2)} &=& \sum_{k=0}^\infty  e_{k} \ketbrad{k} \otimes \left( \ketbrad{1}-\ketbrad{0}\right) \nonumber \\
&&\,+\, f_{k+2}  \ketbra{k}{k+2} \otimes \ketbra{2}{0} + \tilde{f}_{k-2} \ketbra{k}{k-2} \otimes \ketbra{0}{2} \,, \label{ch6/sigma12}
\end{eqnarray}
where
\begin{eqnarray*}
d_{k+1}&=& c_{k+1}\sqrt{k+1} \,,\quad \tilde{d}_{k-1}=c_k \sqrt{k} \,,\\
e_k &=&  c_{k+1}(k+1)  \,,\\
f_{k+2} &=& \frac{1}{\sqrt{2}}c_{k+2}\sqrt{(k+2)(k+1)} \,,\quad 
\tilde{f}_{k-2} = \frac{1}{\sqrt{2}}c_k \sqrt{k(k-1)} \,.
\end{eqnarray*}

We now apply perturbation theory to compute the trace norm $\trnorm{\bar{\sigma}_1-\bar{\sigma}_2}$ in the asymptotic limit $n\to\infty$, up to order $1/n$, using Eqs.~\eqref{appD/barsigma1exp_bis} and \eqref{appD/barsigma2exp_bis}. We start by expressing the trace norm as
%
\begin{equation}\label{ch6/tracenorm col}
\trnorm{\bar{\sigma}_1-\bar{\sigma}_2} \sim\, |\!| A+B/\sqrt{n} + C/n \equiv \Gamma |\!|_1 = \sum_j |\gamma_j| \,,
\end{equation}
where $A=\bar{\sigma}_1 -\bar{\sigma}_2^{(0)}$, $B=-\bar{\sigma}_2^{(1)}$, $C=-\bar{\sigma}_2^{(2)}$, and $\gamma_j$ is the $j$th eigenvalue of $\Gamma$, which admits an expansion of the type $\gamma_j = \gamma_j^{(0)} + \gamma_j^{(1)}/\sqrt{n} + \gamma_j^{(2)}/n$.
The matrix $\Gamma$ belongs to the Hilbert space $\mathcal{H}_\infty \otimes \mathcal{H}_3$, i.e., the first mode is described by the infinite dimensional space generated by the Fock basis, and the second mode by the three-dimensional space spanned by the linearly independent vectors $\{\ket{-\alpha_0},\ket{0},\ket{1}\}$ (we will see that the contribution of $\ket{2}$ vanishes, hence it is not necessary to consider a fourth dimension).
Writing the eigenvalue equation associated to $\gamma_j$ and separating the expansion orders, we obtain the set of equations
\begin{align}
&A \psi_j^{(0)} = \gamma_j^{(0)} \psi_j^{(0)} \,, \label{ch6/pert1}\\
&A \psi_j^{(1)} + B \psi_j^{(0)} = \gamma_j^{(0)} \psi_j^{(1)} +\gamma_j^{(1)} \psi_j^{(0)} \,, \label{ch6/pert2}\\
&A \psi_j^{(2)} + B \psi_j^{(1)} + C \psi_j^{(0)} = \gamma_j^{(0)} \psi_j^{(2)} +\gamma_j^{(1)} \psi_j^{(1)} +\gamma_j^{(2)} \psi_j^{(0)} \,, \label{ch6/pert3}
\end{align}
where $\psi_j$ is the eigenvector associated to $\gamma_j$, which also admits the expansion $\psi_j=\psi_j^{(0)}+\psi_j^{(1)}/\sqrt{n}+\psi_j^{(2)}/n$. Eq.~\eqref{ch6/pert1} tells us that $\gamma_j^{(0)}$ is an eigenvalue of $A$ with associated eigenvector $\psi_j^{(0)}$. We multiply \eqref{ch6/pert2} and \eqref{ch6/pert3} by $\bra{\psi_j^{(0)}}$ to obtain
\begin{eqnarray}
\gamma_j^{(1)} &=& \bra{\psi_j^{(0)}} B \ket{\psi_j^{(0)}} \,, \label{ch6/lambda1} \\
\gamma_j^{(2)} &=& \bra{\psi_j^{(0)}} C \ket{\psi_j^{(0)}} + \sum_{l\neq j} \frac{\left|\! \bra{\psi_j^{(0)}} B \ket{\psi_l^{(0)}} \!\right|^2}{\gamma_j^{(0)}-\gamma_l^{(0)}} \,. \label{ch6/lambda2}
\end{eqnarray}
Note that Eq.~\eqref{ch6/lambda2} assumes that there is no degeneracy in the spectrum of $\Gamma$ at zero order (as we will see, this is indeed the case). From the structure of $A$ we can deduce that the form of its eigenvector $\psi_j^{(0)}$ is
\begin{equation}\label{ch6/psi0}
\ket{\psi_{i,\varepsilon}^{(0)}} = \ket{i} \otimes \ket{v_\varepsilon} \,,
\end{equation}
where we have replaced the index $j$ by the pair of indices $i,\varepsilon$. The index $i$ represents the Fock state $\ket{i}$ in the first mode, and the vectors $\ket{v_\varepsilon}$ are eigenvectors of $\ketbrad{-\alpha_0}-\ketbrad{0}$ and form a basis of $\mathcal{H}_3$ in the second mode. 
Every eigenvalue of $\Gamma$ is now labelled by the pair of indices $i,\varepsilon$, where $i=0,\ldots,\infty$ and $\varepsilon=+,-,0$: the second mode in $A$ has a positive, a negative, and a zero eigenvalue, to which we associate eigenvectors $\ket{v_+}$, $\ket{v_-}$ and $\ket{v_0}$, respectively. It is straightforward to see that the first two are
\begin{equation}\label{ch6/eigvecs}
\ket{v_\pm} = \frac{1}{2} \left( \frac{\ket{-\alpha_0}+\ket{0}}{N_+} \pm \frac{\ket{-\alpha_0}-\ket{0}}{N_-} \right) \,,
\end{equation}
where $N_\pm = \sqrt{1 \pm e^{-|\alpha_0|^2/2}}$. The zero-order eigenvalues of $\Gamma$ with $\varepsilon=\pm$ are
\begin{equation}\label{ch6/lambda0pm}
\gamma_{i,\pm}^{(0)} = \pm c_i \sqrt{1-e^{-|\alpha_0|^2}} \,.
\end{equation}
The third eigenvector $\ket{v_0}$ is orthogonal to the subspace spanned by $\ket{-\alpha_0}$ and $\ket{0}$, and corresponds to the eigenvalue $\gamma_{i,0}^{(0)} = 0$ \footnote{Note that the zero-order eigenvalues $\gamma_{i,\varepsilon}^{(0)}$ are nondegenerate, hence Eq.~\eqref{ch6/lambda2} presents no divergence problems.}.
This eigenvector only plays a role through the overlap $\braket{1}{v_0}$, which arises in Eqs.~\eqref{ch6/lambda1} and \eqref{ch6/lambda2}. We thus do not need its explicit form, but it will suffice to express $\braket{1}{v_0}$ in terms of known overlaps.

From Eqs.~\eqref{ch6/lambda1} and \eqref{ch6/psi0} we readily see that 
$
\gamma_{i,\varepsilon}^{(1)}=0
$.
Using Eqs.~\eqref{ch6/sigma11}, \eqref{ch6/sigma12}, \eqref{ch6/lambda2} and \eqref{ch6/psi0} we can express $\gamma_{i,\varepsilon}^{(2)}$ as
%
%
\begin{eqnarray}
\gamma_{i,\pm}^{(2)} &=&
e_i \left( |\!\braket{0}{v_\pm}\!|^2 - |\!\braket{1}{v_\pm}\!|^2 \right) \nonumber \\
&+& \sum_{\varepsilon} \frac{d_i |\!\braket{0}{v_\pm}\!|^2 |\!\braket{1}{v_\varepsilon}\!|^2}{\gamma_{i,\pm}^{(0)}-\gamma_{i-1,\varepsilon}^{(0)}} + \frac{\tilde{d}_i |\!\braket{1}{v_\pm}\!|^2 |\!\braket{0}{v_\varepsilon}\!|^2}{\gamma_{i,\pm}^{(0)}-\gamma_{i+1,\varepsilon}^{(0)}} \,, \label{ch6/lambda2ov} \\
\gamma_{i,0}^{(2)} &=& 0 \,,\nonumber
\end{eqnarray}
where we have used that, by definition, $\braket{0}{v_0}=\braket{\alpha_0}{v_0}=0$. The overlaps in \eqref{ch6/lambda2ov} are
\begin{eqnarray}
|\!\braket{0}{v_\pm}\!|^2 &=& \frac{1}{2} \left( 1 \mp \sqrt{1-e^{-|\alpha_0|^2}} \right) \,, \label{ch6/ov1}\\
|\!\braket{1}{v_\pm}\!|^2 &=& \frac{|\alpha_0|^2}{2} \frac{1 \pm \sqrt{1-e^{-|\alpha_0|^2}}}{e^{|\alpha_0|^2}-1} \,,\label{ch6/ov2}\\
|\!\braket{1}{v_0}\!|^2 &=& 1-\frac{|\!\braket{1}{-\alpha_0}\!|^2}{1-|\!\braket{0}{-\alpha_0}\!|^2} = 1-\frac{|\alpha_0|^2 e^{-|\alpha_0|^2}}{1-e^{-|\alpha_0|^2}} \,.\label{ch6/ov3}
\end{eqnarray}

Now that we have computed the eigenvalues of $\Gamma$, we are finally in condition to evaluate the sum in the right-hand side of Eq.~\eqref{ch6/tracenorm col}. Incorporating the relevant eigenvalues, given by Eqs.~\eqref{ch6/lambda0pm} and \eqref{ch6/lambda2ov}, it reads
\begin{eqnarray*}
\trnorm{\Gamma} &=& \sum_{i,\varepsilon} \left|\gamma_{i,\varepsilon}^{(0)}+\gamma_{i,\varepsilon}^{(2)}/n\right| \nonumber\\
&=& \sum_{i=0}^\infty \gamma_{i,+}^{(0)} + \frac{1}{n} \gamma_{i,+}^{(2)} - \gamma_{i,-}^{(0)} - \frac{1}{n} \gamma_{i,-}^{(2)} \nonumber\\
&=& \Lambda_+^{(0)} - \Lambda_-^{(0)} + \frac{1}{n} \left( \Lambda_+^{(2)}-\Lambda_-^{(2)} \right) \,,
\end{eqnarray*}
where 
\begin{equation*}
\Lambda_\pm^{(0)} = \sum_{i=0}^\infty \gamma_{i,\pm}^{(0)} = \pm \sqrt{1-e^{-|\alpha_0|^2}}
\end{equation*}
(recall that $\sum_{i=0}^\infty c_i = 1$), and 
\begin{equation}\label{appD/Lambda2pm}
\Lambda_{\pm}^{(2)}=\sum_{i=0}^\infty \gamma_{i,\pm}^{(2)} = \pm \frac{\mu^2 e^{-|\alpha_0|^2/2}}{2\sqrt{e^{|\alpha_0|^2}-1}} \left( 1-\frac{\mu^2+1}{2\mu^2+1} \frac{|\alpha_0|^2 \left(2e^{|\alpha_0|^2}-1\right)}{e^{|\alpha_0|^2}-1} \right) \,.
\end{equation}
\\

\section{\boldmath Conditional probability $p(\beta|\alpha)$, Eq.~\eqref{ch6/heterodyne_prob_ab}}\label{appD/sec:heterodyne_prob}

Given two arbitrary Gaussian states $\rho_A, \rho_B$, the trace of their product is
\begin{equation}\label{ch6/traceAB}
\tr (\rho_A \rho_B) = \frac{2}{\sqrt{\det(V_A+V_B)}} e^{-\delta^T (V_A+V_B)^{-1} \delta} \,,
\end{equation}
where $V_A$ and $V_B$ are their covariance matrices and $\delta$ is the difference of their displacement vectors. For the states $\rho_A \equiv \ketbrad{\sqrt{n}\alpha}$ and $\rho_B \equiv E_{\bar{\beta}}$, we have
\begin{eqnarray*}
V_A &=& \begin{pmatrix} 1 & 0 \\ 0 & 1 \end{pmatrix} \,,\quad 
V_B = R \begin{pmatrix} e^{-2r} & 0 \\ 0 & e^{2r} \end{pmatrix} R^T \,, \\
R &=& \begin{pmatrix} \cos \phi & -\sin \phi \\ \sin \phi & \cos \phi \end{pmatrix} \,, \\
\delta &=& (\sqrt{n} a_1-\bar{b}_1,\sqrt{n} a_2-\bar{b}_2) \,,
\end{eqnarray*}
where $\alpha = a_1 + i a_2$, $\bar{\beta} = \bar{b}_1 + i \bar{b}_2$, $r$ is the squeezing parameter, and $\phi$ indicates the direction of squeezing in the phase space. In terms of $\alpha$ and $\bar{\beta}$, Eq.~\eqref{ch6/traceAB} reads
\begin{equation*}
\tr (\rho_A \rho_B) = \frac{1}{\pi \cosh r} e^{-|\sqrt{n} \alpha - \bar{\beta}|^2-{\rm Re}[(\sqrt{n} \alpha-\bar{\beta})^2 e^{-i 2 \phi}] \tanh r} \,.
\end{equation*}

\section{Trace norm for the E\&D strategy}\label{ch6/sec:eyd_tracenorm}

For assessing the performance of the E\&D strategy, we want to obtain the error probability in discriminating the state $\ketbrad{0}$ and the posterior state $\rho(\beta)$, resulting from a heterodyne estimation of the state of the auxiliary mode that provides the estimate $\beta$.
Under a local Gaussian model around $\alpha_0$ parametrised by the complex variables $u$ and $v$, these states transform into $\ketbrad{-\alpha_0}$ and $\rho(v)$, respectively, where the second is given by
\begin{equation*}
\rho(v)=\int p(u|v) \, |u/\sqrt{n}\rangle\!\langle u/\sqrt{n}| \, d^2u \,,
\end{equation*}
and where $p(u|v)$ is given by Eq.~\eqref{ch6/puv}. The error probability is determined by the trace norm $\trnorm{\ketbrad{-\alpha_0}-\rho(v)}$ [cf. Eq.~\eqref{ch6/tracenorm_eyd}]. To compute it, we first series expand $\rho(v)$ in the limit $n\to\infty$, up to order $1/n$.
We name the appearing integrals of $u,u^*,|u|^2,u^2$, and $(u^*)^2$ over the probability distribution $p(u|v)$ as $I_1,I_1^*,I_2,I_3$, and $I_3^*$, respectively. This allows us to write the trace norm 
as 
\begin{equation*}
\trnorm{\ketbrad{-\alpha_0}-\rho(v)} \sim |\!| A'+B'/\sqrt{n}+C'/n \equiv \Phi |\!|_1 = \sum_\kappa |\lambda_\kappa| \,,
\end{equation*}
where
\begin{eqnarray*}
A' &=& \ketbrad{-\alpha_0} - \ketbrad{0} \,,\\
B' &=& -I_1 \ketbra{1}{0} - I_1^* \ketbra{0}{1} \,,\\
C' &=& -I_2 \left(\ketbrad{1}-\ketbrad{0}\right) - \frac{1}{\sqrt{2}}\left(I_3\ketbra{2}{0}+I_3^*\ketbra{0}{2}\right) \,,
\end{eqnarray*}
and $\lambda_\kappa$ is the $\kappa$th eigenvalue of $\Phi$, which admits the perturbative expansion $\lambda_\kappa = \lambda_{\kappa}^{(0)}+\lambda_{\kappa}^{(1)}/\sqrt{n}+\lambda_{\kappa}^{(2)}/n$, just as its associated eigenvector 
$\varphi_\kappa = \varphi_{\kappa}^{(0)} + \varphi_{\kappa}^{(1)}/\sqrt{n} + \varphi_{\kappa}^{(2)}/n$.
Up to order $1/n$, the matrix $\Phi$ has effective dimension 4 since it belongs to the space spanned by the set of linearly independent vectors $\{\ket{-\alpha_0},\ket{0},\ket{1},\ket{2}\}$. Hence the index $\kappa$ has in this case four possible values, i.e., $\kappa=+,-,3,4$. The zero-order eigenvalues $\lambda_\kappa^{(0)}$, which correspond to the eigenvalues of the rank-2 matrix $A'$, are
\begin{equation*}
\lambda_\pm^{(0)} = \pm \sqrt{1-e^{-\alpha_0^2}} \,,\quad \lambda_{3}^{(0)}=\lambda_{4}^{(0)} = 0 \,
\end{equation*}
(recall that $\alpha_0\in\mathbb{R}$). Their associated eigenvectors are $|\varphi_{\kappa}^{(0)}\rangle = \ket{v_\kappa}$, where $\ket{v_\pm}$ is given by Eq.~\eqref{ch6/eigvecs}, and, by definition, $\braket{v_\kappa}{-\alpha_0}=\braket{v_\kappa}{0}=0$ for $\kappa=3,4$. From analogous expressions to Eqs.~\eqref{ch6/lambda1} and \eqref{ch6/lambda2} we can write the first and second-order eigenvalues as
\begin{eqnarray*}
\lambda_{\kappa}^{(1)} &=& -I_1 \braket{v_\kappa}{1}\braket{0}{v_\kappa} - I_1^* \braket{v_\kappa}{0}\braket{1}{v_\kappa} \,,\\
\lambda_{\kappa}^{(2)} &=& I_2 \left(|\!\braket{v_\kappa}{0}\!|^2 - |\!\braket{v_\kappa}{1}\!|^2 \right) -\frac{1}{\sqrt{2}} \left( I_3 \braket{v_{\kappa}}{2}\braket{0}{v_{\kappa}} + I_3^* \braket{v_{\kappa}}{0}\braket{2}{v_{\kappa}}\right) \nonumber\\
& + &  \sum_{\xi \neq \kappa}  
\left(  |I_1|^2 \frac{ |\!\braket{v_\xi}{1}\!|^2 |\!\braket{v_{\kappa}}{0}\!|^2 + |\!\braket{v_\xi}{0}\!|^2 |\!\braket{v_{\kappa}}{1}\!|^2 }{\lambda_{\kappa}^{(0)}-\lambda_{\xi}^{(0)}} \right.\nonumber\\
&+& \left.\frac{I_1^2 \braket{v_\xi}{1}\braket{v_{\kappa}}{1}\braket{0}{v_{\kappa}}\braket{0}{v_\xi} + (I_1^*)^2 \braket{1}{v_\xi}\braket{1}{v_{\kappa}}\braket{v_{\kappa}}{0}\braket{v_\xi}{0} }{\lambda_{\kappa}^{(0)}-\lambda_{\xi}^{(0)}}
\right) \,.
\end{eqnarray*}
The needed overlaps for computing $\lambda_{\kappa}^{(1)}$ and $\lambda_{\kappa}^{(2)}$ are given by Eqs.~\eqref{ch6/ov1}, \eqref{ch6/ov2}, and
\begin{align}
\braket{v_\pm}{0} &= \frac{1}{2} \left( N_+ \mp N_- \right) \,,\nonumber\\
\braket{v_\pm}{1} &= \frac{1}{2} (-\alpha_0) e^{-\alpha_0^2/2} \left( \frac{1}{N_+} \pm \frac{1}{N_-} \right) \,,\nonumber\\
|\!\braket{v_3}{1}\!|^2 &= 1-\frac{|\!\braket{1}{-\alpha_0}\!|^2}{1-|\!\braket{0}{-\alpha_0}\!|^2-|\!\braket{2}{-\alpha_0}\!|^2} \,, \label{ch6/ov_v3_1}\\
|\!\braket{v_4}{1}\!|^2 &= \frac{|\!\braket{1}{-\alpha_0}\!|^2 |\!\braket{2}{-\alpha_0}\!|^2}{\left(1-|\!\braket{0}{-\alpha_0}\!|^2\right)\left(1-|\!\braket{0}{-\alpha_0}\!|^2-|\!\braket{2}{-\alpha_0}\!|^2\right)} \label{ch6/ov_v3_2}\,.
\end{align}
The expressions for the overlaps \eqref{ch6/ov_v3_1} and \eqref{ch6/ov_v3_2} actually depend on the dimension of the space that we are considering (four in this case), and they are not unique: there are infinitely many possible orientations of the orthogonal pair of vectors $\{\ket{v_3},\ket{v_4}\}$ such that both of them are orthogonal to the plane formed by $\{\ket{-\alpha_0},\ket{0}\}$, which is the only requirement we have. Note, however, that this degeneracy has no effect on the excess risk, thus we are free to choose the particular orientation that, in addition, verifies $\braket{v_3}{2}=0$, yielding the simple expressions \eqref{ch6/ov_v3_1} and \eqref{ch6/ov_v3_2}.

Finally, we write down the trace norm as
\begin{eqnarray}
\trnorm{\Phi} &=& \sum_\kappa |\lambda_{\kappa}^{(0)} + \lambda_{\kappa}^{(1)}/\sqrt{n} + \lambda_{\kappa}^{(2)}/n| \nonumber\\
&=& \lambda_+^{(0)} - \lambda_-^{(0)} + \frac{1}{\sqrt{n}} \left(\lambda_+^{(1)}-\lambda_-^{(1)}\right) \nonumber\\
&& +\; \frac{1}{n} \left(\lambda_+^{(2)}-\lambda_-^{(2)}+|\lambda_3^{(2)}|+|\lambda_4^{(2)}|\right)\,,\label{ch6/tracenorm_eyd_2}
\end{eqnarray}
which we use now to obtain the asymptotic expression for the average error probability, defined in Eq.~\eqref{ch6/perror_eyd}. Recall Eq.~\eqref{ch6/pv} and note that we have to average Eq.~\eqref{ch6/tracenorm_eyd_2} over the probability distribution $p(v)$. Regarding this average, it is worth taking into account the following considerations. First, the $v$-dependence of the eigenvalues comes from $I_1,I_2,I_3$, and its complex conjugates. The integrals needed are given in the last part of Appendix \ref{ch6/sec:gaussian integrals}. Second, because of Eq.~\eqref{appD/intI1}, $\lambda_{\kappa}^{(1)}=0$ and hence the order $1/\sqrt{n}$ term vanishes, as it should. And third, the second-order eigenvalues $\lambda_3^{(2)}$ and $\lambda_4^{(2)}$ are $v$-independent and positive, so we can ignore the absolute values in Eq.~\eqref{ch6/tracenorm_eyd_2}. 
Putting all together, we can express the asymptotic average error probability of the E\&D strategy as
\begin{equation}\label{appD/perror_eyd_final}
P_{\rm e}^{\rm E\&D} \equiv P_{\rm e}^{\rm E\&D}(n\to\infty) \sim \frac{1}{2}\left(1-\sqrt{1-e^{-\alpha_0^2}} + \frac{1}{n} \Delta^{\rm E\&D} \right)\,,
\end{equation}
where
\begin{equation}\label{appD/DeltaEyD}
\Delta^{\rm E\&D} = -\frac{1}{2} \left[\lambda_3^{(2)}+\lambda_4^{(2)} + \int p(v) \left(\lambda_+^{(2)}-\lambda_-^{(2)}\right) dv \right] \,.
\end{equation}

Making use of Eqs.~\eqref{appD/perror_eyd_final} and \eqref{ch6/perror known} we can readily compute the excess risk of the E\&D strategy:

\begin{align}\label{appD/excessrisk_eyd_r}
R^{\rm E\&D}(r) &= n \lim_{\mu\to\infty} \left(P_{\rm e}^{\rm E\&D}-P_{\rm e}^*\right) \nonumber\\
&= \frac{e^{-\alpha_0^2}}
{
16 \sqrt{1-e^{-\alpha_0^2}} \left(e^{\alpha_0^2}-1\right)
}
\left\{  \left[ 4e^{\alpha_0^2} \left(1-e^{\alpha_0^2}\right)\left(\sqrt{1-e^{-\alpha_0^2}}-1\right) \right.\right.\nonumber\\
& \quad\left.\left. +\, \alpha_0^2 \left(4e^{\alpha_0^2}\sqrt{1-e^{-\alpha_0^2}}-2\right) \right] \cosh^2s + \alpha_0^2\sinh(2s)  \right\} \,.
\end{align}
\\

\section{Gaussian integrals}\label{ch6/sec:gaussian integrals}

At many points in Chapter~\ref{ch6_learningcv}, we integrate complex-valued functions over the complex plane, weighted by the bidimensional Gaussian probability distribution $G(u)$. This Section gathers the integrals that we need. Recall that $G(u)$ is defined as
\begin{equation*}
G(u) = \frac{1}{\pi \mu^2} e^{-u^2/\mu^2} \,,\quad u\in\mathbb{C} \,.
\end{equation*}
Expressing $u$ either in polar or Cartesian coordinates in the complex plane, i.e., $u = r e^{i\theta} = u_1+i u_2$, one can readily check that $G(u)$ is normalized:
\begin{eqnarray*}
\int G(u) d^2u &=& \int_0^\infty \int_0^{2\pi} \frac{1}{\pi \mu^2} e^{-r^2/\mu^2} r dr d\theta=1 \,,\\
\int G(u) d^2u &=& \int_{-\infty}^{\infty} \int_{-\infty}^{\infty} \frac{1}{\pi \mu^2} e^{(-u_1^2-u_2^2)/\mu^2} du_1 du_2 = 1 \,.
\end{eqnarray*}
The average of a coherent state $\ketbrad{u}$ over the probability distribution $G(u)$ can be computed by expressing $\ket{u}$ in terms of Fock states, as in Eq.~\eqref{appC/coherentFock}. It gives
\begin{equation}\label{appD/int_coherent}
\int G(u) \ketbrad{u} d^2u = \sum_{k=0}^\infty c_k \ketbrad{k} \,,\quad c_k=\frac{\mu^{2k}}{(\mu^2+1)^{k+1}} \,,
\end{equation}
where $\{\ket{k}\}$ is the Fock basis. Note that the result of averaging a coherent state over $G(u)$ is nothing more than a thermal state with average photon number $\mu^2$.

Variations of Eq.~\eqref{appD/int_coherent} with different complex functions that we use are
%
%
\begin{eqnarray*}
\int G(u) u \ketbrad{u} d^2u &=& \sum_{k=0}^\infty c_{k+1} \sqrt{k+1} \ketbra{k}{k+1} \,, \\
\int G(u) u^* \ketbrad{u} d^2u &=& \sum_{k=0}^\infty c_k \sqrt{k} \ketbra{k}{k-1} \,, \\
\int G(u) |u|^2 \ketbrad{u} d^2u &=& \sum_{k=0}^\infty c_{k+1}  (k+1) \ketbrad{k} \,, \\
\int G(u) u^2 \ketbrad{u} d^2u &=& \sum_{k=0}^\infty c_{k+2} \sqrt{k+2}\sqrt{k+1} \ketbra{k}{k+2} \,, \\
\int G(u) \left(u^*\right)^2 \ketbrad{u} d^2u &=& \sum_{k=0}^\infty c_k \sqrt{k}\sqrt{k-1} \ketbra{k}{k-2} \,,
\end{eqnarray*}
and
\begin{eqnarray}
\int G(u) (u+u^*) d^2u &=& 0 \label{ch6/int u1}\,,\\
\int G(u) (u+u^*)^2 d^2u &=& 2\mu^2 \label{ch6/int u2}\,,\\
\int G(u) |u|^2 d^2u &=& \mu^2 \label{ch6/int u3}\,.
\end{eqnarray}

%

For the computations in Appendix~\ref{ch6/sec:eyd_tracenorm} we also need to perform Gaussian integrals, this time over the probability distribution $p(v)$, defined in Eq.~\eqref{ch6/pv}. We make use of
\begin{eqnarray}
\int p(v) I_1 d^2v &=& \int p(v) I_1^* d^2v = 0 \,,\label{appD/intI1}\\
\int p(v) I_3 d^2v &=& \int p(v) I_3^* d^2v = 0 \,,\label{appD/intI3}\\
\int p(v) I_2 d^2v &=& \mu^2 \,,\nonumber\\
\int p(v) I_1^2 d^2v &=& \int p(v) (I_1^*)^2 d^2v \nonumber\\
&=& \frac{\mu^4 \sinh(2r)}{(2\mu^2+1)\cosh(2r)+2\mu^2(\mu^2+1)+1} \,,\nonumber\\[1em]
\int p(v) |I_1|^2 d^2v &=& \frac{\mu^4 (\cosh(2r)+2\mu^2+1)}{(2\mu^2+1)\cosh(2r)+2\mu^2(\mu^2+1)+1} \,.\nonumber
\end{eqnarray}

\backmatter



\addcontentsline{toc}{chapter}{Bibliography}
\bibliography{library}
\bibliographystyle{bib_style}

%
%
%
%
%
%
%
%
%
%
%



\end{document}